\documentclass[11pt]{article}

\usepackage{fullpage}
\usepackage[top=1in,bottom=1in,left=1in,right=1in]{geometry}
\date{}

\usepackage{multirow}
\usepackage{mathtools}
\usepackage{amssymb} 
\usepackage{amsfonts}
\usepackage{subfig}
\usepackage{amsmath}
\usepackage{hyperref}
\usepackage{bookmark} 
\usepackage{graphicx}
\usepackage{booktabs}
\usepackage{algorithm}
\usepackage{algorithmicx}
\usepackage{algpseudocode}
\newtheorem{problem}{Problem}
\newcommand{\RNum}[1]{\uppercase\expandafter{\romannumeral #1\relax}}
\algrenewcommand\algorithmicrequire{\textbf{Input:}}
\algrenewcommand\algorithmicensure{\textbf{Output:}}
\newcommand{\figurenames}{Figures}

\graphicspath{{./images/}}

\title{SCOUT: simultaneous time \underline{s}egmentation and \underline{co}mm\underline{u}ni\underline{t}y detection\\ in dynamic networks}

\author{
  Yuriy Hulovatyy and Tijana Milenkovi\'{c}\footnote{To whom correspondence should be addressed}\\
\\
 Department of Computer Science and Engineering \\
 Interdisciplinary Center for Network Science and Applications (iCeNSA) \\
 Eck Institute for Global Health\\
 University of Notre Dame, Notre Dame, IN 46556, USA\\ 
 \{yhulovat, tmilenko\}@nd.edu\\
}

\begin{document}

\maketitle

\begin{abstract}
\noindent
Many evolving complex systems can be modeled via dynamic networks. 
An important problem in dynamic network research is community detection, which identifies groups of topologically related nodes. 
Typically, this problem is approached by assuming either that each time point has a distinct community organization or that all time points share one community organization. 
In reality, the truth likely lies between these two extremes, since some time periods can have community organization that evolves while others can have community organization that stays the same. To find the compromise, we consider community detection in the context of the problem of segment detection, which identifies contiguous time periods with consistent network structure. 
Consequently, we formulate a combined problem of segment community detection (SCD), which simultaneously partitions the network into contiguous time segments with consistent community organization and finds this community organization for each segment. 
To solve SCD, we introduce SCOUT, an optimization framework that explicitly considers both segmentation quality and partition quality. SCOUT addresses limitations of existing methods that can be adapted to solve SCD, which typically consider only one of segmentation quality or partition quality. 
In a thorough evaluation, SCOUT outperforms the existing methods in terms of both accuracy and computational complexity.

\end{abstract}

\section{Introduction}
\subsection{Motivation}
Networks (or graphs) are elegant yet powerful abstractions for studying complex systems in various domains, from biological entities to social organizations \cite{newman2010networks}.
Real-world systems evolve over time.
However, until relatively recently, dynamic measurements about their functioning have been unavailable, owing mostly to limitations of technologies for data collection.
Hence, an evolving system has traditionally been analyzed by studying its static network representation, which discards the system's time dimension by combining all of its interacting elements and their connections across multiple times into a single aggregate network.
For example, dynamic cellular functioning has traditionally been modeled as a static protein-protein interaction network that combines biomolecular interactions across different time points and other contexts \cite{berger2013computational, mitra2013integrative}.
However, such an aggregate approach loses important temporal information about the functioning of evolving real-world systems \cite{holme2015modern}.
Analyzing dynamic network representations of evolving systems is crucial for understanding important mechanisms behind various dynamic phenomena such as human aging in the computational biology domain \cite{faisal2014dynamic} or opinion formation in the social network domain \cite{romero2011differences}, especially with the increasing recent availability of temporal real-world data in these and other domains.
The dynamic network representation of an evolving system models its temporal measurement data as a series of snapshots, each of which is a network that encompasses the temporal data observed during the corresponding time interval.
We refer to this snapshot-based representation as a dynamic network.

Approaches for studying dynamic networks can be categorized into: 1) those that extend well-established static network problem formulations and solutions to their dynamic counterparts, and 2) those that consider novel network problems and solutions that arise specifically from the time dimension and are thus native only to the dynamic setting.
A popular problem from category 1 above that is of our interest is \emph{community detection}.
A popular problem from category 2 above that is of our interest is \emph{time segmentation}, or \emph{segment detection} (also known as \emph{change detection}).
We next discuss these two problems.

Community detection aims to study network structure (or topology) from mesoscopic (i.e., intermediate or groups-of-nodes level) perspective,
in contrast to doing so from macroscopic (i.e., global or network level) or microscopic (i.e., local or node level) perspective \cite{fortunato2010community}.
Specifically, the goal of community detection is to identify groups of topologically related (e.g., densely interconnected \cite{newman2004finding, yang2015defining} or topologically similar \cite{milenkovic2008uncovering, solava2012graphlet, hulovatyy2015exploring}) nodes called \emph{communities} (or \emph{clusters}), which are likely to indicate important functional units within the network.
For example, communities can correspond to proteins with similar functions in a biological network or groups of friends in a social network \cite{wiwie2015comparing, fortunato2010community, hric2014community}.
A \emph{partition} is a division of a network into communities, with each node belonging to a single community.
We focus on this mathematical notion of a partition; that is, we consider non-overlapping communities.
Nonetheless, our work can be extended to handle overlapping communities as well.
For an evolving real-world system, community detection in its dynamic network representation is likely to yield additional insights compared to community detection in the system's static network representation \cite{hartmann2014clustering, meng2016multiplex}.
Two extremes of community detection in a dynamic network are:  1) \emph{snapshot clustering} and 2) \emph{consensus clustering}.
On the one hand, snapshot clustering finds a separate partition for each temporal snapshot
 \cite{chi2007evolutionary, fu2009dynamic, mucha2010community, kawadia2012sequential}.
Given the snapshot-level partitions, one can then track their evolution by matching  individual clusters in adjacent snapshots \cite{berger2006framework, palla2007quantifying, rosvall2010mapping, greene2010tracking}.
On the other hand, consensus clustering finds a single partition that fits well all snapshots  \cite{pandit2011detecting, aynaud2011multi, lancichinetti2012consensus, zhang2012common}.
In the real life, community organization most often lies between these two extremes. 
Finding this real life community organization is one of key goals of our study.

Segment detection aims to divide a dynamic network into continuous \emph{segments} (groups of snapshots),
such that the ``border'' between each pair of adjacent segments marks a prominent shift in the network structure \cite{ranshous2015anomaly}.
As a result, all snapshots within a given segment have similar network structure, while every two adjacent segments have snapshots with dissimilar structure.
The set of all segments covering the whole dynamic network is called the \emph{segmentation} of the network.
Time points that separate the segments are called \emph{change points}.
Since change points correspond to shifts in the network structure, they likely indicate functionally important events in the life of the underlying system \cite{ranshous2015anomaly}.
For example, change points can correspond to transitions between different functional states in brain networks or to stock market changes in financial networks  \cite{barnett2016change}. 
Finding change points indicating important structural shifts in the dynamic network is the other key goal of our study.

\begin{figure}[t!]
  \centering\includegraphics[width=0.7\linewidth]{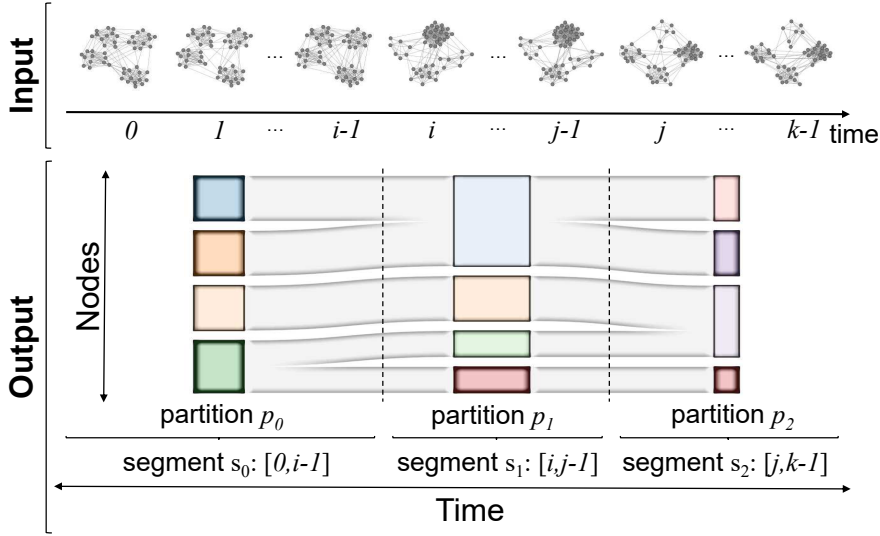}
  \caption[Illustration of SCD]{Illustration of SCD.
}\label{fig:problem}
\end{figure}

There is a connection between community detection and segment detection.
The former aims to partition a dynamic network along the node dimension (by grouping nodes into communities), while the latter does this along the time dimension (by grouping snapshots into segments).
The combination of the two problems, which is our focus and which we refer to as \emph{segment community detection} (\emph{SCD}), can be seen as two-dimensional clustering:
simultaneously grouping snapshots of the dynamic network into segments based on community organization of the snapshots, and 
grouping nodes of the snapshots into communities based on the segments these snapshots belong to
(\figurename~\ref{fig:problem}). 

\begin{figure}[t!]
  \centering\includegraphics[width=0.7\linewidth]{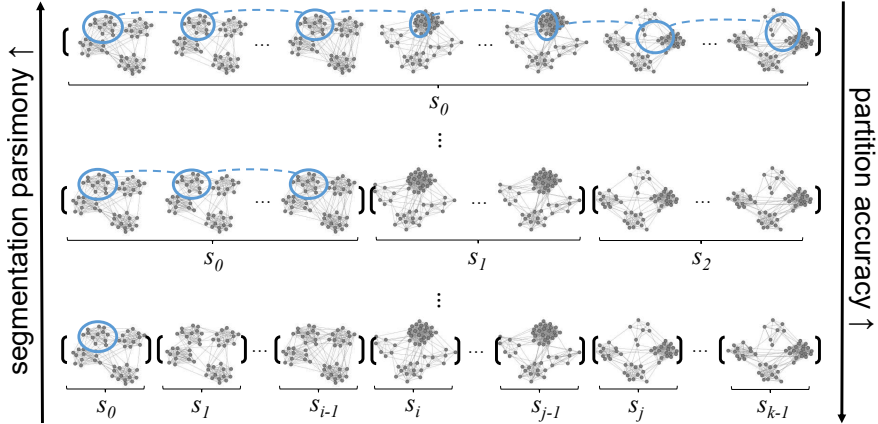}
  \caption[Illustration of how SCD naturally allows for compromising between the two extremes of snapshot clustering and consensus clustering]{
Illustration of how SCD (middle) naturally allows for compromising between the two extremes of snapshot clustering (bottom) and consensus clustering (top), by balancing the number of segments (which we refer to as \emph{segmentation parsimony}; Section~\ref{sec:problem}) and how well a partition fits the corresponding snapshot (which we refer to as \emph{partition accuracy}; Section~\ref{sec:problem}). 
At each of the three horizontal levels, in blue we show the same community across different snapshots within the given segment.   
}\label{fig:balance}
\end{figure}

SCD naturally allows for achieving the goal of compromising between the two extremes of snapshot clustering and consensus clustering to identify the real life community organization.
Namely, while snapshot clustering is set to ``zoom-in'' to the level of individual snapshots and consensus clustering is set to ``zoom-out'' to the level of the whole dynamic network,
segment community detection allows for automatically choosing an appropriate ``zoom level'' by focusing on segments, each  potentially spanning multiple coherent snapshots while still capturing important changes in the community organization (\figurename~\ref{fig:balance}).
As an illustration, consider studying how protein modules evolve with age: it might be more desirable to focus on different stages of the aging process such as infancy, childhood, adolescence, adulthood, etc. \cite{mintz1993life} (via segment community detection) than on each day/month/year of the lifespan (via snapshot clustering) or on the entire lifespan (via consensus clustering).
Similar holds when studying evolution of protein modules with disease (e.g., cancer) progression.

\subsection{Related work}\label{sec:related_work}
Several approaches exist that can be adapted to be able to deal with the SCD problem: GraphScope \cite{sun2007graphscope}, Multi-Step  \cite{aynaud2011multi}, and GHRG \cite{peel2015detecting}.
For a review of how these methods work, see Supplementary Section~\ref{sec:supplement:related_work}.
These existing methods can produce both segments and their corresponding partitions, which indeed is a solution that SCD aims to find.
However, these approaches have the following drawbacks.
\textbf{1)} They generally cannot produce a high-quality solution with respect to both of the segment community detection aspects (i.e., segmentation quality and partition quality), as we will show in Section~\ref{sec:results}. 
\textbf{2)} For each method, either: a) the number of segments can only be determined automatically but not set by the user, or instead b) the number of segments can only be set by the user but not determined automatically.
In applications where some domain expert knowledge on the desired number of segments is available, the user should be able to feed this knowledge into the method by setting the number of segments, but the methods of type ``a'' above (GraphScope and GHRG) cannot handle this. 
On the other hand, in applications where such knowledge is unavailable, the method should be able to determine an appropriate number of segments automatically, but the methods of type ``b'' above (Multi-Step) cannot handle this. 
For a method to be generalizable to both types of applications, the method should be able to handle both automatic as well as user-defined determination of the appropriate number of segments. 
\textbf{3)} Each of the existing methods has a single built-in intuition about what a good segment or partition is.
Hence, each approach could be biased  towards the particular parameters that it implements.
Thus, a more generalizable approach that would offer flexibility in terms of parameter choices is desirable. 
To address these three drawbacks, we introduce \emph{SCOUT}, a new general framework for \underline{s}egment \underline{co}mm\underline{u}ni\underline{t}y detection, as follows.

\subsection{Our contributions}
We propose a novel formulation of the SCD problem as an optimization process that integrates the two aspects (segment detection and community detection) more explicitly than the existing methods.
Also, we propose SCOUT, a general framework for solving the new problem, which addresses the drawbacks of the existing methods:
\textbf{1)} it is capable of producing a high-quality solution with respect to both of the segment community detection aspects;
\textbf{2)} it can handle both automatic and user-defined determination of the appropriate number of segments;
\textbf{3)} it offers high level of flexibility when it comes to the choice of segmentation or partition quality parameters. 

Specifically, SCOUT algorithm consists of three key parts: \emph{objective function} (a measure of what a good SCD solution is),
\emph{consensus clustering} (given a set of  change points, how to find a good partition for each segment), and \emph{search strategy} (how to search through the space of possible change point sets).
We vary choices for each of these three components.
By doing so, we effectively trade-off between different goals, such as between segmentation quality and partition quality, or between accuracy and speed. 

We comprehensively evaluate SCOUT against the existing methods.
We do so on both synthetic and real-world networks of varying sizes.
In particular, because in some domains (such as computational biology) large-scale experimental real-world dynamic network data are not available \cite{berger2013computational, mitra2013integrative, faisal2014dynamic, yoo2015improving}, 
in order to illustrate generalizability of our approach, we first perform evaluation on synthetic dynamic network data. 
For this purpose, we introduce an intuitive model for automatic generation of a synthetic dynamic network of an arbitrary size with known ground truth segmentation as well as community organization, and we perform our evaluation on 20 different synthetic ground truth configurations. 
In addition, we analyze six real-world dynamic networks from domains that do offer such data and that offer such data with some ground truth knowledge embedded into them; these networks span studies of human proximity, communication, and political relationships. 
To evaluate how well each method can reconstruct the ground truth knowledge, we rely on established partition quality and similarity measures as a basis for developing new SCD accuracy measures that can simultaneously
account for both segmentation quality and partition quality.
Interestingly, although the existing approaches can all achieve the same task of SCD, they have not been evaluated against each other to date. 
Hence, our study provides the first ever such evaluation. 
Importantly, we show that SCOUT overall outperforms the existing methods with respect to both segmentation quality and partition quality, while also being more computationally efficient.

\section{Methods}\label{sec:methods}
\subsection{Notations}\label{sec:methods_notation}
A \emph{dynamic network} $D$ is a sequence of $k$ snapshots  $\lbrace G_0, G_1, \dots, G_{k-1} \rbrace$,
where each snapshot $G_i=(V_i,E_i)$ is a static graph capturing network structure during time interval $i$.
A sequence of consecutive snapshots can be grouped into a segment.
Formally, a \emph{segment} $s$ is a sequence of consecutive snapshots $\lbrace G_{i}, G_{i+1}, \dots, G_{j} \rbrace$, $i \leq j$,
with $i$ being its \emph{start time}, $j$ being its \emph{end time}, and $j-i+1$ being its \emph{length}.
A sequence of non-overlapping segments (meaning that each segment in the sequence starts right after the previous one ends) that covers the whole dynamic network (meaning that the first segment in the sequence starts at time $0$ and the last segment in the sequence ends at time $k-1$) forms a segmentation of this network.
Formally, a \emph{segmentation} $S$ is a sequence of $l$ adjacent segments $\lbrace s_0, s_1, \dots, s_{l-1} \rbrace$ such that $s_0$ starts at time $0$ and $s_{l-1}$ ends at time $k-1$. 
We can specify such a segmentation via a set $T=\lbrace t_1, t_2, \dots, t_{l-1} \rbrace$ of $l-1$ time points called \emph{change points}, such that $t_i$ is the start time of segment $s_i$, $i \in [0,l-1]$ (by convention, we  always assume that $t_0=0$).

\subsection{Problem formulation}\label{sec:problem}
Given a dynamic network $D$, the goal of SCD is to simultaneously find a segmentation $S^\star=\lbrace s^\star_0, s^\star_1, \dots, s^\star_{l-1} \rbrace$ (or equivalently a change point set $T^\star=\lbrace t^\star_1, t^\star_2, \dots, t^\star_{l-1} \rbrace$) and a sequence of partitions $P^\star=\lbrace p^\star_0, p^\star_1, \dots, p^\star_{l-1} \rbrace$
such that $S^\star$ identifies important shifts in the community organization of $D$ and each $p^\star_i$ (called \emph{segment partition}) reflects well the community organization of each snapshot within segment $s^\star_i$ (\figurename~\ref{fig:problem}).
Clearly, the output (i.e., the solution) of the SCD problem can be represented as $O^\star=(T^\star, P^\star)$.
Intuitively, in a good output, $T^\star$ should be \emph{parsimonious}
(meaning that it should capture all important shifts in the network with as small as possible number of change points), while $P^\star$ should be \emph{accurate} (meaning that segment partitions should correctly capture community organization of all snapshots within the corresponding segment).
That is, output $O^\star=(T^\star, P^\star)$ should aim to simultaneously satisfy two objectives:
\emph{segmentation parsimony} and \emph{partition accuracy}.
We can now state the problem:

\begin{problem}[SCD]
Given a dynamic network $D=\lbrace G_1, G_2, \dots, G_k \rbrace$,
find a number of segments $l$, a sequence of $l-1$ change points $T^\star=\lbrace t^\star_1, t^\star_2, \dots, t^\star_{l-1} \rbrace$,
and a sequence of $l$ segment partitions $P^\star=\lbrace p^\star_1, p^\star_2, \dots, p^\star_l \rbrace$
such that the output $O^\star=(T^\star,P^\star)$ forms a parsimonious segmentation with accurate segment partitions.
\label{prob:scd}
\end{problem}

In some sense, the two objectives, segmentation parsimony and partition accuracy, are competing with each other.
That is, optimizing one does not necessarily lead to optimizing the other.
For example, at the extreme of snapshot community detection (bottom of \figurename~\ref{fig:balance}), each snapshot is considered to be a separate segment that has its own well-fitting partition, which yields high partition accuracy.
However, such a fine-grained output with the maximum possible number of segments might contain redundancies, because some adjacent snapshots might have similar community organizations.
In this case, segmentation parsimony will be low. 
To optimize (increase) segmentation parsimony, adjacent snapshots with similar community organizations should be grouped together. 
At the other extreme of consensus community detection (top of \figurename~\ref{fig:balance}), all snapshots are grouped together into one segment with a single common segment partition for the whole network, which yields high segmentation parsimony.
However, the single segment partition will have to ``compromise'' between many possibly quite distinct snapshots.
In this case, the segment partition will not be able to fit well all of the distinct snapshots, and consequently, partition accuracy will be low.  
In real-world scenarios, the SCD solution typically lies between these two extremes, but finding such a solution still requires balancing between the two somewhat contradicting goals of optimizing both segment parsimony and partition accuracy. 
We formalize the ways of finding such a solution in Section~\ref{sec:scout}.

Recall from Section~\ref{sec:related_work} the need of being able to find a solution with a user-specified number of segments $l$, in addition to being able to determine this parameter $l$ automatically.
Our current SCD problem formulation (Problem~\ref{prob:scd}) can handle the latter scenario, but we can extend it to handle the former scenario as well. 
Specifically, when finding an SCD solution, in addition to allowing for simultaneously optimizing both aspects of SCD quality (i.e., segmentation parsimony and partition accuracy), we can allow for optimizing only one aspect (partition accuracy) while setting the other one (segmentation parsimony, expressed as the number of segments $l$) as a constraint.
So, we extend the problem formulation by adding to the existing SCD objective from Problem~\ref{prob:scd} the following new objective: given a dynamic network $D$ and the desired number of segments $l$ as input by the user, find an output $O^\star=(T^\star,P^\star)$ with $l$ segments that achieves the highest partition accuracy.
We refer to this new objective as the \emph{constrained SCD problem} (\emph{CSCD}).
We propose SCOUT to solve any of the SCD and CSCD problems, in order to allow for handling both of the above scenarios (automatic vs. user-defined selection of the number of segments $l$, respectively), as follows.

\subsection{Our SCOUT approach}\label{sec:scout}
Given a dynamic network $D$, we aim to find an output $O^\star=(T^\star,P^\star)$ by directly optimizing an \emph{objective function} that measures both segmentation parsimony and partition accuracy (see below for details on how we deal with SCD versus CSCD).
Algorithm~\ref{alg:scout} provides a high-level overview of SCOUT, and Supplementary Section~\ref{sec:supplement:scout} and Supplementary \figurenames~\ref{fig:supplement:exh_search}-\ref{fig:supplement:bu_search} provide further details.
SCOUT has the following five steps.
\textbf{1)} Select the initial change point set as the current change point set $T$ (line 2 in Algorithm~\ref{alg:scout}).
For example, the initial change point set could correspond to a set of all possible snapshot-level segments (bottom-up search) or just one large network-level segment (top-down search).
Given $T$, the method iteratively performs the following steps.
\textbf{2)} Perform \emph{consensus clustering} within each segment $s_i$ to get its corresponding partition $p_i$ (line 7).
In general, the consensus clustering method should aim to obtain the partition set $P$ that maximizes the objective function for $T$. 
Step 2 results in $O=(T,P)$ (line 9).
\textbf{3)} Use a \emph{search strategy} to search for the next change point set that will become the new current change point set $T$ (line 11).
Clearly, the search strategy guides how we explore the space of possible change point sets.
For example, in bottom-up search, the next change point set is obtained by merging two adjacent segments, while in top-down search, the next change point set is obtained by splitting a segment into two.
\textbf{4)} Repeat steps 2 and 3 above until the exploration of the space is finished (corresponding to $T = \varnothing$ in line 3), e.g., until one largest possible network-level segment is reached in bottom-up search or until all possible snapshot-level segments are reached in top-down search. 
\textbf{5)} Choose the best output out of all outputs computed in step 2 as the final output $O^\star=(T^\star,P^\star)$ (line 13).
When solving the SCD problem, the best output is the one maximizing the objective function.
When solving the CSCD problem,  the best output is the one maximizing the objective function while satisfying the constraint (the solution consisting of $l$ segments).
Thus, SCOUT contains three main components: objective function (Supplementary Section~\ref{sec:supplement:scout_obj_func}), consensus clustering (Supplementary Section~\ref{sec:supplement:scout_cons_clust}), and search strategy (Supplementary Section~\ref{sec:supplement:scout_search_strategy}).

\begin{algorithm}[t!]
 \caption{SCOUT overview.
It has three main components: objective function ($\mathbf{ObjectiveFunction}$), consensus clustering ($\mathbf{ConsensusClustering}$), and search strategy ($\mathbf{SearchStrategy}$).
Auxiliary procedure $GetSegmentation$ constructs the segmentation of a dynamic network given a change point set and auxiliary procedure $GetBestOutput$ returns the best (with respect to the objective function) of all considered outputs.
}
\label{alg:scout}
\begin{algorithmic}[1]
\Require $D$
\Ensure $O^\star=(T^\star, P^\star)$
 \State $R \gets \emptyset$
 \State $T \gets \mathbf{SearchStrategy}(\varnothing, R)$
 \While{$T \neq \varnothing$}
 	\State $S  \gets GetSegmentation(T, D)$
 	\State $P \gets \emptyset$
 	\For{$s \in S$}
 		\State $P \gets P \cup \lbrace \mathbf{ConsensusClustering}(s) \rbrace$
 	\EndFor
 	\State $O \gets (T, P)$
 	\State $ R \gets R \cup \lbrace (O, \mathbf{ObjectiveFunction}(O)) \rbrace$
 	\State $T \gets \mathbf{SearchStrategy}(T, R)$
 \EndWhile
 \State \Return{$GetBestOutput(R)$}
\end{algorithmic}
\end{algorithm}

\subsection{Experimental setup}\label{sec:evaluation}

\subsubsection{Methods for comparison}\label{sec:methods_for_comparison}
We compare SCOUT against the three existing approaches: GraphScope, Multi-Step, and GHRG. 
We discuss the methods' parameters that we use in Supplementary Section~\ref{sec:supplement:methods_for_comparison}.

\subsubsection{Datasets}\label{sec:datasets}
We evaluate the methods on two types of networks: \emph{synthetic networks} and \emph{real-world networks}.

\noindent\textbf{Synthetic networks.}
To generate a synthetic dynamic network $D$ with the embedded ground truth $O^{(gt)}=(T^{(gt)},P^{(gt)})$,
we introduce a new dynamic random graph model for this purpose, which we call \emph{segment community generator} (SCG), and which works as follows.
We assume that the following are provided as input by the user: the number of snapshots $k$, the number of segments $l$, the number of nodes in each snapshot $n$, the minimum required number of nodes in each cluster $c_{min}$, and two parameters $c_{in}$ and $c_{out}$ that control intra- and inter-community edge density of the snapshots.
The process of generating a synthetic dynamic network with these parameters contains four steps (Supplementary \figurename~\ref{fig:supplement:synthetic_construction}).
In the first three steps, we generate the ground truth $O^{(gt)}=(T^{(gt)},P^{(gt)})$, and in the last step, we use $O^{(gt)}$ to actually generate snapshots of $D$.
Intuitively, we: 
\textbf{1)} generate the set of change points $T^{(gt)}$ to define segments, 
\textbf{2)} create a special auxiliary graph describing how segment partitions evolve from segment to segment, 
\textbf{3)} use this graph to generate the actual segment partitions $P^{(gt)}$, and 
\textbf{4)} use a stochastic blockmodel  to generate snapshots of $D$, based on the idea that snapshots within the same segment (as defined by $T^{(gt)}$) have the same community organization (as defined by the corresponding segment partition from $P^{(gt)}$).
For details on each step, see Supplementary Section~\ref{sec:supplement:synth_networks} and Supplementary Algorithm~\ref{alg:supplement:synth_step3}.

For our experiments, we generate synthetic dynamic networks with 16 snapshots and 1, 2, 4, 8, and 16 ground truth segments.
We also consider networks of various sizes: 50, 100, 500, and 1000 nodes in each snapshot.
This results in $5 \times 4=20$ different synthetic network configurations.
In each configuration, we set the parameters as follows.
For partition graph $G_P$, we set $c_{min}=5$ when $n \in \{50, 100\}$ and $c_{min}=50$ when $n \in \{500, 1000\}$.
For the stochastic blockmodel, we set $c_{in}=20$ and $c_{out}=4$ \cite{newman2015structure}.
For each synthetic network configuration, we generate 10 random instances in order to account for the randomness in the synthetic network generator. 
This totals to $20 \times 10 = 200$ synthetic networks.

\vspace{0.1cm}\noindent\textbf{Real-world networks.}
Unlike our synthetic networks, real-world networks that we analyze (see below) do not contain the ground truth in the form of $O^{(gt)}=(T^{(gt)},P^{(gt)})$.
The only appropriate ground truth knowledge that we have and that we have only for some of the networks is the set of change points $T^{(gt)}$. 
None of the networks contain the set of segment partitions $P^{(gt)}$ as the ground truth, either because they do not have available any node community structure information whatsoever or because they only have available an inappropriate single \emph{static} community structure for the whole \emph{dynamic} network. 
Nevertheless, we can still evaluate the methods on the real-world networks, by: 1) using evaluation measures that do not rely on the ground truth knowledge, for all real-world networks, and 2) assessing how well the methods can recover the change point set $T^{(gt)}$, for real-world networks that do contain this ground truth knowledge.

We consider six different publicly available real-world dynamic networks.
\textbf{1)} \emph{Hypertext} \cite{isella2011s} network contains information about face-to-face proximity of attendees of the Hypertext 2009 conference.
The nodes correspond to people, and there is an edge between two people if they were close to each other within a given time interval, as measured by wearable radio badges.
This network has $T^{(gt)}$ that corresponds to the list of events from the conference program \cite{isella2011s}.
\textbf{2)} \emph{AMD Hope} \cite{amdhope} network contains information about co-location of attendees of The Last HOPE conference in 2008.
The nodes correspond to people, and there is an edge between two people if they were located in the same room at the same time.
This network has $T^{(gt)}$ that corresponds to the featured/keynote talks and social events  \cite{amdhope}.
\textbf{3)} \emph{High School} \cite{mastrandrea2015contact} network contains information about proximity of students in a  high school during one work week in 2013.
The nodes and edges are added in the same way as in Hypertext network.
This network does not have $T^{(gt)}$.
\textbf{4)} \emph{Reality Mining} \cite{eagle2006reality} network contains information about social interactions of university students and faculty during 2004-2005 academic year.
The nodes correspond to people, and there is an edge between two people if there was a phone call between them in a given time interval.
This network has $T^{(gt)}$ that corresponds to the list of events from the academic calendar \cite{peel2015detecting}.
\textbf{5)} \emph{Enron} \cite{priebe2005scan} network contains information about email communication of employees of the Enron corporation during the 2000-2002 period.
The nodes correspond to people, and there is an edge between two people if there was an email between them in a given time interval.
This network has $T^{(gt)}$ that corresponds to the list of company-related events from the news sources \cite{peel2015detecting}.
\textbf{6)} \emph{Senate} \cite{kawadia2012sequential} network contains information about voting similarities of United States senators during the 1789-2015 period (i.e., for 113 Congresses).
The nodes correspond to states, and there is an edge between two states if the voting similarity between the corresponding senators in a given time interval is high enough \cite{kawadia2012sequential}.
\emph{Senate} network does not have $T^{(gt)}$. 
This is because for this network we cannot use the list of historic events as a formal ground truth change point set, since it is not clear how to objectively select a fixed number of them (i.e., how to determine which events are more important than others and how many of the most important events should be considered). 
For statistics of the real-world networks, see Supplementary Table~\ref{tab:supplement:real_world_data}.

\subsubsection{Evaluation measures}\label{sec:eval_measures}
We evaluate the performance of a given method via:
\emph{network structure-based measures} and \emph{ground truth knowledge-based measures}.

\vspace{0.1cm}\noindent\textbf{Network structure-based measures.}
Here, we measure the quality of the results of a given method with respect to the structure of the input dynamic network $D$, without relying on any ground truth knowledge.
Specifically, we can use one of the objective functions from Supplementary Section~\ref{sec:supplement:scout_obj_func} to measure the quality of the method's $O^\star=(T^\star,P^\star)$.
These objective functions include four $Q_P$  measures of partition quality and two $Q_B$ measures accounting for both segmentation quality and partition quality.
Regarding the four $Q_P$ measures (i.e., modularity, conductance, normalized cut, and average-ODF), in our experiments, all four measures show statistically significantly correlated results with respect to both Pearson and Spearman correlations (with all pairwise $p$-values $ < 10 ^{-49}$).
So, in case of $Q_P$, for brevity, we report results only for modularity.
Regarding the two $Q_B$ measures (i.e., AIC and BIC), we do not evaluate the results with respect to them, since these are the objective functions that SCOUT explicitly aims to optimize, and thus, we want to avoid circular reasoning.

\vspace{0.1cm}\noindent\textbf{Ground truth knowledge-based measures.}
Here, we measure the quality of the results of a given method with respect to the available ground truth knowledge.
We discuss two general ways to achieve this:
\RNum{1}) by measuring similarity of the method's $O^\star=(T^\star,P^\star)$ to the known ground truth $O^{(gt)}=(T^{(gt)},P^{(gt)})$
and \RNum{2}) by evaluating the method's ability to rank time points according to how ``change point-like'' they are.

\noindent\textbf{\RNum{1})}
We introduce three general groups of measures of similarity between $O^\star$ and $O^{(gt)}$:
a) \emph{segmentation similarity} $Sim_T$, focusing only on the segmentation aspect of $O^\star$ and $O^{(gt)}$,
b) \emph{partition similarity} $Sim_P$, focusing only on the partition aspect of  $O^\star$ and $O^{(gt)}$,
and c) \emph{overall similarity} $Sim_B$, focusing simultaneously on both aspects of  $O^\star$ and $O^{(gt)}$.

\begin{figure}
  \centering\includegraphics[width=0.6\linewidth]{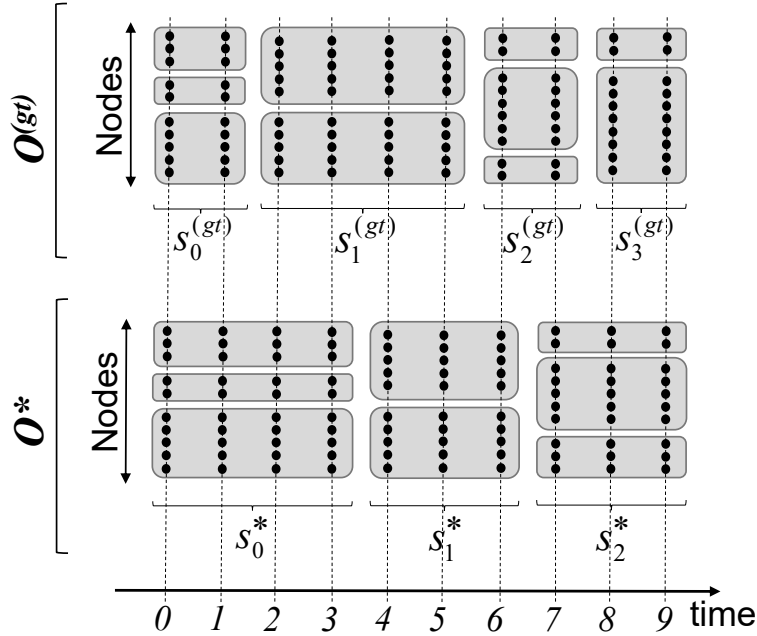}
  \caption[The process of constructing a node-time partition $\mathcal{P}_B$ when computing $Sim_B$]{The process of constructing a node-time partition $\mathcal{P}_B$ when computing $Sim_B$.
  Each black circle corresponds to a node at a given time point.
  Circles on the same horizontal line correspond to a fixed node at different time points.
  Circles on the same vertical line correspond to different nodes at a fixed time point.
 Rectangles illustrate clusters in node-time partitions $\mathcal{P}^\star_B$ and $\mathcal{P}^{(gt)}_B$.
  }\label{fig:overall_sim}
\end{figure}

\textbf{a)}
To measure $Sim_T$ between $O^\star$ and $O^{(gt)}$, intuitively, we first construct for each of them a special time point partition $\mathcal{P}_T$ that captures how the snapshots of $D$ are grouped into segments.
For example, for $O^\star$ and $O^{(gt)}$ in \figurename~\ref{fig:overall_sim}, $\mathcal{P}_T^\star$ contains three clusters ($\{ 0,1,2,3 \}$, $\{ 4,5,6 \}$, and $\{ 7,8,9 \}$) and $\mathcal{P}_T^{(gt)}$ contains four clusters ($\{ 0,1 \}$, $\{ 2,3,4,5 \}$, $\{ 6,7 \}$, and $\{ 8,9 \}$).
Then, we measure similarity between the two resulting time point partitions via an existing partition similarity measure (see below after part ``c'').
For formal details regarding $Sim_T$, see Supplementary Section~\ref{sec:supplement:eval_measures}.

\textbf{b)}
To measure $Sim_P$ between $O^\star$ and $O^{(gt)}$, intuitively, we first measure for each snapshot of $D$ similarity between its corresponding segment partitions in $O^\star$ and $O^{(gt)}$ via an existing partition similarity measure (see below after part ``c'').
For example, for $O^\star$ and $O^{(gt)}$ in \figurename~\ref{fig:overall_sim}, for snapshot $G_0$, we measure similarity between $p^\star_0$ and $p^{(gt)}_0$ (since $G_0$ belongs to the first segment in $O^\star$ and to the first segment in $O^{(gt)}$), while for snapshot $G_2$, we measure similarity between $p^\star_0$ and $p^{(gt)}_1$  (since $G_2$ belongs to the first segment in $O^\star$ and to the second segment in $O^{(gt)}$). 
Then, we average the results over all snapshots.
For formal details regarding $Sim_P$, see Supplementary Section~\ref{sec:supplement:eval_measures}.

\textbf{c)}
To measure $Sim_B$ between $O^\star$ and $O^{(gt)}$, intuitively, we first construct for each of them a special node-time partition $\mathcal{P}_B$ that simultaneously captures how snapshots are grouped by $T$s and how nodes are grouped by $P$s.
For illustrations of node-time partitions of $O^\star$ and $O^{(gt)}$, see \figurename~\ref{fig:overall_sim}.
Then, we measure similarity between the two resulting node-time partitions via an existing partition similarity measure (see below).
For formal details regarding $Sim_B$, see Supplementary Section~\ref{sec:supplement:eval_measures}.

All of $Sim_T$, $Sim_P$, and $Sim_B$ are parameterized with a measure $H$ of similarity between two partitions. 
We test four popular such measures $H$: \textbf{1)}  \emph{Normalized Mutual Information} (\emph{NMI}) \cite{vinh2010information}, \textbf{2)} \emph{Adjusted Mutual Information} (\emph{AMI}) \cite{vinh2010information},
\textbf{3)} \emph{Adjusted Rand Index} (\emph{ARI}) \cite{vinh2010information}, and \textbf{4)} \emph{V-Measure} (\emph{VM}) \cite{rosenberg2007v}.
For details of the above measures, see Supplementary Section~\ref{sec:supplement:eval_measures}.
In our experiments, all four measures $H$ show statistically significantly correlated results with respect to both Pearson and Spearman correlations (with all pairwise $p$-values $ < 10 ^{-239}$).
So, for brevity, we report results only for NMI.

\noindent\textbf{\RNum{2})}
Assessing a given method's ability to detect ground truth change points $T^{(gt)}$ is important in the task of segment detection \cite{ranshous2015anomaly}.
One way to achieve this is via $Sim_T$ from above, which directly compares the given method's change point set $T^\star$ against $T^{(gt)}$.
$Sim_T$ only takes into account time points that were chosen as change points. 
That is, $Sim_T$ does not consider time points that were not chosen as change points, even though some of these time points may have still been good change point candidates. 
Namely, when determining which time points should be change points, a method assigns to each time point a score (or rank) according to how ``change point-like'' the time point is.
So, instead of using ``binary'' information for each time point $t$ as $Sim_T$ does (i.e., either $t \in T^{(gt)}$ or $t \not\in T^{(gt)}$), we can make use of the more complete information on ranking of all time points.
An example of why this would be useful is as follows. 
Even if some ground truth change point $t \in T^{(gt)}$ is not (mistakenly) included into $T^\star$, we still want the method to  rank $t$ higher than some other $t^\prime \not\in T^{(gt)}$. $Sim_T$ would fail to capture this information, so we use an alternative evaluation metric, as follows. 

Having a ranked list of all time points (for details on how we obtain this list for each method, see Supplementary Section~\ref{sec:supplement:eval_measures}), we measure a given method's performance with respect to \emph{change point classification} via three measures:
1) the \emph{area under the precision-recall curve} (\emph{AUPR}),
2) the \emph{maximum F-score},
and 3) the \emph{area under the receiver operator characteristic curve} (\emph{AUROC}).
For details of the above measures, see Supplementary Section~\ref{sec:supplement:eval_measures}.
In our experiments, all three measures show statistically significantly correlated results with respect to both Pearson and Spearman correlations (with all pairwise $p$-values $ < 10^{-64}$).
So, for brevity, we report results only for AUPR.

\section{Results}\label{sec:results}
We compare four different methods (Section~\ref{sec:methods_for_comparison}):
three existing methods (GraphScope, Multi-Step, and GHRG; Section~\ref{sec:related_work}) and our new SCOUT approach (Section~\ref{sec:scout}).
We evaluate the methods on synthetic networks as well as real-world networks (Section~\ref{sec:datasets}).
We evaluate the methods with respect to network structure-based measures and ground truth knowledge-based measures in the task of the SCD problem (Section~\ref{sec:eval_measures}).
As a measure of the former type, we use average snapshot partition quality $Q_P$ based on modularity.
As a measure of the latter type, we use a) similarity of a method's output to the ground truth
and b) change point classification.
For case ``a'' above, we compute segmentation similarity $Sim_T$, partition similarity $Sim_P$, and overall similarity $Sim_B$.
For all of the three similarity measures, we use NMI to measure partition similarity.
For case ``b'' above, we use AUPR.
We measure statistical significance of the improvement of SCOUT over the best of the existing approaches (Supplementary Section~\ref{sec:supplement:methods_stat}).

When we have the complete ground truth information (on both the segmentation aspect and the partition aspect of the SCD problem) available, which is the case for our synthetic networks, we use all of the above measures, but we trust $Sim_B$ the most, since it captures similarity between a given method's solution and the ground truth solution with respect to both SCD aspects.
When we do not have the complete ground truth information (i.e., when we cannot use the two-aspect $Sim_B$), which is the case for our real-world networks, we assess a given method based on the structure-based measure (i.e., $Q_P$ based on modularity) and whichever ground truth knowledge-based measure we \emph{can} compute based on the partial ground truth information about the data.
Since in our case the available ground truth information is the list of change points, for the latter, we can use any measure that captures the segmentation aspect of the solution quality.
Recall that we have two such measures: $Sim_T$ and change point classification (Section~\ref{sec:eval_measures}).
Since we demonstrate in Section~\ref{sec:results_synthetic_t} that the two measures overall yield consistent results on synthetic networks with known ground truth SCD solution, and since per our discussion in Section~\ref{sec:eval_measures} change point classification is theoretically more meaningful than $Sim_T$ as it accounts for ranking of all time points rather than only for the identified change points, for brevity, we focus only on change point classification for real-world networks.

Below, we first discuss the effect of parameter choices on method performance, in order to choose the best parameter values for each method (Section~\ref{sec:results_params}).
Then, we compare the methods on synthetic (Section~\ref{sec:results_synthetic}) and real-world (Section~\ref{sec:results_real_world}) networks.

\subsection{The effect of method parameter choices}\label{sec:results_params}
We perform all experiments from this section on synthetic networks, since they have the known ground truth knowledge embedded into them (Section~\ref{sec:datasets}).
In particular, due to high computational complexity of some of the existing methods and a large number of performed tests, in this section, we use the smallest synthetic data with $50$ nodes per snapshot.
As discussed above, our main criterion for selecting parameters of a given method is overall ground truth similarity $Sim_B$.
Note that GraphScope does not accept any user-specified parameters, and thus we leave it out from consideration in this section.

\vspace{0.1cm}\noindent\textbf{Multi-Step.}
We test the effect on the method's performance of the similarity threshold parameter $\theta$, which determines when to stop the segment merging process  (Supplementary Section~\ref{sec:supplement:methods_for_comparison}).
We find that there is no $\theta$ value that works well for all of the synthetic network configurations with respect to $Sim_B$ (Supplementary \figurename~\ref{fig:supplement:multistep-gl_p__nmi}). 
This is mainly because no single $\theta$ value can reliably estimate the ground truth number of segments across the different configurations (Supplementary \figurename~\ref{fig:supplement:multistep-sgmq__num}). 
Thus,  Multi-Step can be used to reliably solve only the CSCD problem where the number of segments is provided as input.
So, when comparing Multi-Step against other methods in the context of the SCD problem, we instead ask Multi-Step to solve the CSCD problem with the ground truth number of segments given as input.
We refer to this modification of Multi-Step as Multi-Step$^\star$.
This gives Multi-Step an unfair advantage compared to the other methods, but we have to do this in order to include Multi-Step into comparison.

\vspace{0.1cm}\noindent\textbf{GHRG.}
We test the effect on the method's performance of windows size $w$ (Supplementary Section~\ref{sec:supplement:methods_for_comparison}). 
After varying its values, we observe that the value $w=4$ generally leads to the highest $Sim_B$ (Supplementary \figurename~\ref{fig:supplement:ghrg}).
Thus, we use $w=4$ for our experiments.

\vspace{0.1cm}\noindent\textbf{SCOUT.}
We test the effect on the method's performance of a) the objective function,
b) consensus clustering method, and c) search strategy.
We choose $Q_B$ based on BIC as the objective function, sum graph with Walktrap as the consensus clustering method, and the bottom-up search as the search strategy, per our discussion in Supplementary Section~\ref{sec:supplement:results_scout_params} and Supplementary \figurenames~\ref{fig:supplement:scout_q_b}-\ref{fig:supplement:scout_search}.

\subsection{Synthetic networks}\label{sec:results_synthetic}

We next evaluate the methods (under their best parameter values from Section~\ref{sec:results_params}) on synthetic networks, which have the ground truth SCD solution embedded into them.
We consider 20 different synthetic network configurations: five values for the number of segments times four values for the number of nodes per snapshot (Section~\ref{sec:datasets}).
These configurations span the whole ``spectrum'' between the extreme cases of snapshot clustering (where the number of ground truth segments corresponds to the number of snapshots) and consensus clustering (where there is only one ground truth segment corresponding to the whole dynamic network).
For each synthetic network configuration, we generate multiple random network instances (Section~\ref{sec:datasets}) and report results averaged over the multiple instances.

Recall that the main idea behind our synthetic network generation process (snapshots within the same segment having the same community organization) aligns well with the intuition of each of the considered methods.
Thus, we expect all methods to have a fair chance for recovering the ground truth knowledge,
with the exception of Multi-Step, which has an unfair \emph{advantage} over all other methods, per our discussion in Section~\ref{sec:results_params}.
Specifically, recall that we provide the ground truth number of segments as input to Multi-Step.
This \emph{a priori} knowledge gives an unfair advantage to Multi-Step compared to all other methods for all configurations, but this advantage is the most pronounced for the extreme configurations with the minimum and maximum possible numbers of ground truth segments (i.e., with one and 16 segments, respectively; Section~\ref{sec:datasets}).
This is because for these two types of  configurations, the knowledge of the ground truth number of segments guarantees that Multi-Step's solution will have the correct segmentation: given 16 snapshots (which is the size of our synthetic network data), there is only one way to group the 16 snapshots into one segment (the resulting segment will encompass all 16 snapshots) and only one way to group the 16 snapshots into 16 segments (each segment will encompass exactly one of the snapshots).
For the other non-extreme configurations, with more than one but less than 16 segments, while knowing the ground truth number of segments still gives an advantage to Multi-Step (meaning that clearly Multi-Step will produce the correct ground truth number of segments, or equivalently, the correct number of change points), it does not necessarily guarantee that Multi-Step will obtain the correct segmentation (i.e., that the identified change points will be correct).
This is because for these non-extreme configurations, there are multiple ways to group snapshots into the given number of segments.

For each synthetic network, we know the corresponding ground truth segmentation and segment partitions, so we can fully utilize the available ground truth knowledge-based measures.
Below, we start by discussing results when focusing on a single aspect of the SCD problem at a time: first on a segmentation aspect (i.e., $Sim_T$ and change point classification; Section~\ref{sec:results_synthetic_t}) and second on a partition aspect (i.e., $Q_P$ and $Sim_P$; Section~\ref{sec:results_synthetic_p}).
Then, we discuss the results with respect to overall ground truth similarity $Sim_B$  (Section~\ref{sec:results_synthetic_b}).
Recall that $Sim_B$ is the most reliable measure, since its captures both aspects of the SCD problem.
Thus, for $Sim_B$, we also measure the statistical significance of the improvement of SCOUT over the existing methods (Supplementary Section~\ref{sec:supplement:methods_stat}).
Finally, we compare running times of the methods (Section~\ref{sec:results_synthetic_run_time}).

\subsubsection{Segmentation aspect of the solution quality}\label{sec:results_synthetic_t}

\begin{figure}[!htbp]
  \centering\hspace{-0.9cm}\includegraphics[width=0.4\linewidth]{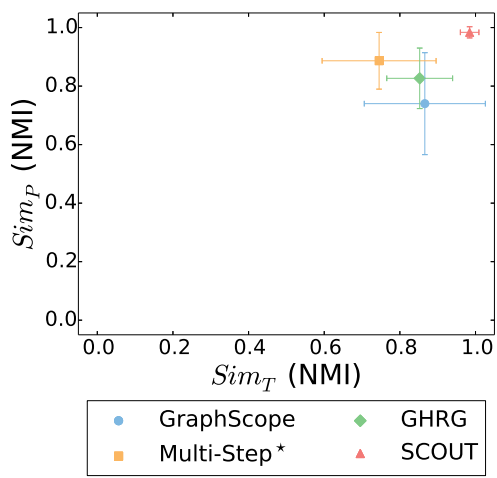}
  \caption[Representative $Sim_T$ and $Sim_P$ scores for synthetic networks with 100 nodes per snapshot and four ground truth segments]{Representative $Sim_T$ and $Sim_P$ scores for synthetic networks with 100 nodes per snapshot (the largest size for which all methods could be run) and four ground truth segments.
The results are averaged over all of the corresponding synthetic network instances.
Equivalent results for the remaining synthetic network configurations are shown in Supplementary \figurename~\ref{fig:supplement:scatter_100}.
}\label{fig:synth_100_scatter_sim_4seg}
\end{figure}

\begin{figure*}[!htbp]
\centering
\subfloat[]{\label{fig:synth_100_aupr}\includegraphics[width=.315\linewidth]{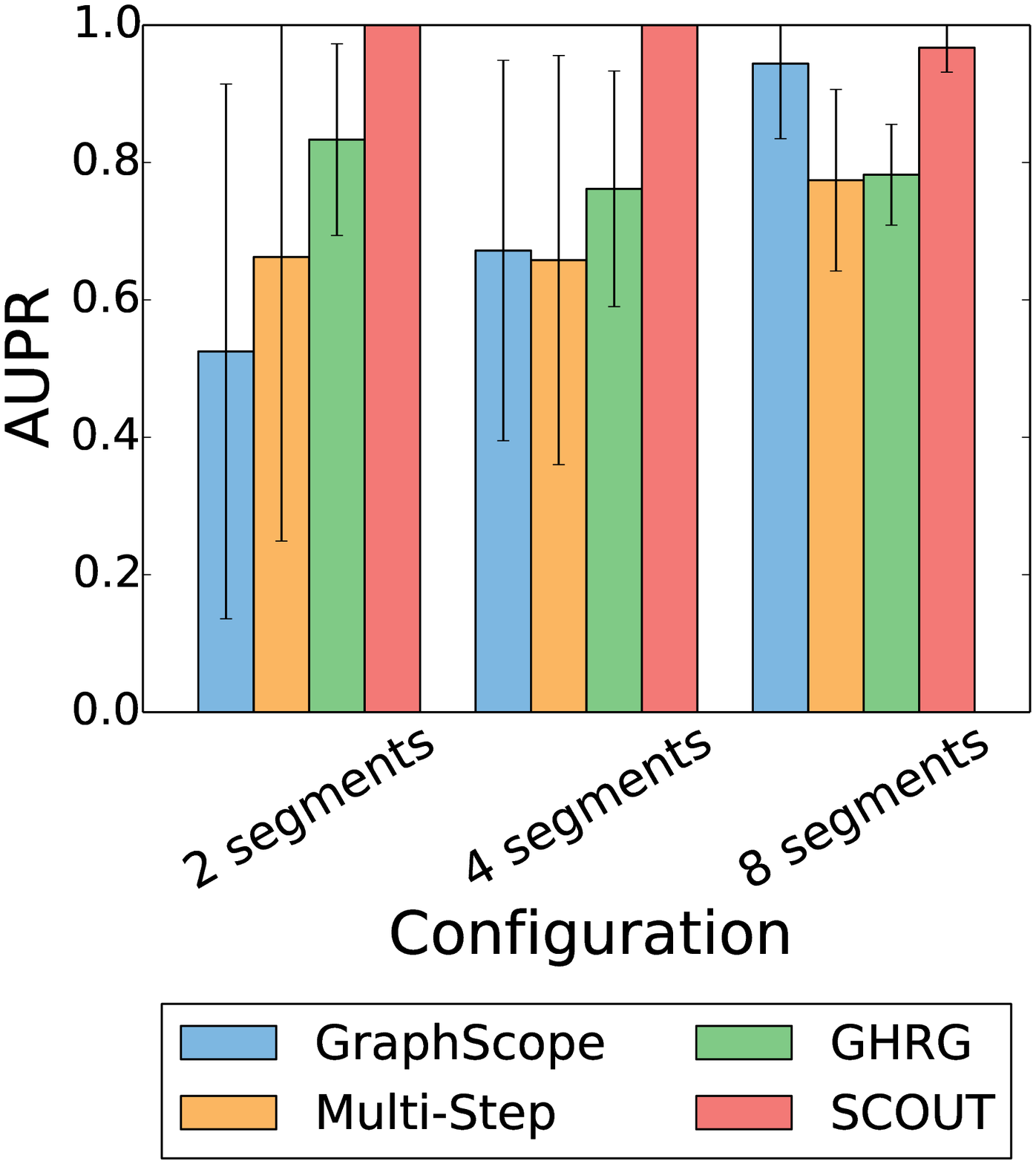}}
\hspace{0.2cm}
\subfloat[]{\label{fig:synth_100_avg_sn__mod}\includegraphics[width=.325\linewidth]{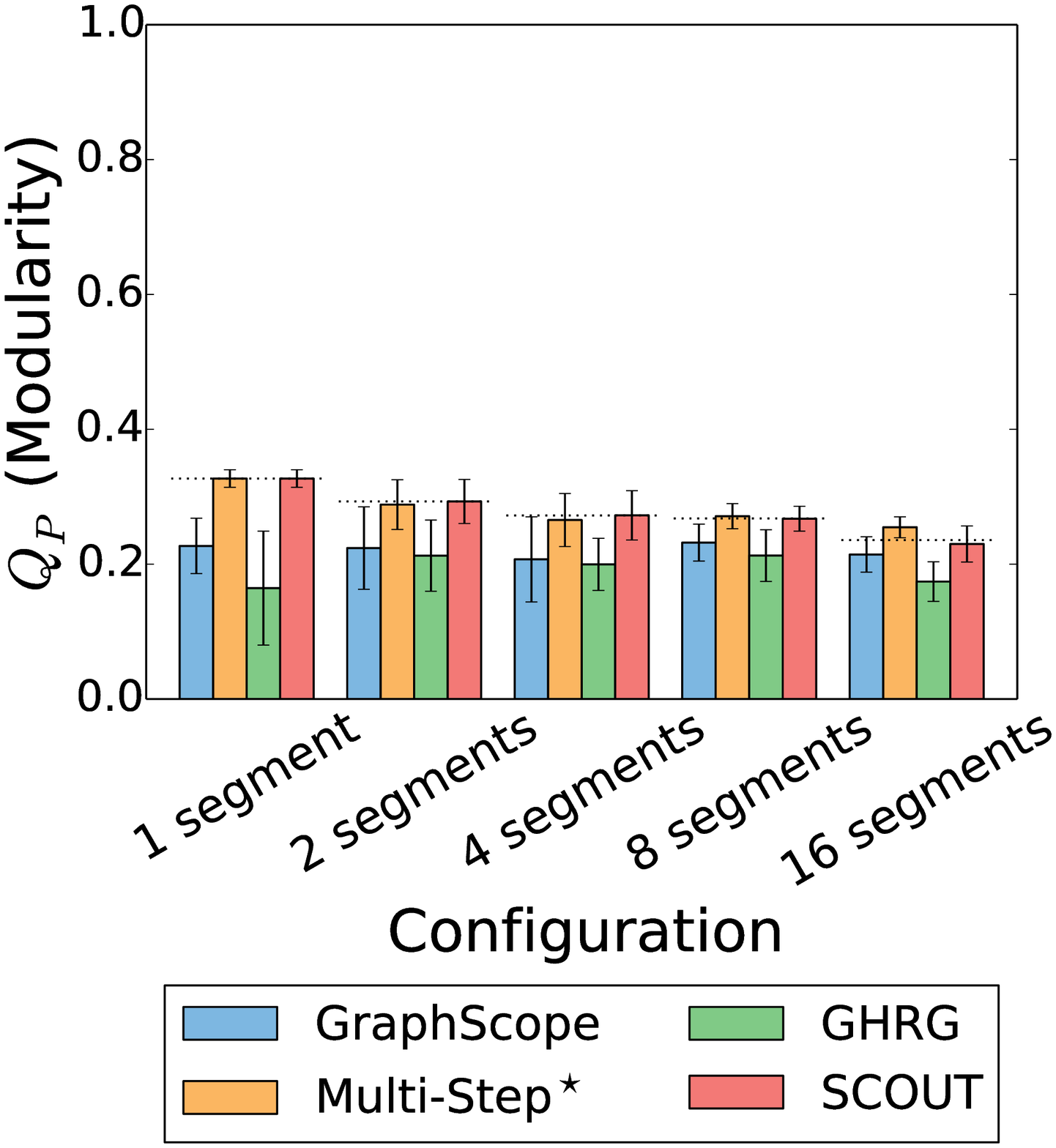}}
\hspace{-0.1cm}
\subfloat[]{\label{fig:synth_100_gl_p__nmi}\includegraphics[width=.325\linewidth]{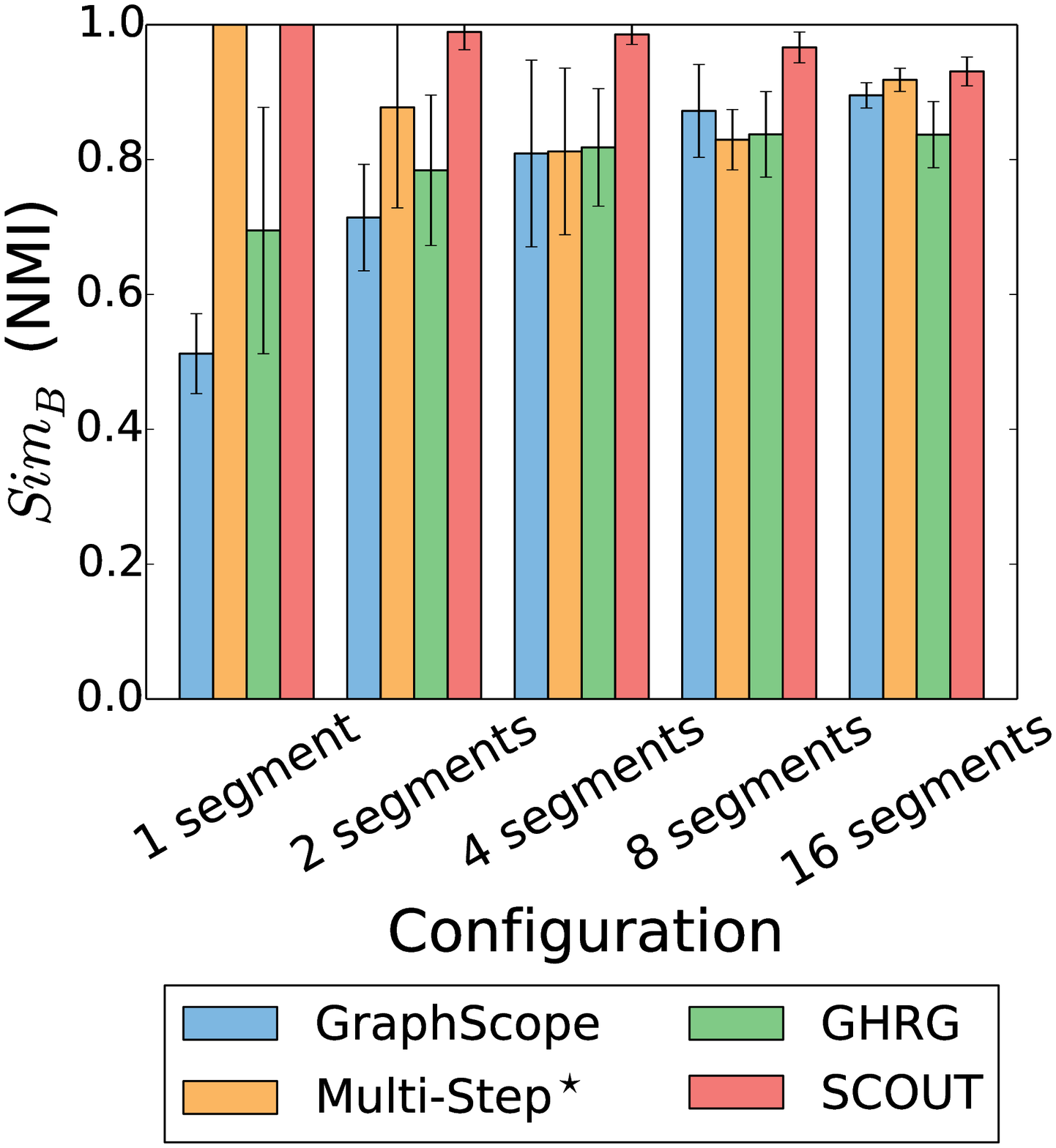}}
  \caption[Representative method comparison  for synthetic networks with 100 nodes per snapshot]{Representative method comparison  for synthetic networks with 100 nodes per snapshot with respect to
\textbf{\protect\subref{fig:synth_100_aupr}} change point classification,
\textbf{\protect\subref{fig:synth_100_avg_sn__mod}} $Q_P$,
 and \textbf{\protect\subref{fig:synth_100_gl_p__nmi}} $Sim_B$.
For a given ground truth configuration, the results are averaged overall all of the corresponding synthetic network instances.
In panel \protect\subref{fig:synth_100_avg_sn__mod}, the dotted lines correspond to the ground truth score.
Note that for panel \protect\subref{fig:synth_100_aupr}, we exclude from consideration the configurations with the minimum and maximum possible numbers of ground truth segments (i.e., one and 16).
We do this because for these configurations, either there are no change points at all (for one segment) or every time point is a change point (for 16 segments), which means that change point classification cannot be performed (Section~\ref{sec:eval_measures}).
Equivalent results for the remaining synthetic network configurations are shown in Supplementary \figurename~\ref{fig:supplement:synth_results}.
}
\label{fig:synth_100}
\end{figure*}

\textbf{For} {\boldmath $Sim_T$}\textbf{,} SCOUT is superior to all other methods, as it achieves the highest scores for 90\% of all synthetic network configurations, while the other methods are relatively comparable to each other (Supplementary \figurename~\ref{fig:supplement:synth_rank_sim_t}).
The remaining 10\%  (i.e., two) of all configurations in which an existing method (in this case, GraphScope) achieves higher scores are configurations with the two largest numbers of nodes per snapshot and with the maximum possible number of segments (\figurename~\ref{fig:synth_100_scatter_sim_4seg} and Supplementary \figurename~\ref{fig:supplement:scatter_100}).
The fact that GraphScope has higher $Sim_T$ for these configurations is not necessarily surprising, for the following reason.
GraphScope generally produces solutions with more segments than the other methods do, frequently overestimating the ground truth number of segments (Supplementary \figurename~\ref{fig:supplement:sgmq__num}). 
Consequently, since for the configurations with the maximum possible number of segments, the most that GraphScope can overestimate is the maximum number of segments itself (i.e., the correct solution), GraphScope is expected to achieve higher $Sim_T$ than the other methods.
Note that when measuring $Sim_T$ for the extreme configurations with the minimum and maximum possible numbers of segments, we exclude Multi-Step from comparison.
This is because, per our discussion from Section~\ref{sec:results_synthetic},
we give Multi-Step an unfair advantage by providing it with the ground truth number of segments as input, which for these extreme configurations means \emph{a priori} knowing the correct segmentation and thus achieving the perfect $Sim_T$ (\figurename~\ref{fig:synth_100_scatter_sim_4seg} and Supplementary \figurename~\ref{fig:supplement:scatter_100}).
Interestingly, for the remaining non-extreme configurations, Multi-Step is always outperformed by SCOUT and at least one of the existing methods  (Supplementary \figurename~\ref{fig:supplement:synth_rank_sim_t}).
Therefore, Multi-Step, which knows the ground truth number of segments \emph{a priori} typically does not yield a high quality segmentation with respect to $Sim_T$, whereas SCOUT does produce a high quality segmentation (and it typically does so better than the other methods) despite not having this prior knowledge.
This is further confirmed by the fact that SCOUT can automatically determine the ground truth number of segments more accurately than the existing methods (Supplementary \figurename~\ref{fig:supplement:sgmq__num}).

\textbf{For change point classification,} SCOUT is superior to all of the existing methods, as it achieves the highest accuracy for 92\% of all synthetic network configurations (Supplementary \figurename~\ref{fig:supplement:synth_rank_aupr}).
Among the existing methods, GHRG is generally superior, followed by GraphScope and Multi-Step  (\figurename~\ref{fig:synth_100_aupr} and Supplementary \figurename~\ref{fig:supplement:synth_aupr}).
In the remaining 8\% of all configurations (which is only one configuration in this case -- the configuration with 500 nodes and eight segments; Supplementary \figurename~\ref{fig:supplement:synth_rank_aupr}) where SCOUT is not superior, an existing method (in this case, GraphScope) achieves only marginally higher score (Supplementary \figurename~\ref{fig:supplement:synth_aupr}).
Overall, the trends with respect to change point classification are similar to those with respect to $Sim_T$, which is not surprising, since both measure the same aspect of the SCD problem.
Note that for change point classification, Multi-Step does not have the unfair advantage over the other methods, as it does for $Sim_T$ above,
since its produced time point ranking depends only on the solutions of the CSCD problem (Supplementary Section~\ref{sec:supplement:eval_measures}).

\subsubsection{Partition aspect of the solution quality}\label{sec:results_synthetic_p}
\textbf{For} {\boldmath $Q_P$}\textbf{,} SCOUT is superior to all other methods, achieving the highest $Q_P$  for 70\% of all synthetic network configurations (Supplementary \figurename~\ref{fig:supplement:synth_rank_avg_sn_mod}).
Among the existing methods, Multi-Step shows the best results,
followed by GraphScope and GHRG that are comparable to each other (\figurename~\ref{fig:synth_100_avg_sn__mod} and Supplementary \ref{fig:supplement:synth_avg_sn__mod}).
Importantly, SCOUT overall outperforms Multi-Step in terms of $Q_P$ despite the fact that Multi-Step explicitly maximizes modularity (which is the basis of $Q_P$; Section~\ref{sec:eval_measures}), while the version of SCOUT under consideration does not rely on $Q_P$ at all (Section~\ref{sec:results_params}).
Note that the configurations on which Multi-Step outperforms SCOUT are mostly those with the maximum possible number of ground truth segments (\figurename~\ref{fig:synth_100_avg_sn__mod} and Supplementary \figurename~\ref{fig:supplement:synth_avg_sn__mod}).
This is not necessarily surprising, since for these 16-segment configurations,  SCOUT can produce a solution with \emph{at most} 16 segments, while Multi-Step is guaranteed to produce the solution with exactly 16 segments (Section~\ref{sec:results}).
That is, intuitively, Multi-Step's solution will have a separate segment partition for each snapshot, and each of those partitions aims to maximize modularity and consequently $Q_P$.
Importantly, for the configurations where Multi-Step outperforms SCOUT, Multi-Step's $Q_P$-based superiority is \emph{not} necessarily an advantage.
This is because Multi-Step achieves higher $Q_P$ scores even compared to $Q_P$ scores of the ground truth solution (\figurename~\ref{fig:synth_100_avg_sn__mod} and Supplementary \figurename~\ref{fig:supplement:synth_avg_sn__mod}).
Thus, even if Multi-Step obtains the highest $Q_P$, its partitions might not necessarily be closer to the ground truth than SCOUT's partitions, as we justify next.

\textbf{For} {\boldmath $Sim_P$}\textbf{,} SCOUT is superior to all other methods, as it achieves the highest $Sim_P$ score for 100\% of all synthetic network configurations (Supplementary \figurename~\ref{fig:supplement:synth_rank_sim_p}).
The other methods are relatively comparable to each other, with slight superiority of Multi-Step over the other two methods  (\figurename~\ref{fig:synth_100_scatter_sim_4seg} and Supplementary \figurename~\ref{fig:supplement:scatter_100}).
Interestingly, trends with respect to $Sim_P$ are not always consistent with those for $Q_P$, even though the two measure the same aspect of the SCD problem. 
For example, for the configuration with 100 nodes per snapshot and 16 ground truth segments, even though Multi-Step achieves the highest $Q_P$ score (\figurename~\ref{fig:synth_100_avg_sn__mod}), it is the worst-performing method in terms of $Sim_P$ (Supplementary \figurename~\ref{fig:supplement:scatter_100}).
The difference in trends between $Q_P$ and $Sim_P$ is not necessarily suprising, since modularity is known not to always be able to capture well the ground truth communities \cite{yang2015defining}.

\subsubsection{Overall solution quality}\label{sec:results_synthetic_b}

\begin{figure}[!tbp]
\centering
\begin{minipage}{0.5\linewidth}
\includegraphics[width=\linewidth, keepaspectratio=true]{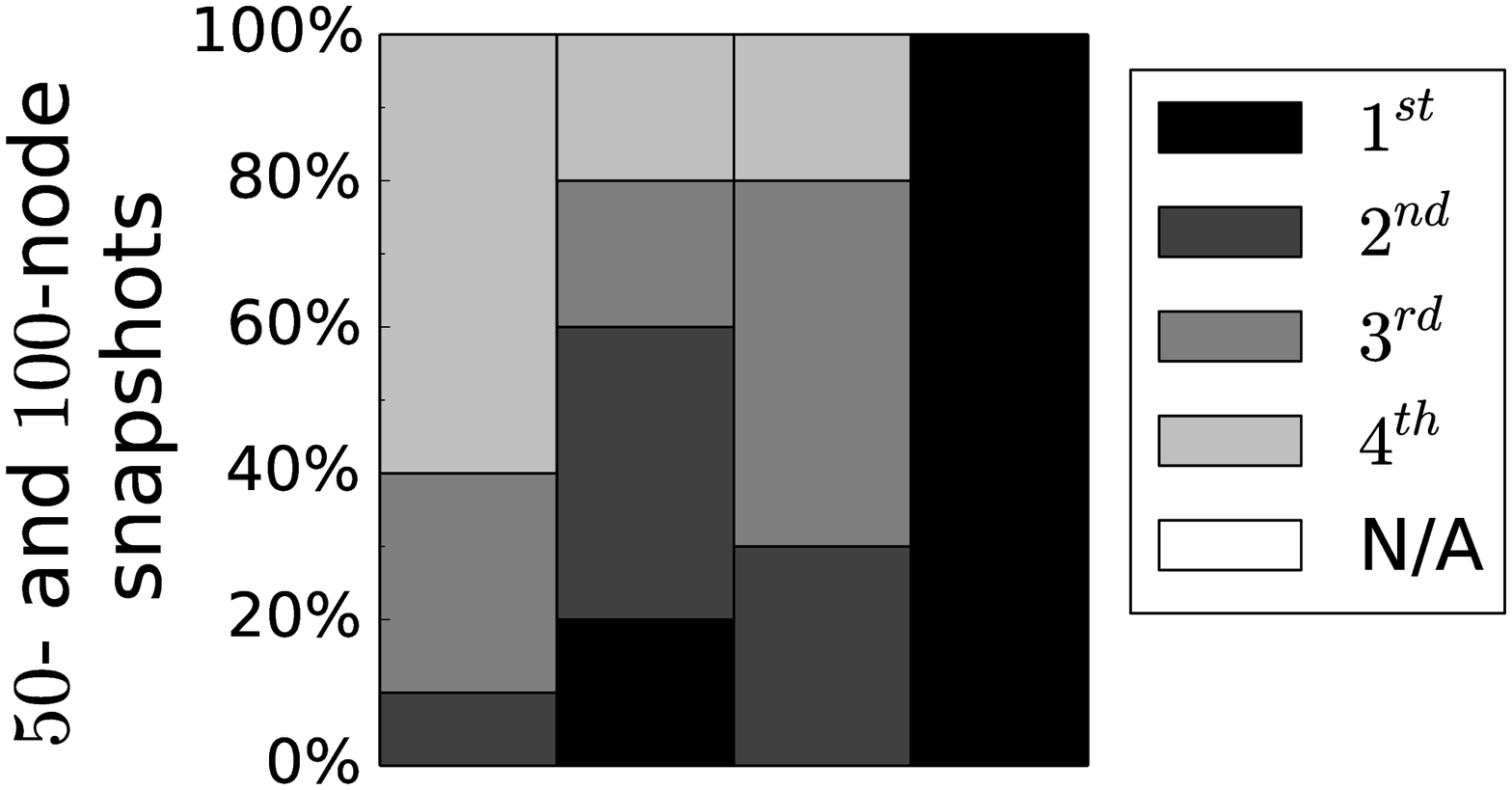}
\includegraphics[width=0.73\linewidth, keepaspectratio=true]{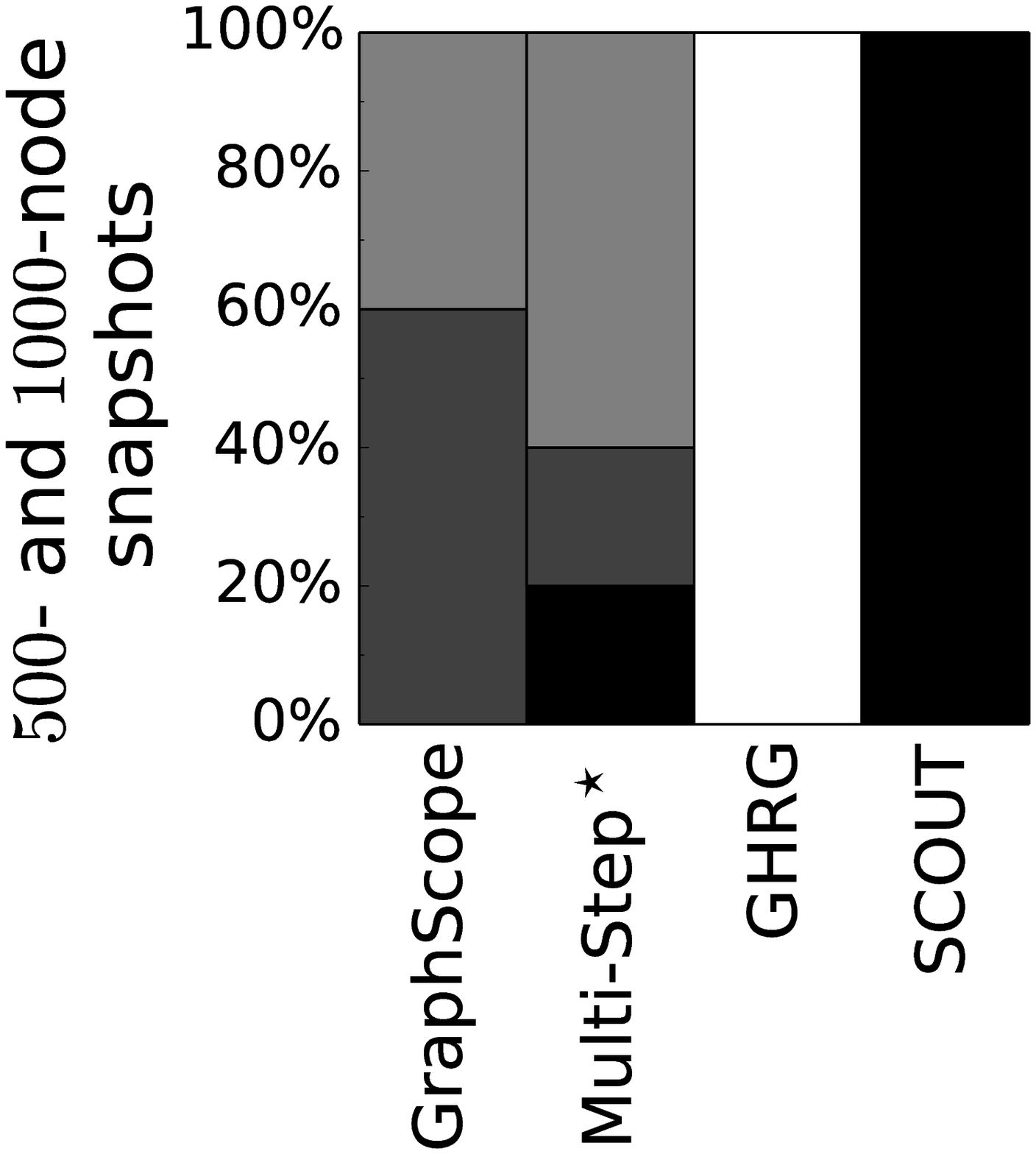}
\end{minipage} 
 \caption[Rankings of the methods for synthetic networks with respect to $Sim_B$]{Rankings of the methods for synthetic networks with respect to $Sim_B$.
Since GHRG could not be run for the larger networks, the results are split into those for the configurations with 50 and 100 nodes per snapshot (top) and those for the configurations with 500 and 1000 nodes per snapshot (bottom).
The rankings are computed as follows. 
For each synthetic network configuration, we compare the four methods' $Sim_B$ scores (averages over all instances of the given configuration) to identify the first, second, third, and fourth best method; ties are allowed, in which case, two methods would be assigned the same rank. 
Then, we summarize these results over all considered synthetic network configurations by measuring, for each method ($x$-axis), how many times the given method is ranked as  the first, second, third, and fourth best method (expressed as the percentage of all considered configurations; $y$-axis).
``N/A'' indicates that the given method could not be run (which is the case for GHRG for the larger networks). 
The figure can intuitively be interpreted as follows: the darker the bar of a given method, the better its performance.
}
\label{fig:synth_rank_sim_b}
\end{figure}

\begin{table}[!tbp]
\centering
{\renewcommand{\arraystretch}{1.25}
\caption[Statistical significance of the superiority of SCOUT over the best existing method in terms of $Sim_B$ on synthetic networks]{Statistical significance of the superiority of SCOUT over the best existing method in terms of $Sim_B$ on synthetic networks. 
``Configuration'' columns describe the synthetic network configurations in terms of the number of nodes per snapshot and the number of segments.
``Average $Sim_B$'' columns contain $Sim_B$ scores of SCOUT and the best of all existing methods (shown in the parentheses), where the scores are averaged over all corresponding synthetic network instances; the highest $Sim_B$ score for a given configuration (i.e., in the given row) is shown in bold. 
The ``$p$-value'' column shows the statistical significance of the difference between $Sim_B$ scores of SCOUT and the best existing method.
``N/A'' means that the scores are identical.
$p$-values less than 0.05 are shown in bold, $p$-values less than 0.01 are shown with one star, and $p$-values less than 0.001 are shown with two stars.
}
\label{tab:synth_stats_sim_b}
\begin{tabular}{@{\extracolsep{4pt}}@{\hskip5pt}cccll}
\toprule
\multicolumn{2}{c}{Configuration} & \multicolumn{2}{c}{Average $Sim_B$}  & \multicolumn{1}{c}{\multirow{2}{*}{$p$-value}}\\
\cline{1-2} \cline{3-4}
\# of nodes & \# of segments & SCOUT & \multicolumn{1}{c}{Best existing method} & \\
\midrule
 \multirow{5}{*}{50} & 1 & \textbf{1.000} & \textbf{1.000} (Multi-Step$^\star$) & \multicolumn{1}{c}{N/A}\\ 
  & 2 & \textbf{0.991} & 0.941  (Multi-Step$^\star$) & 1.548E-01 \\ 
  & 4 & \textbf{0.995} & 0.866  (GHRG) & \textbf{2.075E-03$^\star$} \\ 
  & 8 & \textbf{0.938} & 0.845  (GHRG) & \textbf{4.889E-04$^{\star\star}$} \\
  & 16 & \textbf{0.978} & 0.956  (Multi-Step$^\star$) & \textbf{3.876E-04$^{\star\star}$} \\
\hline
 \multirow{5}{*}{100} & 1 & \textbf{1.000} & \textbf{1.000} (Multi-Step$^\star$) & \multicolumn{1}{c}{N/A} \\ 
  & 2 & \textbf{0.989} & 0.877  (Multi-Step$^\star$) & \textbf{4.643E-02} \\ 
  & 4 & \textbf{0.986} & 0.818  (GHRG) & \textbf{4.045E-04$^{\star\star}$} \\ 
  & 8 & \textbf{0.966} & 0.872  (GraphScope) & \textbf{1.691E-03$^\star$} \\
  & 16 & \textbf{0.931} & 0.918  (Multi-Step$^\star$) & \textbf{3.311E-02} \\
\hline
 \multirow{5}{*}{500} & 1 & \textbf{1.000} & \textbf{1.000}  (Multi-Step$^\star$) & \multicolumn{1}{c}{N/A} \\ 
  & 2 & \textbf{0.953} & 0.757  (GraphScope) & \textbf{1.408E-05$^{\star\star}$} \\ 
  & 4 & \textbf{0.982} & 0.792  (GraphScope) & \textbf{8.709E-06$^{\star\star}$} \\ 
  & 8 & \textbf{0.960} & 0.820  (GraphScope) & \textbf{2.047E-05$^{\star\star}$} \\
  & 16 & \textbf{0.933} & 0.884  (Multi-Step$^\star$) & \textbf{1.628E-05$^{\star\star}$} \\
\hline
 \multirow{5}{*}{1000} & 1 & \textbf{1.000} & \textbf{1.000}  (Multi-Step$^\star$) & \multicolumn{1}{c}{N/A} \\ 
  & 2 & \textbf{0.971} & 0.710  (Multi-Step$^\star$) & \textbf{4.666E-04$^{\star\star}$} \\ 
  & 4 & \textbf{0.937} & 0.659  (GraphScope) & \textbf{1.677E-04$^{\star\star}$} \\ 
  & 8 & \textbf{0.902} & 0.721  (GraphScope) & \textbf{3.058E-06$^{\star\star}$} \\
  & 16 & \textbf{0.841} & 0.763  (GraphScope) & \textbf{6.030E-06$^{\star\star}$} \\
\bottomrule
  \end{tabular}
  }  
\end{table}

For $Sim_B$, SCOUT outperforms the other methods, as it achieves the highest score for 100\% of all synthetic network configurations (\figurename~\ref{fig:synth_rank_sim_b}).
The other methods are comparable to each other  (\figurename~\ref{fig:synth_100_gl_p__nmi} and Supplementary \figurename~\ref{fig:supplement:synth_gl_p__nmi}).
Intuitively, the trends with respect to $Sim_B$ seem to follow the trends with respect to segmentation (Section~\ref{sec:results_synthetic_t}) and partition (Section~\ref{sec:results_synthetic_p}) quality aspects of the SCD problem, which is not surprising given that $Sim_B$ captures both of these aspects.
When we measure the statistical significance of the improvement of SCOUT over the existing methods, we find that SCOUT statistically significantly improves upon the best existing method in 75\%, 65\%, and 55\% of all cases at $p$-value threshold of 0.05, 0.01, and 0.001, respectively (Table~\ref{tab:synth_stats_sim_b}).
Thus, in most of the cases, SCOUT not only improves upon the existing methods but also its improvement is statistically significant.
Note that the above percentages could not be perfect, since for 20\% of all configurations (namely, the four configurations with the minimum number of ground truth segments), in addition to SCOUT that achieves the perfect $Sim_B$, Multi-Step also (unfairly, per our above discussion) achieves the perfect $Sim_B$ and is thus comparable to SCOUT.

\begin{figure}[!tbp]
  \centering\hspace{-0.9cm}\includegraphics[width=0.4\linewidth]{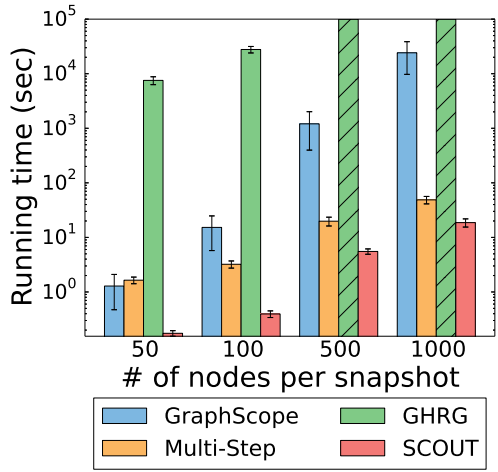}
  \caption[Running times of the methods for synthetic networks]{Running times of the methods for synthetic networks (logarithmic scale).
The results are grouped by the number of nodes per snapshot.
For each number of nodes, running times are averaged over the corresponding numbers of segments and synthetic network instances (i.e., over $5 \times 10 = 50$ runs).
The striped bars mean that the method could not finish within the allowed time. 
  }\label{fig:synthetic_run_time}
\end{figure}

\subsubsection{Running time}\label{sec:results_synthetic_run_time}
SCOUT has the lowest running time of all methods, over all synthetic network configurations  (\figurename~\ref{fig:synthetic_run_time}).
It is followed by Multi-Step, GraphScope, and GHRG, respectively.
Note that GHRG, even when parallelized, cannot be run for the larger networks due to its high computational complexity.

\subsection{Real-world networks}\label{sec:results_real_world}

We next evaluate the methods on real-world networks.
Recall that we consider six real-world networks (Section~\ref{sec:datasets}).
Since the complete ground truth knowledge (i.e., both change points and segment partitions) is unavailable for any of these networks, we perform evaluation based on $Q_P$ and change point classification (Section~\ref{sec:results}).

We discuss first the segmentation aspect of the SCD problem (change point classification; Section~\ref{sec:results_real_world_t}) and second the partition aspect of the SCD problem ($Q_P$; Section~\ref{sec:results_real_world_p}).
Third, we compare running times of the methods (Section~\ref{sec:results_real_world_run_time}).

\subsubsection{Segmentation aspect of the solution quality}\label{sec:results_real_world_t}

\begin{figure*}[!tbp]
\centering
\subfloat[]{\label{fig:real_world_aupr}\includegraphics[width=.4\linewidth]{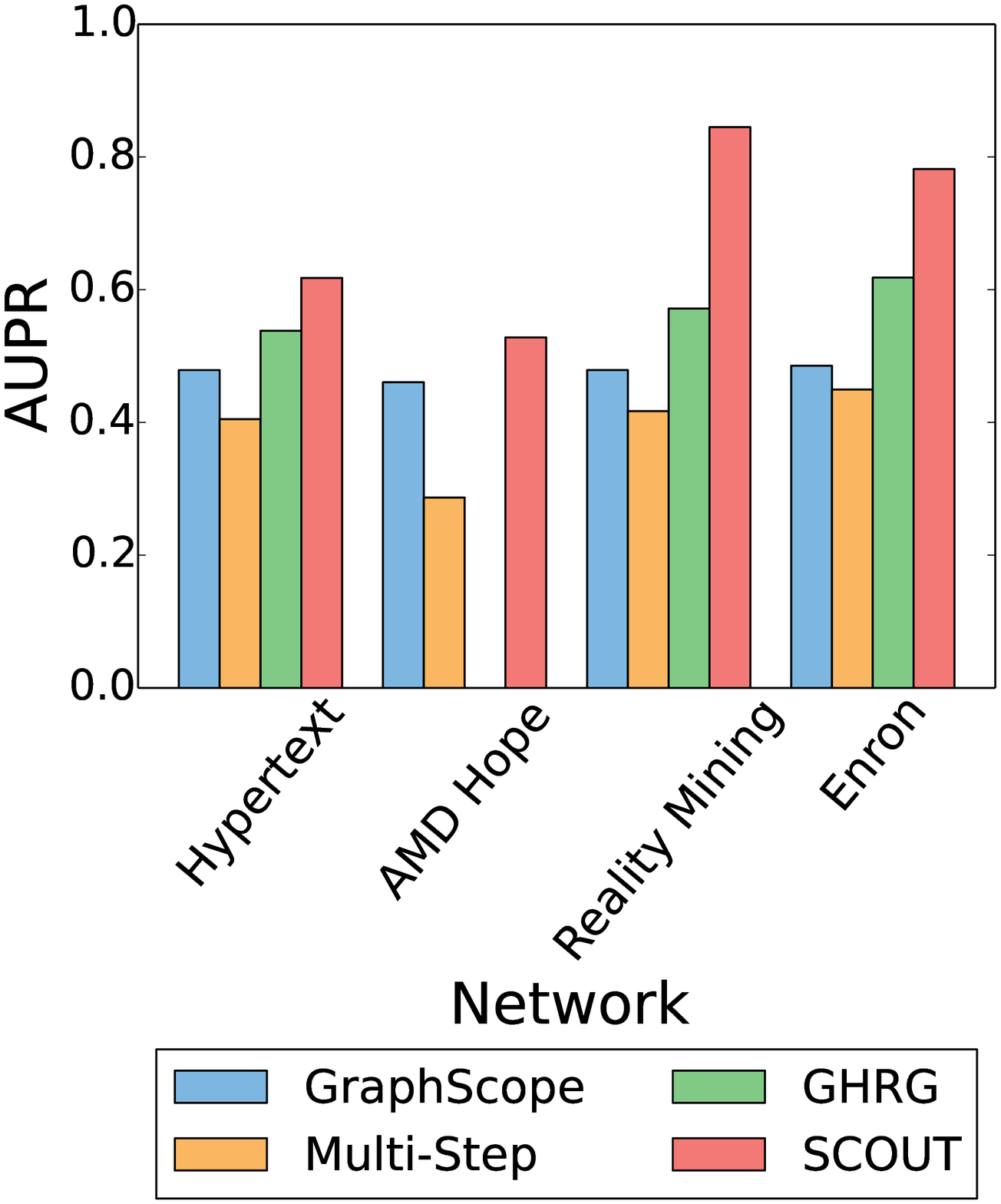}}
\hspace{0.2cm}
\subfloat[]{\label{fig:real_world_avg_sn__mod}\includegraphics[width=.4\linewidth]{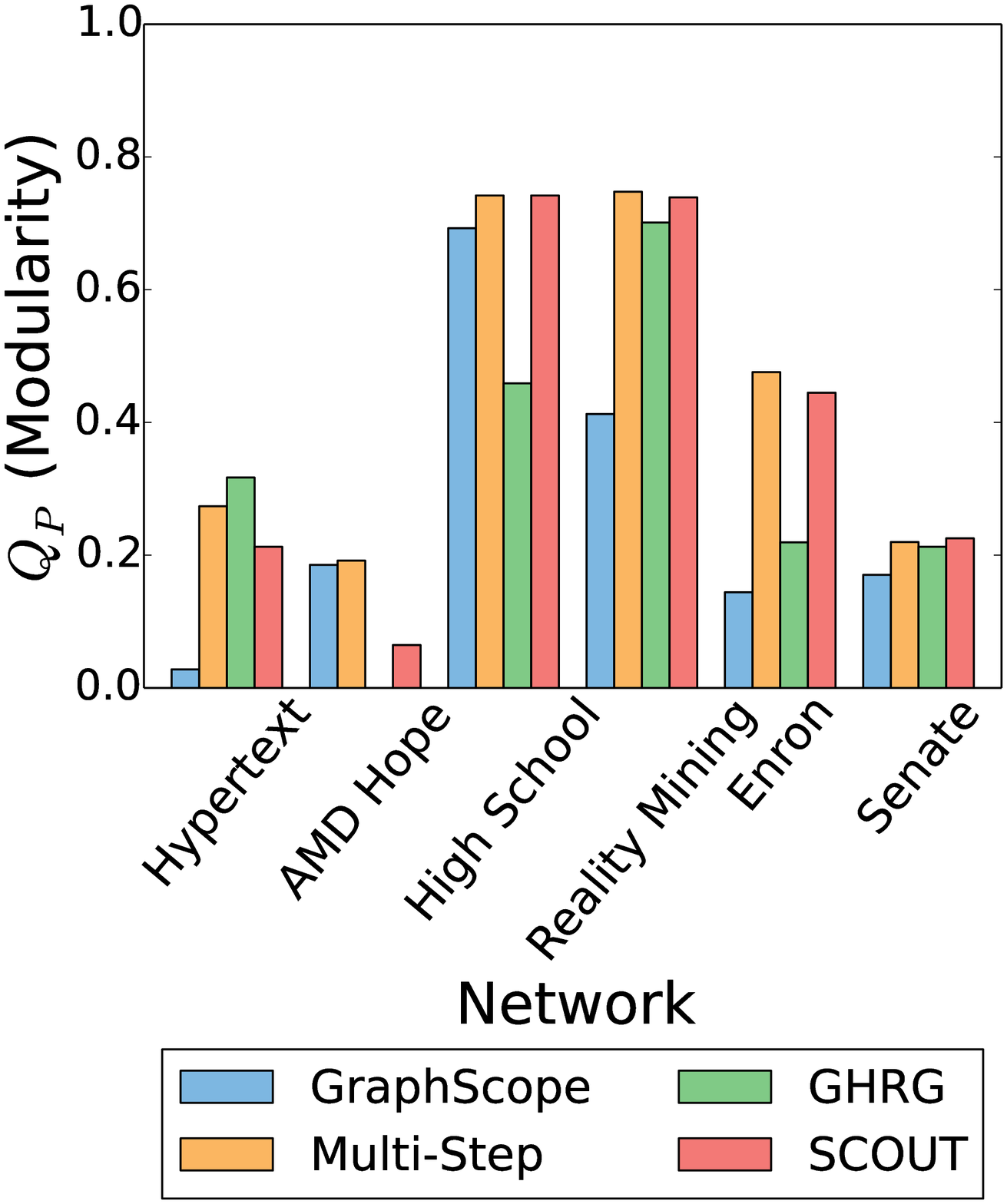}}
  \caption[Method comparison  for real-world networks]{Method comparison  for real-world networks with respect to
\textbf{\protect\subref{fig:real_world_aupr}} change point classification and
\textbf{\protect\subref{fig:real_world_avg_sn__mod}} $Q_P$.
In panel \protect\subref{fig:real_world_aupr}, only networks with known ground truth change points are shown, since otherwise change point classification cannot be computed.
GHRG could not be run on \emph{AMD Hope} network due to its high computational complexity.
}
\label{fig:real_world}
\end{figure*}

For change point classification, SCOUT is superior to all of the existing methods, since it achieves the highest accuracy for all considered real-world networks, and it is followed by GHRG, GraphScope, and Multi-Step, respectively (\figurename~\ref{fig:real_world_aupr}).
This method ranking is consistent with that for synthetic networks (Section~\ref{sec:results_real_world_t}).
Recall that we have formal lists of change points only for four of the six networks (Section~\ref{sec:datasets}), and thus the above change point classification is performed only on those four networks.
However, we can still intuitively (i.e., informally) discuss segmentation results of the methods for the remaining two networks, \emph{High School} and \emph{Senate}, as follows.

Regarding \emph{High School} network, recall that this network captures proximity of students in a high school (Section~\ref{sec:datasets}).
Intuitively, for this network, we do not expect large-scale changes in the students' interaction patterns over time
(meaning that we expect very few change points, if any, i.e., very few segments, possibly only one), since students typically interact with other students from the same classes \cite{mastrandrea2015contact}.
Consistent with this intuition, SCOUT (as well as GraphScope and Multi-Step) detects only one segment for \emph{High School} network (Supplementary \figurename~\ref{fig:supplement:real_world_sgmq__num}).
Moreover, SCOUT (as well as Multi-Step) produces the partition for this single segment that perfectly matches the (static) partition of students according to their classes \cite{mastrandrea2015contact}.
Hence, it is encouraging that SCOUT (as well as Multi-Step) captures the above intuition about the expected dynamics and structure of \emph{High School} network.

\begin{figure*}[!tbp]
  \centering\includegraphics[width=\linewidth]{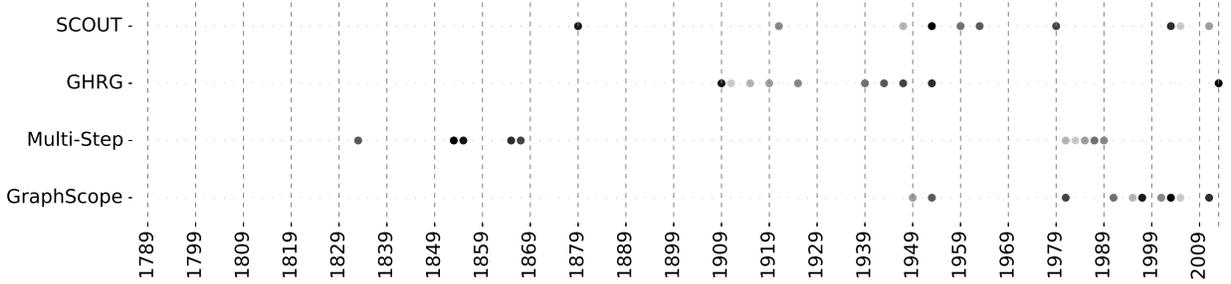}
  \caption[The top 10 highest ranked time points by each method for \emph{Senate} network]{The top 10 highest ranked time points by each method for \emph{Senate} network.
  Time points (expressed in years) are shown on the $x$-axis.
  Methods are shown on the $y$-axis.
  Each circle corresponds to one of the top 10 ranked time points, and the darker its color, the higher its rank.
  }\label{fig:senate_changepoints}
\end{figure*}

Regarding \emph{Senate} network, for a given method, we identify its 10 top-ranked ``change point''-like time points (Section~\ref{sec:eval_measures}).
Interestingly, the lists of top ranked time points produced by the different methods have little overlap (\figurename~\ref{fig:senate_changepoints}).
Specifically, given the four methods and 10 identified points per method, if all methods combined identified only 10 distinct time points, this would mean that the methods produced identical results. 
On the other hand, if all methods combined identified all $4 \times 10 = 40$ possible distinct time points, this would mean that the methods produced completely different results with no time point in the overlap of any two methods. 
In our case, the four methods combined identify 33 out of all 40 possible distinct time points (i.e., 82.5\% of them), which means that their results are quite complementary. 
This is further supported by the fact that there is only one time point that is identified by more than two methods (namely, the 83$^{rd}$ Congress in 1953, which is among the top 10 ranked time points of SCOUT, GraphScope, and GHRG).
We aim to empirically evaluate whether these top ranked points correspond to some important historical events. 
If so, this would further validate the given method. 
This evaluation needs to be performed qualitatively (rather than quantitatively, as has been done so far), since it is hard to determine the ranking of all historical events in terms of their importance and consequently to correlate this ranking with the methods' ranking of the time points.
Because of this, and because the resulting qualitative evaluation is time consuming, while we illustrate the top 10 ranked change points for each method (\figurename~\ref{fig:senate_changepoints}), we do not focus here on comparing the different methods. 
Instead, we focus on discussing SCOUT's results only, to at least intuitively assess the meaningfulness of its results.
SCOUT's top four time points (1953, 1879, 2003, and 1979, respectively) correspond to Congresses with shifts in the structure of the Senate's majority between the Democratic and Republican parties.
SCOUT's next three time points correspond to the 86$^{th}$, 88$^{th}$, and 67$^{th}$ Congress, respectively. 
The first two brought major civil rights acts (Civil Rights Act of 1960 and Civil Rights Act of 1964, respectively), and during the third one, ``Teapot Dome'' Scandal occurred, which is considered one of the most significant investigations in the history of the Senate.\footnote{\url{http://www.senate.gov/history/1921.htm}}
SCOUT's remaining three of the top 10 ranked time points correspond to divided Congresses:
the 112$^{th}$ Congress that almost lead to government shutdown,\footnote{\url{https://en.wikipedia.org/wiki/112th_United_States_Congress}} plus the $80^{th}$ Congress and the 109$^{th}$ Congress, both of which were nicknamed as ``do-nothing''.\footnote{\url{https://en.wikipedia.org/wiki/109th_United_States_Congress}}
Overall, it is encouraging that SCOUT identifies as likely change points those time points that correspond to important historical events.

\subsubsection{Partition aspect of the solution quality}\label{sec:results_real_world_p}
For $Q_P$, with the exception of \emph{Hypertext} and \emph{AMD Hope} networks, SCOUT and Multi-Step are comparable, and they outperform both GraphScope and GHRG (\figurename~\ref{fig:real_world_avg_sn__mod}); this is the same trend as for synthetic networks (Section~\ref{sec:results_synthetic_p}).
For \emph{Hypertext} network, SCOUT is outperformed by GHRG and Multi-Step, respectively (\figurename~\ref{fig:real_world_avg_sn__mod}).
For \emph{AMD Hope} network, SCOUT is outperformed by Multi-Step and GraphScope, respectively  (\figurename~\ref{fig:real_world_avg_sn__mod}).  
These results for \emph{Hypertext} and \emph{AMD Hope} networks are not necessarily surprising, for the following reason.
Different methods can produce solutions with different numbers of segments.
In particular, for these two networks, GHRG and GraphScope produce more segments than SCOUT and Multi-Step (Supplementary \figurename~\ref{fig:supplement:real_world_sgmq__num}).
Recall from Section~\ref{sec:problem} that the more segments exist in a solution, the easier it is for this solution to obtain a high partition quality score (i.e., $Q_P$).
Hence, a direct comparison of $Q_P$ scores of the solutions with different numbers of segments may not necessarily provide a realistic view of the methods' performance.
As an illustration, consider comparing some two methods: if method 1 has a slightly higher $Q_P$ score than method 2, but it also achieves this score with ten times as many segments as method 2, does it mean that method 1 has a better partition accuracy than method 2?
Probably not.
Thus, ideally, we would compare $Q_P$ scores of the solutions with equal numbers of segments.

\begin{figure}[!tbp]
  \centering
\subfloat[\emph{High School} ]{\label{fig:high_school_avg_sn__mod_cscd}\includegraphics[width=0.4\linewidth]{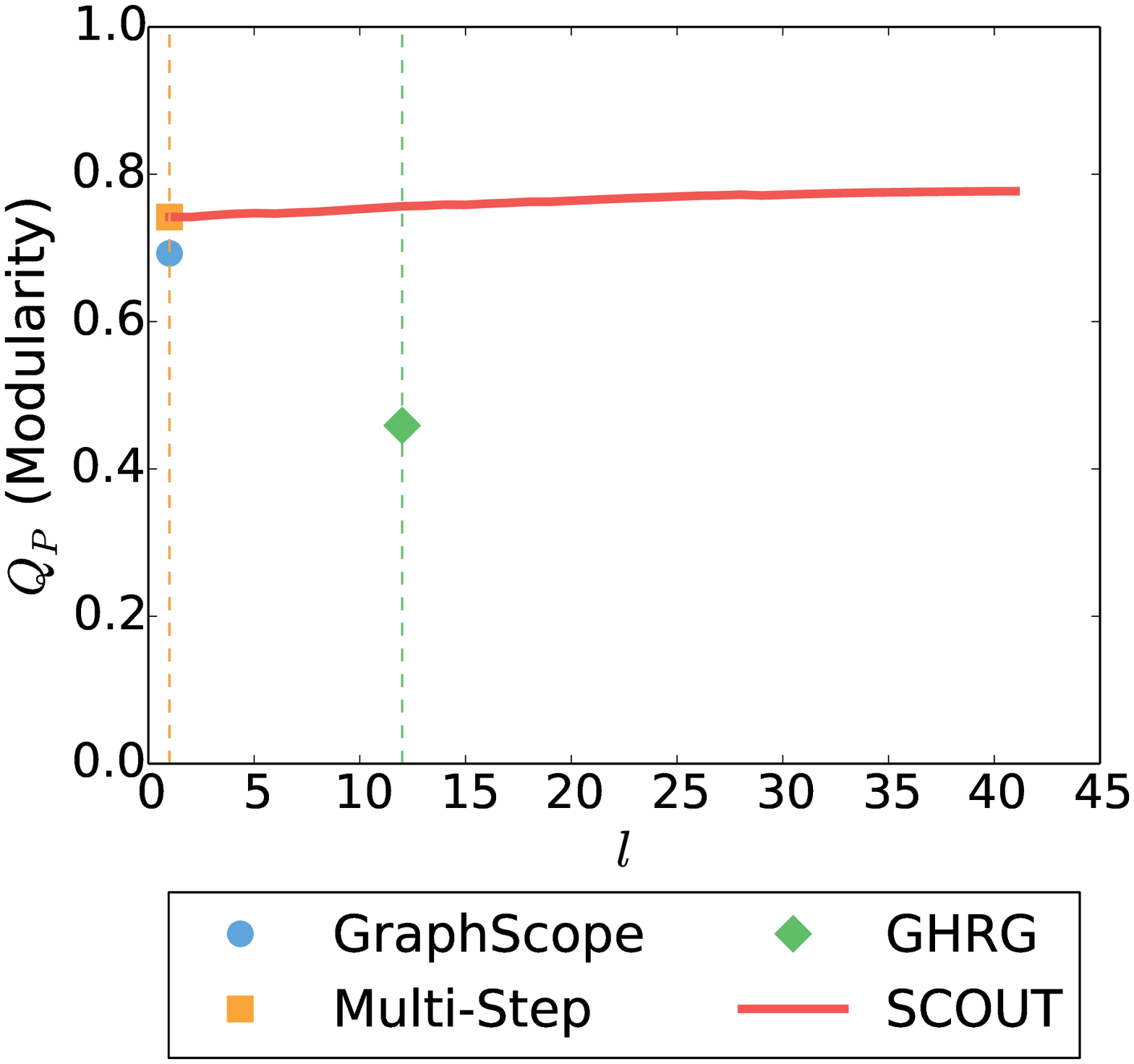}}
\hspace{0.2cm}
\subfloat[\emph{Senate}]{\label{fig:senate_avg_sn__mod_cscd}\includegraphics[width=0.4\linewidth]{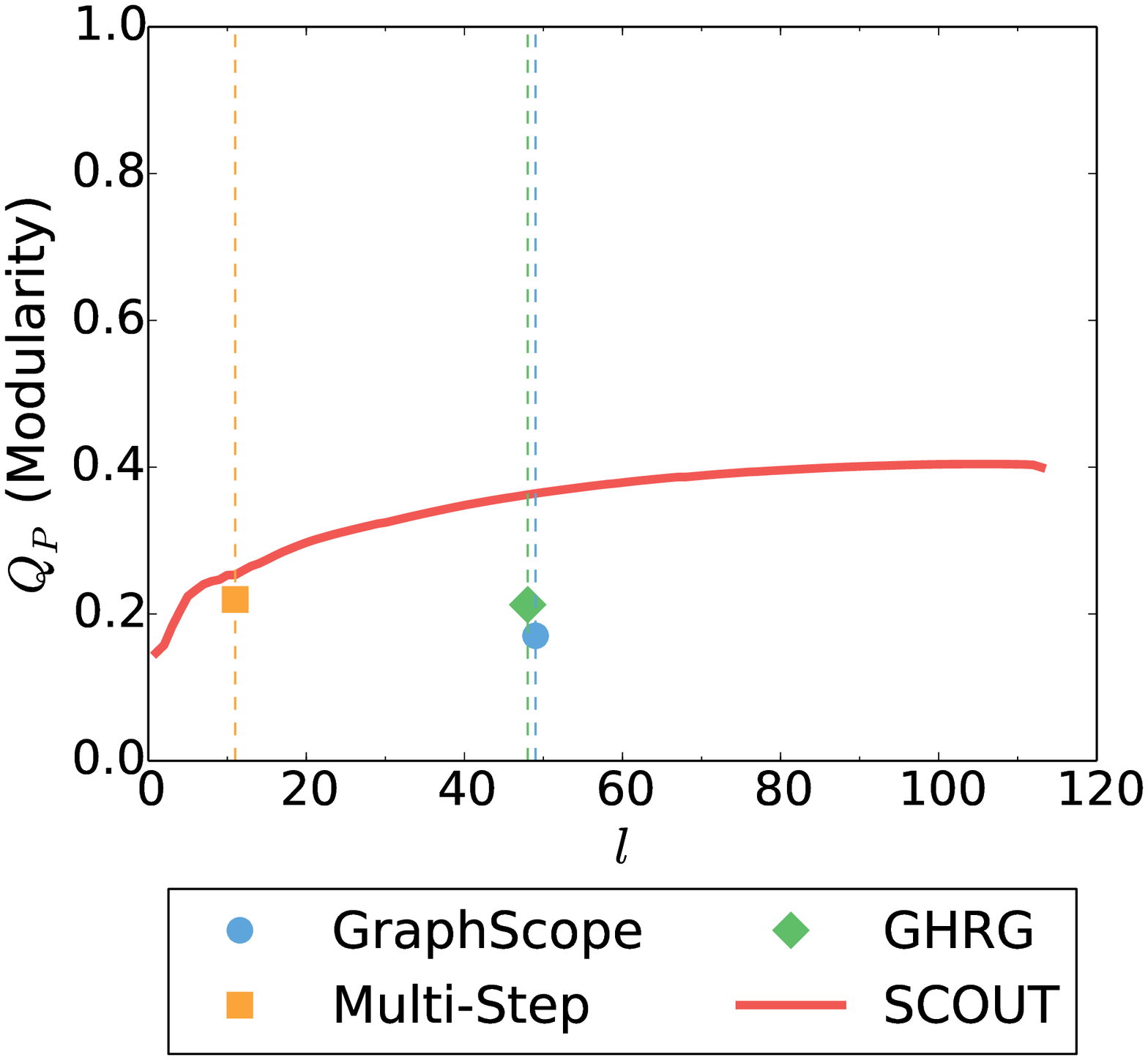}}
  \caption[Representative $Q_P$ scores of SCOUT's solutions for different numbers of segments $l$ and the solutions of the existing methods for \emph{High School} network and \emph{Senate} network]{Representative $Q_P$ scores of  1) SCOUT's solutions for different numbers of segments $l$ and 2) the solutions of the existing methods, for
\textbf{\protect\subref{fig:high_school_avg_sn__mod_cscd}} \emph{High School} network
and  \textbf{\protect\subref{fig:senate_avg_sn__mod_cscd}} \emph{Senate} network.
For SCOUT, the line shows its $Q_P$ score when solving the CSCD problem while varying the number of segments. 
For each of the existing methods, the mark shows $Q_P$ score of the given method's solution, with the position of the mark along the $x$-axis corresponding to the number of segments $l$ in the solution.
Equivalent results for the remaining real-world networks are shown in Supplementary \figurename~\ref{fig:supplement:real_world_avg_sn__mod_cscd}.
  }\label{fig:real_world_avg_sn__mod_cscd}
\end{figure}

For this reason, since SCOUT is capable of producing a solution with not only an automatically determined but also user-provided number of segments (i.e., since it can solve both SCD and CSCD problems; Section~\ref{sec:problem}), we compare $Q_P$ score of each existing method and $Q_P$ score of SCOUT when solving the $Q_P$-based CSCD problem and producing a solution with the same number of segments as the solution of the given existing method.
In this way, we avoid the bias arising from the fact that the two compared methods might have different numbers of segments.
According to this evaluation, SCOUT outperforms all methods (\figurename~\ref{fig:real_world_avg_sn__mod_cscd} and Supplementary \figurename~\ref{fig:supplement:real_world_avg_sn__mod_cscd}).

The shape of the $Q_P$-curve as a function of the number of segments $l$ could provide insights into the dynamics of the network in question.
Even though the $x$-axis of the curve does not correspond to time, and thus it cannot tell us when changes in community organization (if any) occur, the fact that the $x$-axis corresponds to $l$ can intuitively tell us something about the number of such changes and their scale.
Namely, on the one hand, if $Q_P$ increases slowly (or does not increase at all) as $l$ increases, this could mean that the community organization of the network does not change a lot with increase in the number of segments, and thus, the increase in the number of segments in unnecessary.
For example, this is the case for \emph{High school} network (\figurename~\ref{fig:high_school_avg_sn__mod_cscd}), which agrees with our discussion in Section~\ref{sec:results_real_world_t}.
On the other hand, if $Q_P$ increases drastically as $l$ increases,  this could mean that the community organization of the network indeed changes a lot with increase in the number of segments, and thus, the increase in the number of segments is justified. 
For example, this is the case for \emph{Senate} network (\figurename~\ref{fig:senate_avg_sn__mod_cscd}), which agrees with our discussion in Section~\ref{sec:results_real_world_t}.

\begin{figure}[!tbp]
  \centering\hspace{-0.9cm}\includegraphics[width=0.4\linewidth]{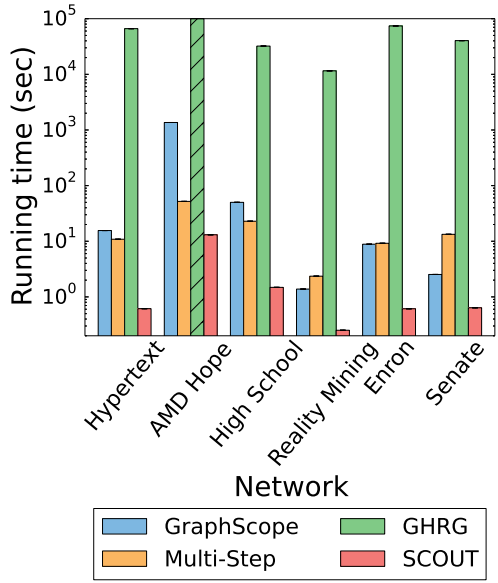}
  \caption[Running times of the methods for real-world networks]{Running times of the methods for real-world networks (logarithmic scale).
The striped bars mean that the method could not finish within the allowed time. 
  }\label{fig:real_world_run_time}
\end{figure}

\subsubsection{Running time}\label{sec:results_real_world_run_time}
Just as for synthetic networks, SCOUT has the lowest running time of all methods, over all real-world networks  (\figurename~\ref{fig:real_world_run_time}).
Again, GHRG is the slowest among all considered methods, which means that it cannot be run for the larger networks due to its high computational complexity.

\section{Conclusions}
We study the problem of community detection in dynamic networks.
To capture the intuition of a compromise between the two extremes of snapshot clustering and consensus clustering, we combine community detection with the problem of segment detection to formulate a new problem of SCD. 
To address the drawbacks of the existing methods that can be employed to solve the SCD problem, we introduce SCOUT. 
To comprehensively evaluate SCOUT against the existing methods, we introduce a synthetic network generator that produces a dynamic network with the known ground truth segments and their community organization, where by varying the model parameters, different synthetic dynamic network configurations can be obtained.
To quantify the performance of a given method, we introduce new measures of SCD quality.
We perform our experiments on a variety of synthetic as well as real-world networks.
We demonstrate that SCOUT outperforms the existing methods with respect to both segmentation aspect and partition aspect of the SCD problem.
At the same time, SCOUT is more computationally efficient than the existing methods.
Ultimately, we show that the SCD problem and SCOUT in particular is a useful framework for studying community organization of dynamic networks, as it can identify both when communities evolve by identifying change points and how communities look like at each stage of their evolution by identifying segment partitions.
The solution of the SCD problem  provides a concise yet informative description of the dynamic network from the perspective of its community organization.

Our work has several potential future directions.
From the methodological perspective, SCOUT could be extended to different problem settings, such as dealing with weighted networks or overlapping communities.
From the application perspective, an important problem in dynamic network analysis is to choose a meaningful time scale for defining network snapshots.
Usually, the time scale is chosen so that each snapshot is assumed to have the same duration (e.g., one week), and the duration is determined empirically to fit the context of the given application. 
Instead, the output of the SCD problem could provide a systematic way for defining snapshots.
Namely, the smallest meaningful traditional empirical equal-length snapshots would be used define the initial dynamic network. 
Then, this network would be given as input to SCOUT to group the small snapshots with consistent community organization into larger segments. 
Finally, the time interval of each segment would correspond to a new, more meaningful snapshot, and collection of all such new snapshots would form a new, more meaningful dynamic network.
In this way, each snapshot of the new network would capture the period during which community organization is consistent.
Moreover, the duration of different snapshots could be different.
These newly constructed snapshots (i.e., the new dynamic network) could then be used as input to various methods for dynamic network analysis,
which could improve the quality of results compared to using the same methods on the traditionally determined empirical same-length snapshots (i.e., on the initial network that was given as input to SCOUT).

\section*{Acknowledgments}
This work was supported by the National Science Foundation (CAREER CCF-1452795 and CCF-1319469).

\clearpage
\appendix
\graphicspath{{./supplement/images/}}

\floatname{algorithm}{Supplementary Algorithm}
\renewcommand{\figurename}{Supplementary Figure}
\renewcommand{\tablename}{Supplementary Table}
\setcounter{table}{0}
\renewcommand{\thetable}{S\arabic{table}}%
\setcounter{figure}{0}
\renewcommand{\thefigure}{S\arabic{figure}}%
\setcounter{algorithm}{0}
\renewcommand{\thealgorithm}{S\arabic{algorithm}}%
\renewcommand{\thesection}{S\arabic{section}}%

\part*{Supplementary information}
\addcontentsline{toc}{part}{Supplementary information} 
\section{Related work}\label{sec:supplement:related_work}
Here, we expand our discussion from Section~\ref{sec:related_work} in the main paper and discuss the three existing methods, GraphScope, Multi-Step, and GHRG, which can deal with the SCD problem.

\textbf{GraphScope} \cite{sun2007graphscope, ferlez2008monitoring, duan2009community} works as follows.
The first snapshot becomes the current segment.
Given the current segment, the method iteratively examines the next snapshot in the temporal sequence to determine whether: 1) the community organization of the snapshot in question matches well the community organization of the current segment, and thus, the snapshot should be added to the current segment (this simply extends the current segment for the next iteration),
or instead 2) the community organization of the snapshot does not match well the community organization of the current segment, and thus, the snapshot should begin a new segment (which becomes the current segment for the next iteration).
Community organizations of the given snapshot and the current segment are obtained and their match is measured via the minimum description length (MDL) principle.

\textbf{Multi-Step} \cite{aynaud2011multi} uses an agglomerative hierarchical clustering approach as follows. 
Each snapshot starts as a singleton segment.
Then, in every iteration, the most similar (in terms of community organization) pair of segments are combined.
Specifically, the level of similarity between two segments quantifies how well the community organization (i.e., the partition) of the first segment fits the second segment, and also how well the partition of the second segment fits the first segment.
Here, the quality of the fit of a partition to a segment is based on average modularity (Supplementary Section~\ref{sec:supplement:scout_obj_func}), and a partition for the given segment is detected by greedily maximizing average modularity via a modification of Louvain algorithm for static community detection (Supplementary Section~\ref{sec:supplement:scout_cons_clust}). 
The output of the above iterative Multi-Step procedure is a hierarchical tree with snapshots as leaves. However, it is not clear how to automatically cut the tree to obtain segments and their corresponding change points. 
As such, Multi-Step is suitable when the desired number of segments is provided as input.

\textbf{GHRG} \cite{peel2015detecting} considers a fixed-length sliding window of the most recent snapshots and uses a statistical test to evaluate whether: 1) within the window, the snapshots before and after a given time point originate from different community organization-related models, and thus, this time point should be declared as a change point, or instead 2) all snapshots within the window come from the same model, and thus, there is no change point in that window.
As its community organization-related model, GHRG uses generalized hierarchical random graphs.

\section{Methods}
\subsection{Our SCOUT approach}\label{sec:supplement:scout}
Here, we expand our discussion from Section~\ref{sec:scout} in the main paper
on the three main components of SCOUT: objective function (Supplementary Section~\ref{sec:supplement:scout_obj_func}), consensus clustering (Supplementary Section~\ref{sec:supplement:scout_cons_clust}), and search strategy (Supplementary Section~\ref{sec:supplement:scout_search_strategy}).

\subsubsection{Objective function}\label{sec:supplement:scout_obj_func}
For the CSCD problem, in which segmentation parsimony is fixed, an objective function $Q$ should measure partition accuracy of an output $O$. 
For the SCD problem, an objective function $Q$ should measure both segmentation parsimony and partition accuracy.
We organize the rest of this section as follows.
\RNum{1}) We discuss the group of objective functions $Q_P$ that measure only partition accuracy.
\RNum{2}) We discuss the group of objective functions $Q_B$ that measure both aspects of the output quality.
\RNum{3}) We discuss how to use the above two groups of objective functions to solve the CSCD and SCD problems.

\vspace{0.1cm}\noindent\textbf{\RNum{1})} To measure only partition accuracy, we define $Q_P$ as the \emph{average snapshot partition quality}:
\begin{equation} \label{eq:supplement:q_p}
Q_P(O, D) = \frac{1}{k} \sum_{i=0}^{l-1} \sum_{j=t_i}^{t_{i+1}-1} F(p_i,G_j),
\end{equation}
where $F$ measures the fit of partition $p_i$ to snapshot $G_j$.
Since there is no one universally accepted measure $F$ of how well a given partition $p$ fits a given snapshot $G=(V,E)$,
we test four popular such measures $F$ \cite{yang2015defining}.
Let $|p|$ be the number of clusters in partition $p$.
For a given cluster $c \in p$, let $n_c$ be the number of its nodes, let $m_c$ be the number of its internal edges, and let $b_c$ be the number of its boundary edges (edges between the nodes in $c$ and the nodes in $V \setminus c$).
We consider the following choices of $F$:
\textbf{1)} \emph{Modularity} \cite{newman2004finding}:
$F(p,G)=\frac{1}{2|E|} \sum_{c \in p} (m_c - \mathbb{E}(m_c))$, where $\mathbb{E}(m_c)$ is the expected number of $c$'s internal edges under a configuration model (a random model with the same degree distribution as $G$).
Intuitively, a partition is of high quality with respect to modularity if its clusters are denser than at random.
The higher the modularity score, the better the partition accuracy.
The remaining three measures are based on the intuition that in a good partition, clusters should have more inside than boundary edges.
\textbf{2)} \emph{Conductance} \cite{cheng2010uncovering}:
$F(p,G)=\frac{1}{|p|} \sum_{c \in p} \frac{b_c}{2m_c + n_c}$.
\textbf{3)} \emph{Normalized Cut} \cite{shi2000normalized}:
$F(p,G)=\frac{1}{|p|} \sum_{c \in p} (\frac{b_c}{2m_c + n_c} + \frac{b_c}{2(m-m_c) + n_c})$.
\textbf{4)} \emph{Average-ODF} \cite{flake2000efficient}:
$F(p,G)=\frac{1}{|p|} \sum_{c \in p} \frac{1}{|n_c|} \sum_{u \in c} \frac{|\lbrace |(u,v) \in E| v \not\in c \rbrace|}{d_u}$, where $d_u$ is the degree of node $u$.
Because for the last three measures, the lower the score, the better the partition accuracy, and because SCOUT aims to \emph{maximize} (rather than minimize) its objective function, SCOUT uses $F^\prime(p,G)=1-F(p,G)$ instead of $F$ in its objective function for these three measures.

\vspace{0.1cm}\noindent\textbf{\RNum{2})} To simultaneously measure both segmentation parsimony and partition accuracy, we define $Q_B$ based on the \emph{model selection problem} \cite{kadane2004methods}.
Intuitively, given some $O=(T,P)$ for a dynamic network $D$, if we use $O$ as a generative model for creating a dynamic network, how well does this model $O$ fit $D$? 
On the one hand, the more complex the model (intuitively, the more segments there are in $O$, i.e., the lower the segmentation parsimony, and also, the more clusters there are in each segment partition), the more likely it is that we will observe a high fit (as measured by the likelihood of $D$ given $O$, which mostly reflects partition accuracy).
On the other hand, the less complex the model, the more likely it is that we will observe a low fit.
Given a set of $O$s under consideration (see below), the goal of the model selection problem is to choose $O^\star$ that optimizes some measure of quality over all such $O$s. 
This measure of quality should balance between the \emph{goodness of the fit} of $O$ to $D$ (mostly partition accuracy) and the \emph{complexity of the model} $O$ (mostly segmentation parsimony).

To solve the model selection problem, we test two popular approaches \cite{kadane2004methods,grunwald2000model}:
\textbf{1)} \emph{Akaike Information Criterion} (\emph{AIC}) \cite{akaike1998information} and \textbf{2)} \emph{Bayesian Information Criterion} (\emph{BIC}) \cite{schwarz1978estimating}.
Both approaches compute the goodness of the fit in the same way.
They also compute the complexity of the model in the same way.
However, the two approaches differ in how they penalize the objective function by the complexity of the model.
We define $Q_B$ using AIC or BIC as follows:
\begin{equation} \label{eq:supplement:q_ic}
Q_B(O, D)=\ell(D|O) - w(D)N_{p}(O).
\end{equation}
In the above formula, the goodness of the fit is measured via $\ell(D|O)$, the log-likelihood of $D$ given $O$ (see below).
The complexity of the model is measured via $N_{p}(O)$, the \emph{number of parameters} in $O$ (see below).
The above two quantities, $\ell(D|O)$ and $N_{p}(O)$, are balanced via \emph{penalty weight} $w(D)$.
For AIC, $w(D) = 1$.
For BIC, $w(D) = \frac{1}{2}\log N_{o}(D)$, where $N_{o}(D)$ is the \emph{number of observations} in $D$ (how ``large'' $D$ is; see below).
In general, $w(D)$ is larger in BIC than in AIC, which means that BIC penalizes complex models more heavily than AIC.
Intuitively, in our case, this means that BIC prefers outputs with smaller numbers of segments than AIC.

Next, we discuss how to compute $\ell(D|O)$, $N_{p}(O)$, and $N_{o}(D)$.

\textbf{\emph{To compute $\boldsymbol{\ell(D|O)}$}}, we assume that each segment $s_i$ is independent of the others, and thus $\ell(D|O)$ is just the sum of log-likelihoods of the individual segments $\ell(s_i|p_i)$:
\begin{equation} \label{eq:supplement:ll_d_seg}
\ell(D|O) = \sum_{i=0}^{l-1} \ell(s_i|p_i),
\end{equation}
where segmentation $S=\{s_0, s_1, \dots, s_l \}$ is determined by change point set $T$ of $O$.
To compute $\ell(s_i|p_i)$, we assume that $s_i$ has an associated \emph{stochastic blockmodel} (see below) and each snapshot $G_j$ within segment $s_i$ is independent given this blockmodel.
A stochastic blockmodel is a generative model where probability of an edge is determined by the cluster memberships of its endpoints \cite{karrer2011stochastic}.
The blockmodel contains two parts: a partition $p$ and a \emph{stochastic block matrix} $\theta$ of size $|p| \times |p|$, where $\theta_{c_u,c_v}$ is the probability of an edge between two nodes $u,v$ from clusters $c_u, c_v \in p$, respectively.
The blockmodel associated with $s_i$ is based on the corresponding segment partition $p_i$ and has the stochastic block matrix $\theta^{(i)}$ (see below).
Thus, $\ell(s_i|p_i)$ is just the sum of log-likelihoods of the individual snapshots $\ell(G_j|\hat{\theta}^{(i)},p_i)$:
\begin{equation} \label{eq:supplement:ll_seg_sn}
\ell(s_i|O) = \sum_{j=t_i}^{t_{i+1}-1} \ell(G_j|\hat{\theta}^{(i)}, p_i),
\end{equation}
where $\hat{\theta}^{(i)}$ is the maximum likelihood estimator of $\theta_{c_u,c_v}^{(i)}$.
That is, $\hat{\theta}^{(i)}_{c_u,c_v}$ is computed as the fraction of the actual and the maximum possible numbers of edges between nodes in cluster $c_u$ and nodes in cluster $c_v$ across all snapshots $G_j$ of segment $s_i$:
\begin{equation} \label{eq:supplement:sbm_thetas}
\hat{\theta}^{(i)}_{c_u,c_v} = \frac{\sum_{j=t_i}^{t_{i+1}-1} m^{(j)}_{c_uc_v}}{\sum_{j=t_i}^{t_{i+1}-1} n^{(j)}_{c_uc_v}}.
\end{equation}
In the above formula, $m^{(j)}_{c_uc_v}$ is the number of edges in $G_j$ between nodes in cluster $c_u$ and nodes in cluster $c_v$, and $n^{(j)}_{c_uc_v}$ is the maximum possible number of such edges.
If $c_u \neq c_v$, then $n^{(j)}_{c_uc_v}=n^{(j)}_{c_u}n^{(j)}_{c_v}$, where $n^{(j)}_{c_u}$ and $n^{(j)}_{c_v}$ are the numbers of nodes from $G_j$ that are in clusters $c_u$ and $c_v$, respectively.
If $c_u=c_v$, then $n^{(j)}_{c_uc_v}=\binom{n^{(j)}_{c_u}}{2}$.
To compute $\ell(G_j|\hat{\theta}^{(i)}, p_i)$, the log-likelihood of $G_j=(V_j,E_j)$ given $\hat{\theta}^{(i)}$ and $p_i$, because we are using a stochastic blockmodel,
we assume that an edge between each pair of nodes $u,v \in V_j$ is independent of others and its probability is based on the cluster memberships $c_u,c_v \in p_i$ of $u, v$, respectively.
Thus, $\ell(G_j|\hat{\theta}^{(i)}, p_i)$ is just the sum of log-likelihoods of individual edges and non-edges observed in $G_j$:
\begin{equation} \label{eq:supplement:ll_sn}
\ell(G_j|\hat{\theta}^{(i)}, p_i) = \sum_{(u,v) \in E_j} \log{\hat{\theta}^{(i)}_{c_u,c_v}} 
 + \sum_{(u,v) \not\in E_j} \log(1-\hat{\theta}^{(i)}_{c_u,c_v}).
\end{equation}
By combining Equations \ref{eq:supplement:ll_seg_sn} -- \ref{eq:supplement:ll_sn}, we can compute $\ell(D|O)$ in Equation \ref{eq:supplement:ll_d_seg}.

\textbf{\emph{To compute $\boldsymbol{N_{p}(O)}$}}, we count the number of values in $\hat{\theta}^{(i)}$s across all segments $s_i, i \in [0,l-1]$.
For a given segment $s_i$, we have one value in $\hat{\theta}^{(i)}$ for each pair of clusters in $p_i$ (including a cluster with itself), so, in total:
\begin{equation} \label{eq:supplement:n_p}
N_{p}(O)=\sum_{i=0}^{l-1} (\binom{|p_i|}{2} + |p_i|)=\sum_{i=0}^{l-1} \frac{|p_i|(|p_i|+1)}{2}.
\end{equation}

\textbf{\emph{To compute $\boldsymbol{N_{o}(D)}$}}, we count the number of node pairs in all snapshots $G_j=(V_j,E_j)$ in $D$ (Equation~\ref{eq:supplement:ll_sn}):
\begin{equation} \label{eq:supplement:n_o}
N_{o}(D)=\sum_{j=0}^{k-1} \binom{|V_j|}{2}=\sum_{j=0}^{k-1} \frac{|V_j|(|V_j|-1)}{2}.
\end{equation}

By combining Equations \ref{eq:supplement:ll_d_seg}, \ref{eq:supplement:n_p}, and, for BIC, \ref{eq:supplement:n_o}, we can compute $Q_B(O, D)$ in Equation \ref{eq:supplement:q_ic}.

\vspace{0.1cm}\noindent\textbf{\RNum{3})} Given some consensus clustering method and search strategy (see below), and given the above two groups of objective functions, $Q_P$ and $Q_B$, we now discuss how to solve the CSCD and SCD problems.

\textbf{\emph{To solve the CSCD problem}}, we pick as $O^\star$ a solution with the desired number of segments $l$ that maximizes $Q \in \{Q_P, Q_B\}$:
\begin{equation}\label{eq:supplement:o_cscd}
O^\star = \underset{|T|=l-1, O \in R}{\mathrm{argmax}} Q(O,D),
\end{equation}
where $T$ is the change point set of $O$ (recall that we need $l-1$ change points to produce $l$ segments)
and $R$ is the set of the considered outputs (note that this set is determined by the search strategy; see below).
Here, $Q$ can measure either only partition accuracy (i.e., $Q_P$) or both aspects of the SCD problem (i.e., $Q_B$).

\textbf{\emph{To solve the SCD problem}}, we first solve the CSCD problem $\forall l \in [1,k]$ using $Q_P$ or $Q_B$ as described above, and then we pick as $O^\star$ one of these $k$ solutions that maximizes $Q_B$.
Let $R^\star = \{O^\star_{(i)}|i \in [1,k]\}$, where $O^\star_{(i)}$ is the solution of the CSCD problem with $i$ segments (Equation~\ref{eq:supplement:o_cscd}).
Given $R^\star$, we select $O^\star$ as follows:
\begin{equation} \label{eq:supplement:o_scd}
O^\star = \underset{O \in R^\star}{\mathrm{argmax}} \; Q_B(O,D).
\end{equation}
Note that if we use the same $Q_B$ when constructing $R^\star$ (Equation~\ref{eq:supplement:o_cscd}) and when selecting  $O^\star$ from $R^\star$ (Equation~\ref{eq:supplement:o_scd}), the described procedure for solving the SCD problem is equivalent to directly aiming to find $O^\star$ with the optimal value of $Q_B$.

\subsubsection{Consensus clustering}\label{sec:supplement:scout_cons_clust}
Given change point set $T$, we obtain the set of segment partitions $P$ by applying consensus clustering to each segment.
That is, for each segment $s_i$, we aim to find a single partition $p_i$ that works well for all snapshots in $s_i$.
Note that if $s_i$ contains only one snapshot, consensus clustering is equivalent to simple static network clustering, since there are no multiple snapshots to compute consensus for;
yet, for consistency, we still refer to such clustering process as consensus clustering.
Intuitively, the chosen consensus clustering method should align with the objective function, 
meaning that, for a given change point set $T$, consensus clustering should aim to find the set $P$ of segment partitions that maximize the objective function $Q$.
We consider three consensus clustering methods: \emph{sum graph} \cite{aynaud2011multi}, \emph{Average-Louvain} \cite{aynaud2011multi}, and \emph{consensus matrix} \cite{lancichinetti2012consensus}.

\vspace{0.1cm}\noindent\textbf{Sum graph.}
An intuitive way to perform consensus clustering for a given segment $s_i$ is to first construct a special graph that ``summarizes'' the topology of all snapshots in $s_i$ and then find community organization in this ``summary'' graph under the hypothesis that this organization will fit well all snapshots in $s_i$.
Here, we construct this ``summary'' graph for $s_i$ simply as a \emph{sum graph}, a weighted graph whose adjacency matrix is the sum of the adjacency matrices of all snapshots in $s_i$ \cite{aynaud2011multi}.
Then, we use a static community detection method that can handle weighted graphs to find a partition in this sum graph.
We test seven popular static community detection methods \cite{fortunato2010community}:
\textbf{1)} \emph{Fast Modularity} \cite{clauset2004finding}: the method starts with each node as a singleton community, and then at every iteration it merges two communities to greadily optimize modularity.
\textbf{2)} \emph{Label Propagation} \cite{raghavan2007near}: the method starts with each node as a singleton community (referred to as a \emph{label}), and then at every iteration each node adopts the label used by the majority of its neighbors.
\textbf{3)} \emph{Leading Eigenvector} \cite{newman2006finding}: the method optimizes modularity based on the eigenspectrum of a modularity matrix (a matrix analogous to graph Laplacian in graph partitioning).
\textbf{4)} \emph{Infomap} \cite{rosvall2010map}: the method aims to find a partition minimizing the expected description length of a random walker trajectory.
\textbf{5)} \emph{Walktrap} \cite{pons2005computing}: the method finds a partition based on the intuition that short random walks tend to get ``trapped''  in the same community, since, intuitively, there are many edges pointing inside the community and only few pointing outside.
\textbf{6)} \emph{Louvain} \cite{blondel2008fast}: the method starts with each node as a singleton community and then repeatedly performs two phases: greedily optimizing modularity by moving nodes to neighboring communities and constructing a new graph with communities as nodes.
\textbf{7)} \emph{Stabilized Louvain} \cite{aynaud2010static}: a modification of Louvain algorithm for snapshot clustering that aims to produce stable partitions (i.e., prevent two snapshots with similar topologies from having dissimilar partitions); to achieve stability, the method clusters a snapshot at time $t$ via Louvain algorithm initialized with the partition obtained for the snapshot at time $t-1$.

\vspace{0.1cm}\noindent\textbf{Average-Louvain.}
This method aims to find a segment partition $p_i$ that maximizes average modularity over all snapshots in $s_i$  \cite{aynaud2011multi}.
To achieve this, the method uses a modification of Louvain algorithm for static community detection (see above).
Recall that Louvain method contains two phases.
In Average-Louvain, the first phase is modified so that the modularity gain of each move is computed as the average gain of this move across all snapshots in the given segment.
The second phase, constructing a network of communities, is modified so that the same transformation is performed independently on all snapshots within the given segment. 
Thus, all snapshots have the same partition, which becomes $p_i$.

\vspace{0.1cm}\noindent\textbf{Consensus matrix.}
This method aims to find a segment partition $p_i$ directly from the partitions of snapshots in $s_i$ \cite{lancichinetti2012consensus}. 
That is, given individual snapshot partitions as input, the method computes a \emph{consensus matrix} $M$ based on the co-occurrence of nodes in clusters of the input partitions.
Specifically, entry $M_{ij}$ of this matrix indicates the fraction of the input partitions in which nodes $i$ and $j$ are in the same cluster.
Matrix $M$, which can be thought of as a weighted graph, can then be clustered by some static community detection method to produce a consensus partition.
To compute snapshot partitions as well as to cluster $M$, we use the same static community detection methods as for the sum graph approach above.

\subsubsection{Search strategy}\label{sec:supplement:scout_search_strategy}
We test three strategies for exploring the space of possible change point sets: the \emph{exhaustive search}, \emph{top-down search}, and \emph{bottom-up search}.
Each strategy first produces one best solution for each possible number of segments for the CSCD problem (Equation~\ref{eq:supplement:o_cscd}), which are then used to solve the SCD problem (Equation~\ref{eq:supplement:o_scd}).
The first strategy is aimed at producing a globally optimal solution at the expense of larger running time, while the last two are heuristics aimed at producing a good solution in a faster manner.
Below, for each strategy, we discuss how the strategy works and its ``conceptual'' computational complexity. By ``conceptual'', we mean that we express the running time of a given strategy in terms of the number of times that consensus clustering is performed.
We do this because: 1) performing consensus clustering is SCOUT's most computationally intensive step whose running time dominates all other steps, and 2) we vary consensus clustering methods within SCOUT, and thus, we account only for the number of times that consensus clustering is performed, since the actual computational complexity of performing each consensus clustering depends on the chosen clustering method.

\vspace{0.1cm}\noindent\textbf{Exhaustive search.}
This strategy aims to find a globally optimal solution under the chosen consensus clustering method by exhaustively searching through the space of all possible $T$s.
There are $\binom{k-1}{l-1}$ ways to group all $k$ snapshots of $D$ into $l \in [1,k]$ segments.
Thus, for all $l$s, the exhaustive search needs to explore the total of $\sum_{i=1}^{k} \binom{k-1}{i-1}=2^{k-1}$ different segmentations (or, equivalently, change point sets).

\begin{figure}[h]
  \centering\includegraphics[width=0.6\linewidth]{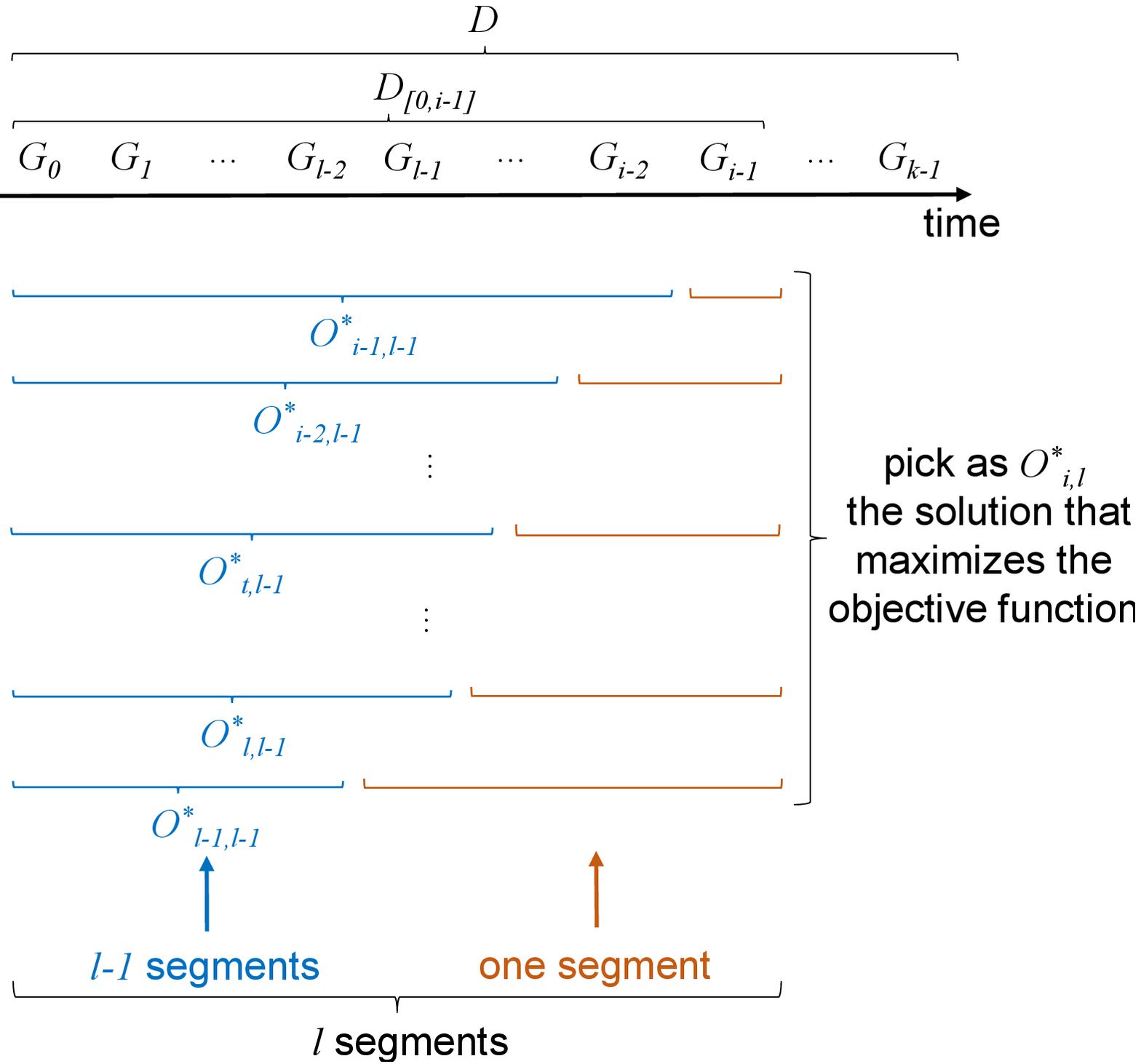}
  \caption[The procedure of obtaining solution $O^\star_{i,l}$ for $D_{[0,i-1]}$ that has $l$ segments during the $i^{th}$ iteration of the exhaustive search]{
The procedure of obtaining solution $O^\star_{i,l}$ for $D_{[0,i-1]}$ that has $l$ segments during the $i^{th}$ iteration of the exhaustive search.
For this, we construct a set of candidate solutions (shown in rows) and pick as $O^\star_{i,l}$ the one solution from this set that maximizes the objective function.
Each candidate solution is obtained by combining:
1) solution $O^\star_{t,l-1}$ that has $l-1$ segments, which was obtained in one of the previous iterations (shown in blue), and
2) a solution with one segment, whose only segment partition is obtained during the current iteration (shown in orange).
  }\label{fig:supplement:exh_search}
\end{figure}

To reduce the computational complexity, we use dynamic programming, as follows.
The search contains $k$ iterations.
Consider the $i^{th}$ iteration ($i \in [1,k]$).
Let $D_{[q,r]}=\lbrace G_q, G_{q+1}, \dots, G_{r-1}, G_{r} \rbrace$ be all consecutive snapshots of $D$ from time $q$ to time $r$, inclusively.
The goal of the $i^{th}$ iteration is to solve the SCD problem for $D_{[0,i-1]}$ (i.e., for the first $i$ snapshots of $D$).
For $i=k$, this means obtaining the solution for the whole network $D=D_{[0,k-1]}$.
Recall from Supplementary  Section~\ref{sec:supplement:scout_obj_func} that in order to solve the SCD problem, we first need to solve the CSCD problem for each possible number of segments (Equation~\ref{eq:supplement:o_scd}).
That is, in the $i^{th}$ iteration, $\forall l \in [1,i]$, we need to find the optimal solution $O^\star_{i,l}$ for $D_{[0,i-1]}$ that has $l$ segments.
Next, we discuss how to find such $O^\star_{i,l}$. 
Any solution $O$ for $D_{[0,i-1]}$ that has $l$ segments can be split into two parts with respect to start time $t$ of its last segment: 
\textbf{1)} the part with the first $l-1$ segments, which can be thought of as a solution for prefix $D_{[0, t-1]}$ of $D_{[0,i-1]}$ and
\textbf{2)} the part with the last $l^{th}$ segment, which can be thought of as a solution for suffix $D_{[t, i-1]}$ of $D_{[0,i-1]}$.
Conversely, solution $O$ can be constructed by combining the above two parts.
Now, since  $O^\star_{i,l}$ is the optimal solution (with respect to the given consensus clustering method), its corresponding two parts should be optimal too.
That is, the first part should be the optimal solution with $l-1$ segments for $D_{[0,t-1]}$ (which is exactly $O^\star_{t,l-1}$, and which is known from the earlier $t^{(th)}$ iteration),
and the second part should be the optimal solution with one segment for $D_{[t, i-1]}$ (which is a solution with just one segment, whose only segment partition can be obtained by performing consensus clustering of $D_{[t, i-1]}$).
So, we know the first part of $O^\star_{i,l}$ from one of the previous iterations and can compute its second part in the current iteration.
However, in order to actually construct $O^\star_{i,l}$ from the above two parts, we need to know start time $t$ of its last segment.
If $l=1$, there is only one value for $t$ (namely, $t=0$), since the $l^{th}$ (i.e., the only) segment should encompass the whole $D_{[0,i-1]}$.
If $l \in [2,i]$, $t$ can take any value from $l-1$ (in which case the $l-1$ segments in the first part are all singletons) to $i-1$ (in which case the $l^{th}$ segment in the second part is a singleton).
So, for $l \in [2,i]$, to find $t$, we simply test all of its possible values and pick the one that produces the solution that maximizes the objective function (\figurename~\ref{fig:supplement:exh_search}).

Given the above procedure, we next discuss its computational complexity.
Consider the $i^{th}$  iteration ($i \in [1,k]$).
In this iteration, we need to find $O^\star_{i,l}$ for each $l \in [1,i]$.
Recall from the above paragraph that each $O^\star_{i,l}$ is constructed from two parts,
and only for the second part, corresponding to its last segment and starting at some time point $t$, we do not know the corresponding segment partition and thus need to obtain this partition in the current iteration.
For $l=1$, we test only one value of $t$ ($t=0$), and for $l \in [2,i]$, we test all values of $t$ from $l-1$ to $i-1$.
For each $t$, we need to obtain consensus partition for suffix $D_{[t,i-1]}$ of $D_{[0,i-1]}$.
Clearly, different values of $l$ can deal with the same $t$, i.e., the same suffix,
and for a given suffix we need to obtain its consensus partition only once.
Overall, there are $i$ distinct values of $t$, from $0$ to $i-1$,
and thus $i$ distinct suffixes of $D_{[0,i-1]}$ for which we need to obtain consensus partitions.
Thus, in the $i^{th}$ iteration, we need to perform consensus clustering for the total of $i$ times, once for each such suffix.
Therefore, for all $k$ iterations, we perform consensus clustering $\sum_{i=1}^{k} i=\frac{k^2+k}{2}=O(k^2)$ times.

\begin{figure}[th]
  \centering\includegraphics[width=0.6\linewidth]{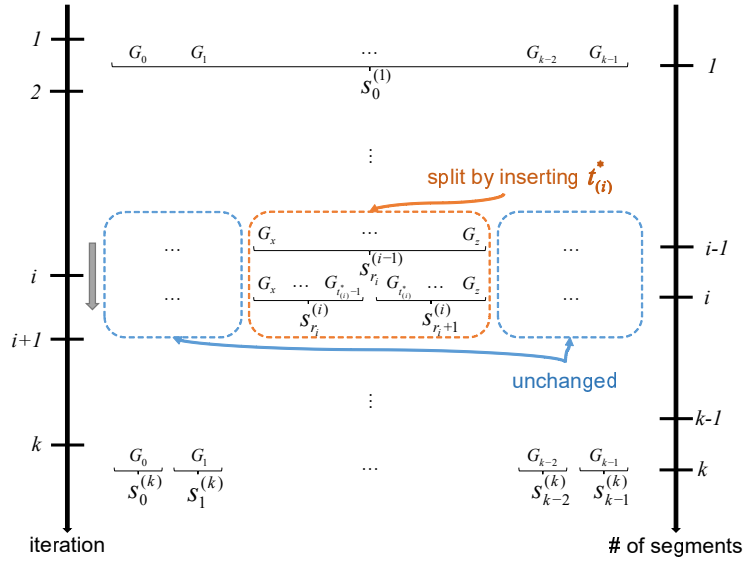}
  \caption[The top-down search]{The top-down search.
  The procedure starts with one segment encompassing the whole network and then iteratively splits one of the current segments into two.
  For the $i^{th}$ iteration, the orange box indicates the segment that was split as a result of this iteration,
  and blue boxes indicate all other, unchanged segments.
  }\label{fig:supplement:td_search}
\end{figure}

\vspace{0.1cm}\noindent\textbf{Top-down search.}
This strategy aims to find a good solution by greedily searching through the set of possible $T$s in a top-down manner.
The search contains $k$ iterations.
We start with one segment of length $k$ and at each subsequent iteration split one of the existing segments into two parts in a locally optimal way with respect to the chosen objective function, until we reach $k$ singleton segments (\figurename~\ref{fig:supplement:td_search}).
Since the search starts with one segment and since in each iteration the number of segments is increased by one, the solution obtained in the $i^{th}$ iteration is the solution for the CSCD problem with $i$ segments.
Hence, after $k$ iterations, we have one solution for the CSCD problem for each possible number of segments, which can be used to solve the SCD problem (Supplementary Section~\ref{sec:supplement:scout_obj_func}).
More specifically, the top-down search works as follows.
In the first iteration, we have only one segment, and we perform consensus clustering once for this segment.
Then, at the start of the $i^{th}$ iteration ($i \in [2,k]$), we have $i-1$ segments from the previous iteration (denoted as $\{ s^{(i-1)}_0, s^{(i-1)}_1, \dots, s^{(i-1)}_{i-2} \}$), and we aim to split one of these $i-1$ segments into two parts by inserting a new change point $t^\star_{(i)}$, in order to produce $i$ segments.
There are $k-i+1$ candidate time points $t^\prime$ for $t^\star_{(i)}$: the total of $k$ time points minus $t_0=0$  and minus $i-2$ change points selected in the previous iterations.
Out of these candidates, we choose the one that maximizes gain (or minimizes loss) in our objective function.

To reduce the computational complexity, we show that we can reuse in each iteration the results from the previous iterations.
In the $i^{th}$ iteration ($i \in [2,k]$), each candidate time point $t^\prime$ leads to splitting some current segment $s^{(i-1)}_j$ into two parts.
So, for a given $t$, we need to know two segment partitions: one for the first half of segment $s^{(i-1)}_j$ (that ends at time $t-1$) and one for the second half of segment $s^{(i-1)}_j$ (that starts at time $t$).
We do not necessarily need to compute these segment partitions in the current iteration, since we can reuse the results from the previous iterations, as follows.
Consider the next $(i+1)^{st}$ iteration.
Let $s^{(i)}_{r_i}$ be the segment that was split in the $i^{th}$ iteration (by inserting $t^\star_{(i)}$).
In the $(i+1)^{st}$ iteration, all segments except the two resulted from splitting $s^{(i-1)}_{r_i}$ are the same as in the $i^{th}$ iteration (\figurename~\ref{fig:supplement:td_search}).
Thus, when testing candidate time points inside these unchanged segments, we can just reuse segment partitions from the previous iterations.
So, in the $(i+1)^{st}$ iteration, we only need to perform consensus clusterings for those candidate time points that are within the two newly created segments (\figurename~\ref{fig:supplement:td_search}).
Such time points are all time points inside the segment that was split in the $i^{th}$ iteration except its start point and except already taken $t^\star_{(i)}$.

The above described reuse of previous consensus clusterings generally allows for reducing the complexity compared to the exhaustive search.
However, in the worst case, the complexity of the top-down search is still $O(k^2)$: if each new change point $t^\star_{(i)}$ is selected as the earliest one (i.e., if $t^\star_{(i)}=i-1$), the overall number of performed consensus clusterings is $1 + 2(k-1) + \sum_{i=3}^{k} (k-i+1) =  \sum_{i=1}^{k} (k-i+1) = (k^2+k)/2 = O(k^2)$.
Nevertheless, even though this is the same theoretic complexity as for the exhaustive search above, in practice, the top-down search is faster (Supplementary Section~\ref{sec:supplement:results_scout_params}).

\vspace{0.1cm}\noindent\textbf{Bottom-up search.}
This strategy aims to find a good solution by greedily searching through the set of possible $T$s in a bottom-up manner.
The search contains $k$ iterations.
We start with $k$ singleton segments and at each subsequent iteration merge two existing adjacent segments in a locally optimal way (with respect to the chosen objective function), until we reach one large segment of length $k$.
Since the search starts with $k$ segments and since in each iteration the number of segments is decreased by one, the solution obtained in the $i^{th}$ iteration is the solution for the CSCD problem with $k-i+1$ segments.
Hence, after $k$ iterations, we have one solution for the CSCD problem for each possible number of segments, which can be used to solve the SCD problem (Supplementary Section~\ref{sec:supplement:scout_obj_func}).
More specifically, the bottom-up search works as follows.
In the first iteration, we perform consensus clustering $k$ times, once for each segment.
Then, at the start of the $i^{th}$ iteration ($i \in [2,k]$), we have $k-i+2$ segments from the previous iteration (denoted as $\{ s^{(i-1)}_0, s^{(i-1)}_1, \dots, s^{(i-1)}_{k-i+1} \}$), and we aim to merge some two adjacent segments $(s^{(i)}_{r_i}, s^{(i)}_{r_i+1})$ from these $k-i+2$ segments in order to produce $k-i+1$ segments.
There are $k-i+1$ candidate segment pairs $(s^{(i)}_j, s^{(i)}_{j+1})$  (or, equivalently, $k-i+1$ candidate change points to be removed), since there are $k-i+2$ segments and since we consider only adjacent segment pairs.
Out of these candidates, we choose the one that maximizes gain (or minimizes loss) in our objective function.

\begin{figure}
  \centering\includegraphics[width=0.6\linewidth]{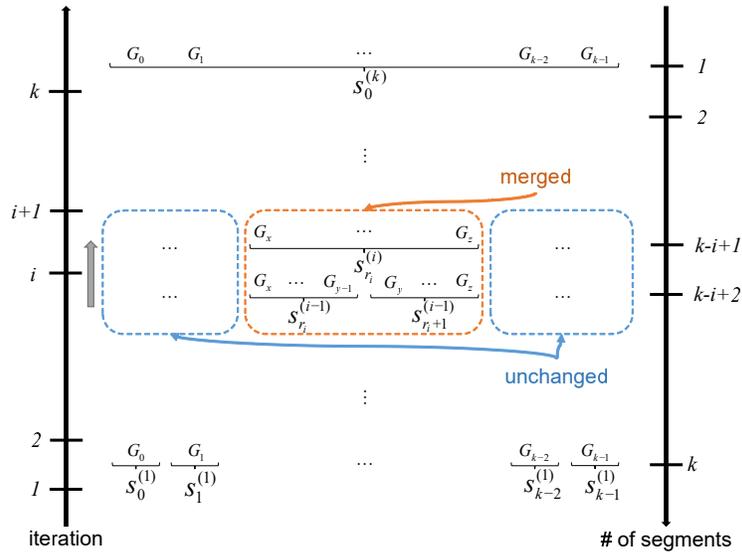}
  \caption[The bottom-up search]{The bottom-up search.
  The procedure starts with $k$ singleton segments and then iteratively merges two of the current adjacent segments into one.
  For the $i^{th}$ iteration, the orange box indicates the segment that was created as a result of this iteration,
  and blue boxes indicate all other, unchanged segments.
  }\label{fig:supplement:bu_search}
\end{figure}

To reduce the computational complexity, we next show that we can reuse in each iteration the results from the previous iterations.
In the $i^{th}$ iteration ($i \in [2,k]$), each candidate segment pair $(s^{(i)}_j, s^{(i)}_{j+1})$ leads to merging segments $s^{(i)}_j$ and $s^{(i)}_{j+1}$.
So, for a given segment pair $(s^{(i)}_j, s^{(i)}_{j+1})$, we need to know one segment partition for the merged segment $s^{(i)}_q \cup s^{(i)}_{q+1}$.
We do not necessarily need to compute these segment partitions in the current iteration, since we can reuse the results from the previous iterations, as follows.
Consider the next $(i+1)^{st}$ iteration.
In the $(i+1)^{st}$ iteration, all segments except the newly created one (i.e., the segment resulting from merging the two segments chosen in the $i^{th}$ iteration)
are the same as in the $i^{th}$ iteration (\figurename~\ref{fig:supplement:bu_search}).
Thus, when testing candidate segment pairs not involving the new segment, we can just reuse segment partitions from the previous iteration.
So, in the $(i+1)^{st}$ iteration, we only need to perform consensus clusterings for those candidate segment pairs that involve the new segment.
There are at most two such segment pairs, since we consider only adjacent segments.

In the first iteration, we perform consensus clustering $k$ times (once for each snapshot).
In the second iteration, we perform consensus clustering $k-1$ times (once for each pair of adjacent snapshots).
For all subsequent iterations, as discussed above, we perform consensus clustering at most twice. 
Therefore, for all $k$ iterations, we need to perform consensus clustering at most $k + (k-1) + \sum_{i=3}^{k} 2 = 4k - 5 = O(k)$ times.

\subsection{Experimental setup}

\subsubsection{Methods for comparison}\label{sec:supplement:methods_for_comparison}
Here, we expand our discussion from Section~\ref{sec:methods_for_comparison} in the main paper and discuss the methods that we use in our experiments and their parameters.

\textbf{GraphScope} does not accept any user-defined parameters.
Note that GraphScope was originally designed to work only with bipartite graphs, and thus it produces two separate partitions.
Hence, in order to handle unipartite graphs such as the data from our study, we constrain GraphScope to produce only one partition \cite{sun2007graphscope}. 
The method can solve only the SCD problem.
We use a publicly available implementation of GraphScope \cite{sarovios2014graphscope}.

\textbf{Multi-Step} performs an agglomerative clustering of the snapshots, merging them into segments to produce a hierarchical tree.
To get a solution for the CSCD problem, we cut the tree at the level that results in the desired number of segments.
To get a solution for the SCD problem, we first test Multi-Step's suggested procedure of cutting the tree at the level above which the highest segment similarity is negative.
We note that this procedure is used as a way to stop the merging process early in order to prevent meaningless merges, rather than as a way to select the best segmentation \cite{aynaud2011multi}.
Importantly, as we show in Section~\ref{sec:results} in the main paper, this procedure consistently underestimates the number of segments that actually exist in the data.
To address this, we introduce a user-specified segment similarity threshold $\theta$, and instead of stopping the merging process as soon as the highest similarity becomes negative, we instead stop this process as soon as the highest similarity becomes less than $\theta$.
That is, in Multi-Step's default procedure, $\theta=0$.
To give Multi-Step the best-case advantage, we vary $\theta$ from $0$ to $2$ in increments of $0.1$ (note that since the similarity between two segments is computed as the sum of modularities of the two segments, and since the maximum value of modularity for any segment is $1$, the maximum possible segment similarity value is $2$; Supplementary Section~\ref{sec:supplement:related_work}).
However, we find that the optimal threshold $\theta$ is network-specific, and a threshold that works well for one network may not work well for other networks (Section~\ref{sec:methods_for_comparison} in the main paper).
Note that by trying different values of $\theta$, essentially, we allow Multi-Step to try solutions with different numbers of segments.
Thus, when the ground truth number of segments is known (see below), we simply provide this information as input to Multi-Step.
That is, instead of using Multi-Step to solve the SCD problem, we use it to solve the CSCD problem where we set $l$ to match the ground truth number of segments.
This is not fair to the other methods (including SCOUT), which aim to solve the full SCD problem (and thus automatically find $l$ that ideally matches the ground truth value).
Yet, this is what we have to do in order to include Multi-Step into the comparison, since this method can reliably solve only the CSCD problem. 
We use a publicly available implementation of Multi-Step \cite{aynaud2011multi}.

\textbf{GHRG} relies on a sliding window approach with the length of the window $w$ being a user-defined parameter.
We test $w \in \lbrace 4,8,12 \rbrace$.
For each segment, GHRG results in a generalized hierarchical tree model instead of a partition.
To obtain a partition from this model, we cut the tree in a way that maximizes modularity \cite{blondel2008fast}. 
The method can solve only the SCD problem.
We use a publicly available implementation of GHRG \cite{peel2015detecting}.

\textbf{SCOUT} contains three main components: objective function, consensus clustering, and search strategy.
We test different choices for these components: two objective functions (with four choices for $Q_P$ and two choices for $Q_B$; Supplementary Section~\ref{sec:supplement:scout_obj_func}), three consensus clustering approaches (with seven choices for each of sum graph and consensus matrix and one choice for Average-Louvain; Supplementary Section~\ref{sec:supplement:scout_cons_clust}), and three search strategies (Supplementary Section~\ref{sec:supplement:scout_search_strategy}).
To allow for experimenting with the different parameter choices, our initial SCOUT implementation focuses on flexibility (to allow for easily testing various parameter choices for the method's components) rather than on running time.
However, once we finalize the most optimal (i.e., accurate yet efficient) choice of the parameters (Supplementary Section~\ref{sec:supplement:results_scout_params}), we develop a faster parallel SCOUT implementation tailored for the selected parameters and aimed at reducing the running time.
So, when comparing SCOUT against the other approaches, we use its latter fast implementation.
As discussed in Supplementary Section~\ref{sec:supplement:scout_obj_func}, SCOUT can solve both the CSCD and the SCD problems.

\subsubsection{Datasets}\label{sec:supplement:synth_networks}
\begin{figure}
  \centering\includegraphics[width=0.75\linewidth]{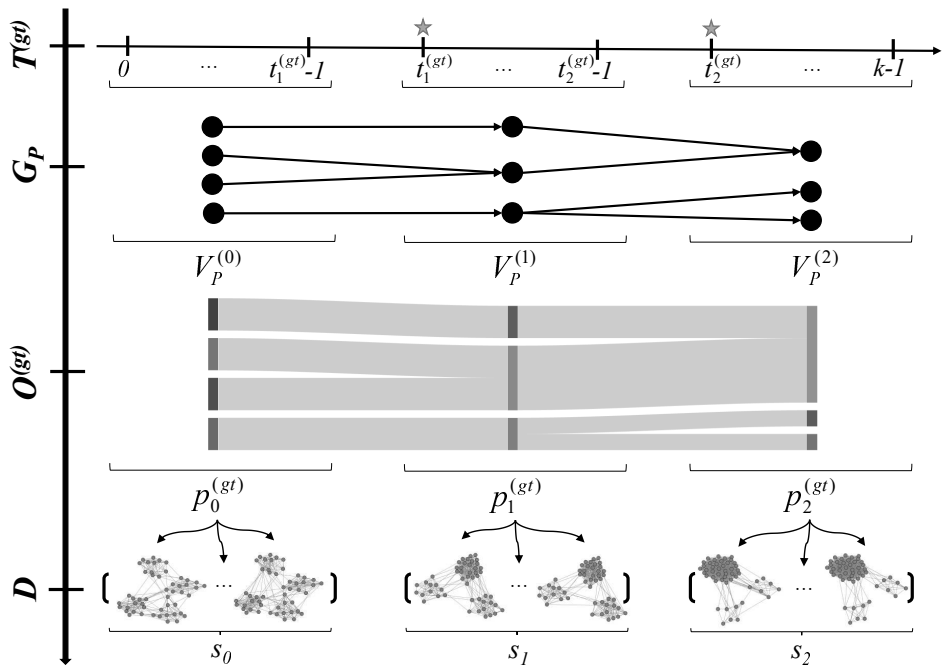}
  \caption[The process of constructing synthetic dynamic network $D$ with known ground truth solution $O^{(gt)}=(T^{(gt)},P^{(gt)})$]{The process of constructing synthetic dynamic network $D$ with known ground truth solution $O^{(gt)}=(T^{(gt)},P^{(gt)})$.
The four steps of the process, described in the text, are illustrated from top to bottom.
 }\label{fig:supplement:synthetic_construction}
\end{figure}

Here, we expand our discussion from Section~\ref{sec:datasets} in the main paper
and describe in detail the four steps of synthetic network generation (\figurename~\ref{fig:supplement:synthetic_construction}).
Recall that our model has the following parameters:
the number of snapshots $k$, the number of segments $l$, the number of nodes in each snapshot $n$, the minimum possible number of nodes in each cluster $c_{min}$, and the two parameters $c_{in}$ and $c_{out}$ controlling intra- and inter-community edge density of the snapshots.
As described in the main paper, the goals of the steps are: 
1) create $T^{(gt)}$, 
2) create a special auxiliary graph $G_P$, 
3) use $G_P$ to create $P^{(gt)}$, and 
4) use $T^{(gt)}$ and $P^{(gt)}$ to generate $D$.
The details of the four steps are as follows:

\textbf{1)} To create a segmentation with $l$ segments, we randomly sample $l-1$ change points $T^{(gt)}$ from $[1,k-1]$.
For example, in \figurename~\ref{fig:supplement:synthetic_construction}, we select two change points $t^{(gt)}_1$ and $t^{(gt)}_2$ to create three segments.
Thus, at the end of this step, we have the change point set $T^{(gt)}$.

\textbf{2)} In this and the next step, we aim to generate segment partitions $P^{(gt)}$, as follows.
To achieve this, we use a special \emph{partition graph} $G_P$.
In this step, we intuitively define $G_P$ and describe  how we generate $G_P$ and in the next step we describe how use $G_P$ to create $P^{(gt)}$.

We need to create $G_P=(V_P, E_P)$ as a weighted directed $l$-partite graph  (i.e., $V_P=V^{(0)}_P \cup V^{(1)}_P \cup \dots \cup V^{(l-1)}_P$) with every edge in $E_P$ having form $(u,v)$, where $u \in V^{(i-1)}_P$ and $v \in V^{(i)}_P$ for some $i \in [1, l-1]$.
To avoid confusion between nodes of $G_P$ and nodes of $D$, we refer to nodes of $G_P$ as \emph{supernodes}.
Intuitively, once we generate $G_P$ at the end of this step, each of its supernode sets  $V^{(i)}_P$ will correspond to segment partition $p^{(gt)}_i$ of $O^{(gt)}$, with each supernode of $V^{(i)}_P$ corresponding to some cluster of $p^{(gt)}_i$.
For example, in \figurename~\ref{fig:supplement:synthetic_construction}, $V^{(0)}_P$ has four supernodes, so $p^{(gt)}_0$ will have four clusters.
Each edge between two supernodes of $G_P$ intuitively means that the two clusters corresponding to these two supernodes have shared members (see the next step for more details).

We construct $G_P$ starting with empty sets $V_P$ and $E_P$.
To create $V_P$, we add a random number $r_i \in [2, \lfloor n/c_{min} \rfloor]$ of supernodes to each $V^{(i)}_P \subset V_P$, $i \in [0,l-1]$.
Recall from above that  $c_{min}$ is the user-defined minimum required number of nodes in a cluster.
Thus, the limits for $r_i$ mean that we want segment partition corresponding to $V^{(i)}_P$ to have at least two clusters, while at the same time we want to prevent it from having too many clusters.
To create $E_P$, we randomly create edges between each pair of adjacent sets $V^{(i-1)}_P$ and $V^{(i)}_P$ ($i \in [1, l-1]$) as follows.
For each edge, we randomly pick its two endpoints: one supernode in $V^{(i-1)}_P$ and one supernode in $V^{(i)}_P$.
When adding edges to $E_P$, we aim to satisfy the following three conditions: \textbf{a)} the edges do not form a perfect matching between $V^{(i-1)}_P$ and $V^{(i)}_P$ (i.e., segment partitions $p^{(gt)}_{i-1}$ and $p^{(gt)}_{i}$ are not identical), \textbf{b)} each supernode in $V^{(i-1)}_P$ has at least one outgoing edge and each supernode in $V^{(i)}_P$ has at least one incoming edge  (i.e., clusters cannot appear/disappear),
and \textbf{c)} for each edge $e=(u,v)$, assuming $d^{+}$ and $d^{-}$ is the outdegree and indegree of a supernode, respectively, one of the following holds: (i) $d^{+}(u)=1$ and $d^{-}(v)>1$ (i.e., the cluster corresponding to $u$ merges with at least one other cluster into the cluster corresponding to $v$; e.g., the top right edge in \figurename~\ref{fig:supplement:synthetic_construction}), (ii) $d^{+}(u)>1$ and $d^{-}(v)=1$ (i.e., the cluster corresponding to $u$ is split into several clusters including the cluster corresponding to $v$; e.g., the bottom right edge in \figurename~\ref{fig:supplement:synthetic_construction}),
or (iii) $d^{+}(u)=d^{-}(v)=1$ (i.e., the cluster corresponding to $u$ and the cluster corresponding to $u$ have the same members; e.g., the top left edge in \figurename~\ref{fig:supplement:synthetic_construction}).
Intuitively, for two adjacent segment partitions, edges satisfying (i) correspond to merges of clusters, edges satisfying (ii) correspond to splits of cluster, and edges satisfying  (iii) correspond to unchanged clusters. 
Note that in general, the condition (iii) can be relaxed; we include it to provide further coherence between individual clusters so that their changes can be described via three simple events (merge, split, and continuation).

\textbf{3)} Using $G_P$ from the previous step, we now generate the set of segment partitions $P^{(gt)}$, as shown in Supplementary Algorithm~\ref{alg:supplement:synth_step3}.
Intuitively, each edge in $G_P$ means that some two clusters share members.
We construct segment partitions one by one, starting from the partition of the first segment.
For each segment partition, we determine the membership of a given cluster based on the incoming edges of the corresponding supernode and the previous segments partition.
At the end of this step, we have the complete $O^{(gt)}=(T^{(gt)},P^{(gt)})$.

\begin{algorithm}[th]
 \caption{Step 3 of our synthetic network generation process.
Auxiliary procedure $RandomPartition$ randomly partitions a given set into a given number of clusters and
auxiliary procedure $Outneighbors$ returns the list of the outneighbors of a given supernode.
}
\label{alg:supplement:synth_step3}
\begin{algorithmic}[1]
\Require $G_P$
\Ensure $P^{(gt)}$
 \State $V \gets [0, n-1]$
 \State $p^{(gt)}_{0} \gets RandomPartition(V, |V^{(0)}_P|)$
 \For{$i \in [1, l-1]$}
   \For{$v \in V^{(i)}_P$}
     \State $c_{(v)} \gets \emptyset$
   \EndFor
   \For{$u \in V^{(i-1)}_P$}
     \State $p^{(temp)} \gets  RandomPartition(c_{(u)},  d^{+}(u))$
     \State $j \gets 0$
     \For{$c^{(temp)} \in p^{(temp)}$}
        \State $c_{(Outneighbors(u)[j])} \gets c_{(Outneighbors(u)[j])} \cup p^{(temp)}$
        \State $j \gets j+1$
     \EndFor
   \EndFor
   \State $p^{(gt)}_{i} \gets \{ c_{(v)} | v \in V^{(i)}_P \}$
 \EndFor
 \State $P^{(gt)} \gets \{ p^{(gt)}_{i} | i \in [0, l-1]\}$
 \State \Return{$P^{(gt)}$}
\end{algorithmic}
\end{algorithm}

\textbf{4)} We use $O^{(gt)}=(T^{(gt)},P^{(gt)})$ to generate snapshots of $D$ as follows.
Intuitively, we aim to generate each snapshot based solely on the corresponding segment partition.
Thus, to generate a snapshot $G_j$ of a given segment $s_i$, we use the stochastic blockmodel with partition $p^{(gt)}_i$ and stochastic block matrix $\theta$  (Supplementary Section~\ref{sec:supplement:scout_obj_func}).
We use the same  $\theta$ for all segments, with $\theta_{c_uc_v}=c_{in}/n$ if $c_u \neq c_v$ and
$\theta_{c_uc_v}=c_{in}/n$ otherwise.
So, to generate a snapshot $G_j$ of segment $s_i$, for each pair of nodes in $G_j$, we independently place an edge between the two nodes with probability $c_{in}/n$ if they are in the same cluster in $p_i$ and with probability $c_{out}/n$ otherwise.
Note that even though all snapshots in segment $s_i$ are created based on the same segment partition $p_i$, they still likely differ from each other due to randomness in the stochastic blockmodel.
At the end of this step, we have all snapshots $G_j$ of $D$, and, moreover, the structure of $D$ reflects the ground truth $O^{(gt)}=(T^{(gt)},P^{(gt)})$.

\subsubsection{Evaluation measures}\label{sec:supplement:eval_measures}
Here, we expand our discussion from Section~\ref{sec:eval_measures} in the main paper.

\noindent\textbf{\RNum{1})}
We start by discussing the three output similarity measures.

\noindent\textbf{\emph{Segmentation similarity} {\boldmath $Sim_T$}.}
Formally, for a given $O=(T,P)$, we construct its time point partition $\mathcal{P}_T$ as a partition of the set $[0,k-1]$ into $l$ clusters $c_i, i \in [0,l-1]$, where cluster $c_i=\{ t_i, t_i+1, \dots, t_{i+1}-1 \}$.
Then, to compute $Sim_T$ between $O^\star$ and $O^{(gt)}$, we measure similarity of their corresponding time point partitions $\mathcal{P}_T^\star$ and $\mathcal{P}_T^{(gt)}$, as follows:
\begin{equation} \label{eq:supplement:sim_t}
Sim_T(O^\star, O^{(gt)}) = H(\mathcal{P}_T^\star,\mathcal{P}_T^{(gt)}),
\end{equation}
where $H$ can be any partition similarity measure.

\noindent\textbf{\emph{Partition similarity} {\boldmath $Sim_P$}.}
Formally, to compute $Sim_P$ between $O^\star$ and $O^{(gt)}$, we introduce \emph{average snapshot partition similarity}, as follows:
\begin{equation} \label{eq:supplement:sim_p}
Sim_P(O^\star, O^{(gt)}) = \frac{1}{k} \sum_{j=0}^{k-1} H(p^\star_{seg(j, T^\star)},p^{(gt)}_{seg(j,T^{(gt)})}),
\end{equation}
where $H$ can be any partition similarity measure and $seg(j, T)$ is the function that returns the index of the segment containing snapshot $G_j$ under the segmentation induced by change point set $T$ (i.e., $seg(j, T)=i \iff G_j \in s_i$).
Note that even though $Sim_P$ focuses on the partition aspect of the SCD problem, it still implicitly relies on the segmentation aspect via the above $seg$ function.

\noindent\textbf{\emph{Overall similarity} {\boldmath $Sim_B$}.}
Formally, for a given $O=(T,P)$, we construct a node-time partition  $\mathcal{P}_B$ as a partition of the set $\{ (u,t) |  u \in V_t, t \in [0,k-1] \}$ into $\sum_{i=0}^{l-1} |p_i|$ clusters.
Two node-time pairs $(u_1,t_1)$ and $(u_2,t_2)$ are clustered together in $\mathcal{P}_B$ if their time points $t_1$ and $t_2$ belong to the same segment (i.e., if $t_1$ and $t_2$ are in the same cluster in $\mathcal{P}_T$, or $seg(t_1, T)=seg(t_2, T)$) and if their nodes $u_1$ and $u_2$ belong to the same cluster in the corresponding segment partition (i.e., if $u_1$ and $u_2$ are in the same cluster in $p_{seg(t_1, T)}$).
Then, to compute $Sim_B$ between $O^\star$ and $O^{(gt)}$, we measure similarity between their corresponding node-time partitions $\mathcal{P}_B^\star$ and $\mathcal{P}_B^{(gt)}$, as follows:
\begin{equation} \label{eq:supplement:sim_b}
Sim_B(O^\star, O^{(gt)}) = H(\mathcal{P}_B^\star,\mathcal{P}_B^{(gt)}),
\end{equation}
where $H$ can be any partition similarity measure. 

Next, we describe the four partition similarity measures $H$ that we use:
\textbf{1)} \emph{Normalized Mutual Information} (\emph{NMI}) \cite{vinh2010information} -- a measure of similarity based on the mutual information (MI), normalized to have values in $[0,1]$.
\textbf{2)} \emph{Adjusted Mutual Information} (\emph{AMI}) \cite{vinh2010information} -- an adjusted for chance version of MI.
\textbf{3)} \emph{Adjusted Rand Index} (\emph{ARI}) \cite{vinh2010information} -- an adjusted for chance version of the Rand Index, a measure  of similarity based on counting pairs of observations assigned to the same cluster or different clusters in two partitions.
For AMI and ARI, the adjustment for chance means correction for chance agreement between two partitions \cite{vinh2010information}.
\textbf{4)} \emph{V-Measure} (\emph{VM}) \cite{rosenberg2007v} -- the harmonic mean of \emph{homogeneity}
(whether the first partition groups together \emph{only} those objects that are grouped together in the second partition)
and \emph{completeness} (whether the first partition groups together \emph{all} those objects that are grouped together in the second partition).
Note that for all of the above measures, a higher value means higher similarity, with two identical partitions having similarity of one.

\noindent\textbf{\RNum{2})}
Next, we describe the three classification accuracy measures that we use:
\textbf{1)} $\forall i \in [1,k]$,  \emph{precision} is the fraction of the top ranked $i$ time points that are ground truth change points (i.e., that belong to $T^{(gt)}$), and \emph{recall} is the fraction of all ground truth change points that are among the top ranked $i$ time points.  
To summarize the values of precision and recall over all possible values of $i$, we compute the \emph{area under the precision-recall curve} (\emph{AUPR}).
\textbf{2)} $\forall i \in [1,k]$, \emph{F-score} is the harmonic mean of precision and recall (that is, F-score balances the two quantities).
We report the \emph{maximum F-score} over all values of $i$. 
\textbf{3)}  $\forall i \in [1,k]$, \emph{sensitivity} is equal to recall and \emph{specificity} is the fraction of ground truth non-change points (corresponding to the complement of $T^{(gt)}$) that are not among the top ranked $i$ time points.
To summarize the values of sensitivity and specificity over all possible values of $i$, we compute the \emph{area under the receiver operator characteristic curve} (\emph{AUROC}).

Finally, we discuss how to obtain the ranked list of all time points for each of the considered methods.
For each method, to get the ranked list, we compute the score (see below) for each time point $t \in [1,k-1]$, such that the time points with lower scores are ranked higher (i.e., are more ``change point-like'').
Note that here we exclude from consideration time point $t_0=0$, because, by definition, for any method, $t_0$ always denotes the start of the first segment, and hence including it into comparison does not provide any method-specific information.

We compute the ranking for each method as follows.
If a method is capable of solving the CSCD problem, we obtain the ranking by using the solutions with all possible numbers of segments for the CSCD problem.
Specifically, let $O^\star_{(i)}=(T^\star_{(i)}, P^\star_{(i)})$ be the solution with $i$ segments for the CSCD problem.
Intuitively, if a given time point $t$ is a change point in $O^\star_{(i)}$ (i.e., if $t \in T^\star_{(i)}$), this means that the method, when asked to select only $i-1$ change points (i.e., to produce $i$ segments), chooses $t$ as one of these $i-1$ change points.
Hence, to capture the intuition that more ``change point-like'' time points appear in the solutions with smaller number of segments, we compute the score of a given time point $t$ as the smallest number of segments for which $t$ appears as a change point in the corresponding CSCD solution (i.e., $score(t)=\min \{ i | i \in [1,k], t \in T^\star_{(i)} \}$).
So, the highest ranked time point will appear in the solution with two segments (for which only one time point is selected as change point), while the lowest ranked time point will appear only in the solution with $k$ segments (for which all time points are selected as change point).
Since among the considered methods only Multi-Step and SCOUT can solve the CSCD problem (Supplementary Section~\ref{sec:supplement:methods_for_comparison}), the above procedure can only be used for these two methods.
For the remaining two methods, GraphScope and GHRG, we use alternative strategies for extracting their ranked lists, as follows.
GraphScope, at each step, marks a time point $t$ as a change point if the community organization of $G_t$ does not match well the community organization of the current segment (Supplementary Section~\ref{sec:supplement:related_work}).
The match is measured via the MDL principle, as the difference of the cost of encoding the current segment and $G_t$ together and the cost of encoding them separately.
Intuitively, the smaller the difference, the ``cheaper'' it is to add $G_t$ to the current segment, and when the difference is negative, $G_t$ is added to the current segment.
Thus, we use the difference of the encoding costs as the score of $t$.
GHRG performs a statistical test at each step to determine whether the current window contains a change point (Supplementary Section~\ref{sec:supplement:related_work}).
A change point is detected if its corresponding $p$-value is smaller than the chosen threshold.
Intuitively, the smaller the $p$-value, the more confident the method is that the given time point $t$ is a change point.
Thus, we use the $p$-value as the score of $t$.

\subsubsection{Statistical significance of two methods' performance difference}\label{sec:supplement:methods_stat}
Given a synthetic network configuration, evaluation measure, and a pair of methods, we compute the statistical significance of the difference between the performance of the two methods as follows.
For each method, we create a list containing performance scores of the method for all network instances of the synthetic network configuration.
Since we know which score in a given list corresponds to which network instance, we use \emph{paired} $t$-test to compute the statistical significance of the difference between the two lists produced by the two methods.
There are four possible outcomes:
\textbf{1)} method 1 outperforms method 2, and the improvement is statistically significant,
\textbf{2)} method 1 outperforms method 2, but the improvement is not statistically significant,
\textbf{3)} method 2 outperforms method 1, but the improvement is not statistically significant, and
\textbf{4)} method 2 outperforms method 1, and the improvement is statistically significant
(note that outcomes 2 and 3 also cover the case when the two methods' scores are tied).
By statistically significant, we mean that the $p$-value that results from paired $t$-test is below a threshold.
We test three $p$-value thresholds: 0.05, 0.01, and 0.001.

\section{Results: the effect of method parameter choices}\label{sec:supplement:results_scout_params}
Here, we expand our discussion from Section~\ref{sec:results_params} in the main paper and discuss the effect of SCOUT parameters.
We test the effect on the method's performance of a) the objective function,
b) consensus clustering method, and c) search strategy.

\begin{figure}
\centering
\subfloat[]{\label{fig:supplement:scout_q_b-gl_p__nmi}\includegraphics[width=.46\linewidth]{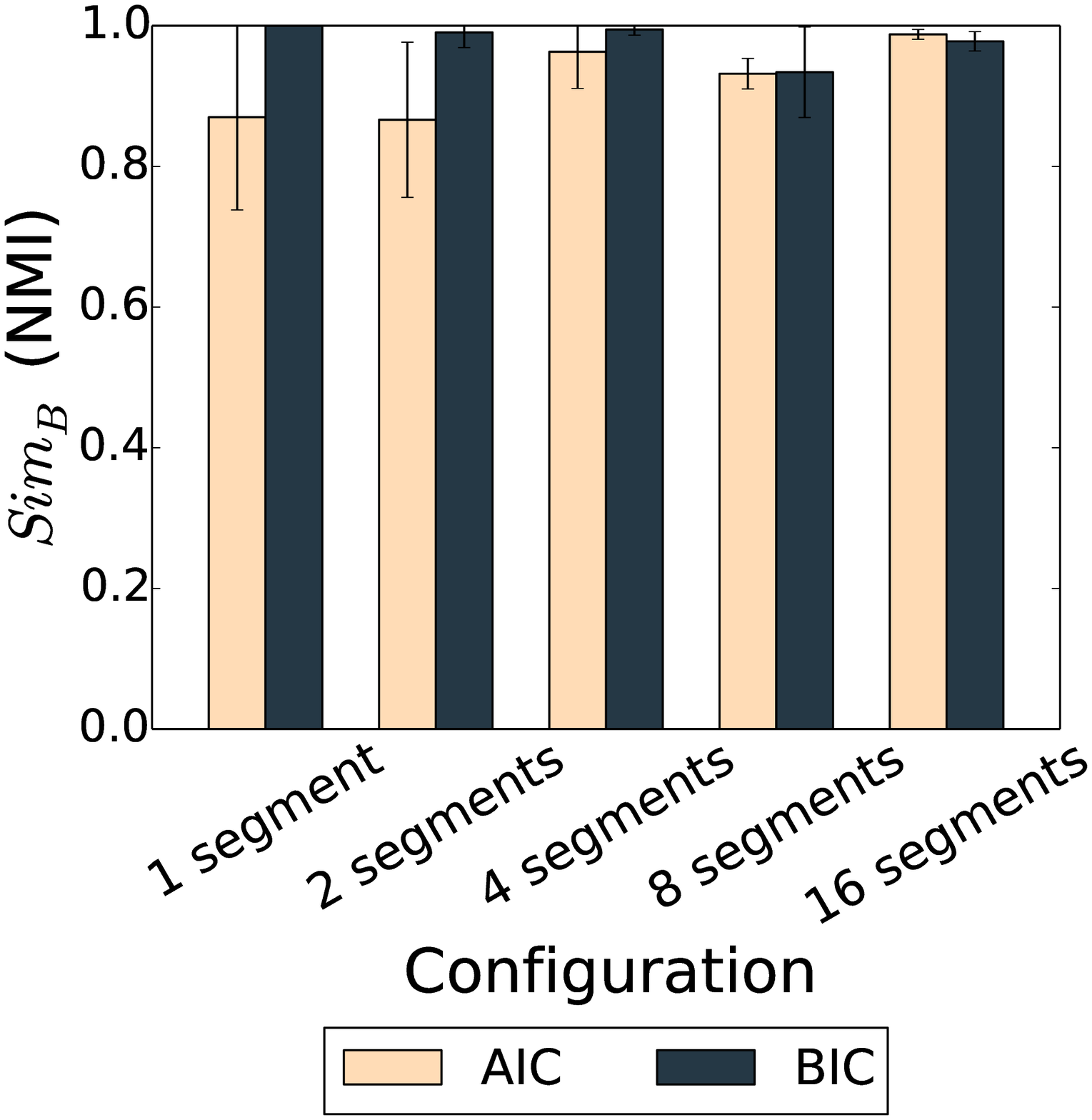}}
\hspace{0.2cm}
\subfloat[]{\vspace{1mm}\label{fig:supplement:scout_q_b-sgmq__num}\includegraphics[width=.45\linewidth]{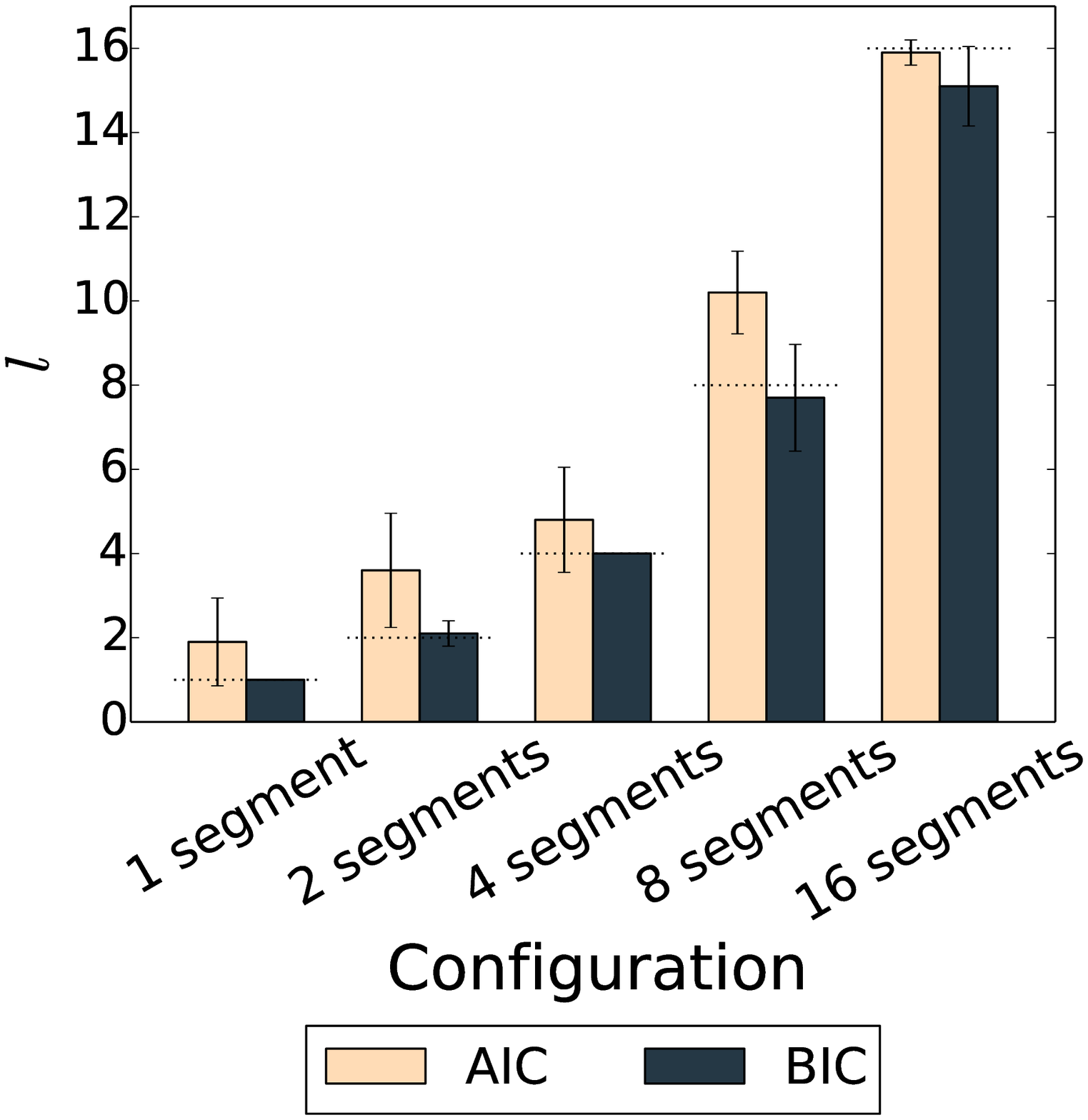}}
  \caption[The effect of $Q_B$ choice for SCOUT when choosing the optimal number of segments]{The effect of $Q_B$ choice for SCOUT when choosing the optimal number of segments, in terms of
\textbf{\protect\subref{fig:supplement:scout_q_b-gl_p__nmi}} $Sim_B$
 and \textbf{\protect\subref{fig:supplement:scout_q_b-sgmq__num}}  the number of segments $l$, 
as we vary the ground truth synthetic network configuration ($x$-axis).
Here, in all cases, we fix consensus clustering method as sum graph with Walktrap and search strategy as exhaustive search.
In panel \protect\subref{fig:supplement:scout_q_b-sgmq__num}, for a given ground truth configuration, the dotted line corresponds to the ground truth number of segments.
}
\label{fig:supplement:scout_q_b}
\end{figure}

\begin{figure}
\centering
\includegraphics[width=.45\linewidth]{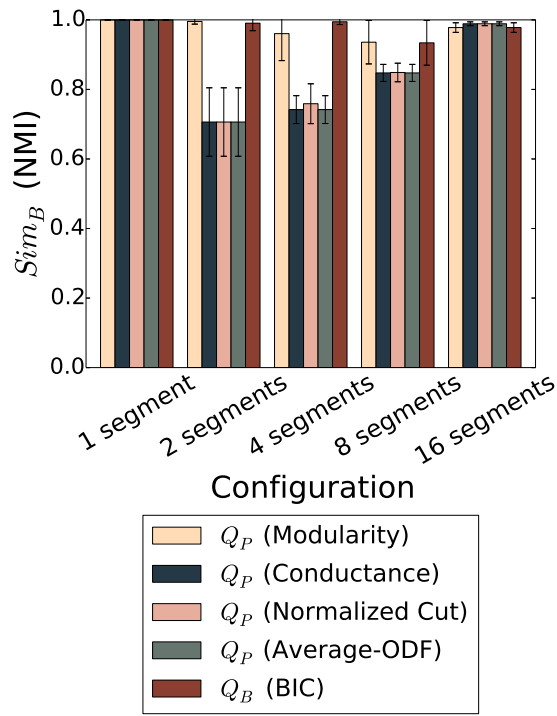}
  \caption[The effect of $Q$ choice for SCOUT when producing the best solution for each possible number of segments]{The effect of $Q$ choice for SCOUT when producing the best solution for each possible number of segments, in terms of $Sim_B$,
as we vary the ground truth synthetic network configuration ($x$-axis).
Note that when using $Q_P$ (the first four series), we still need to choose the optimal number of segments, which we do using $Q_B$ (i.e.,  BIC).
Here, in all cases, we fix consensus clustering method as sum graph with Walktrap, and we fix search strategy as exhaustive search.
}
\label{fig:supplement:scout_q_p}
\end{figure}

\textbf{a)} Objective function is used twice (Supplementary Section~\ref{sec:supplement:scout_obj_func}): 1) when computing the best solution for each possible number of segments (Equation~\ref{eq:supplement:o_cscd}), and 2) when choosing among these best solutions the final one (Equation~\ref{eq:supplement:o_scd}).
The objective functions can differ between the two cases:
in case ``1'', we can use any $Q_P$ or any $Q_B$, and in case ``2'', we have to use a $Q_B$ (Supplementary Section~\ref{sec:supplement:scout_obj_func}).
Thus, since SCOUT uses $Q_B$ up to two times, while it uses $Q_P$ up to one time, we first test the effect of $Q_B$.
Recall that we evaluate two $Q_B$ measures: one based on BIC and the other based on AIC (Supplementary Section~\ref{sec:supplement:scout_obj_func}). 
In general, BIC results in higher $Sim_B$ compared to AIC (\figurename~\ref{fig:supplement:scout_q_b-gl_p__nmi}).
The reason for this is that AIC produces more segments than BIC, usually overestimating the ground truth number of segments (\figurename~\ref{fig:supplement:scout_q_b-sgmq__num}).
Recall that this behavior of AIC is not surprising (Supplementary Section~\ref{sec:supplement:scout_obj_func}).
So, we focus on $Q_B$ based on BIC.
This gives us the choice of $Q_B$ for case ``2''.
For case ``1'', we can use $Q_B$ based on BIC or one of the four $Q_P$s (based on modularity, conductance, normalized cut, or average-ODF; Supplementary Section~\ref{sec:supplement:scout_obj_func}). 
Hence, we next test the effect of $Q_P$ versus $Q_B$ in case ``1''.
Out of all $Q_P$s, modularity generally leads to the highest $Sim_B$ (\figurename~\ref{fig:supplement:scout_q_p}).
However, the best results in terms of $Sim_B$ are achieved when using $Q_B$ based on BIC and not $Q_P$ based on modularity (\figurename~\ref{fig:supplement:scout_q_p}).
So, whether we are considering case ``1'' or case ``2'', $Q_B$ based on BIC overall outperforms all other tested objective functions.
Thus, we focus on $Q_B$ based on BIC as SCOUT's objective function.

\begin{figure}
\centering
\subfloat[]{\label{fig:supplement:scout_cons_sum-gl_p__nmi}\includegraphics[width=.45\linewidth]{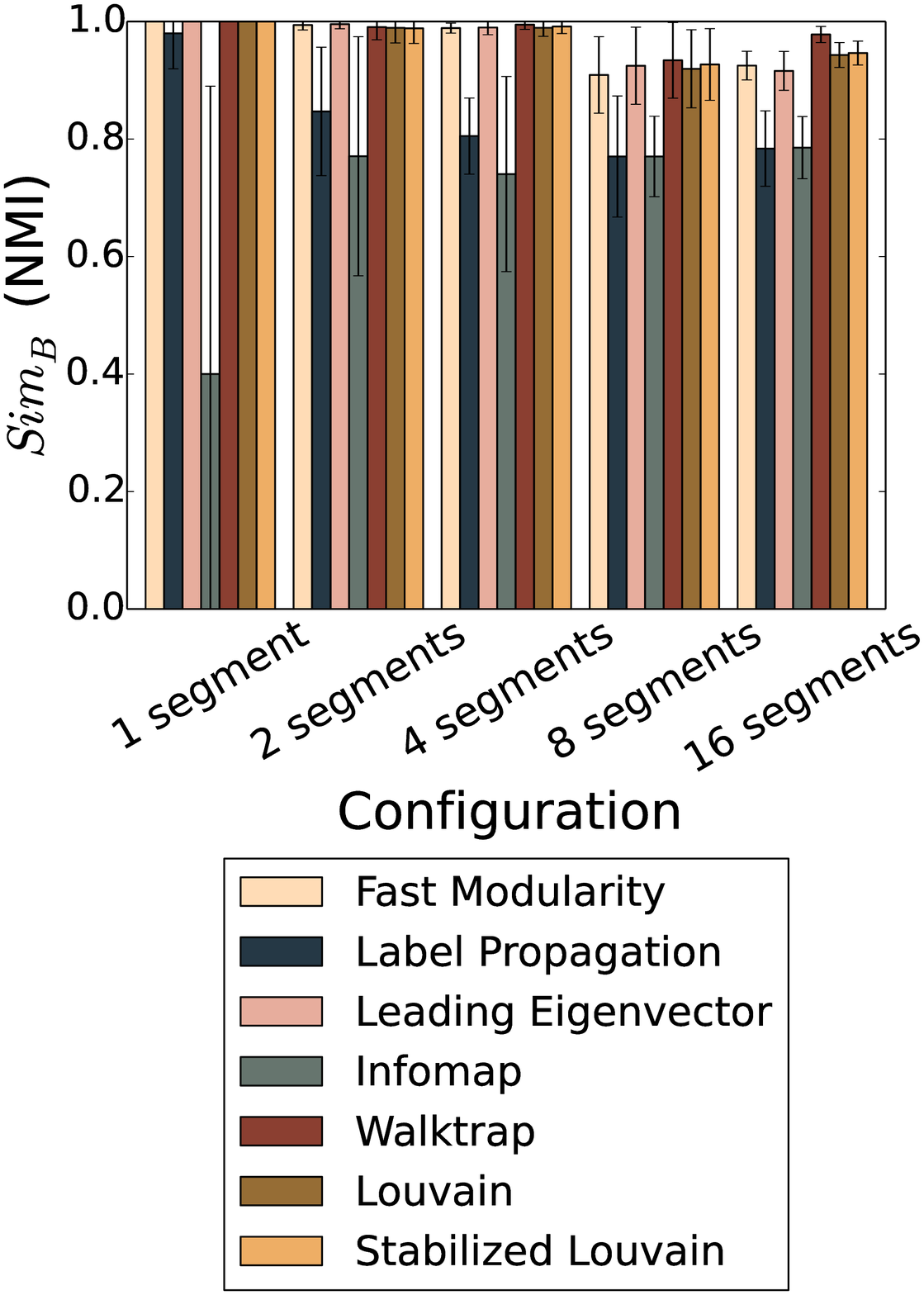}}
\subfloat[]{\label{fig:supplement:scout_cons_sum-run_time}\includegraphics[width=.45\linewidth]{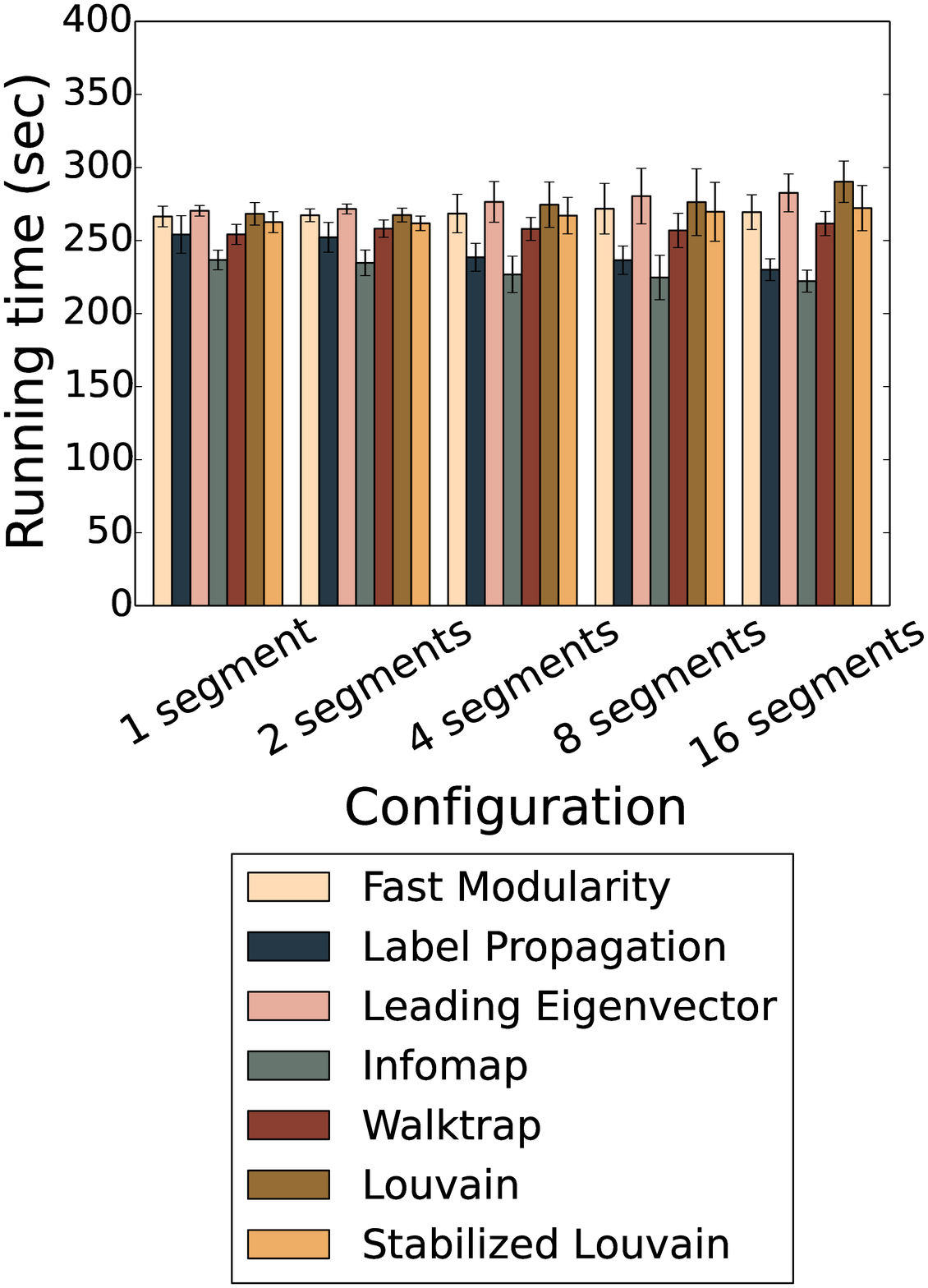}}
  \caption[The effect of the choice of static clustering method in the sum graph consensus clustering method for SCOUT]{The effect of the choice of static clustering method in the sum graph consensus clustering method for SCOUT, in terms of
 \textbf{\protect\subref{fig:supplement:scout_cons_sum-gl_p__nmi}} $Sim_B$
and \textbf{\protect\subref{fig:supplement:scout_cons_sum-run_time}} running time,
as we vary the ground truth synthetic network configuration ($x$-axis).
Here, in all cases, we fix search strategy as exhaustive search.
}
\label{fig:supplement:scout_cons_sum}
\end{figure}

\begin{figure}
\centering
\subfloat[]{\label{fig:supplement:scout_cons-gl_p__nmi}\includegraphics[width=.45\linewidth]{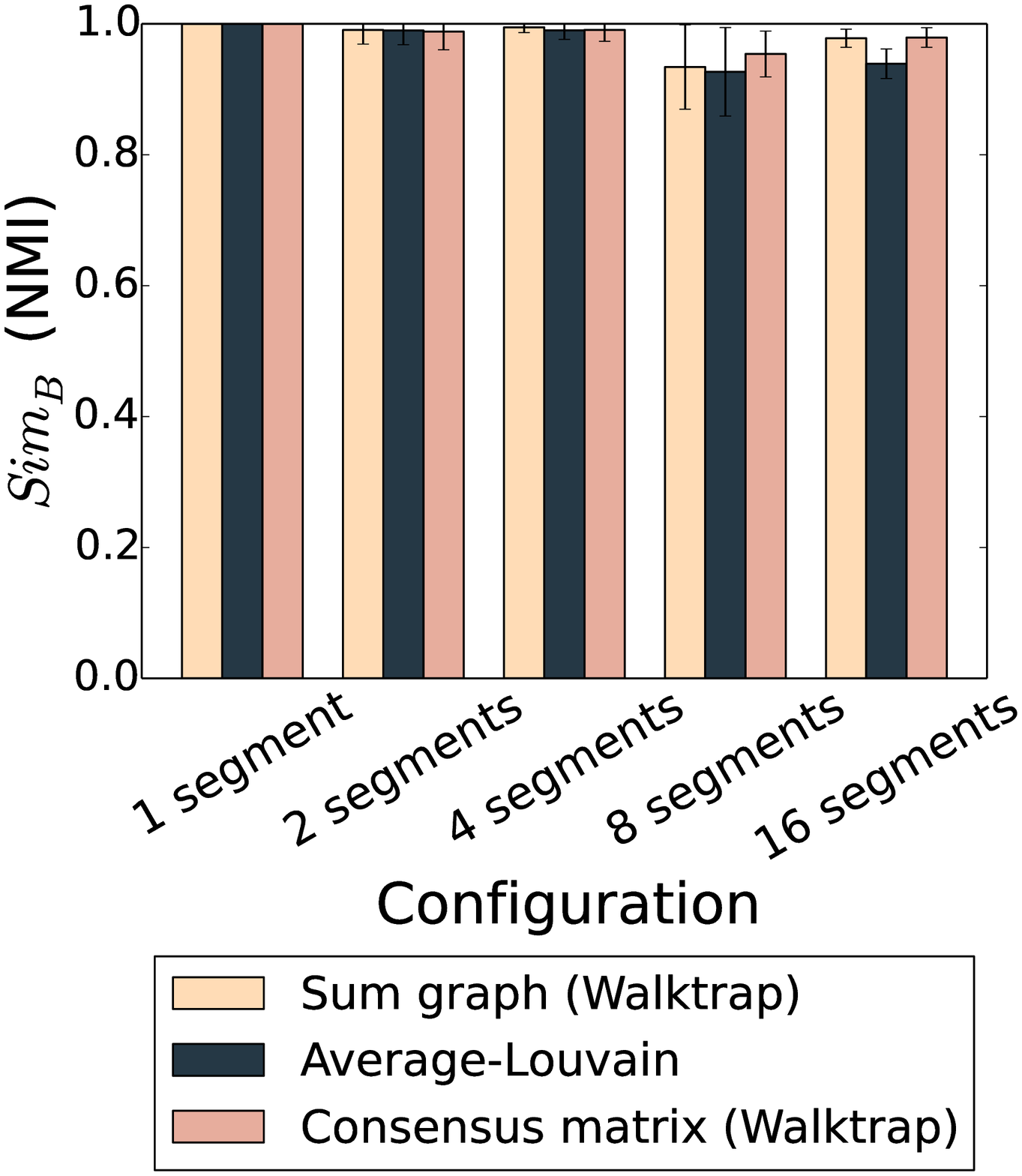}}
\hspace{0.2cm}
\subfloat[]{\label{fig:supplement:scout_cons-run_time}\includegraphics[width=.45\linewidth]{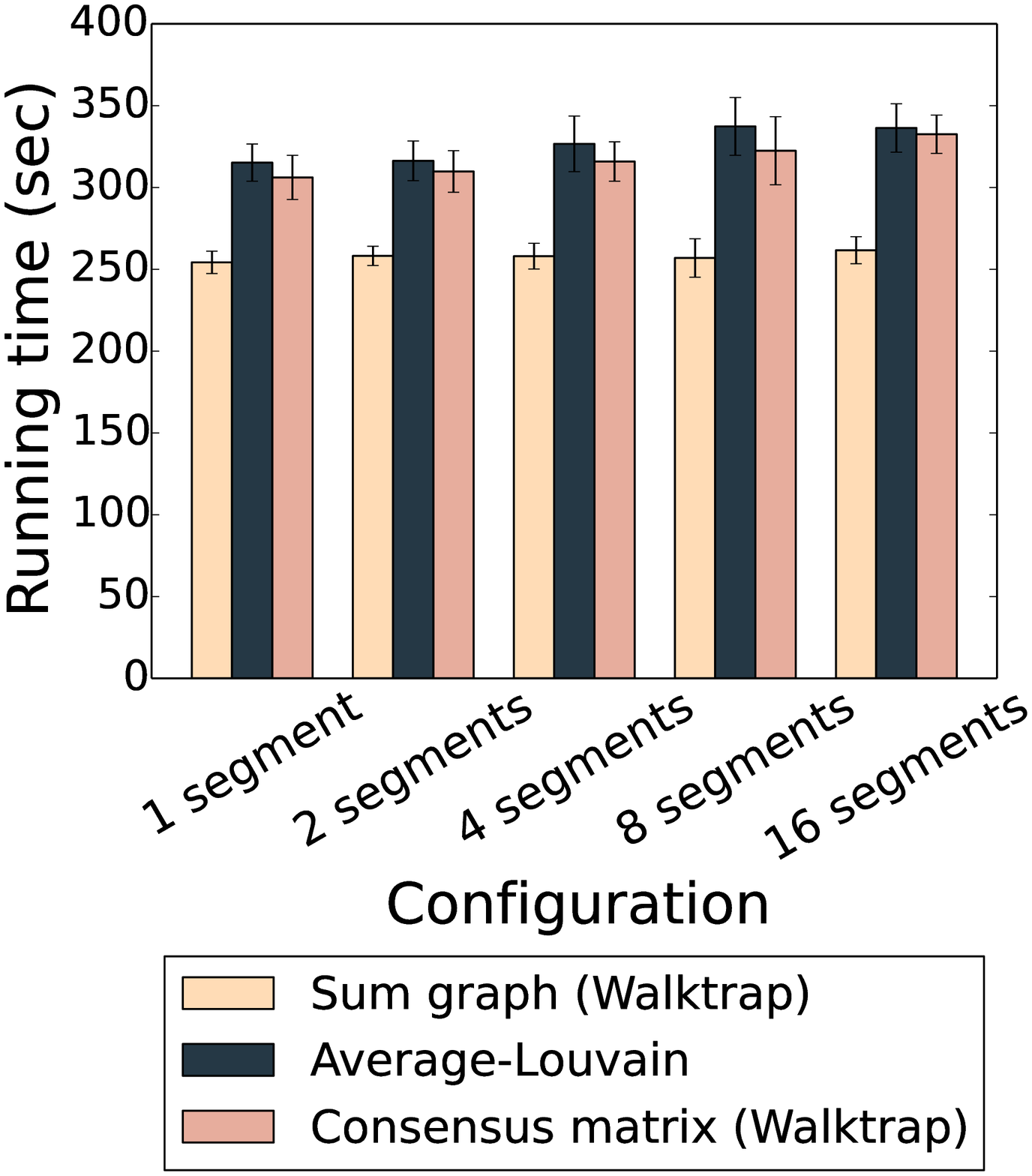}}
  \caption[The effect of the choice of consensus clustering method for SCOUT]{The effect of the choice of consensus clustering method for SCOUT, in terms of
 \textbf{\protect\subref{fig:supplement:scout_cons-gl_p__nmi}} $Sim_B$
and \textbf{\protect\subref{fig:supplement:scout_cons-run_time}} running time,
as we vary the ground truth synthetic network configuration ($x$-axis).
Here, in all cases, we fix search strategy as exhaustive search.
}
\label{fig:supplement:scout_cons}
\end{figure}

\textbf{b)} Consensus clustering is used to produce segment partitions, given a segmentation.
We use three general types of consensus clustering approaches: sum graph, Average-Louvain, and consensus matrix (Supplementary Section~\ref{sec:supplement:scout_cons_clust}).
Recall that the sum graph and consensus matrix approaches are parameterized with the static clustering method (Supplementary Section~\ref{sec:supplement:scout_cons_clust}).
So, before we compare the above three general types of approaches, we first aim to choose the best static clustering method for sum graph and consensus matrix approaches.
We evaluate seven static clustering methods: Fast Modularity, Label Propagation, Leading Eigenvector, Infomap, Walktrap, Louvain, and Stabilized Louvain (Supplementary Section~\ref{sec:supplement:scout_cons_clust}).
We find that generally Walktrap works the best in terms of $Sim_B$ while having comparable running time (\figurename~\ref{fig:supplement:scout_cons_sum}).
So, we focus on Walktrap as the static clustering method for both sum graph and consensus matrix approaches.
Next, we compare the three general approach types.
In terms of $Sim_B$, all three approaches lead to comparable results (\figurename~\ref{fig:supplement:scout_cons-gl_p__nmi}), even though Average-Louvain and consensus matrix are more sophisticated compared to sum graph and thus would be expected to be superior.
In terms of the running time, sum graph is the fastest of the three approaches (\figurename~\ref{fig:supplement:scout_cons-run_time}).
Thus, we focus on sum graph with Walktrap as SCOUT's consensus clustering approach.

\begin{figure}
\centering
\subfloat[]{\label{fig:supplement:scout_search-gl_p__nmi}\includegraphics[width=.45\linewidth]{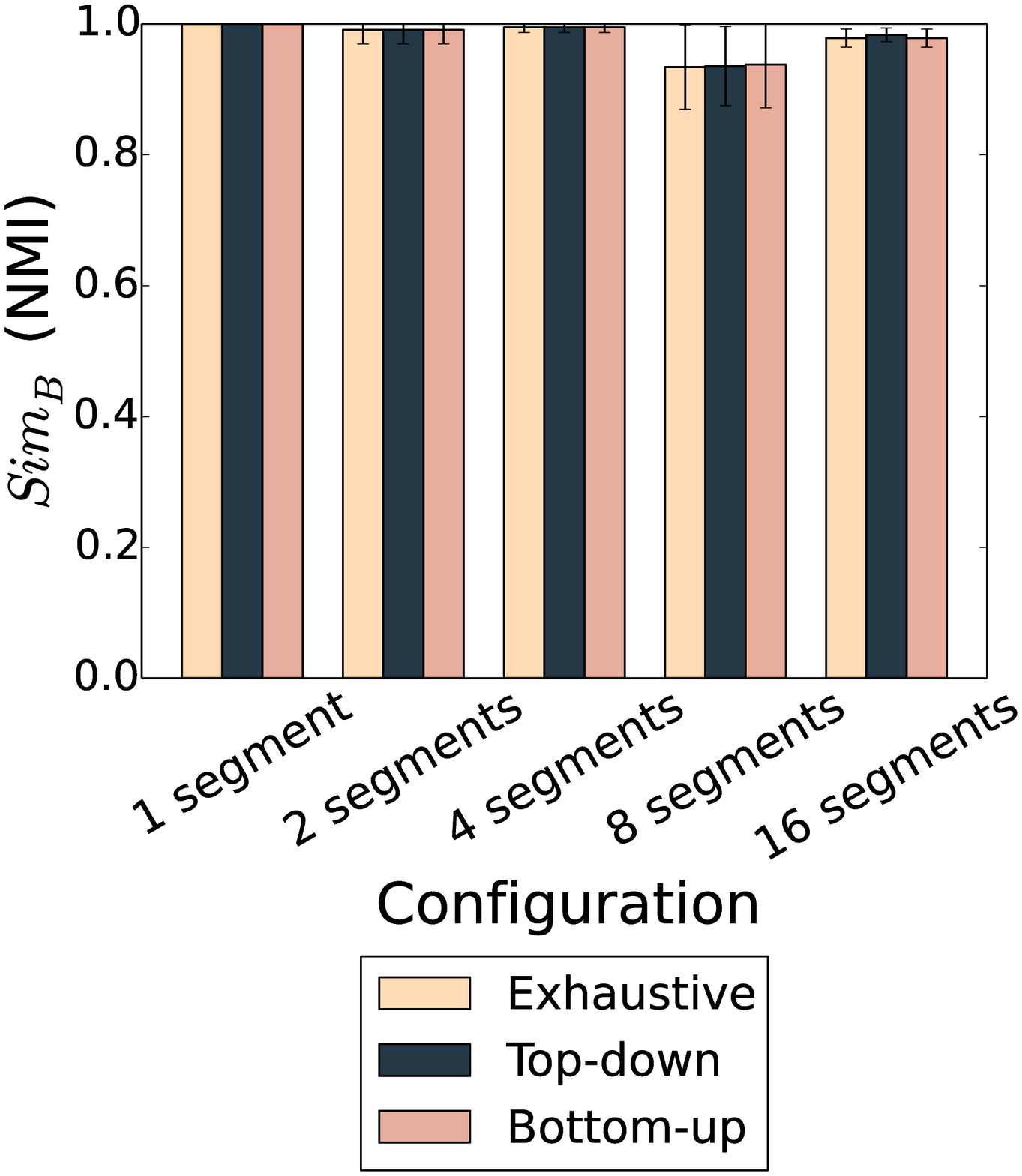}}
\hspace{0.2cm}
\subfloat[]{\label{fig:supplement:scout_search-run_time}\includegraphics[width=.45\linewidth]{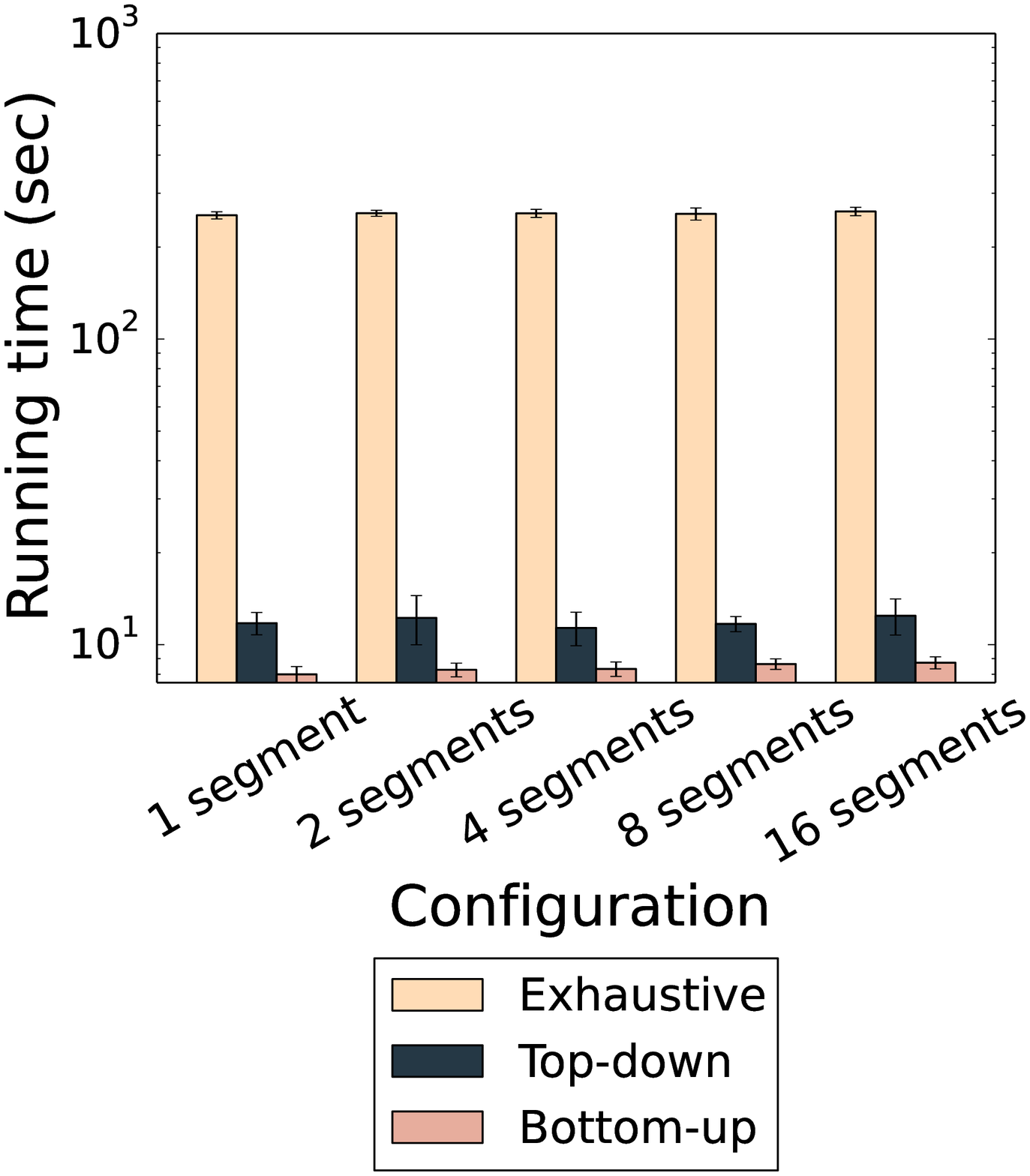}}
  \caption[The effect of the choice of search strategy for SCOUT]{The effect of the choice of search strategy for SCOUT, in terms of
 \textbf{\protect\subref{fig:supplement:scout_search-gl_p__nmi}} $Sim_B$
and \textbf{\protect\subref{fig:supplement:scout_search-run_time}} running time  (logarithmic scale),
as we vary the ground truth synthetic network configuration ($x$-axis).
}
\label{fig:supplement:scout_search}
\end{figure}

\textbf{c)} Search strategy is used to determine how SCOUT searches through the space of possible segmentations.
We evaluate three strategies: the exhaustive search, top-down search, and bottom-up search (Supplementary Section~\ref{sec:supplement:scout_search_strategy}).
In terms of $Sim_B$, all three strategies lead to comparable results (\figurename~\ref{fig:supplement:scout_search-gl_p__nmi}).
In terms of the running time, bottom-up search is the fastest one (it is an order of magnitude faster than the exhaustive search and somewhat faster than the top-down search; \figurename~\ref{fig:supplement:scout_search-run_time}).
Thus, we focus on the bottom-up search as SCOUT's search strategy.

\clearpage
\section*{Supplementary Figures}
\addcontentsline{toc}{section}{Supplementary Figures} 

\clearpage

\begin{figure}
\centering
\subfloat[]{\label{fig:supplement:multistep-gl_p__nmi}\includegraphics[width=.46\linewidth]{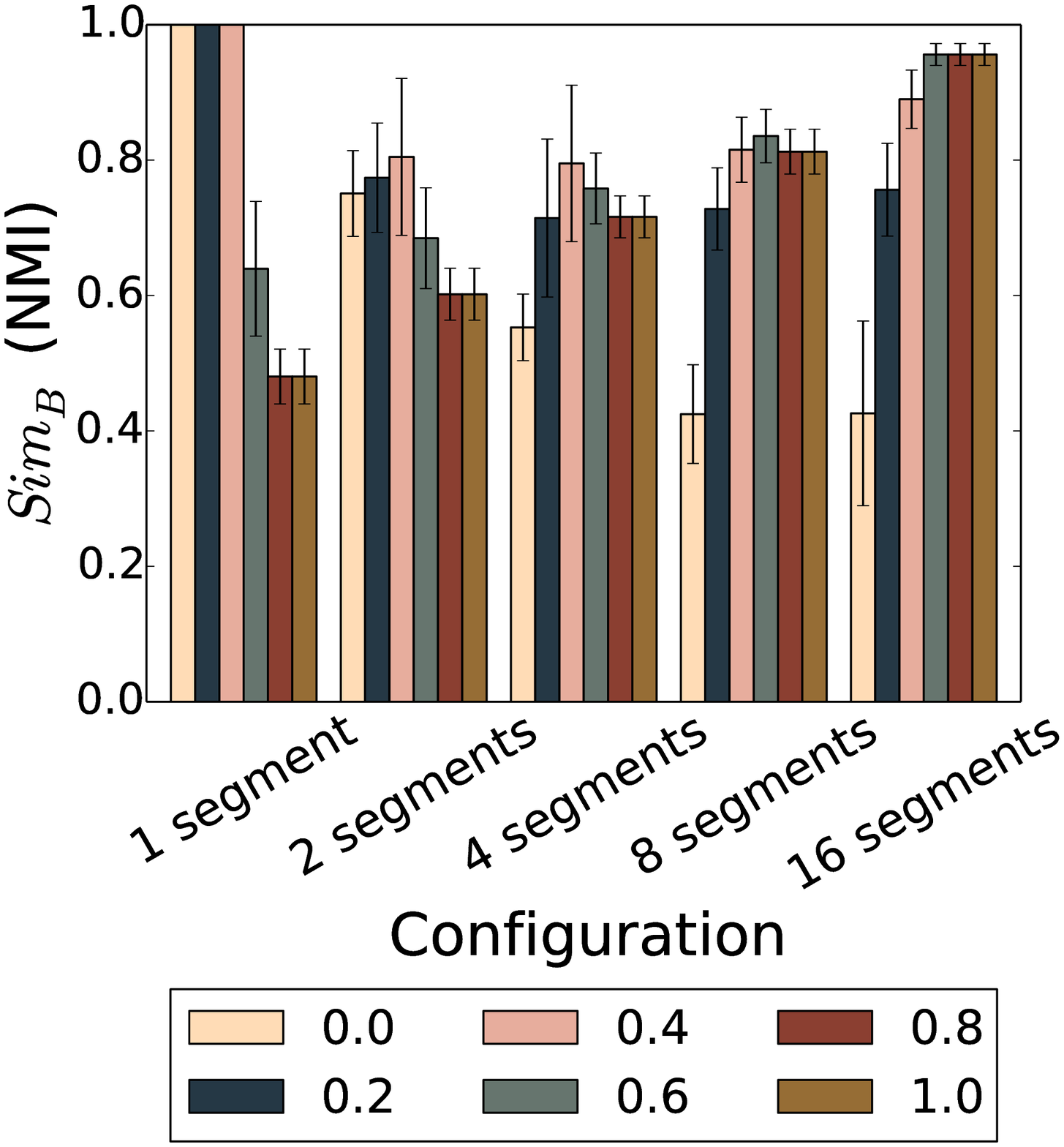}}
\hspace{0.2cm}
\subfloat[]{\vspace{1mm}\label{fig:supplement:multistep-sgmq__num}\includegraphics[width=.45\linewidth]{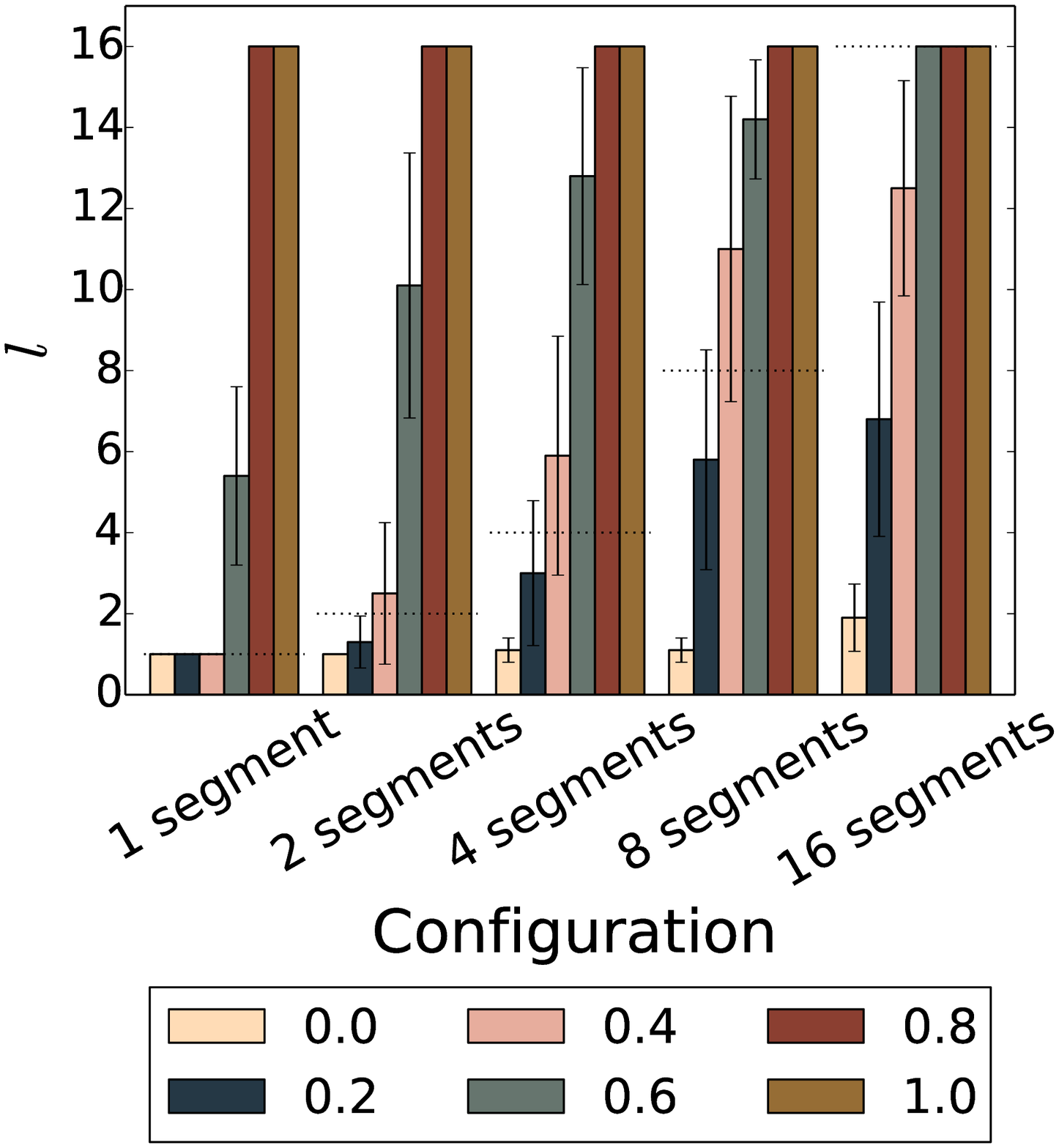}}
\caption[The effect of $\theta$ value for Multi-Step]{The effect of $\theta$ value for Multi-Step, in terms of
  \textbf{\protect\subref{fig:supplement:multistep-gl_p__nmi}} $Sim_B$ 
  and \textbf{\protect\subref{fig:supplement:multistep-sgmq__num}} the number of segments $l$,
as we vary the ground truth synthetic network configuration ($x$-axis).
Note that we run the analysis for $\theta \in [0,2]$ in increments of $0.1$.
However, since the results for $\theta \geq 1.0$ are all the same, we show the results only for $\theta \leq 1.0$, and for visual clarity, we show the results in increments of $0.2$.
In panel \protect\subref{fig:supplement:multistep-sgmq__num}, for a given ground truth configuration, the dotted line corresponds to the ground truth number of segments. 
  }
\label{fig:supplement:multistep}
\end{figure}

\begin{figure}
\centering
\includegraphics[width=.45\linewidth]{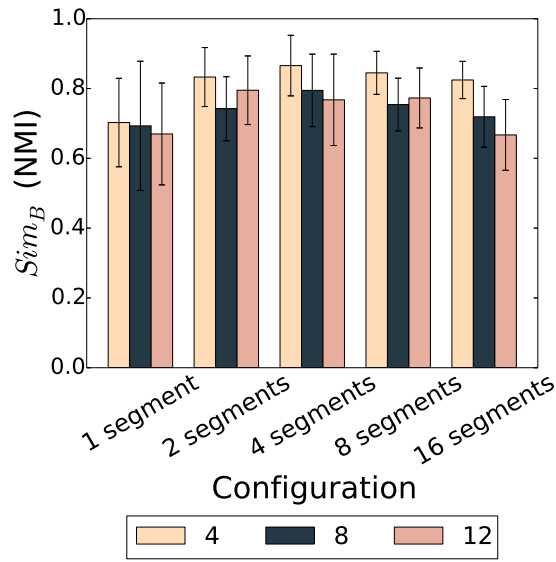}
\caption[The effect of $w$ value for GHRG]{The effect of $w$ value for GHRG, in terms of $Sim_B$,
as we vary the ground truth synthetic network configuration ($x$-axis).
}
\label{fig:supplement:ghrg}
\end{figure}

\clearpage

\clearpage

\begin{figure*}
\centering
\subfloat[$Sim_T$]{
\begin{minipage}[t]{0.315\linewidth}
\includegraphics[width=\linewidth]{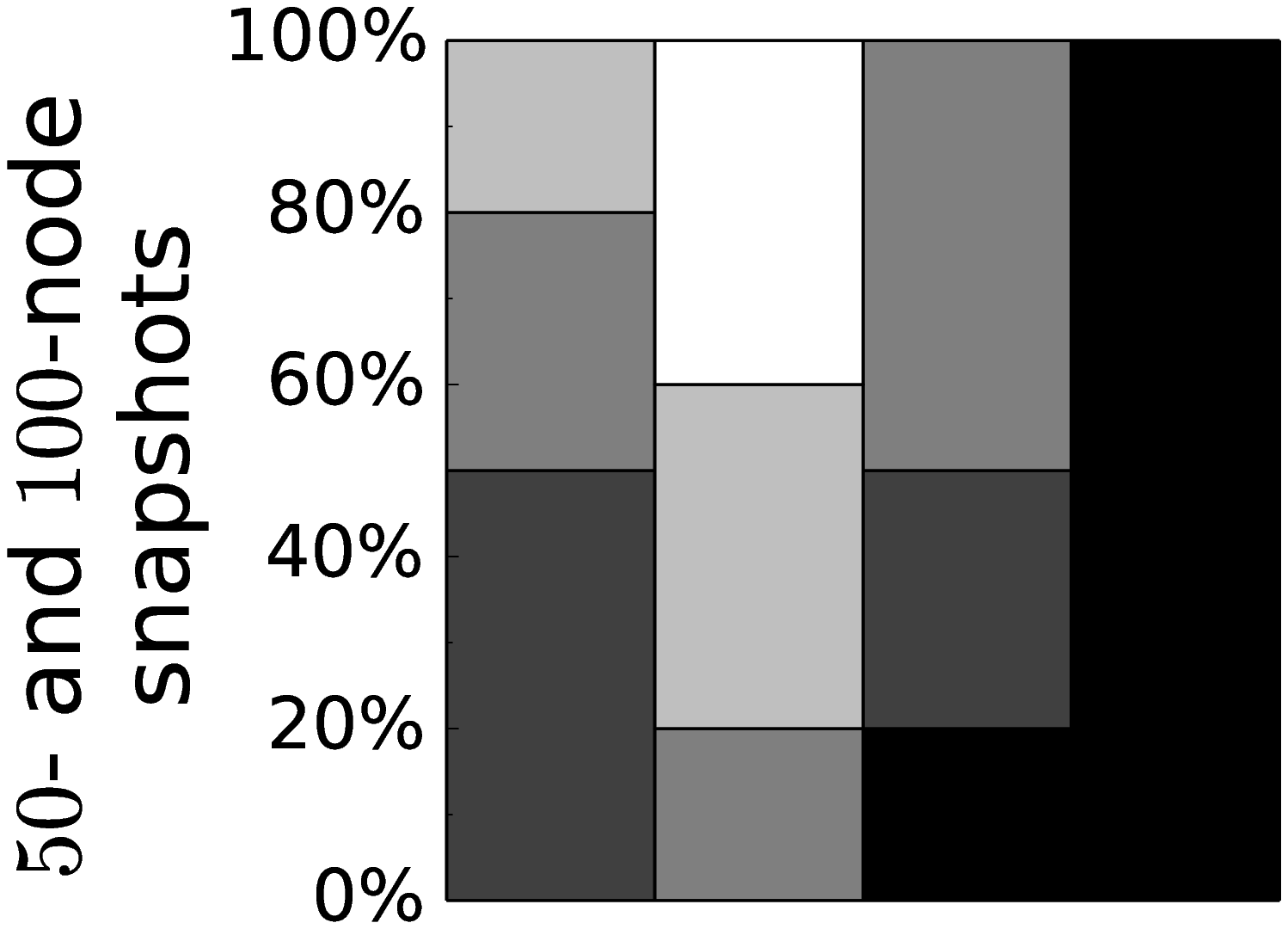}
\includegraphics[width=\linewidth]{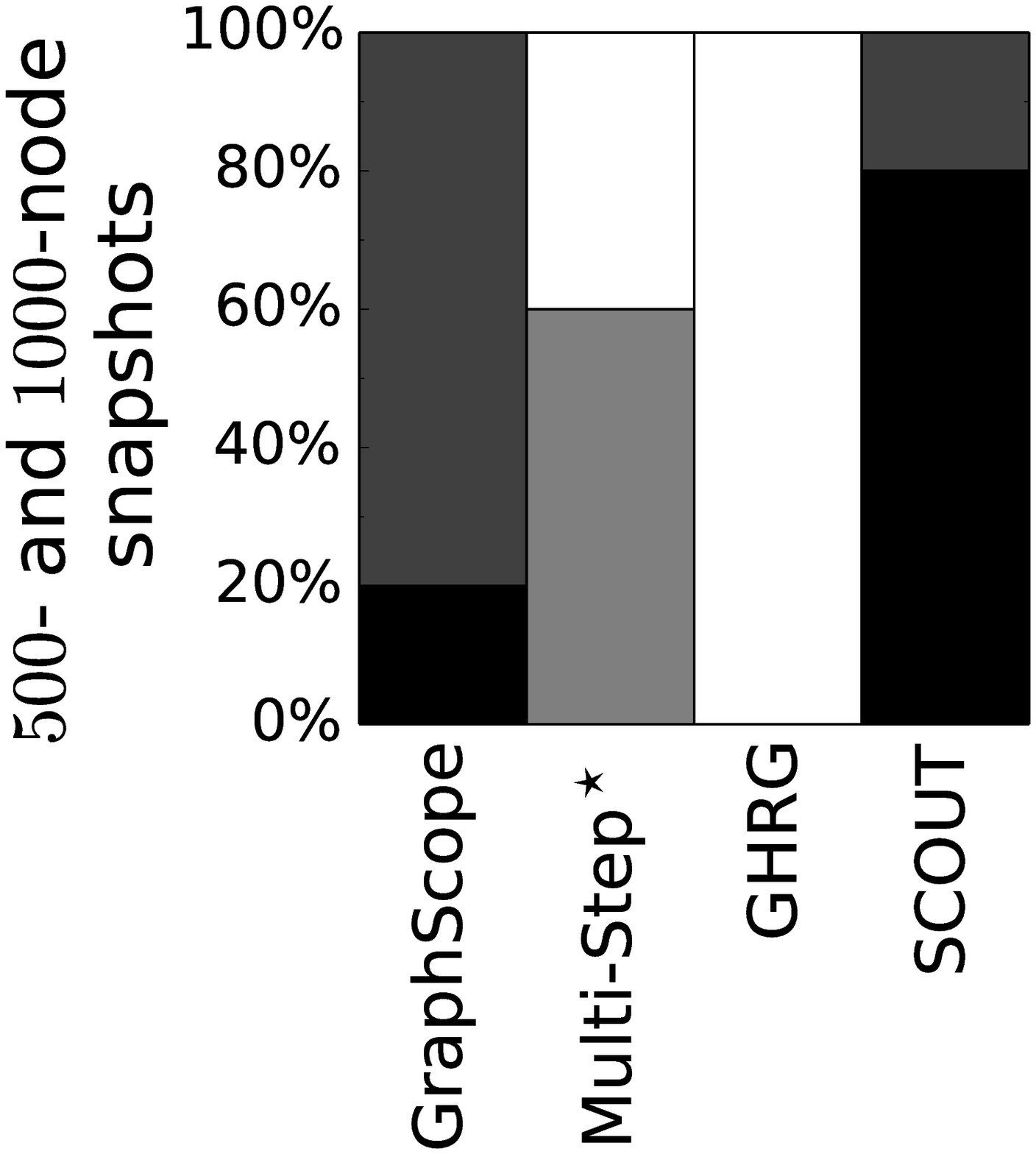}
\label{fig:supplement:synth_rank_sim_t}
\end{minipage} 
}
\subfloat[Change point classification]{
\begin{minipage}[t]{0.38\linewidth}
\includegraphics[width=\linewidth]{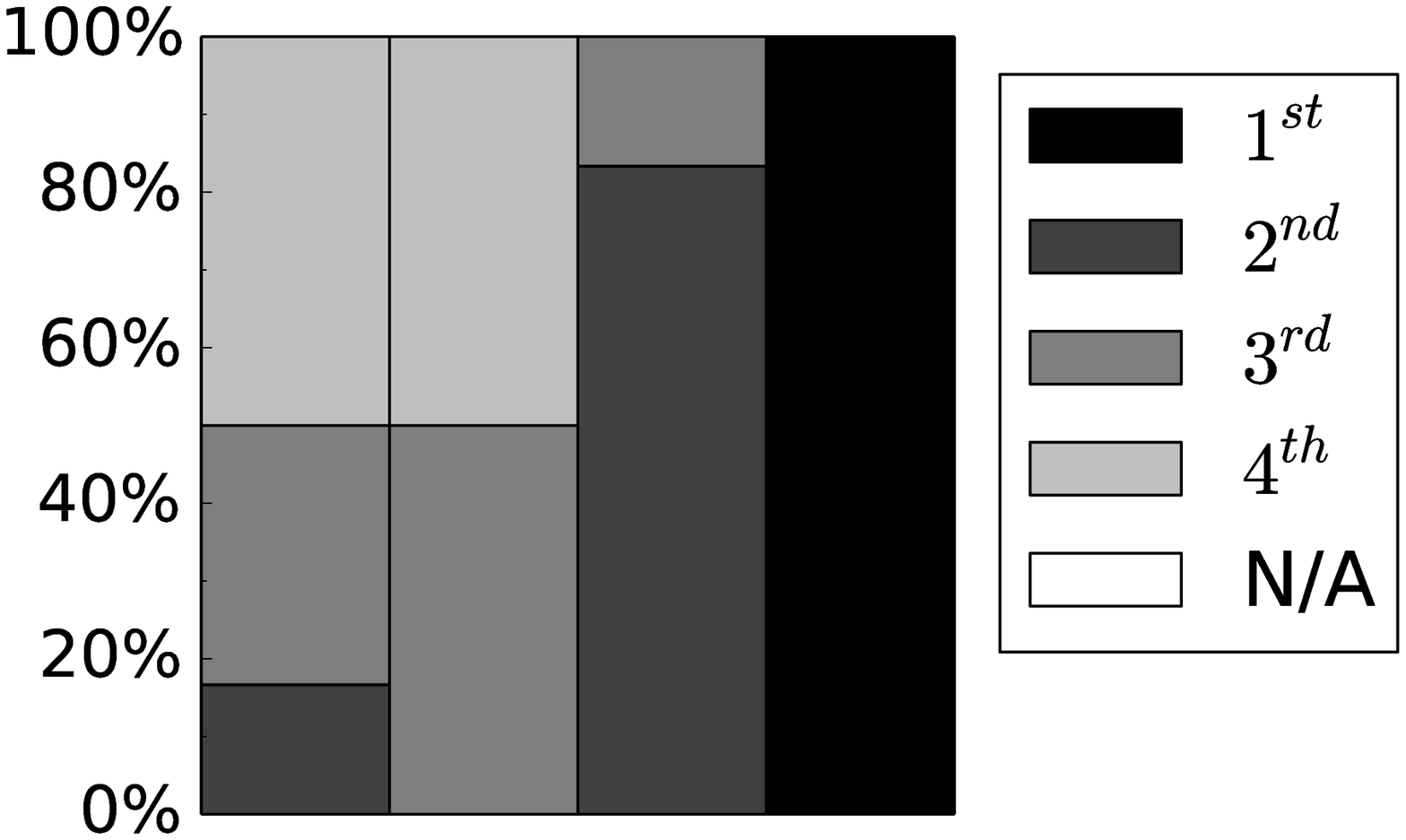}
\includegraphics[width=0.695\linewidth]{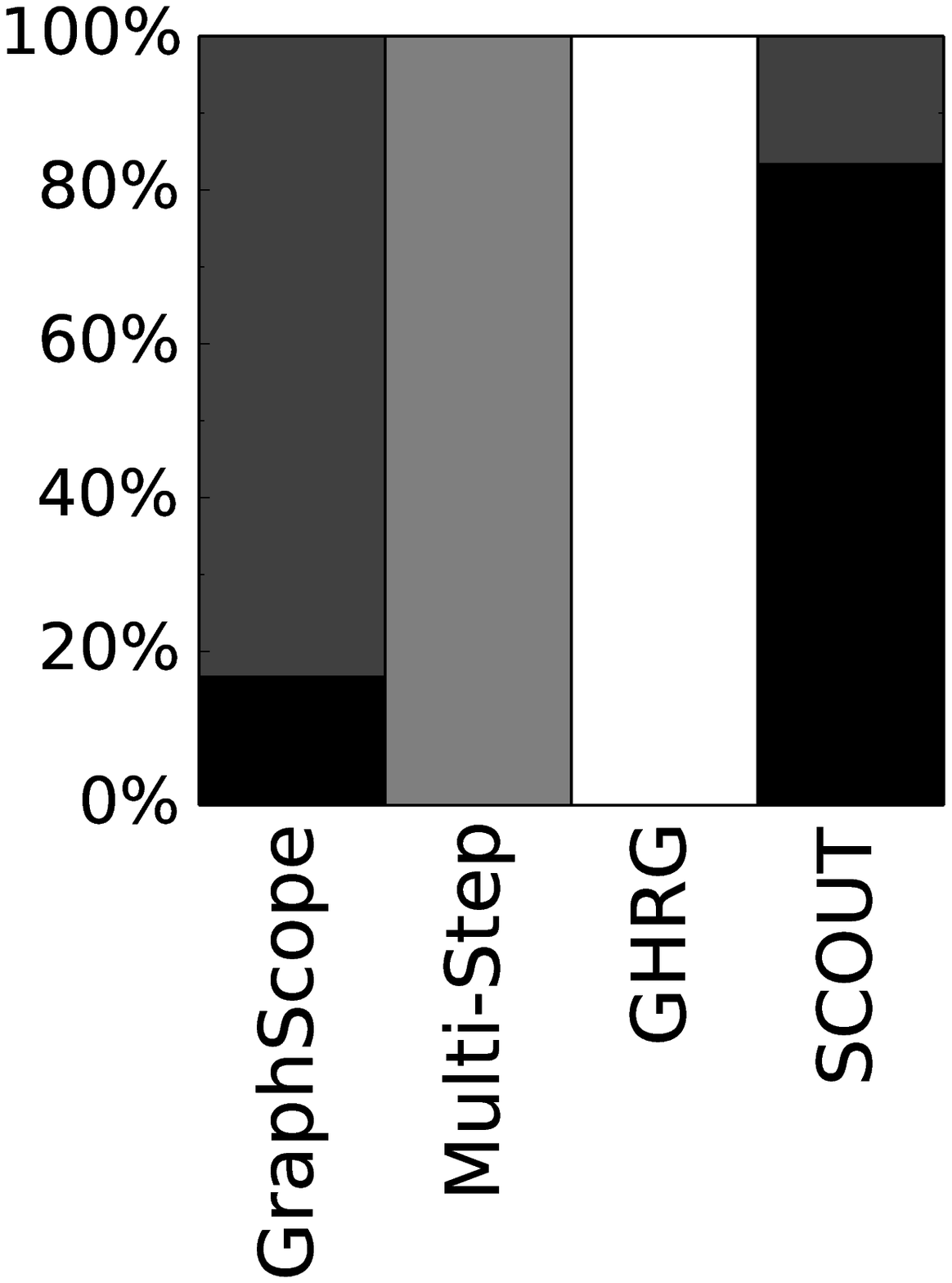}
\vspace{5mm} 
\label{fig:supplement:synth_rank_aupr}
\end{minipage} 
}
\caption[Rankings of the methods with respect to $Sim_T$ and change point classification]{Rankings of the methods with respect to 
\textbf{\protect\subref{fig:supplement:synth_rank_sim_t}} $Sim_T$ and
\textbf{\protect\subref{fig:supplement:synth_rank_aupr}} change point classification.
Since GHRG could not be run for the larger networks, the results are split into those for the configurations with 50 and 100 nodes per snapshot (top) and those for the configurations with 500 and 1000 nodes per snapshot (bottom).
Note that for change point classification, we exclude from consideration configurations with the minimum (i.e., one) and maximum (i.e., 16) possible numbers of ground truth segments.
This is because for these configurations, either there are no change points at all (i.e., for one segment) or every time point is a change point (i.e., for 16 segments), so change point classification cannot be performed.
The rankings are computed as follows. 
For each synthetic network configuration, we compare the four methods' scores (average scores over all instances of the given configuration) to identify the first, second, third, and fourth best method; ties are allowed, in which case, two methods would be assigned the same rank. 
Then, we summarize these results over all considered synthetic network configurations by measuring, for each method ($x$-axis), how many times the given method is ranked as  the first, second, third, and fourth best method (expressed as the percentage of all considered configurations; $y$-axis).
``N/A'' indicates that either a given method could not be run (which is the case for GHRG for the larger networks in both panels) or it was excluded from the consideration (which is the case for Multi-Step in panel \protect\subref{fig:supplement:synth_rank_sim_t}).
The reason we exclude Multi-Step from the consideration in certain configurations in panel \protect\subref{fig:supplement:synth_rank_sim_t}  (namely, those with the minimum and maximum possible numbers of segments) is as follows.
Since we provide Multi-Step with the ground truth number of segments as input and since there is only one possible segmentation with minimum or maximum possible number of segments, Multi-Step trivially returns the perfect $Sim_T$ score for these extreme configurations.
The figure can intuitively be interpreted as follows: the darker the bar of a given method, the better its performance.
}
\end{figure*}

\begin{figure}
\centering
\resizebox{\linewidth}{!}{
\subfloat[1 segment]{
\begin{minipage}[t]{0.23\linewidth}
\includegraphics[width=\linewidth]{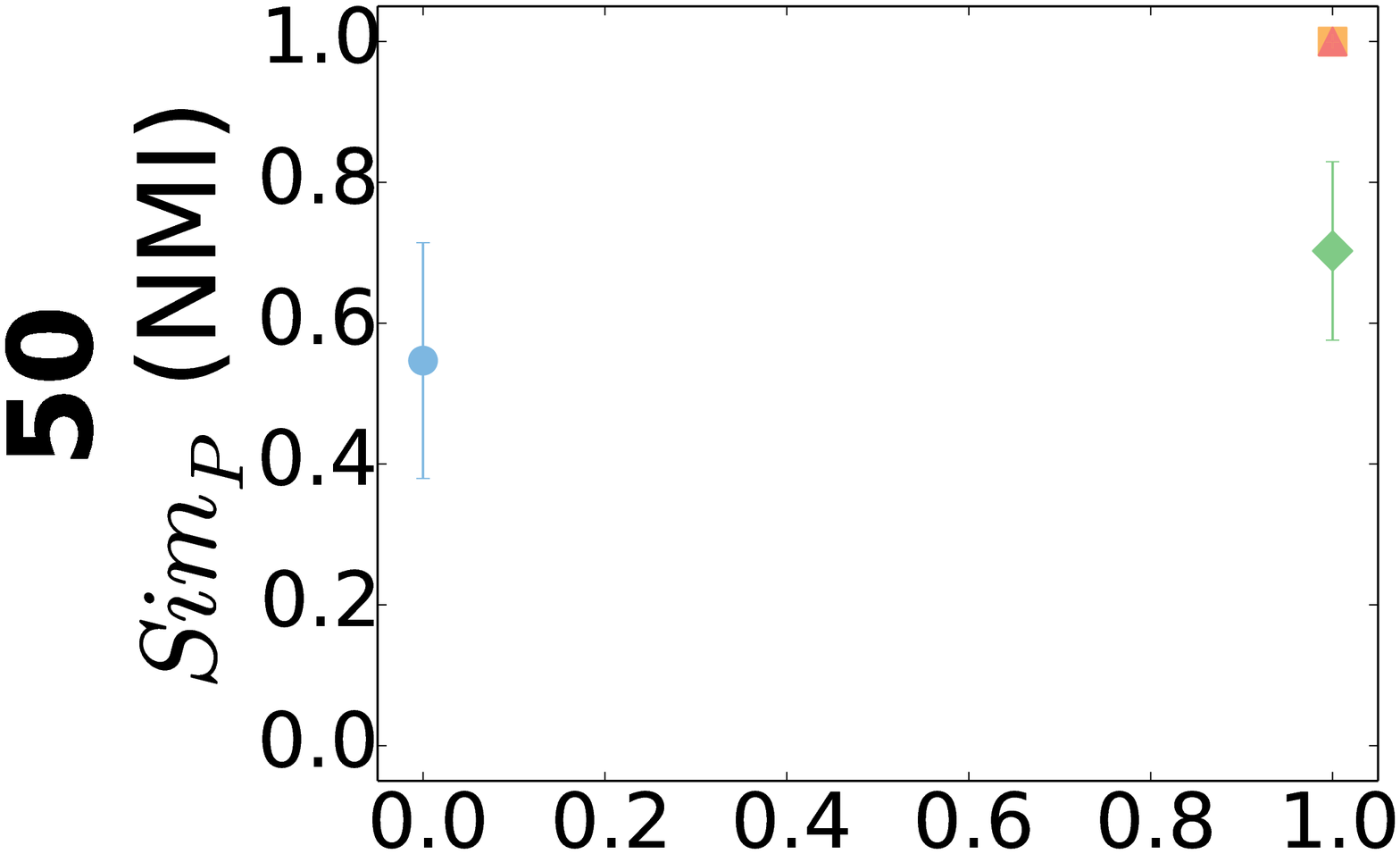}
\includegraphics[width=\linewidth]{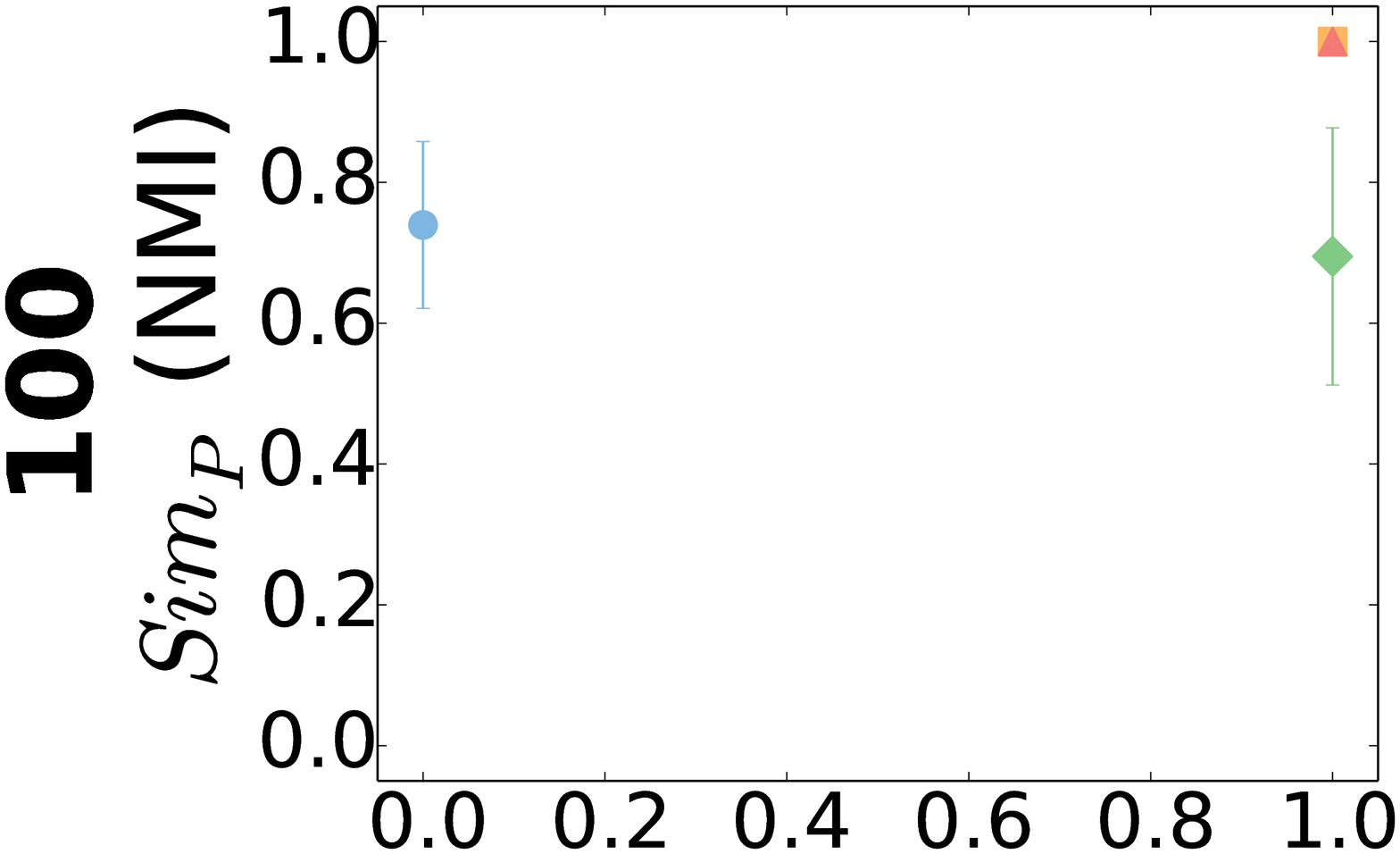}
\includegraphics[width=\linewidth]{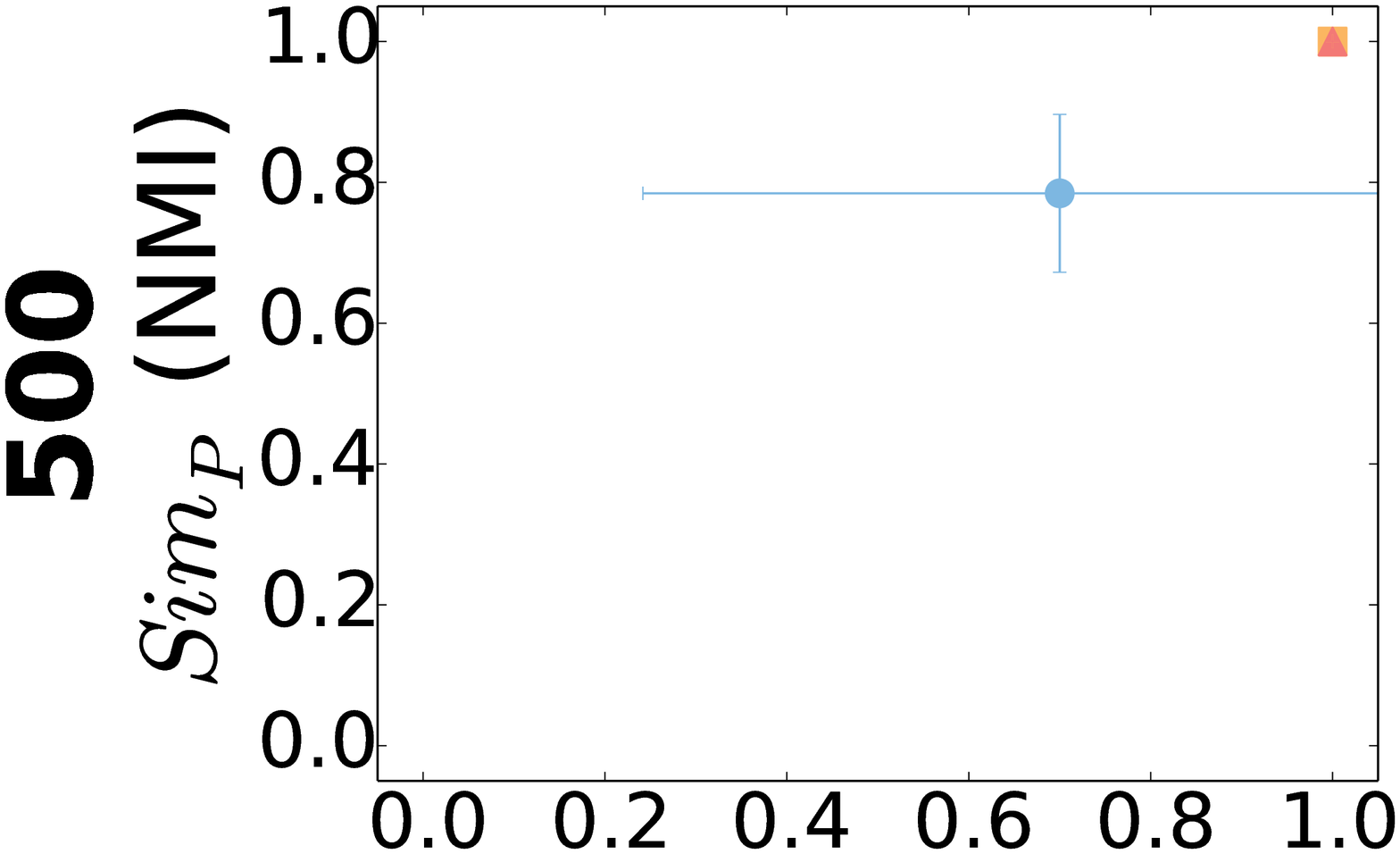}
\includegraphics[width=\linewidth]{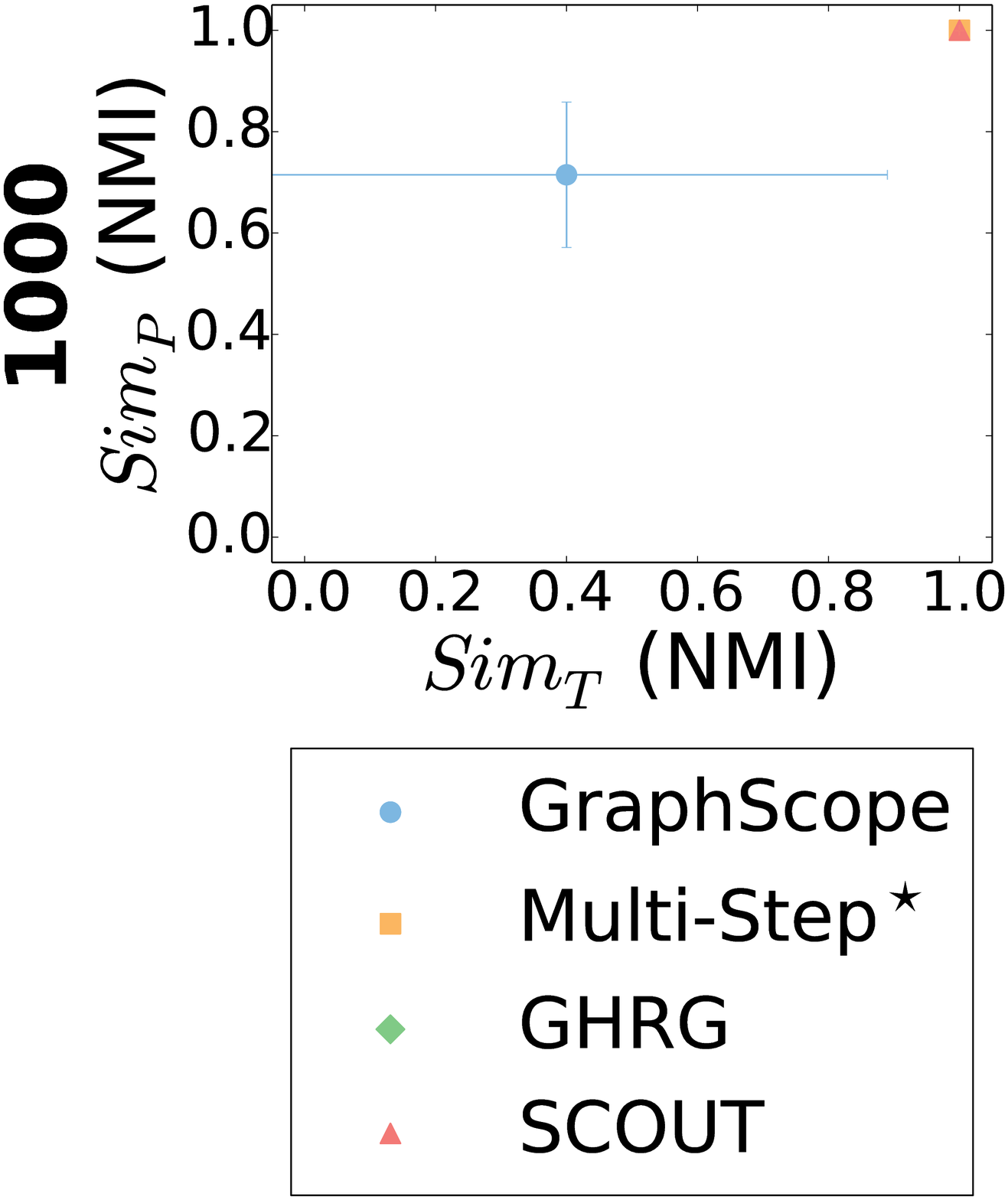}
\label{fig:supplement:scatter_1segment}
\end{minipage} 
}
\subfloat[2 segments]{
\begin{minipage}[t]{0.19\linewidth}
\includegraphics[width=\linewidth]{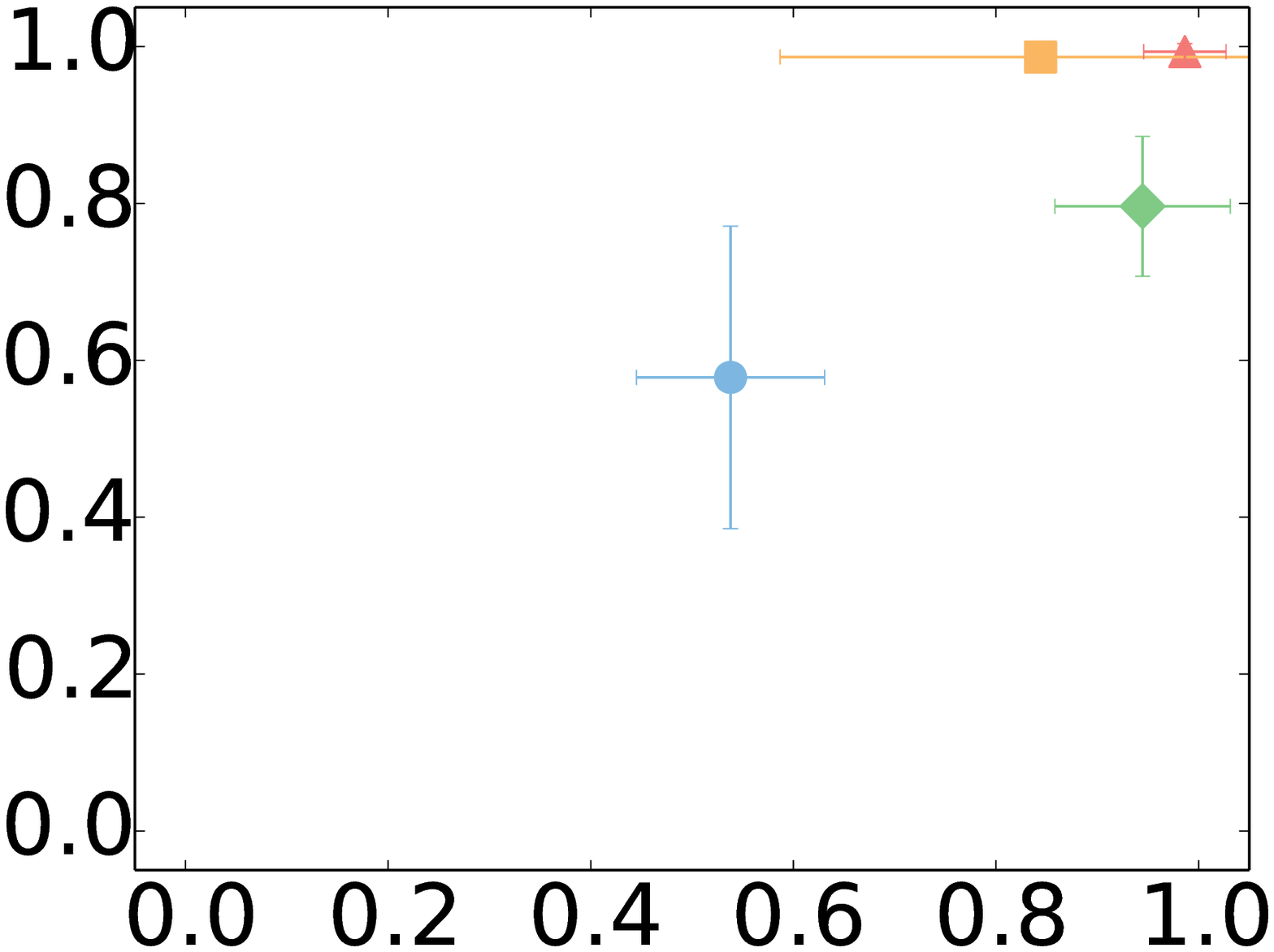}
\includegraphics[width=\linewidth]{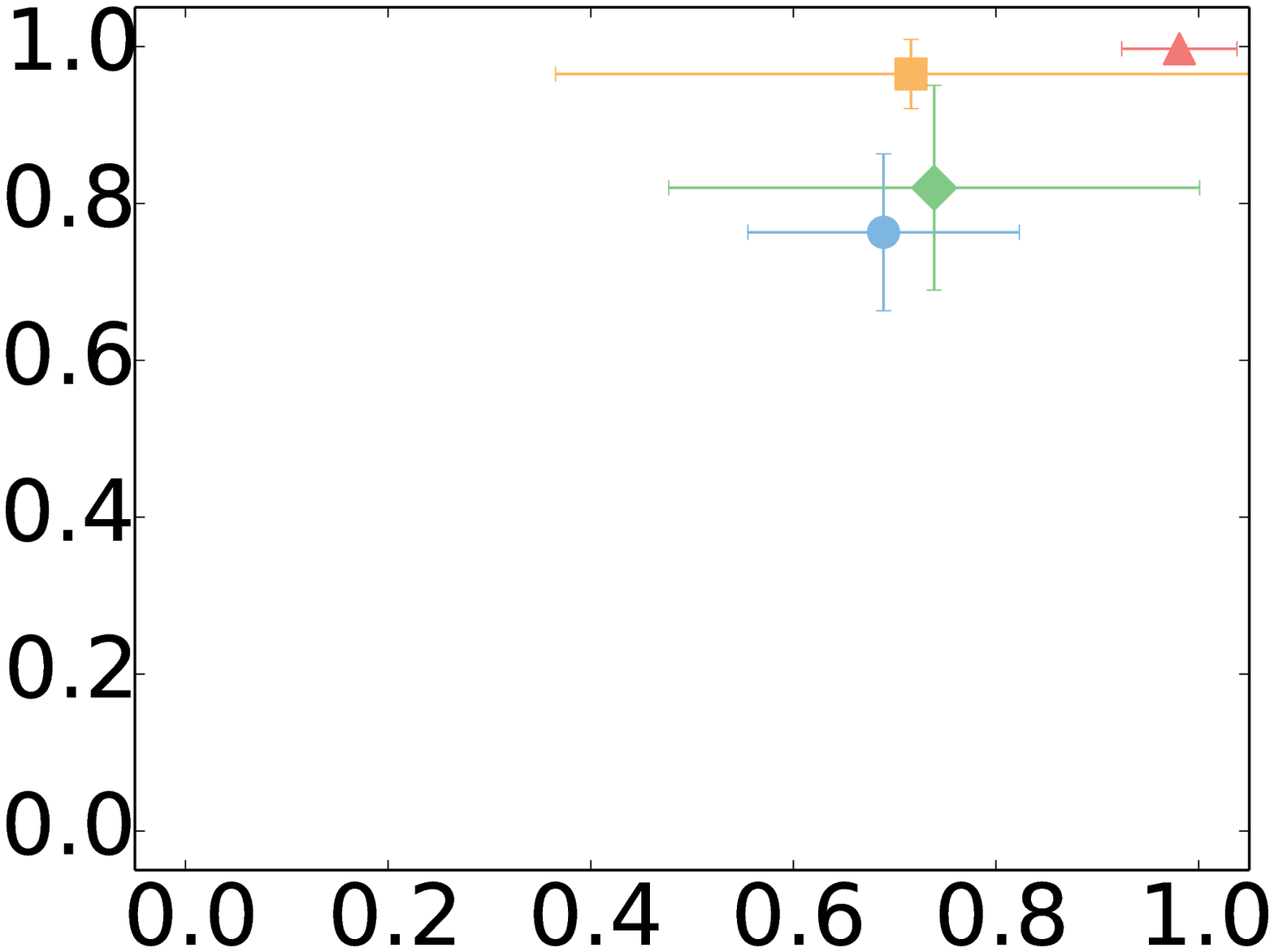}
\includegraphics[width=\linewidth]{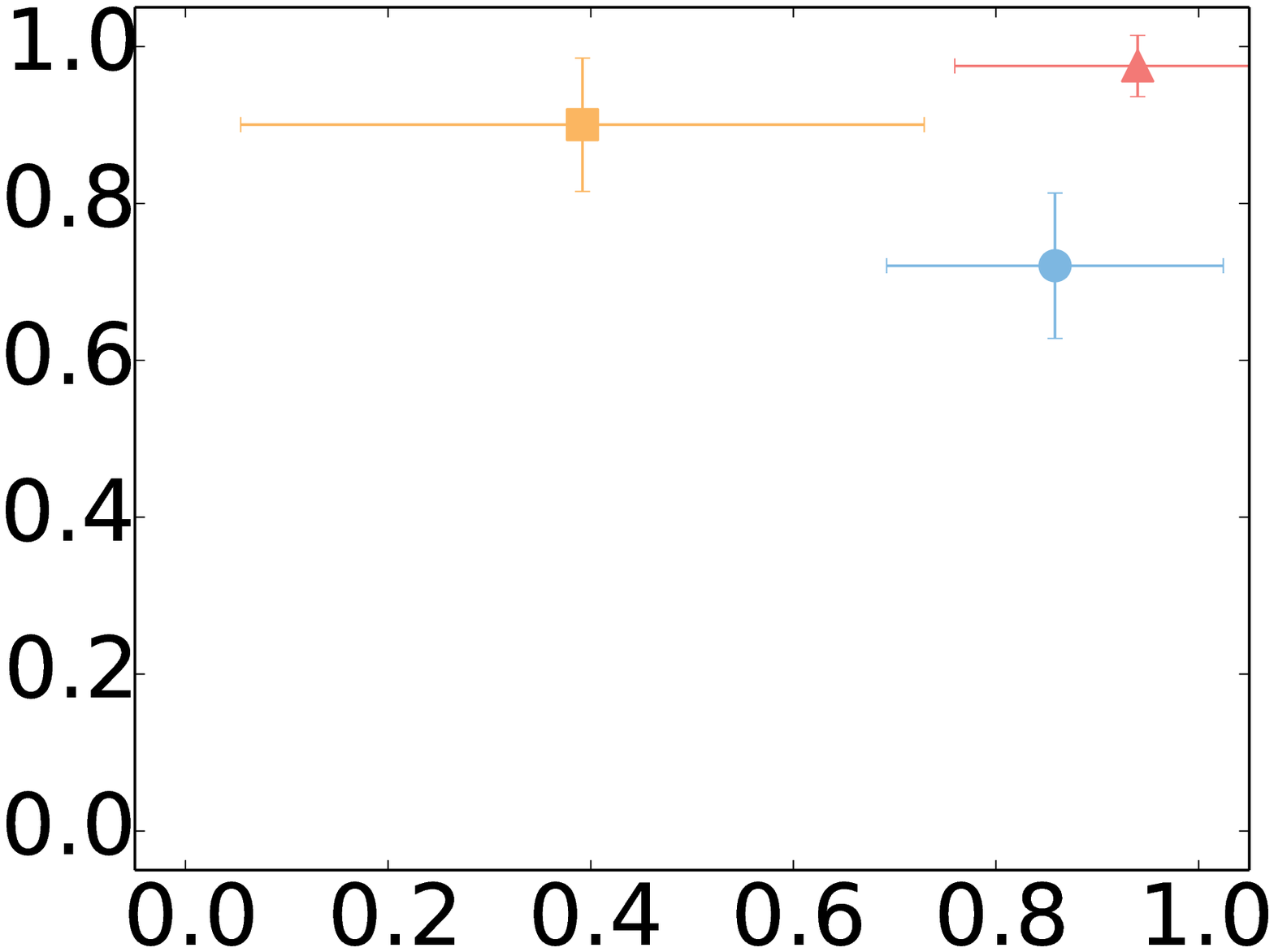}
\includegraphics[width=\linewidth]{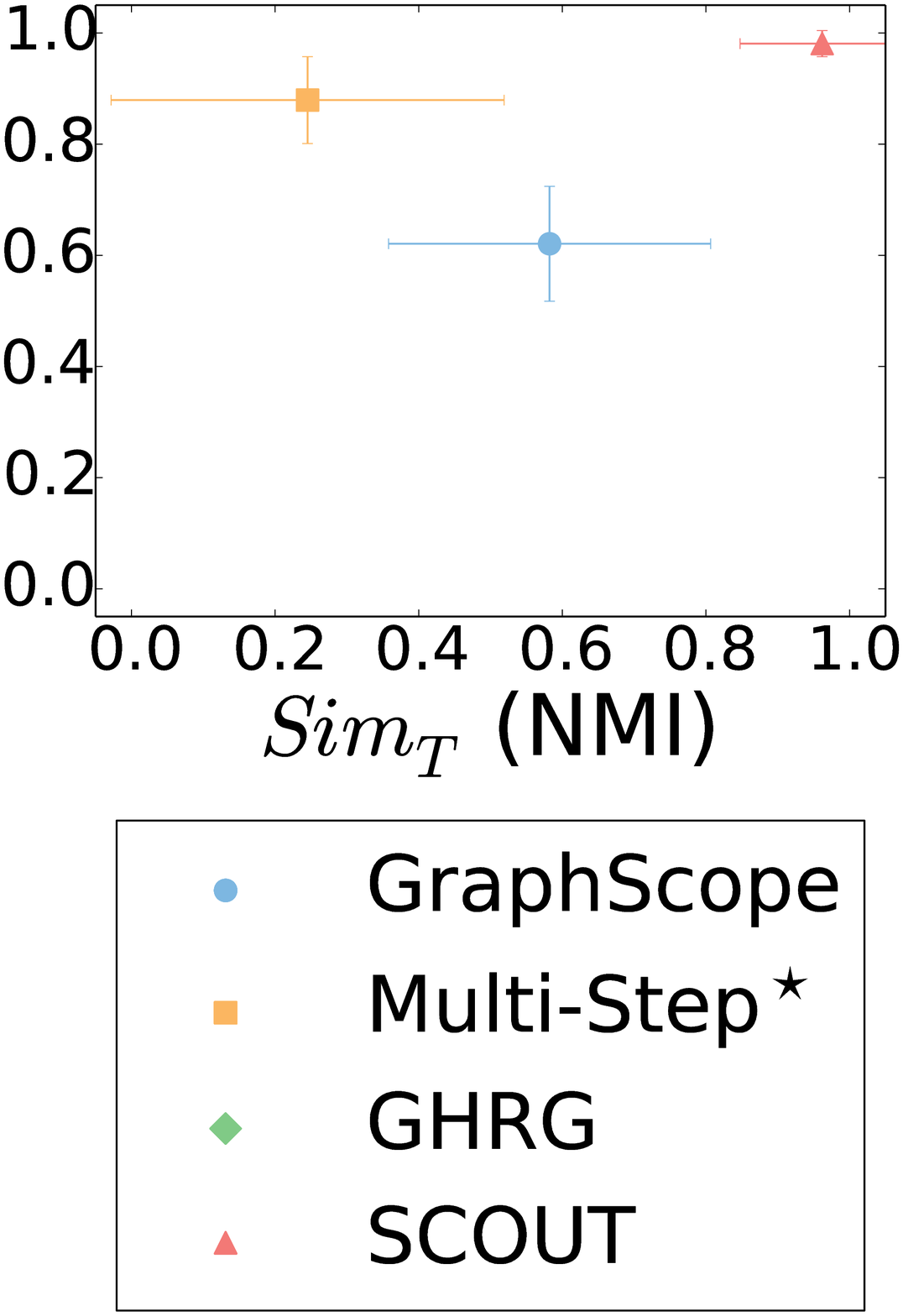}
\label{fig:supplement:scatter_2segments}
\end{minipage} 
}
\subfloat[4 segments]{
\begin{minipage}[t]{0.19\linewidth}
\includegraphics[width=\linewidth]{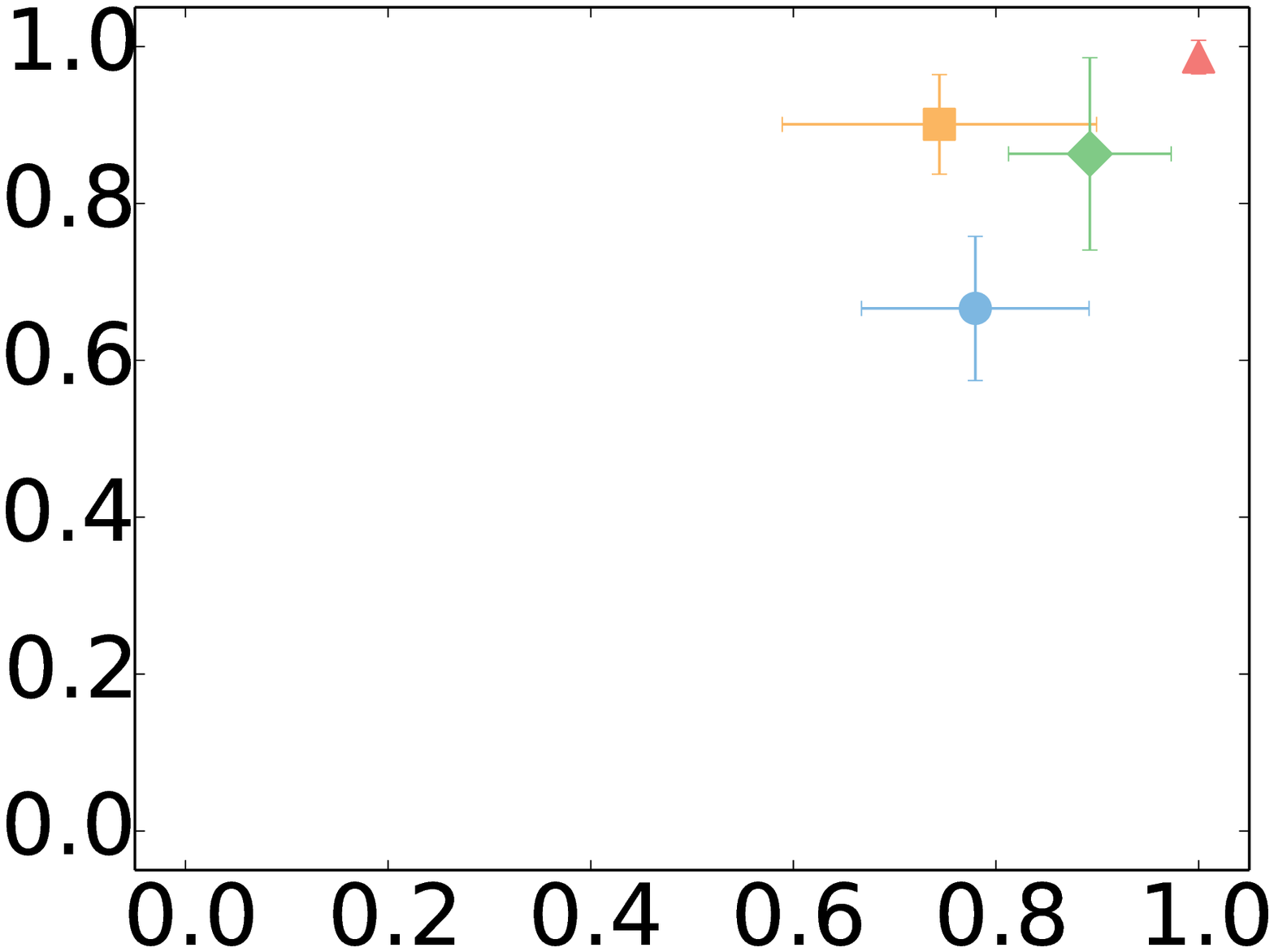}
\includegraphics[width=\linewidth]{synthetic_100/4_segments_avg_seg_t__nmi_-vs-_avg_sn_p__nmi}
\includegraphics[width=\linewidth]{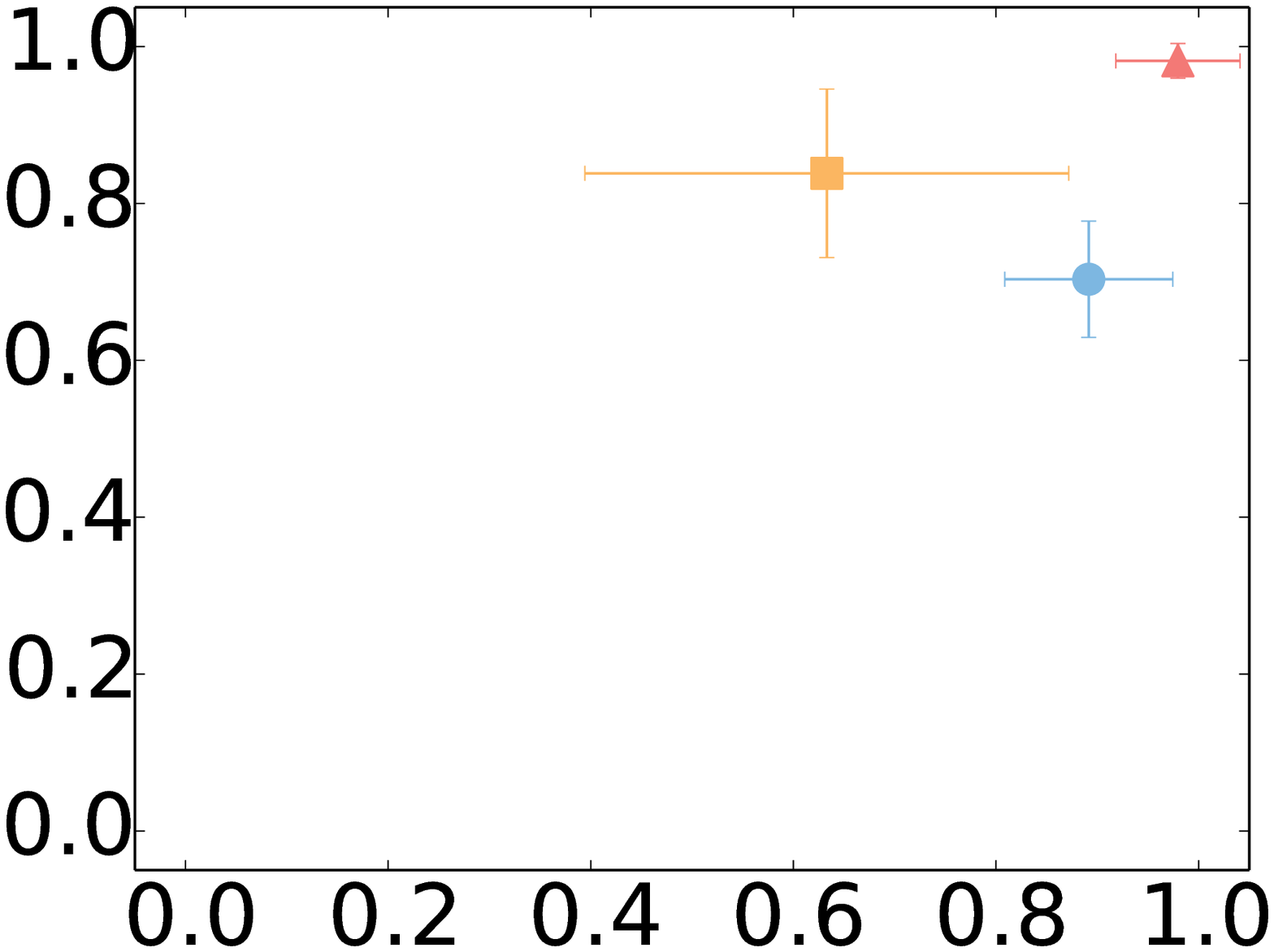}
\includegraphics[width=\linewidth]{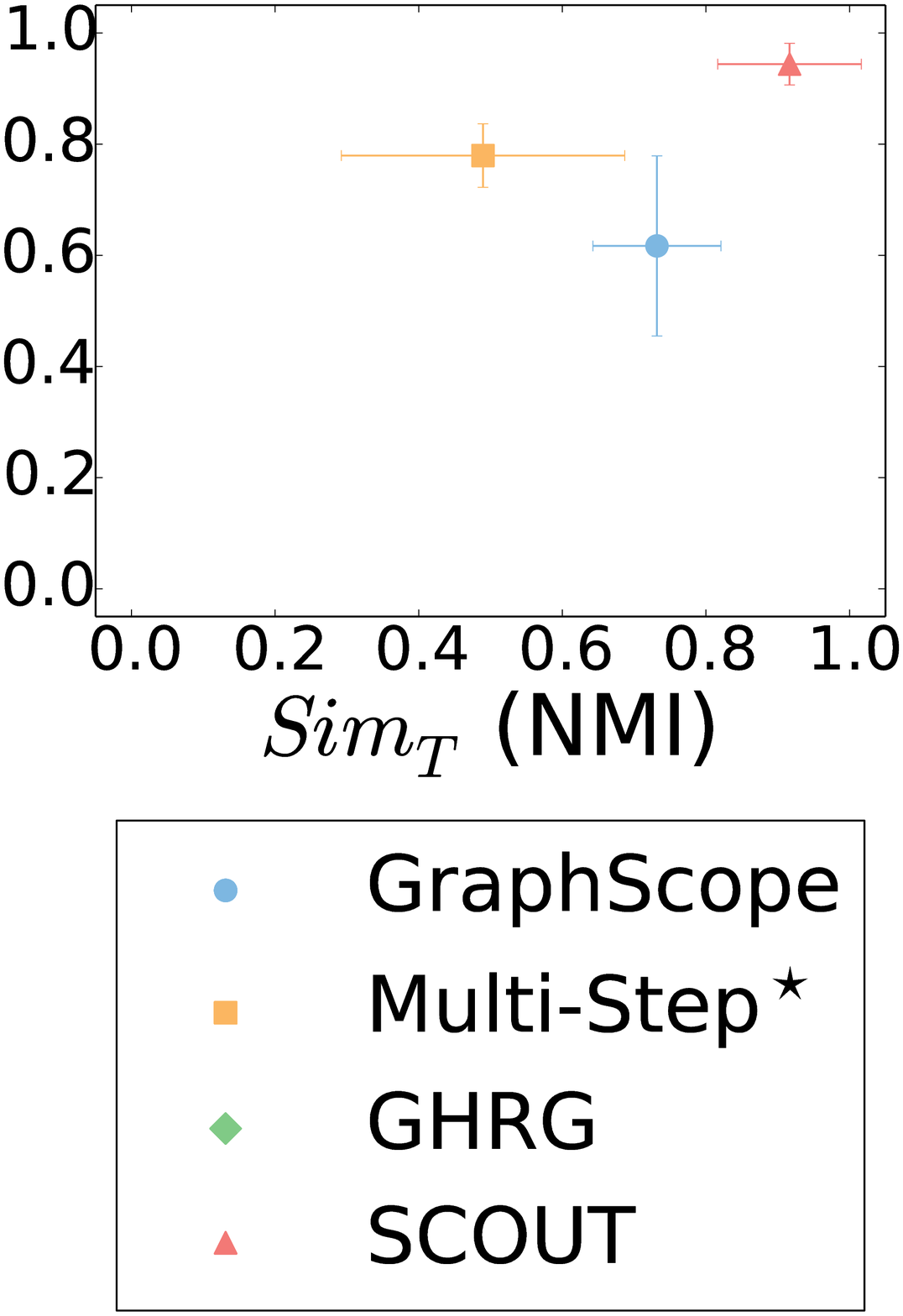}
\label{fig:supplement:scatter_4segments}
\end{minipage} 
}
\subfloat[8 segments]{
\begin{minipage}[t]{0.19\linewidth}
\includegraphics[width=\linewidth]{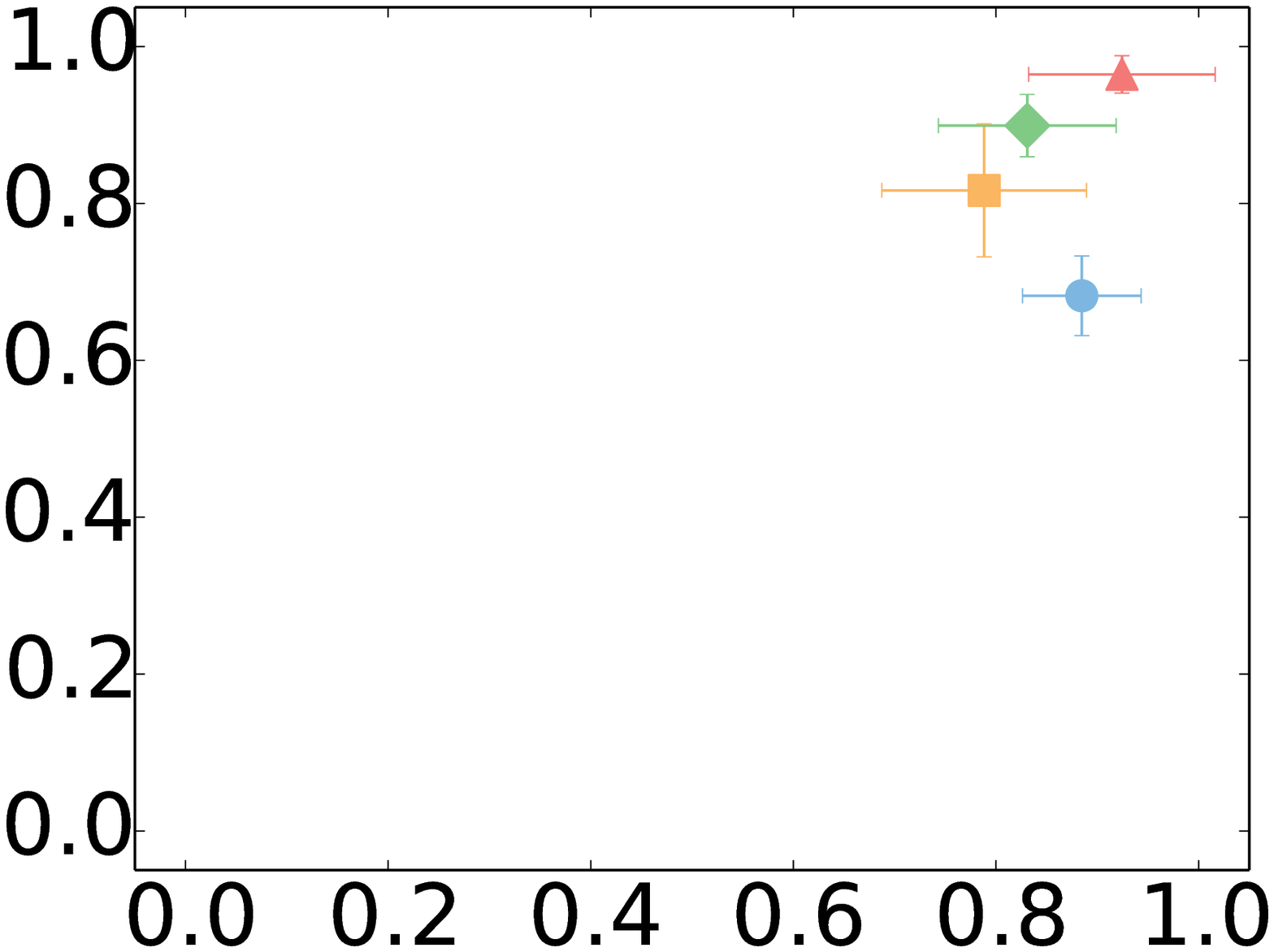}
\includegraphics[width=\linewidth]{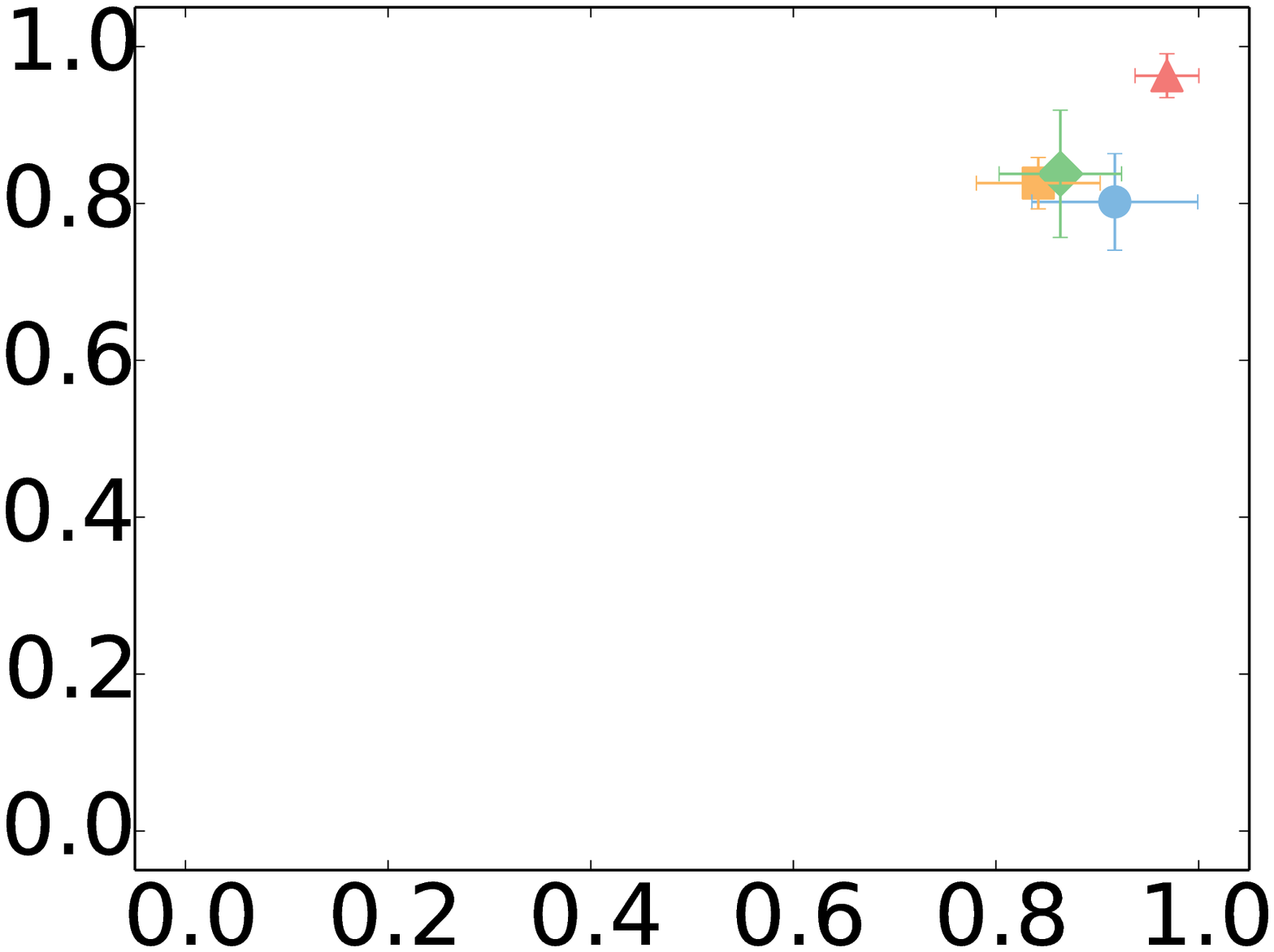}
\includegraphics[width=\linewidth]{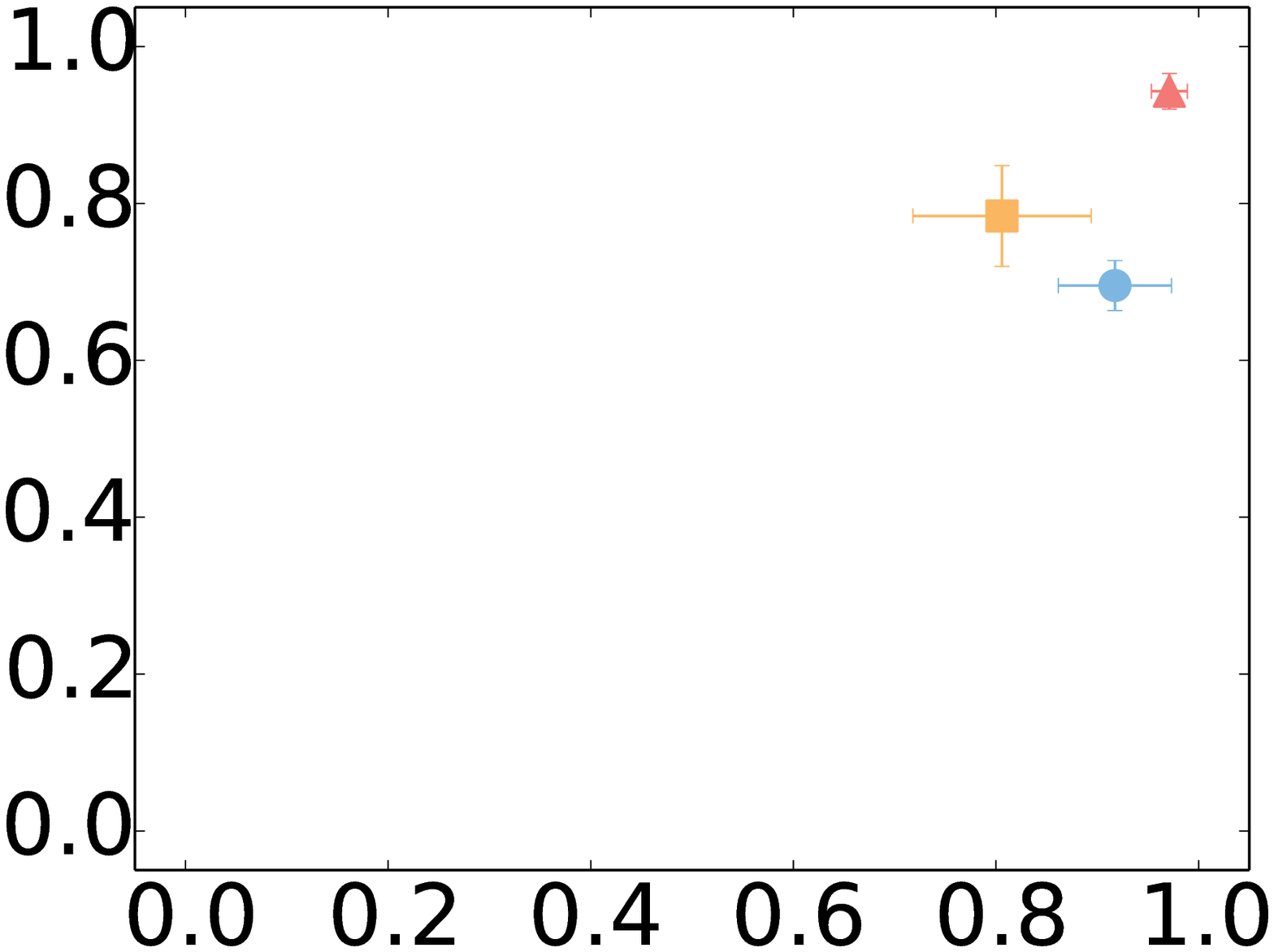}
\includegraphics[width=\linewidth]{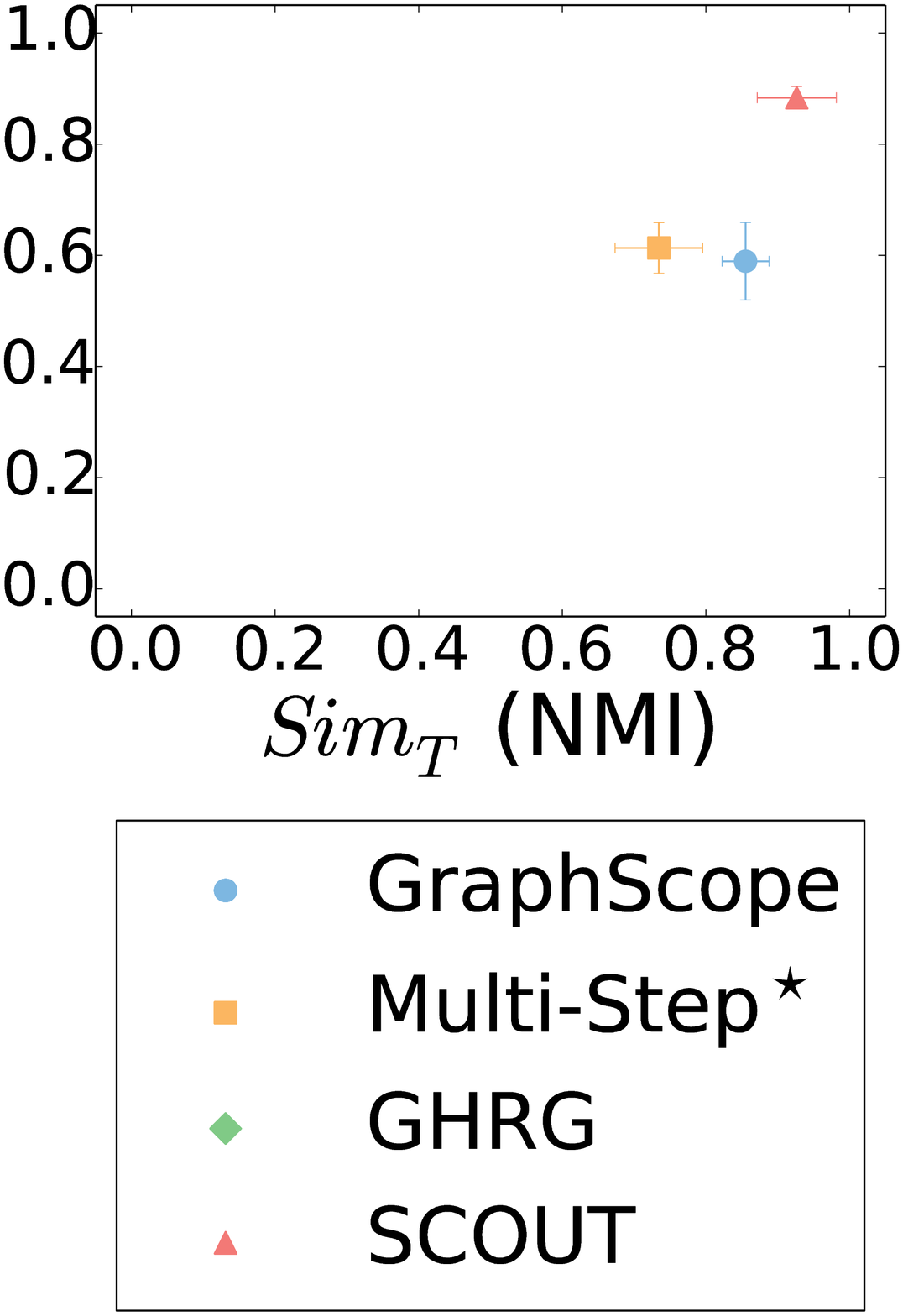}
\label{fig:supplement:scatter_8segments}
\end{minipage} 
}
\subfloat[16 segments]{
\begin{minipage}[t]{0.19\linewidth}
\includegraphics[width=\linewidth]{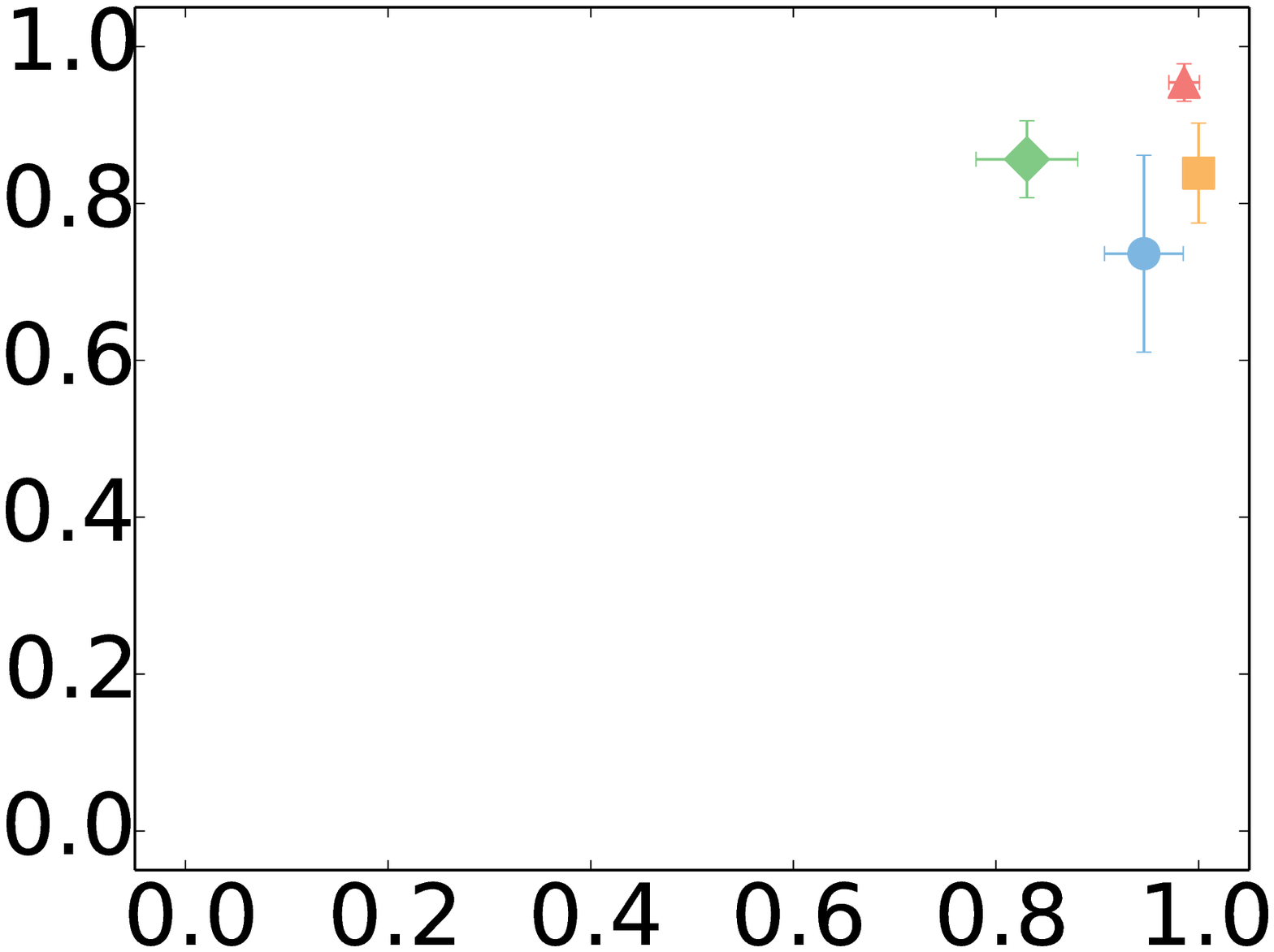}
\includegraphics[width=\linewidth]{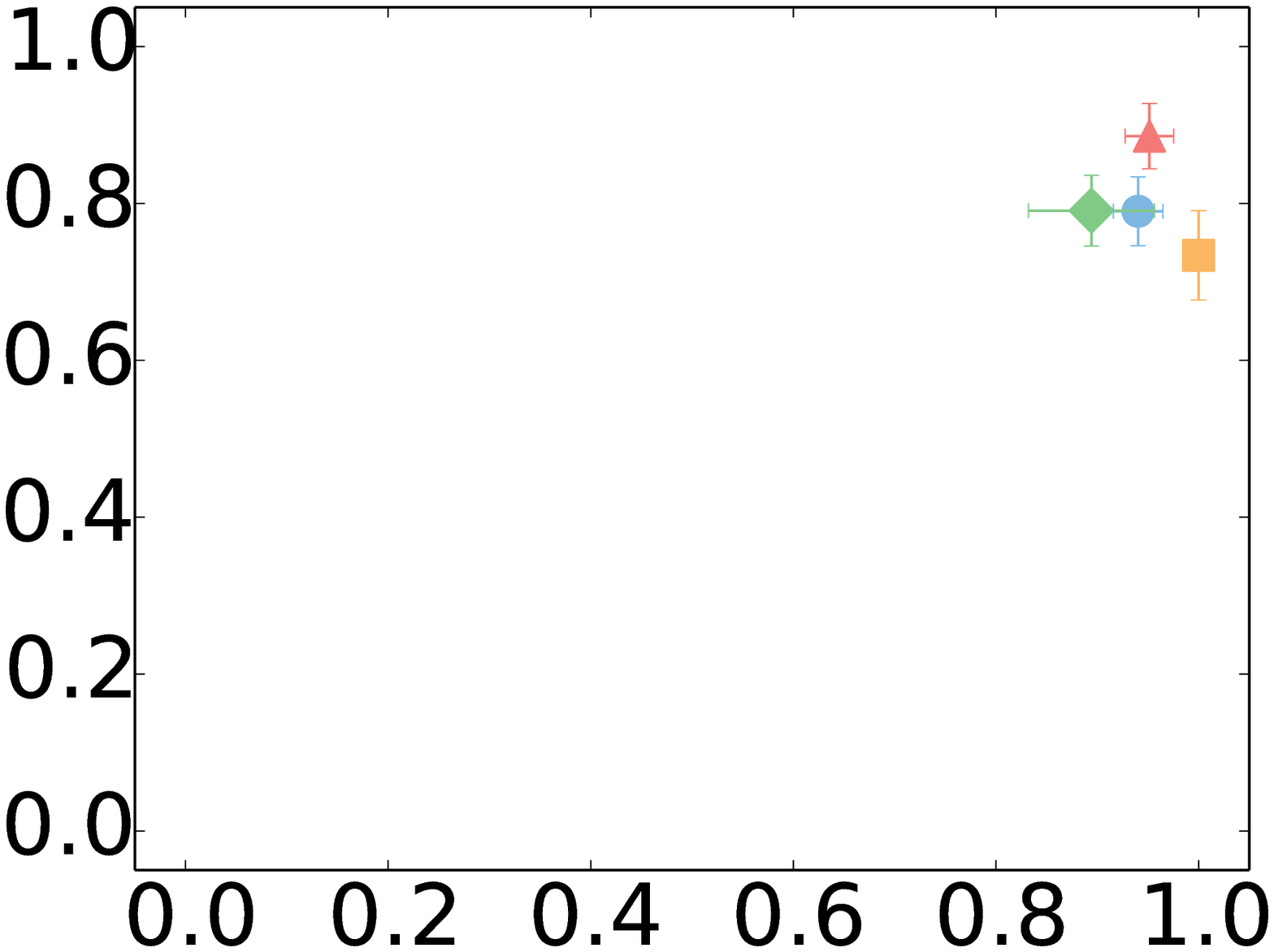}
\includegraphics[width=\linewidth]{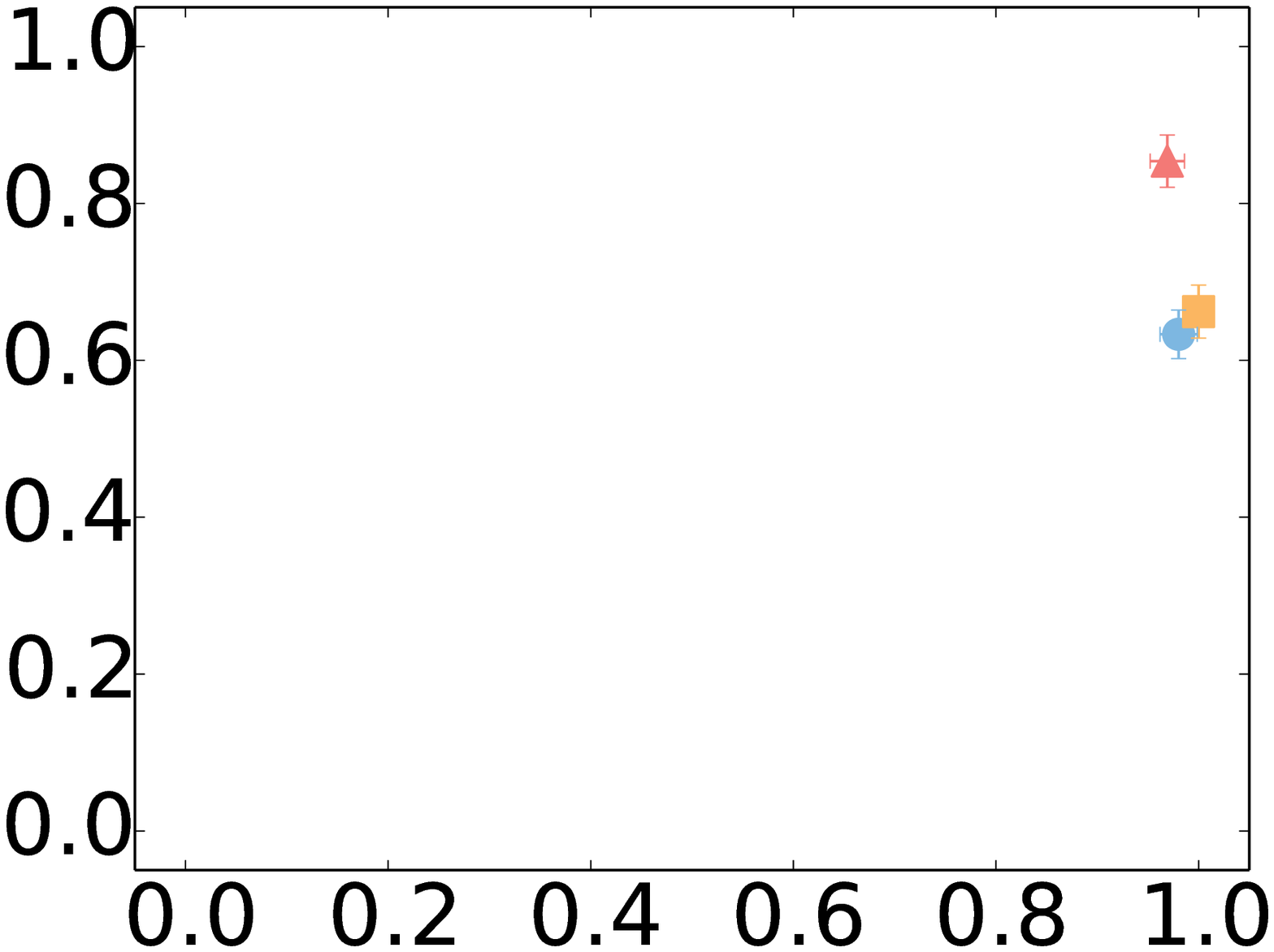}
\includegraphics[width=\linewidth]{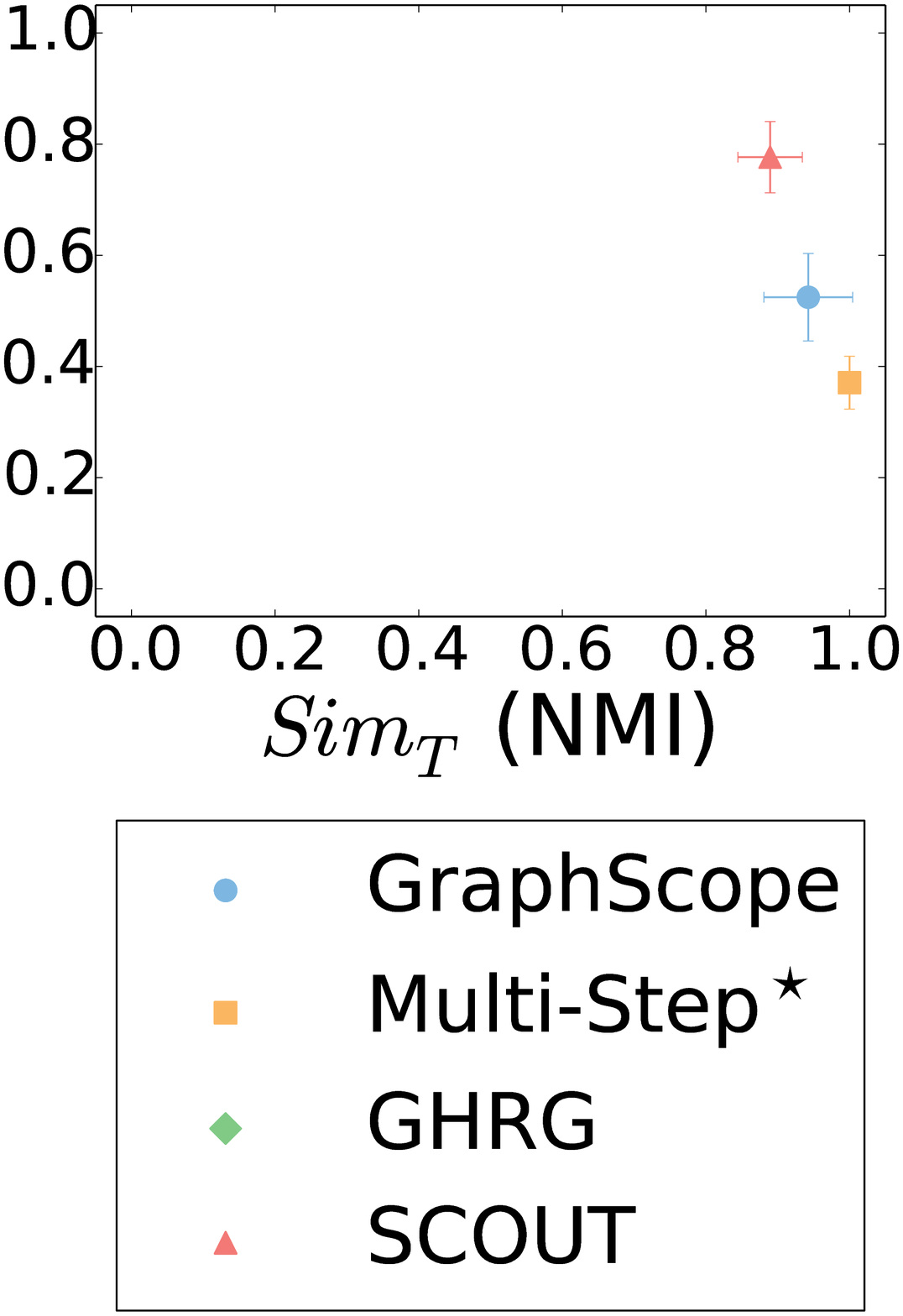}
\label{fig:supplement:scatter_16segments}
\end{minipage} 
}
}
  \caption[$Sim_T$ and $Sim_P$ scores for all 20 synthetic network configurations]{$Sim_T$ and $Sim_P$ scores for all 20 synthetic network configurations.
The rows correspond to four different numbers of nodes per snapshot, with the number of nodes increasing from top to bottom. 
The columns correspond to five different numbers of ground truth segments:
\textbf{\protect\subref{fig:supplement:scatter_1segment}} one ground truth segment,
 \textbf{\protect\subref{fig:supplement:scatter_2segments}}  two ground truth segments,
 \textbf{\protect\subref{fig:supplement:scatter_4segments}} four ground truth segments,
\textbf{\protect\subref{fig:supplement:scatter_8segments}} eight ground truth segments,
and \textbf{\protect\subref{fig:supplement:scatter_16segments}} 16 ground truth segments.
In each panel, for each method, the results are averaged over all of the corresponding synthetic network instances.
}
\label{fig:supplement:scatter_100}
\end{figure}

\begin{figure}
\centering
\subfloat[$50$-node snapshots]{\label{fig:supplement:sgmq__num_50}\includegraphics[width=.37\linewidth]{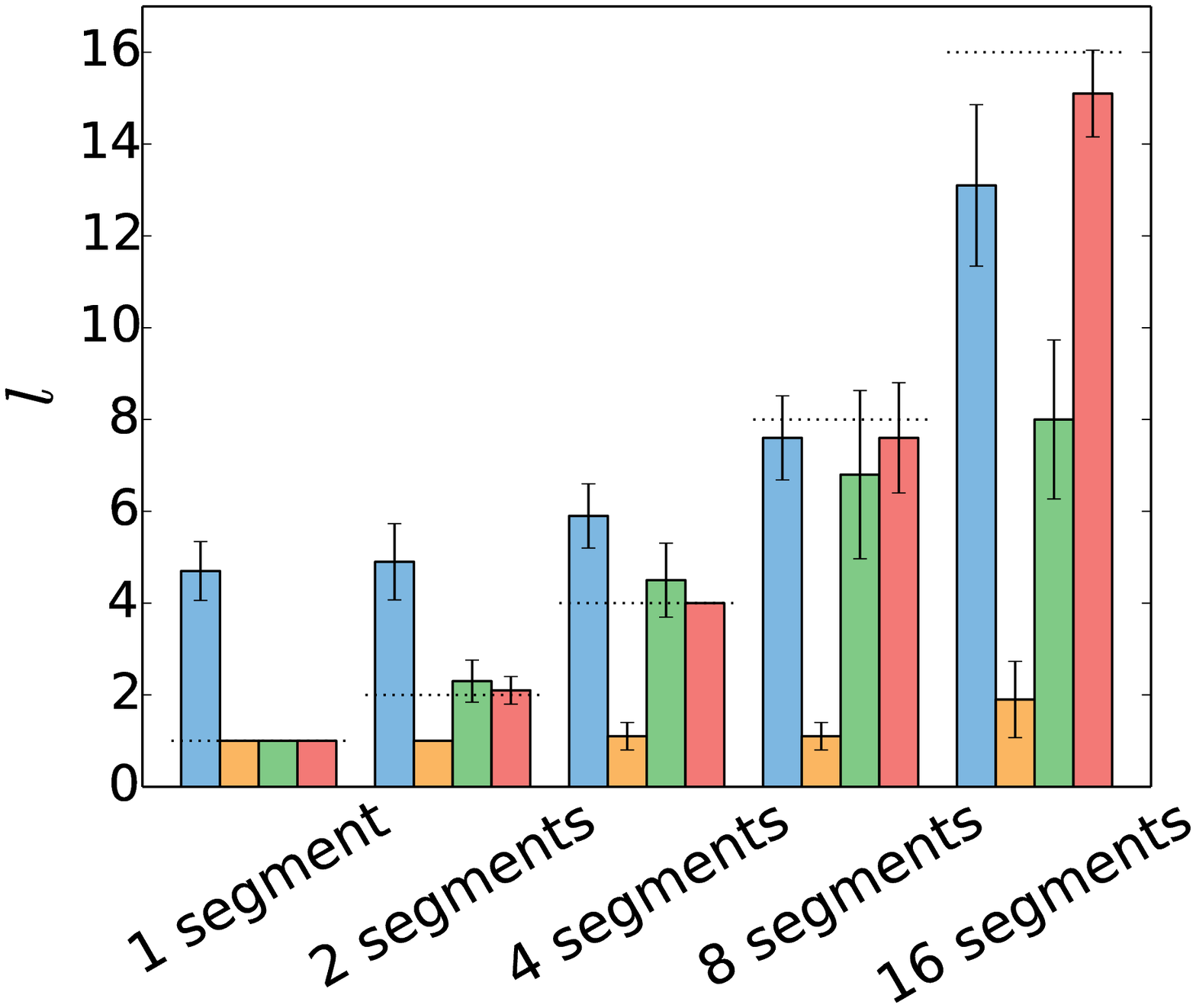}}
\hspace{0.2cm}
\subfloat[$100$-node snapshots]{\label{fig:supplement:sgmq__num_100}\includegraphics[width=.35\linewidth]{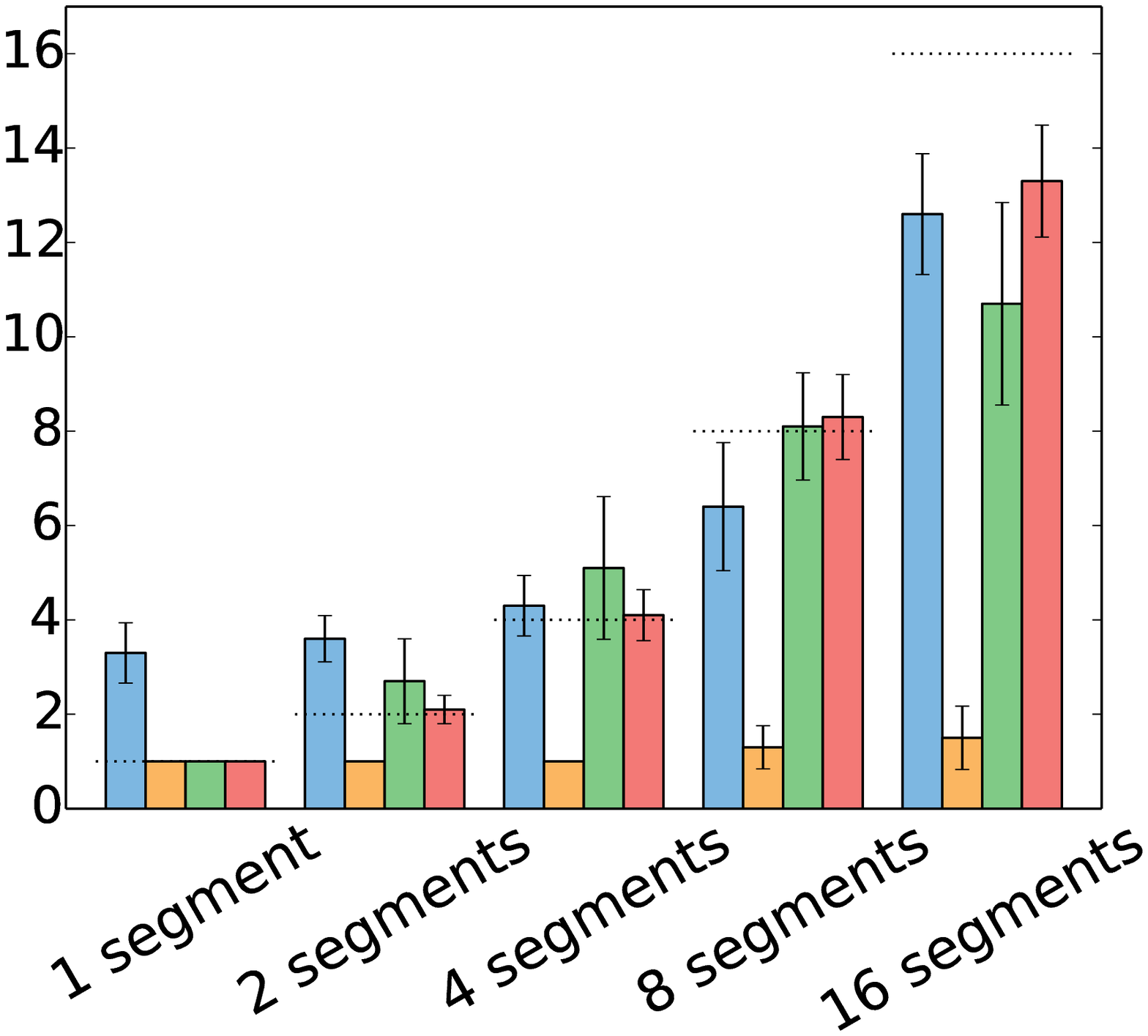}}
\\
\subfloat[$500$-node snapshots]{\label{fig:supplement:sgmq__num_500}\includegraphics[width=.37\linewidth]{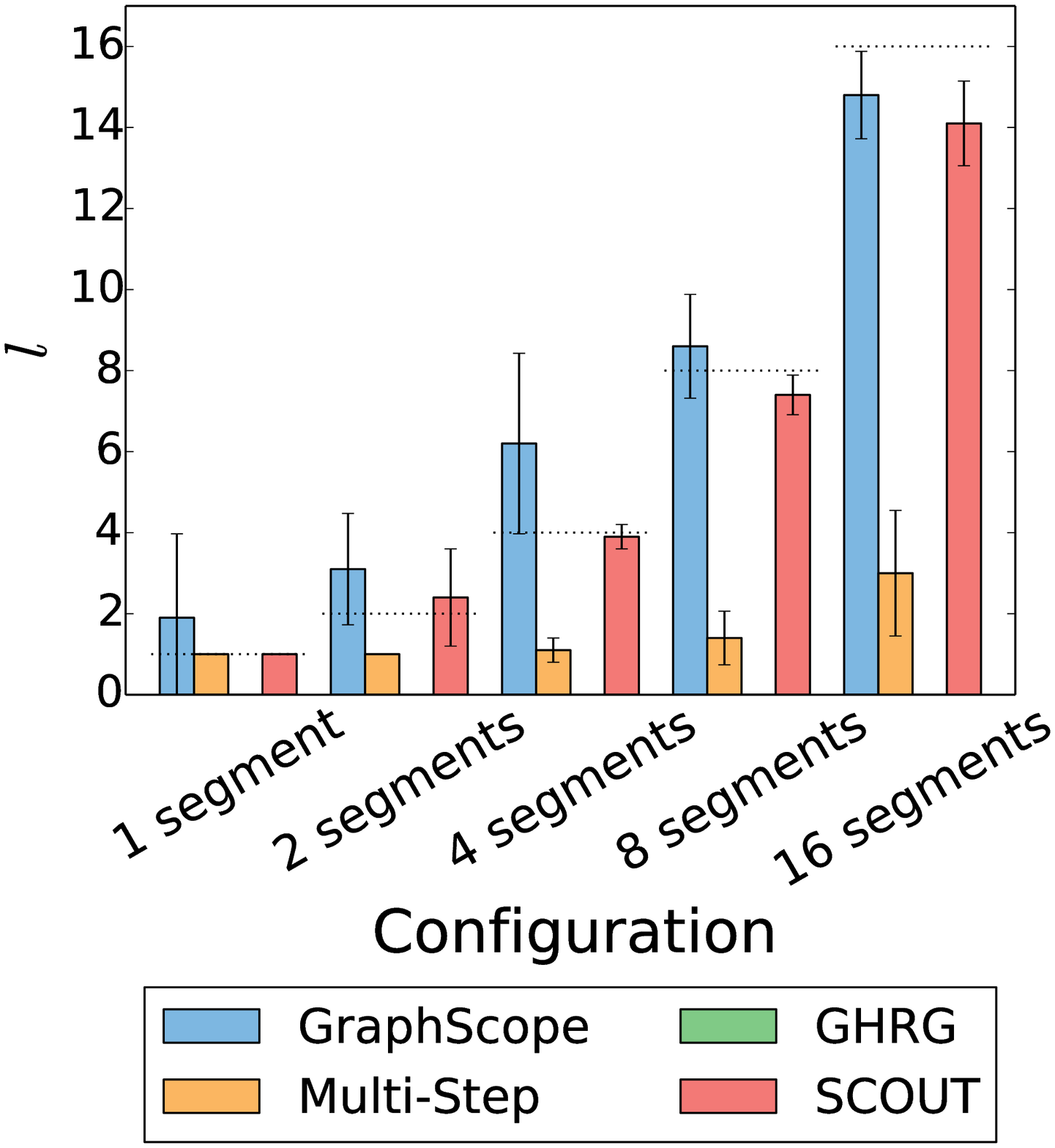}}
\hspace{0.2cm}
\subfloat[$1000$-node snapshots]{\label{fig:supplement:sgmq__num_1000}\includegraphics[width=.35\linewidth]{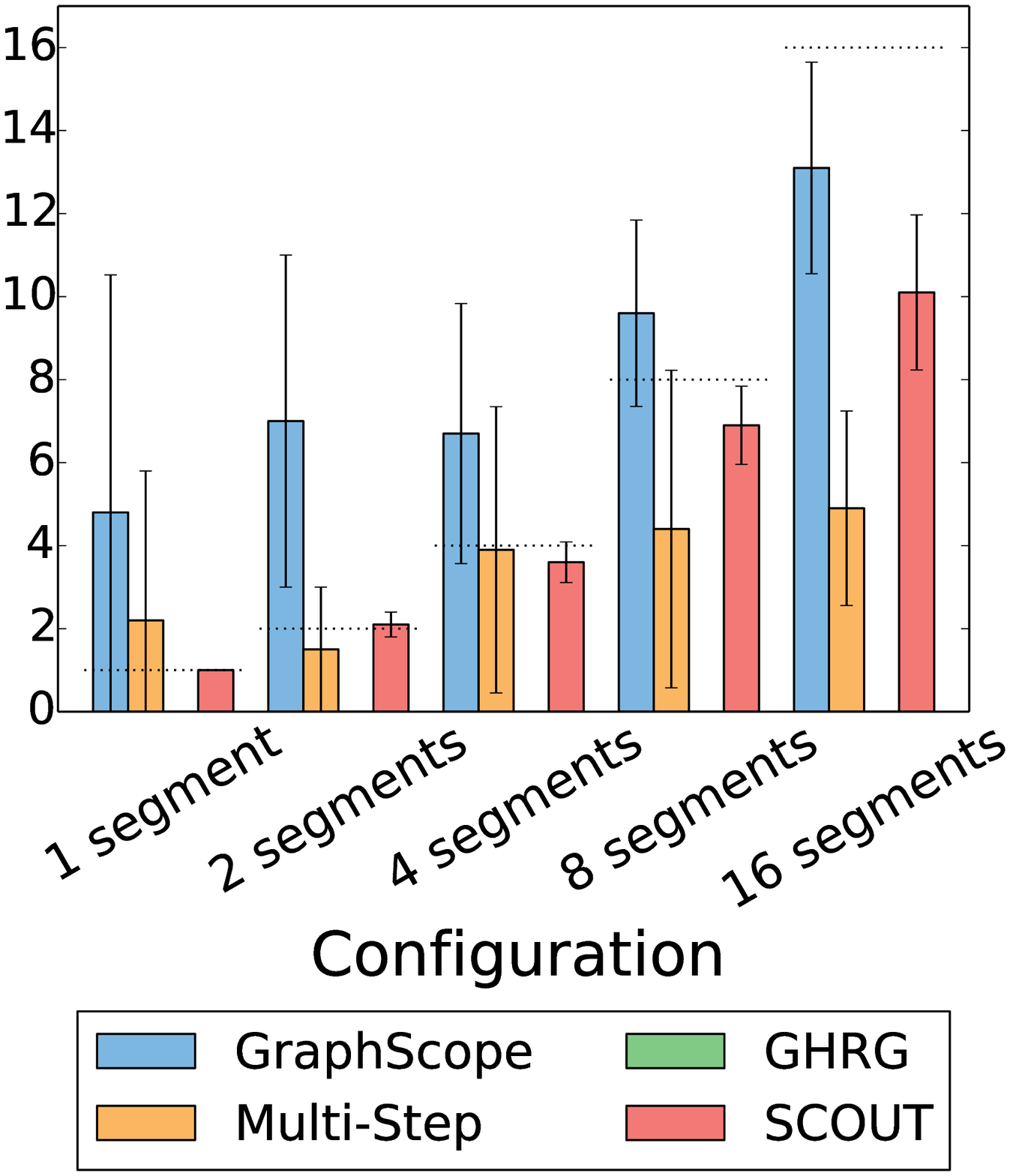}}
  \caption[The number of segments $l$ in the solutions produced by the methods for synthetic networks]{The number of segments $l$ in the solutions produced by the methods for synthetic networks with
\textbf{\protect\subref{fig:supplement:sgmq__num_50}} $50$-node snapshots,
 \textbf{\protect\subref{fig:supplement:sgmq__num_100}}  $100$-node snapshots,
\textbf{\protect\subref{fig:supplement:sgmq__num_500}} $500$-node snapshots,
and \textbf{\protect\subref{fig:supplement:sgmq__num_1000}} $1000$-node snapshots.
  In each panel, the results are grouped by the number of ground truth segments and averaged over all of the corresponding synthetic network instances.
  A given dotted line corresponds to the ground truth number of segments in the given synthetic network configuration.
  Note that for Multi-Step, we use the default parameters, since when using Multi-Step$^\star$ that is provided with the number of ground truth segments as input, this method trivially returns the correct number of ground truth segments.
}
\label{fig:supplement:sgmq__num}
\end{figure}

\begin{figure*}
\centering
\resizebox{\linewidth}{!}{
\subfloat[Change point classification]{
\begin{minipage}[t]{0.32\linewidth}
\includegraphics[width=\linewidth]{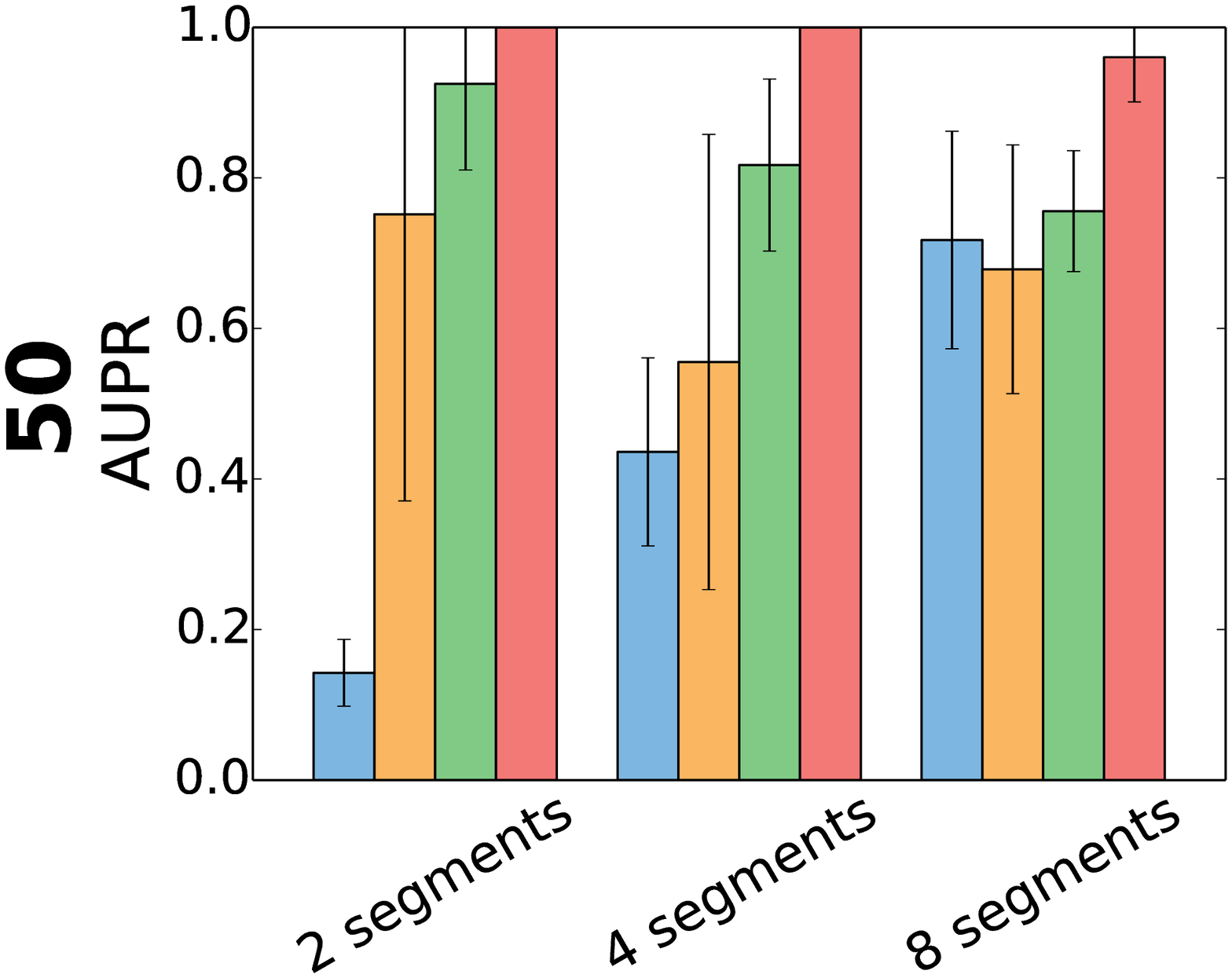}
\includegraphics[width=\linewidth]{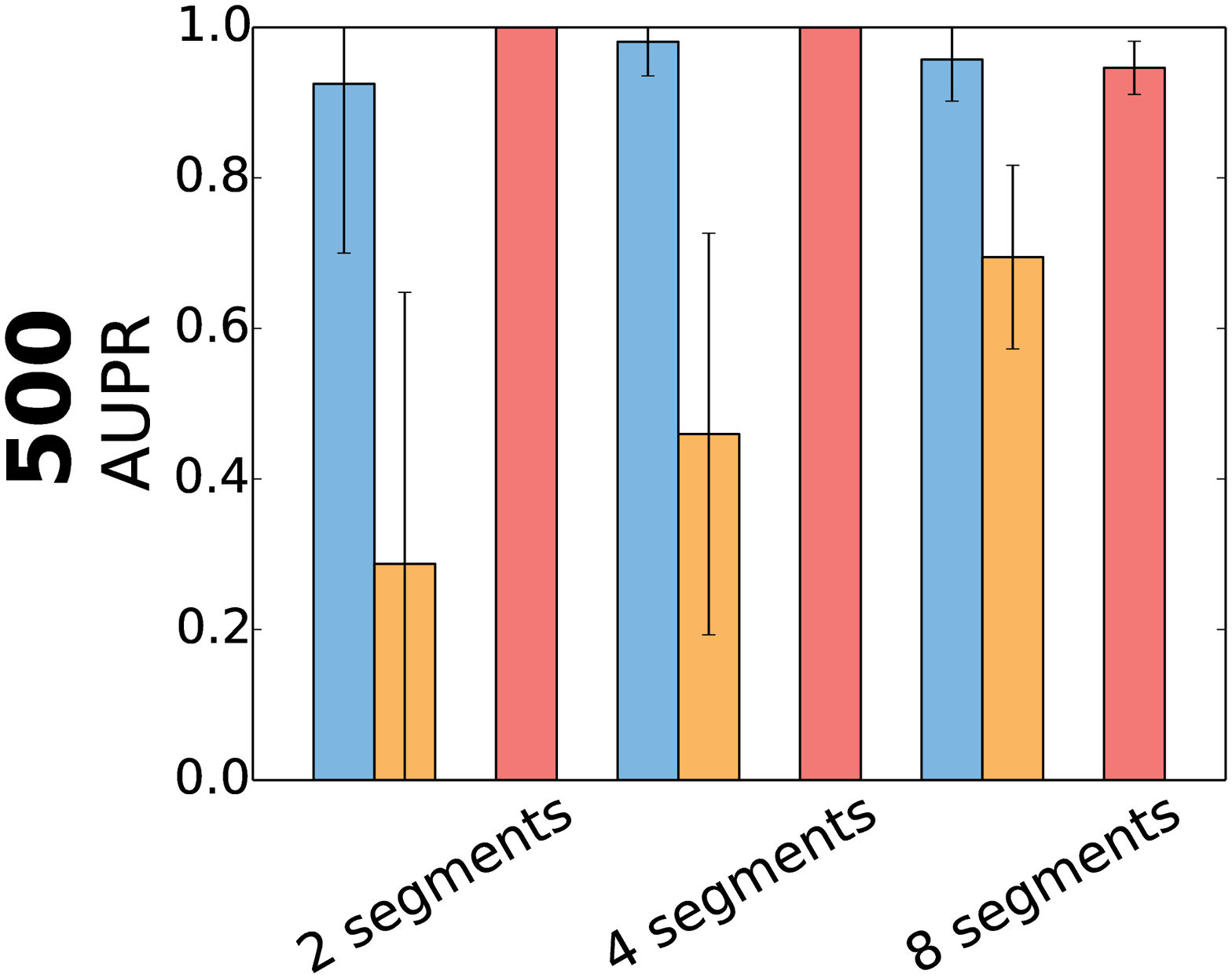}
\includegraphics[width=\linewidth]{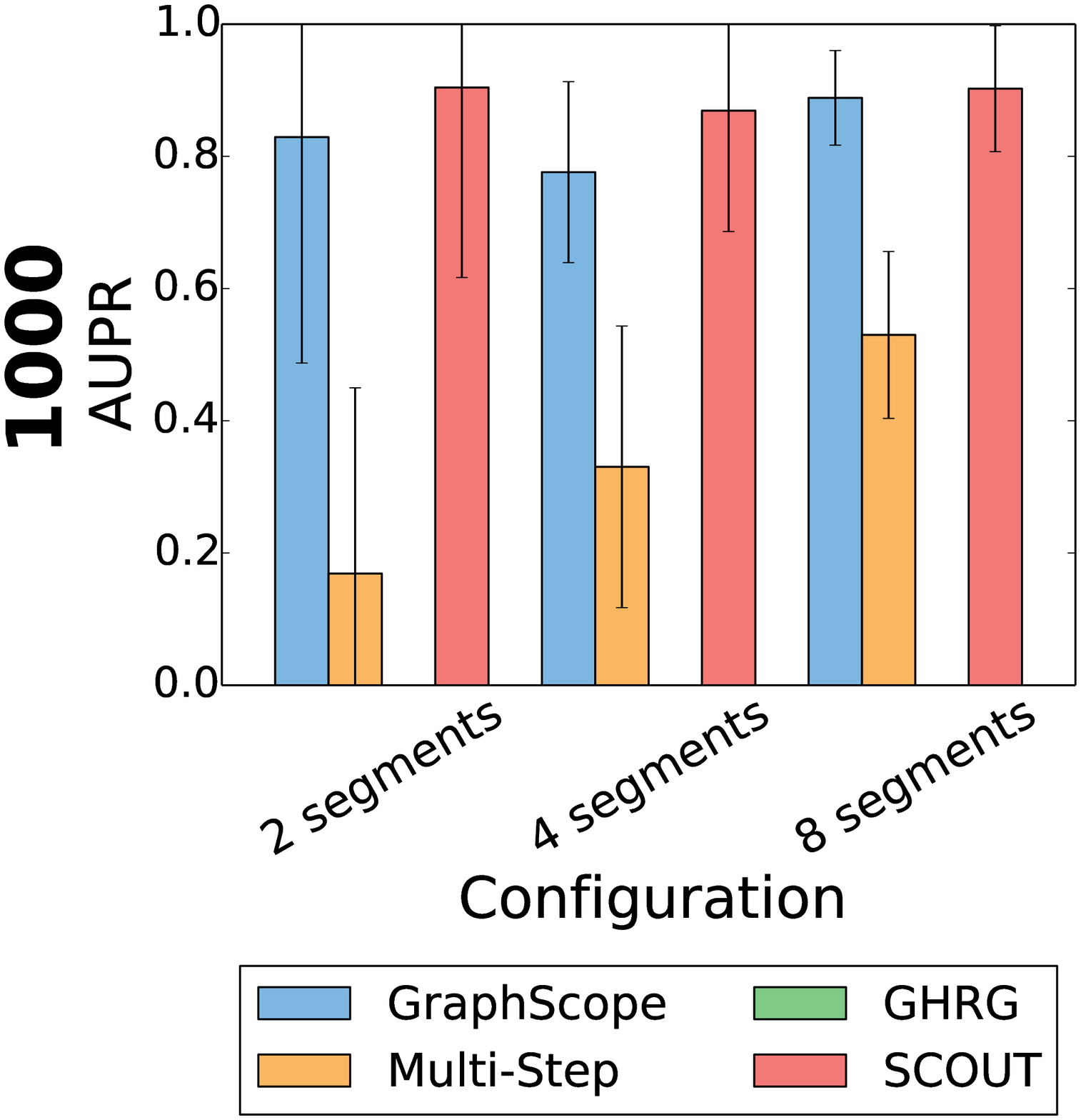}
\label{fig:supplement:synth_aupr}
\end{minipage} 
}
\subfloat[Partition quality $Q_P$]{
\begin{minipage}[t]{0.307\linewidth}
\includegraphics[width=\linewidth]{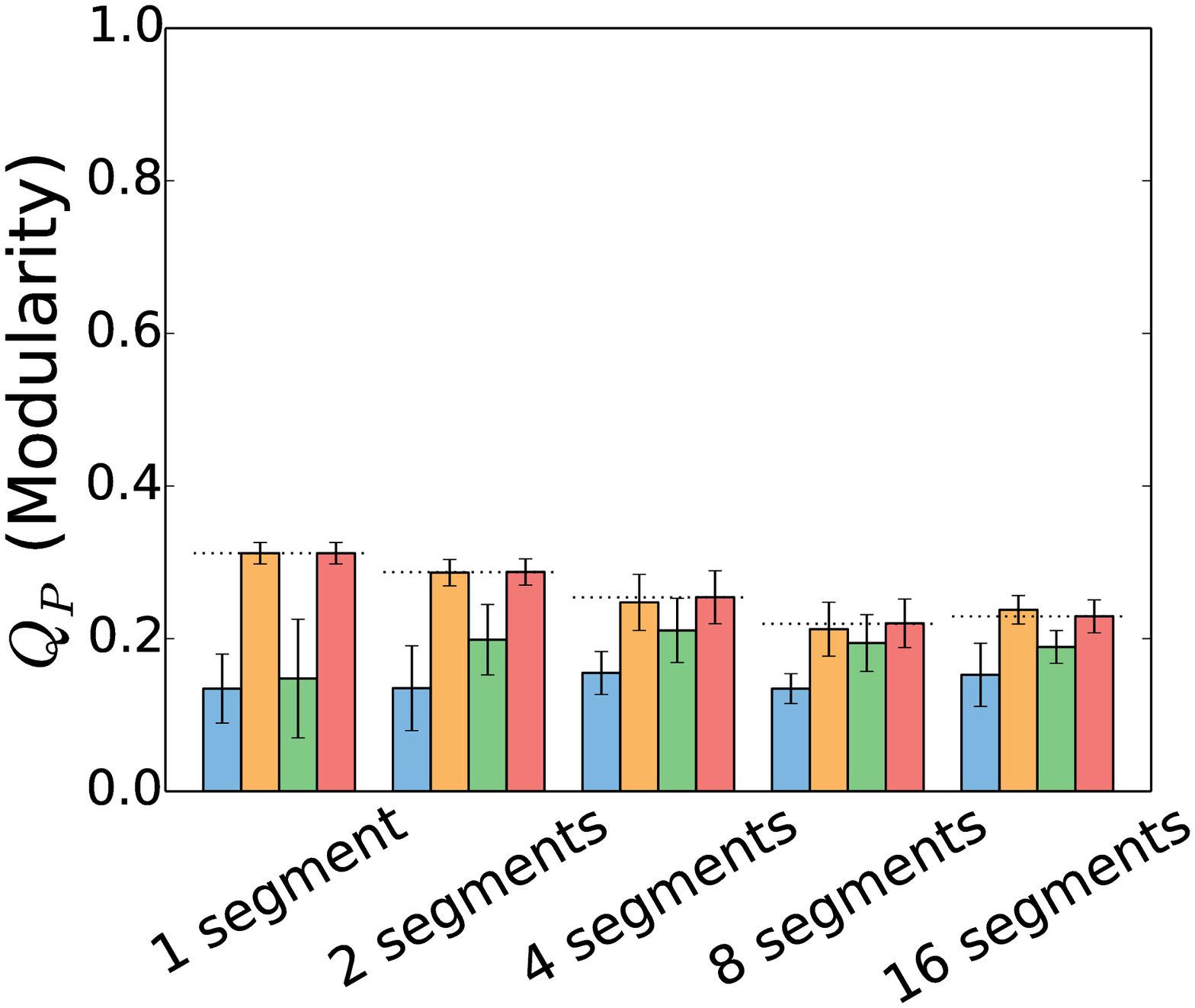}
\includegraphics[width=\linewidth]{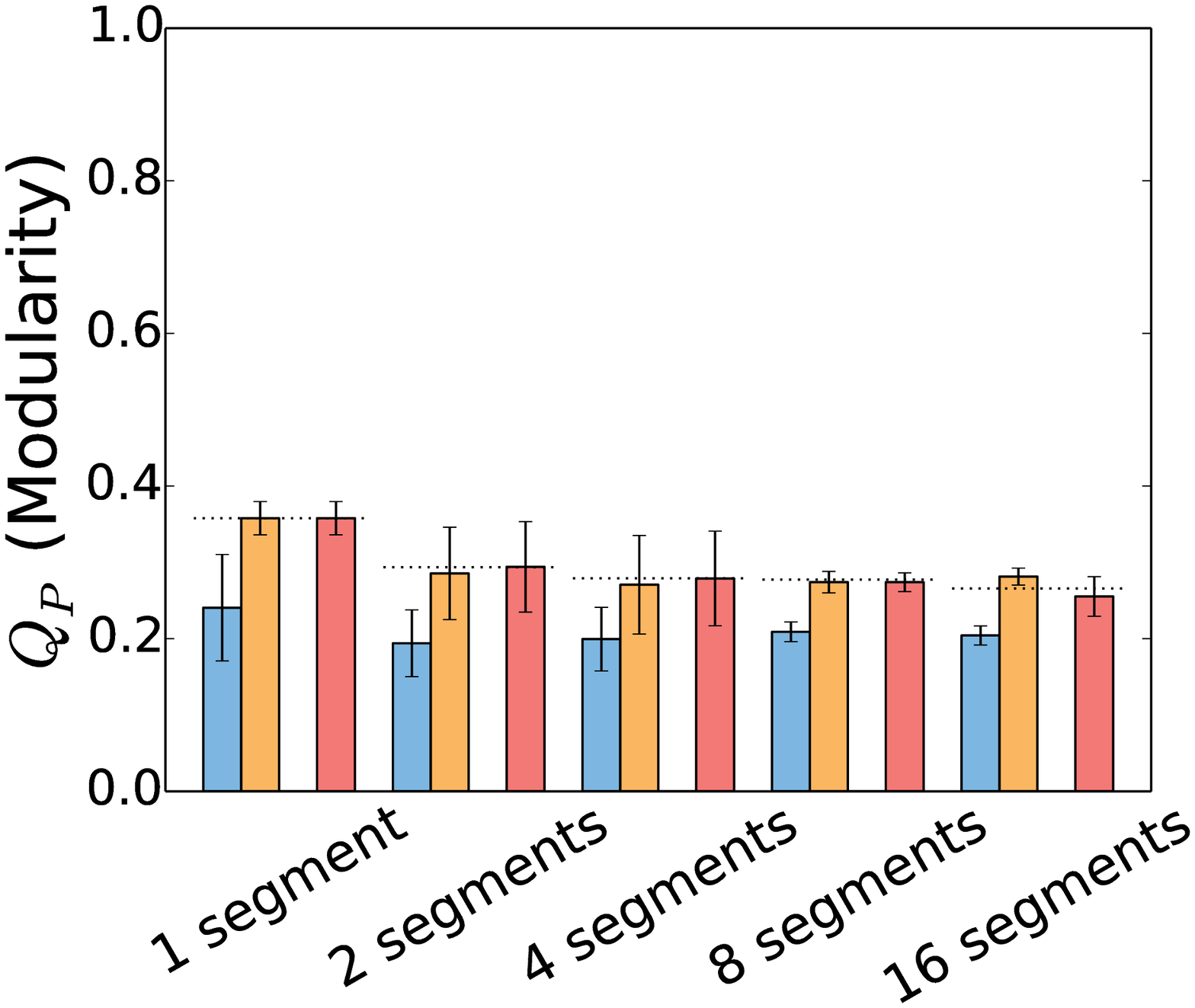}
\includegraphics[width=\linewidth]{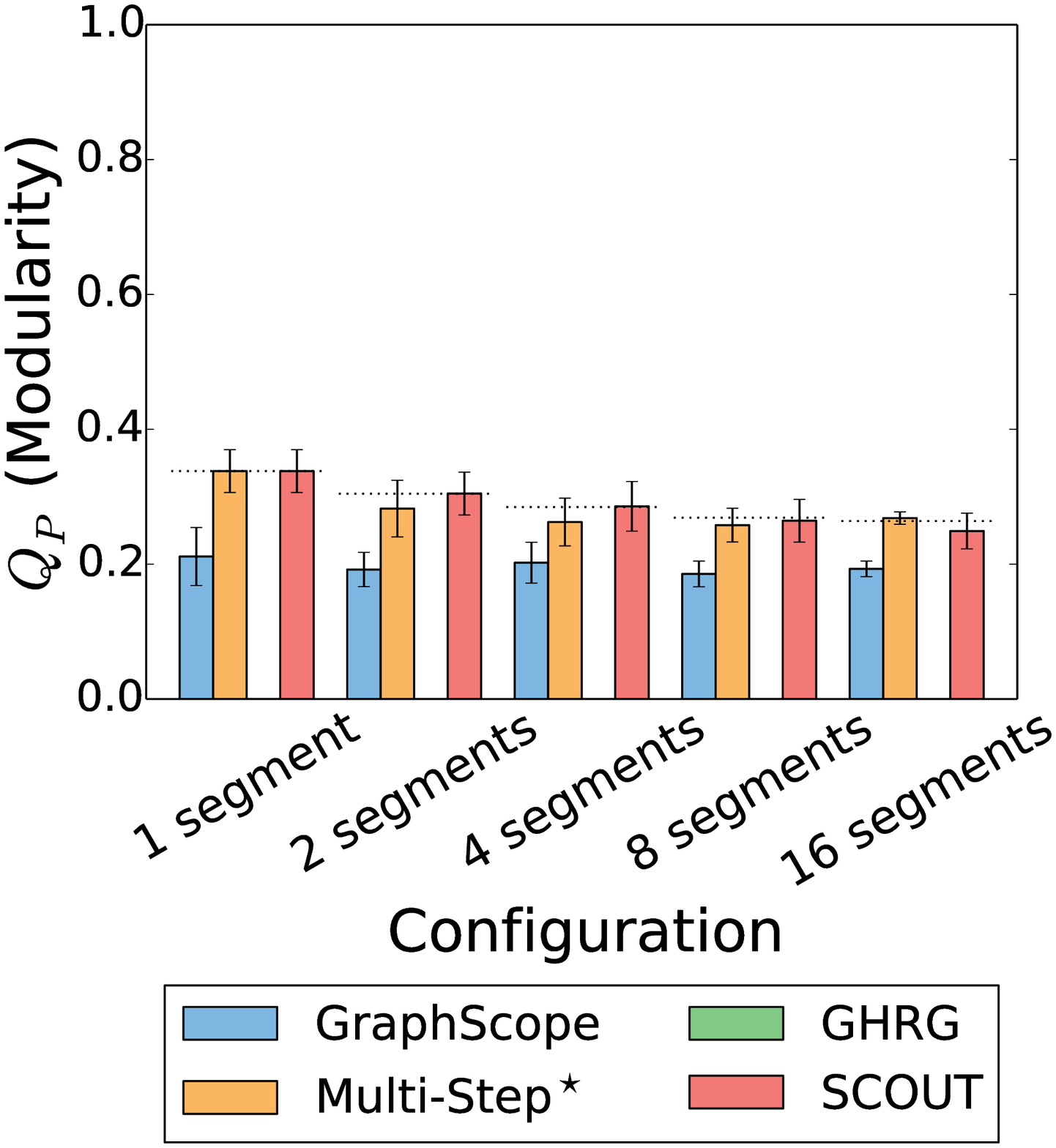}
\label{fig:supplement:synth_avg_sn__mod}
\end{minipage} 
}
\hspace{-0.3cm}
\subfloat[Overall similarity $Sim_B$]{
\begin{minipage}[t]{0.307\linewidth}
\includegraphics[width=\linewidth]{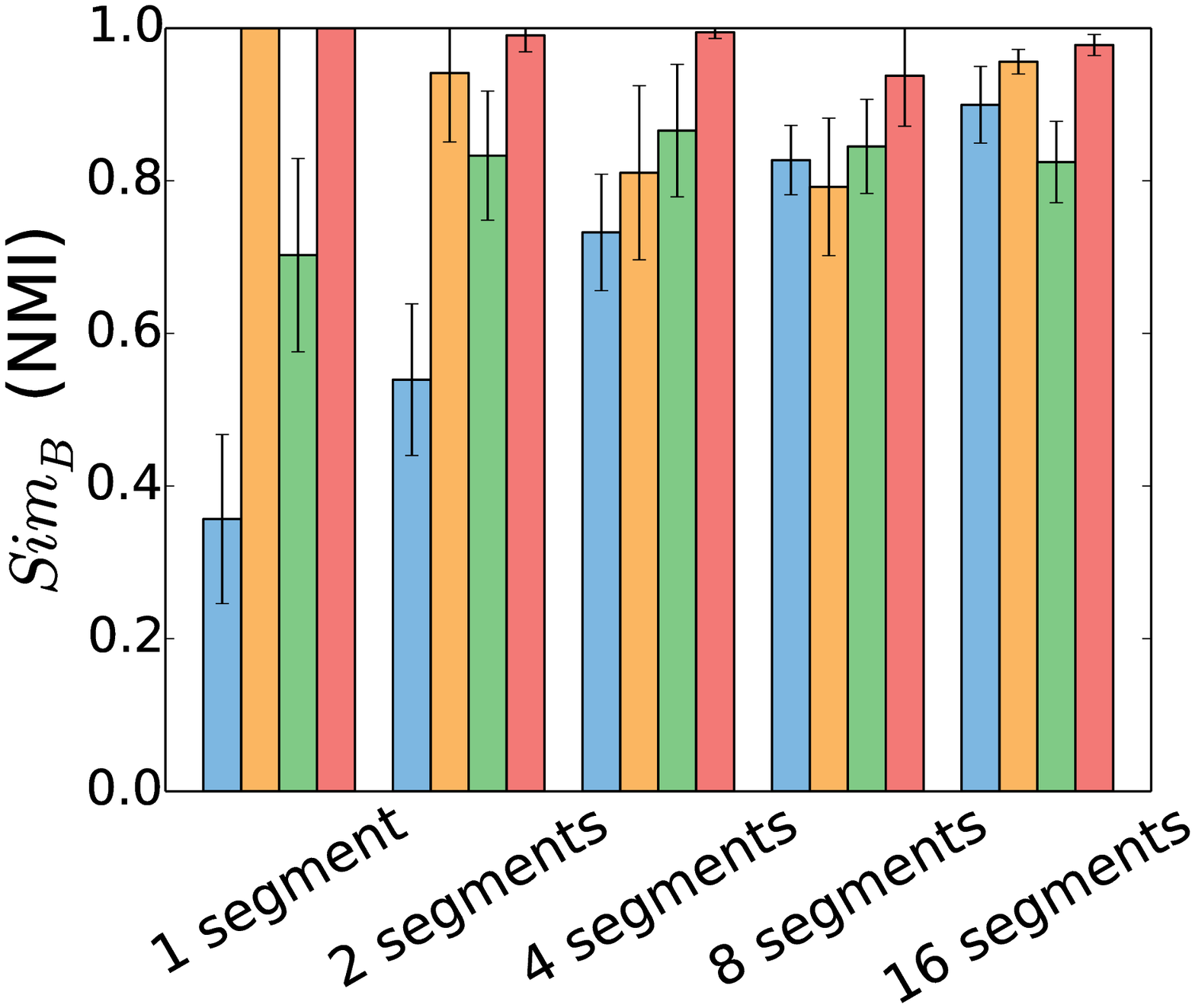}
\includegraphics[width=\linewidth]{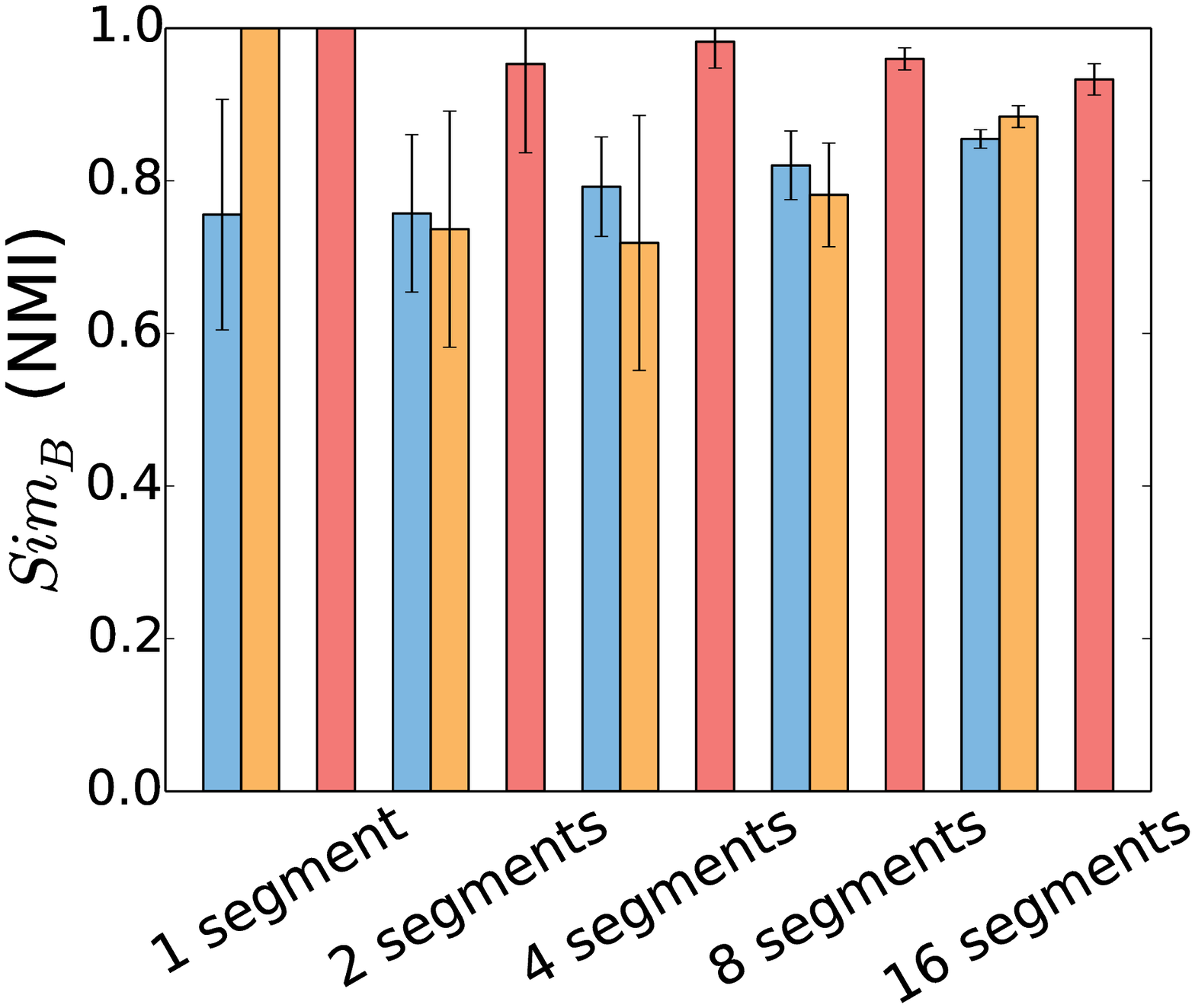}
\includegraphics[width=\linewidth]{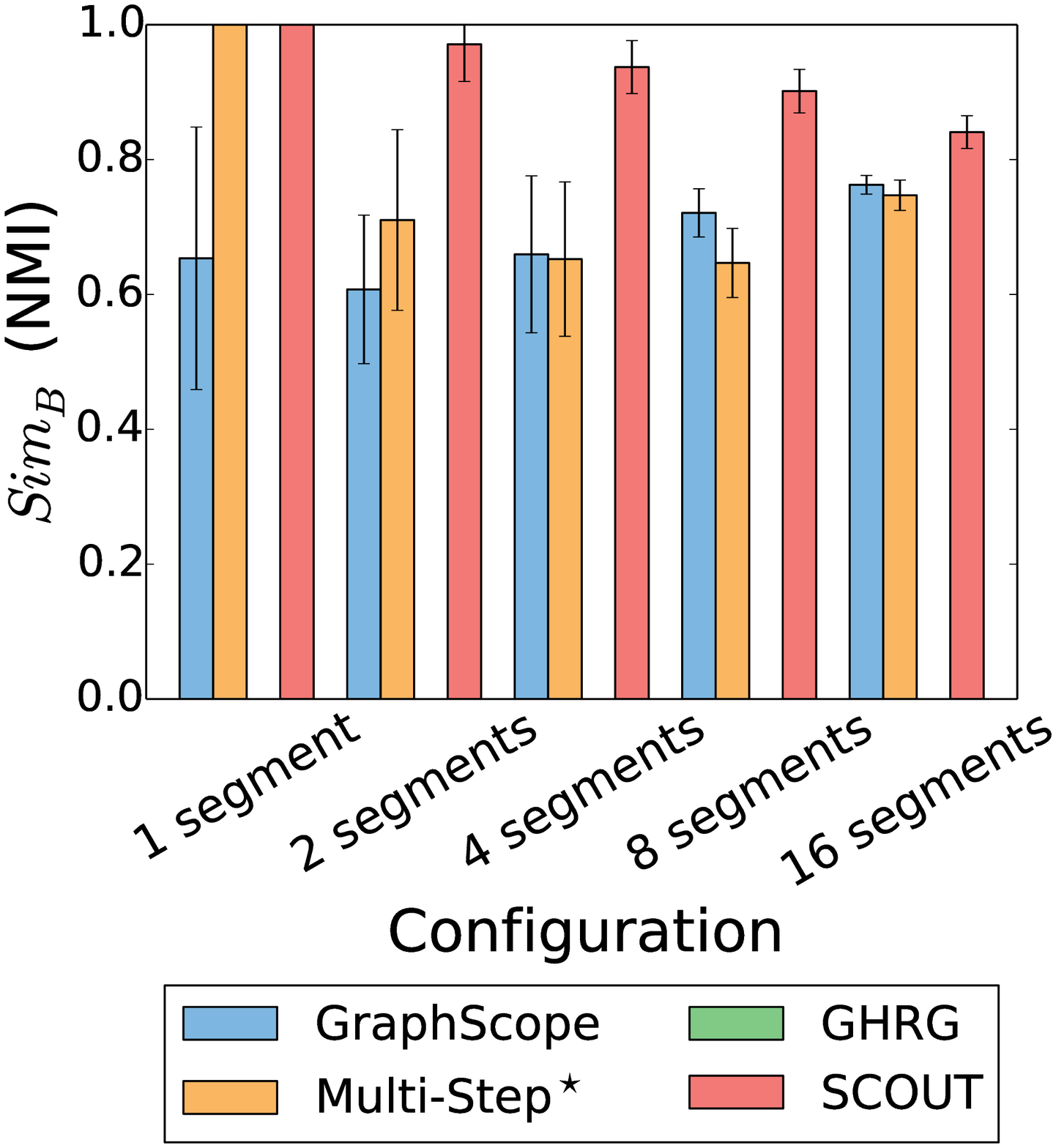}
\label{fig:supplement:synth_gl_p__nmi}
\end{minipage} 
}
}
\caption[Method comparison  for synthetic networks with 50, 500, and 1000 nodes per snapshot]{Method comparison  for synthetic networks with 50, 500, and 1000 nodes per snapshot (shown from top to bottom) with respect to
\textbf{\protect\subref{fig:supplement:synth_aupr}} change point classification,
\textbf{\protect\subref{fig:supplement:synth_avg_sn__mod}} $Q_P$,
and \textbf{\protect\subref{fig:supplement:synth_gl_p__nmi}} $Sim_B$. 
For a given ground truth configuration, the results are averaged over all of the corresponding synthetic network instances.
In panel \protect\subref{fig:supplement:synth_avg_sn__mod}, the dotted lines correspond to the ground truth score.
Note that for panel \protect\subref{fig:supplement:synth_aupr}, we exclude from consideration the configurations with the minimum and maximum possible numbers of ground truth segments.
We do this because for these configurations, either there are no change points at all (for one segment) or every time point is a change point (for 16 segments), which means that change point classification cannot be performed.
GHRG could not be run for two largest network sizes due to its high computational complexity.
Equivalent results for synthetic networks with 100 nodes per snapshot are shown in Figure~\ref{fig:synth_100} in the main paper.
}
\label{fig:supplement:synth_results}
\end{figure*}

\begin{figure*}
\centering
\subfloat[$Q_P$]{
\begin{minipage}[t]{0.315\linewidth}
\includegraphics[width=\linewidth]{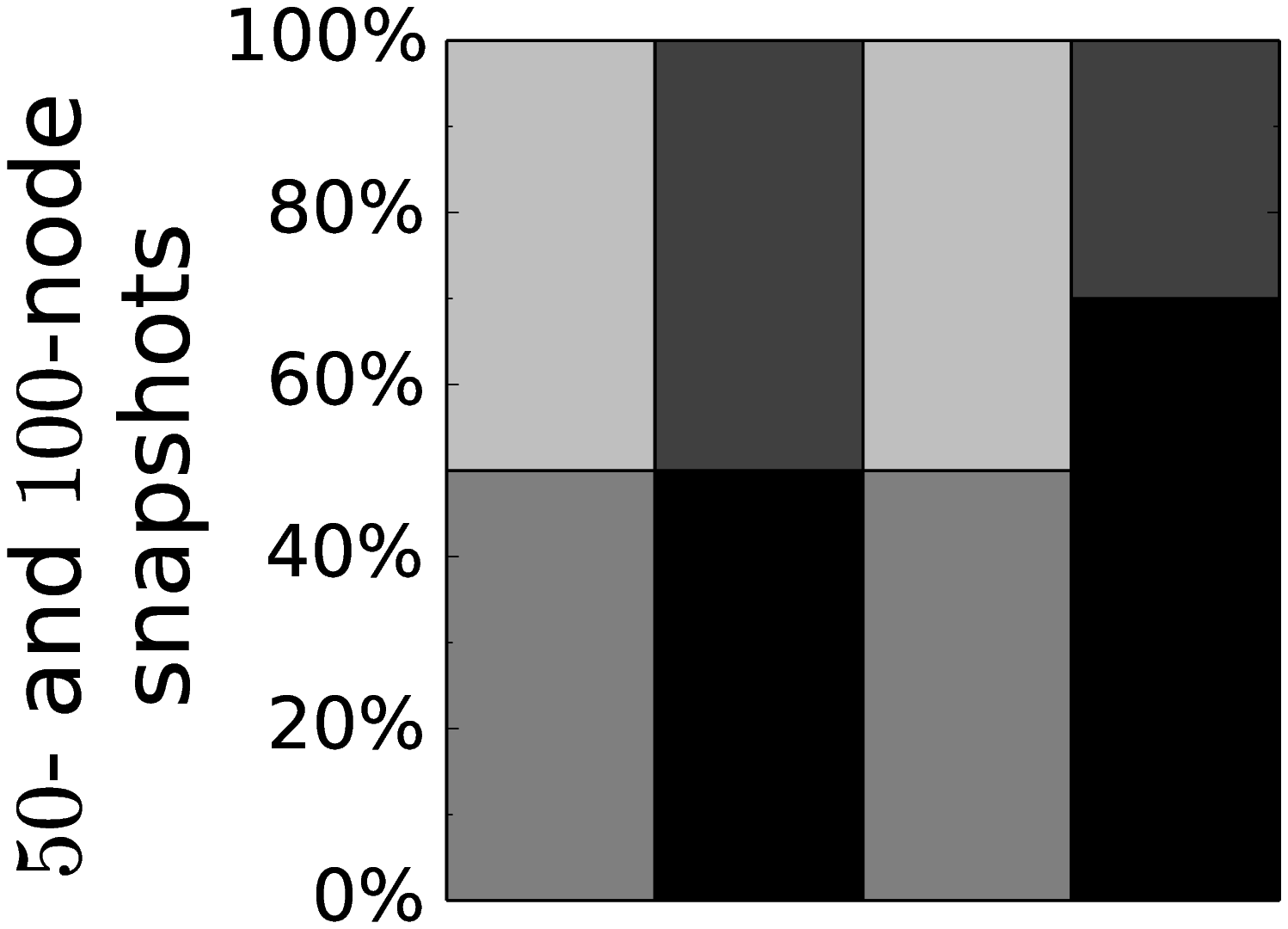}
\includegraphics[width=\linewidth]{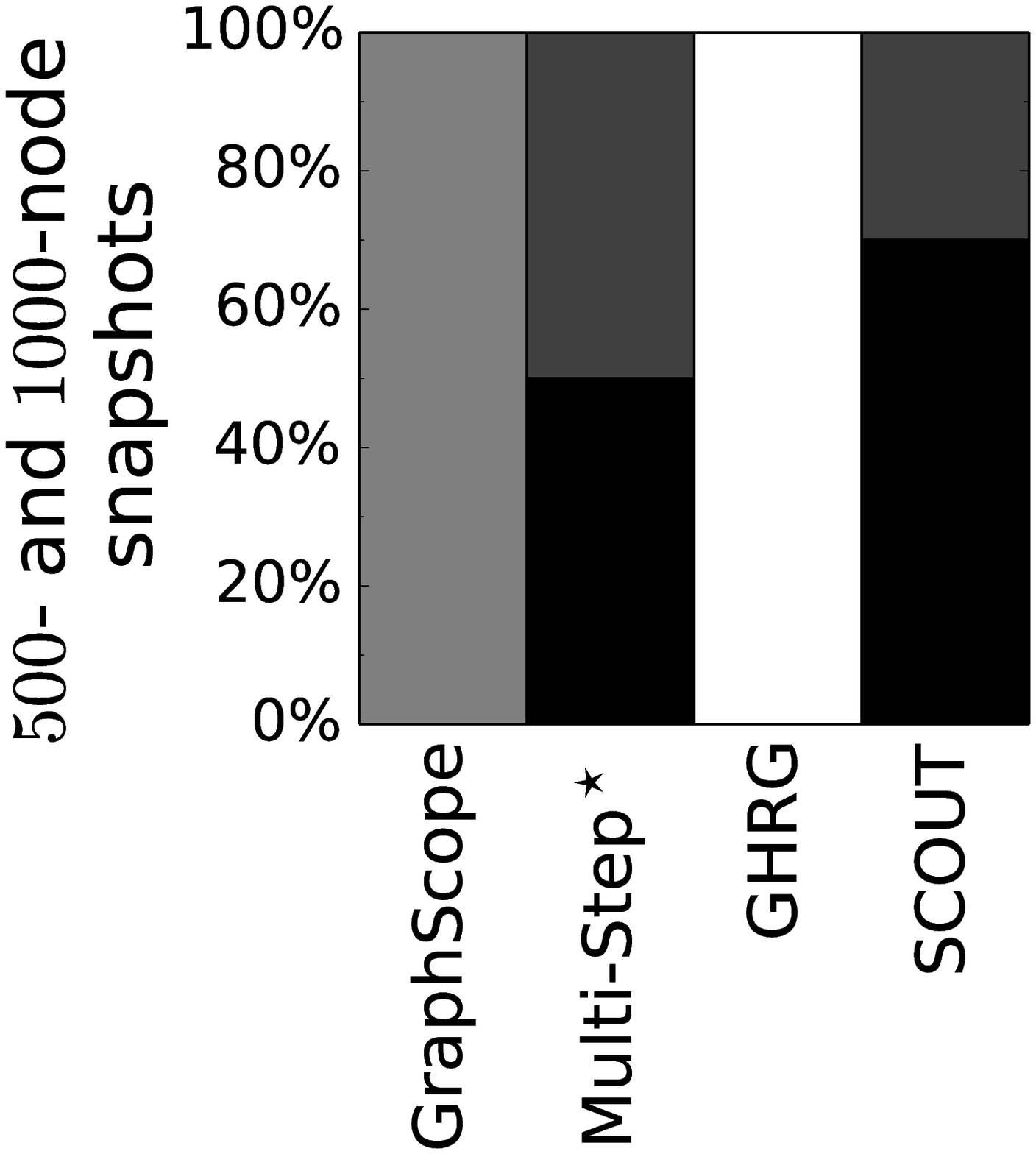}
\label{fig:supplement:synth_rank_avg_sn_mod}
\end{minipage} 
}
\subfloat[$Sim_P$]{
\begin{minipage}[t]{0.38\linewidth}
\includegraphics[width=\linewidth]{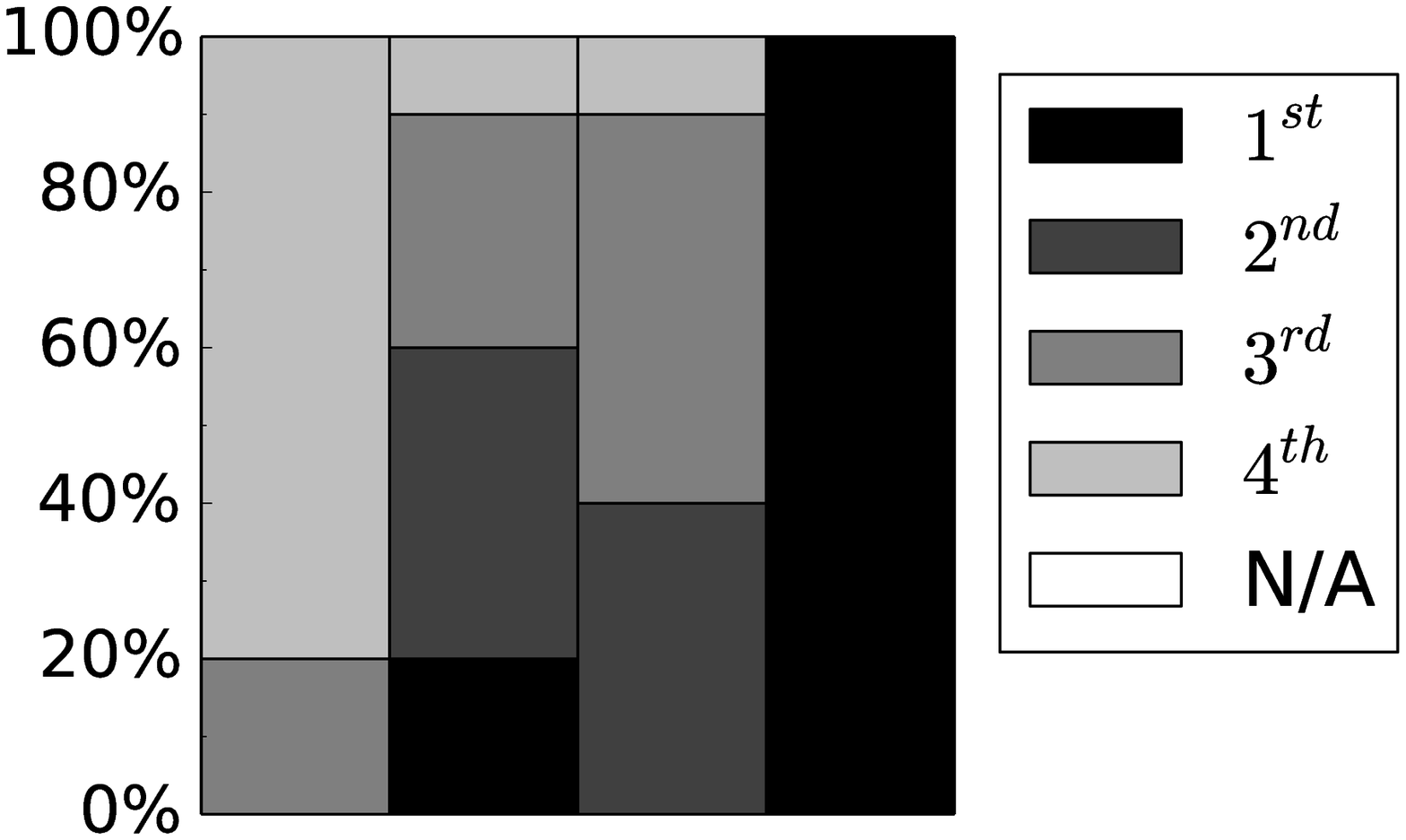}
\includegraphics[width=0.695\linewidth]{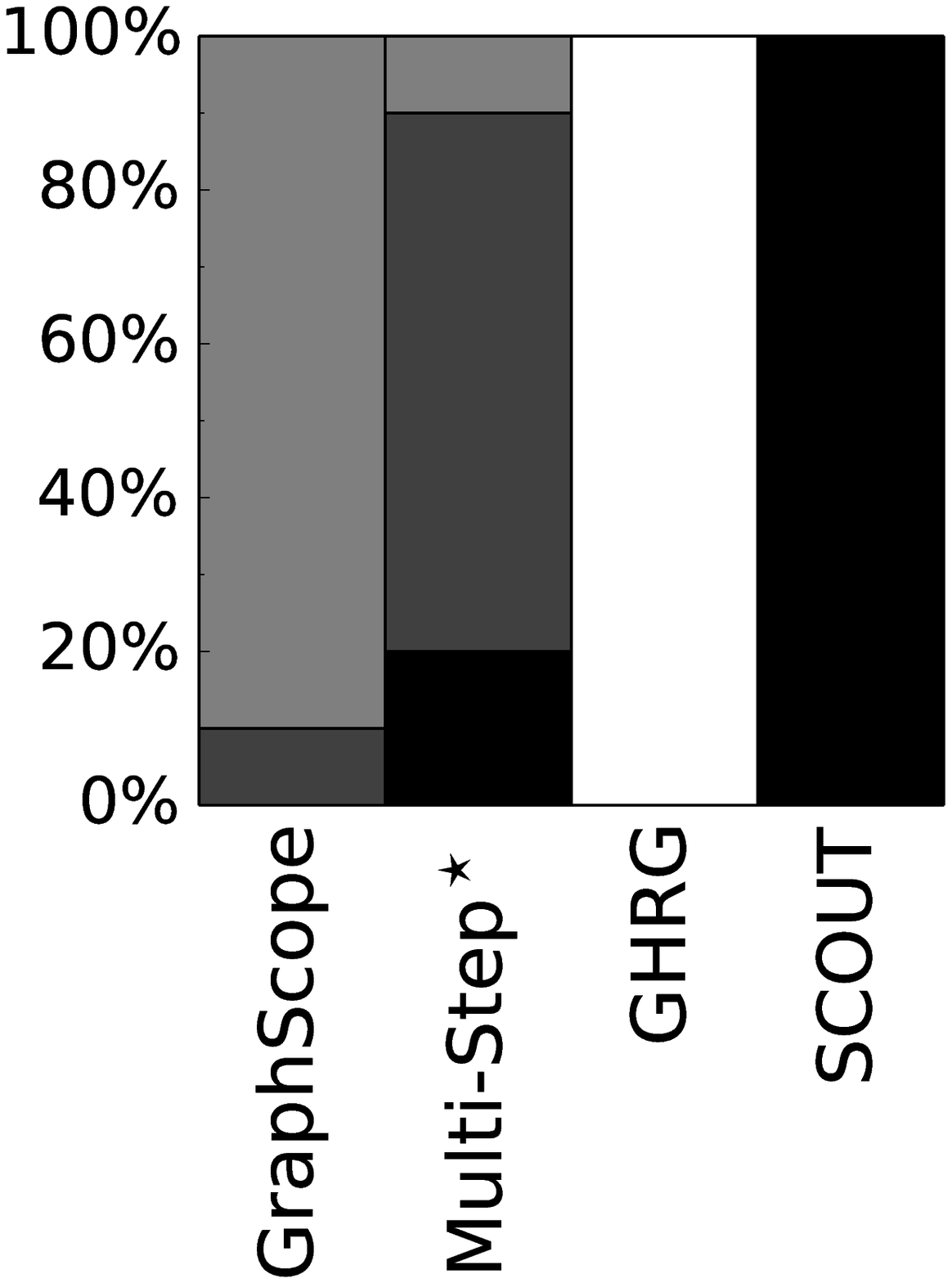}
\vspace{5mm} 
\label{fig:supplement:synth_rank_sim_p}
\end{minipage} 
}
\caption[Rankings of the methods with respect to $Q_P$ and $Sim_P$]{Rankings of the methods with respect to 
\textbf{\protect\subref{fig:supplement:synth_rank_avg_sn_mod}} $Q_P$ and
\textbf{\protect\subref{fig:supplement:synth_rank_sim_p}} $Sim_P$. 
Since GHRG could not be run for the larger networks, the results are split into those for the configurations with 50 and 100 nodes per snapshot (top) and those for the configurations with 500 and 1000 nodes per snapshot (bottom).
The rankings are computed as follows. 
For each synthetic network configuration, we compare the four methods' scores (average scores over all instances of the given configuration) to identify the first, second, third, and fourth best method; ties are allowed, in which case, two methods would be assigned the same rank. 
Then, we summarize these results over all considered synthetic network configurations by measuring, for each method ($x$-axis), how many times the given method is ranked as  the first, second, third, and fourth best method (expressed as the percentage of all considered configurations; $y$-axis).
``N/A'' indicates that the given method could not be run (which is the case for GHRG for the larger networks). 
The figure can intuitively be interpreted as follows: the darker the bar of a given method, the better its performance.
}
\end{figure*}

\begin{figure}
  \centering\includegraphics[width=0.5\linewidth]{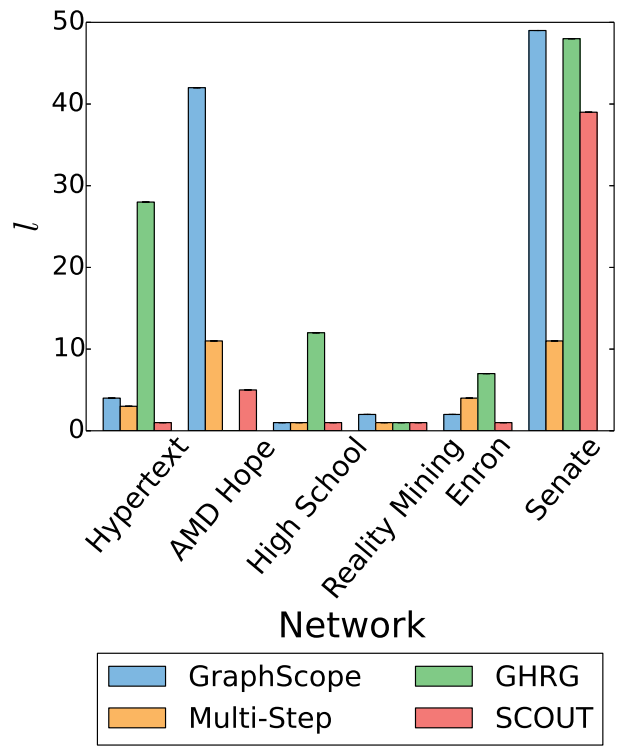}
  \caption{The number of segments $l$ in the solutions produced by the methods for real-world networks.
  }\label{fig:supplement:real_world_sgmq__num}
\end{figure}

\begin{figure*}
\centering
\subfloat[\emph{Hypertext}]{\label{fig:supplement:real_world_hypertext_avg_sn__mod_cscd}\includegraphics[width=0.447\linewidth]{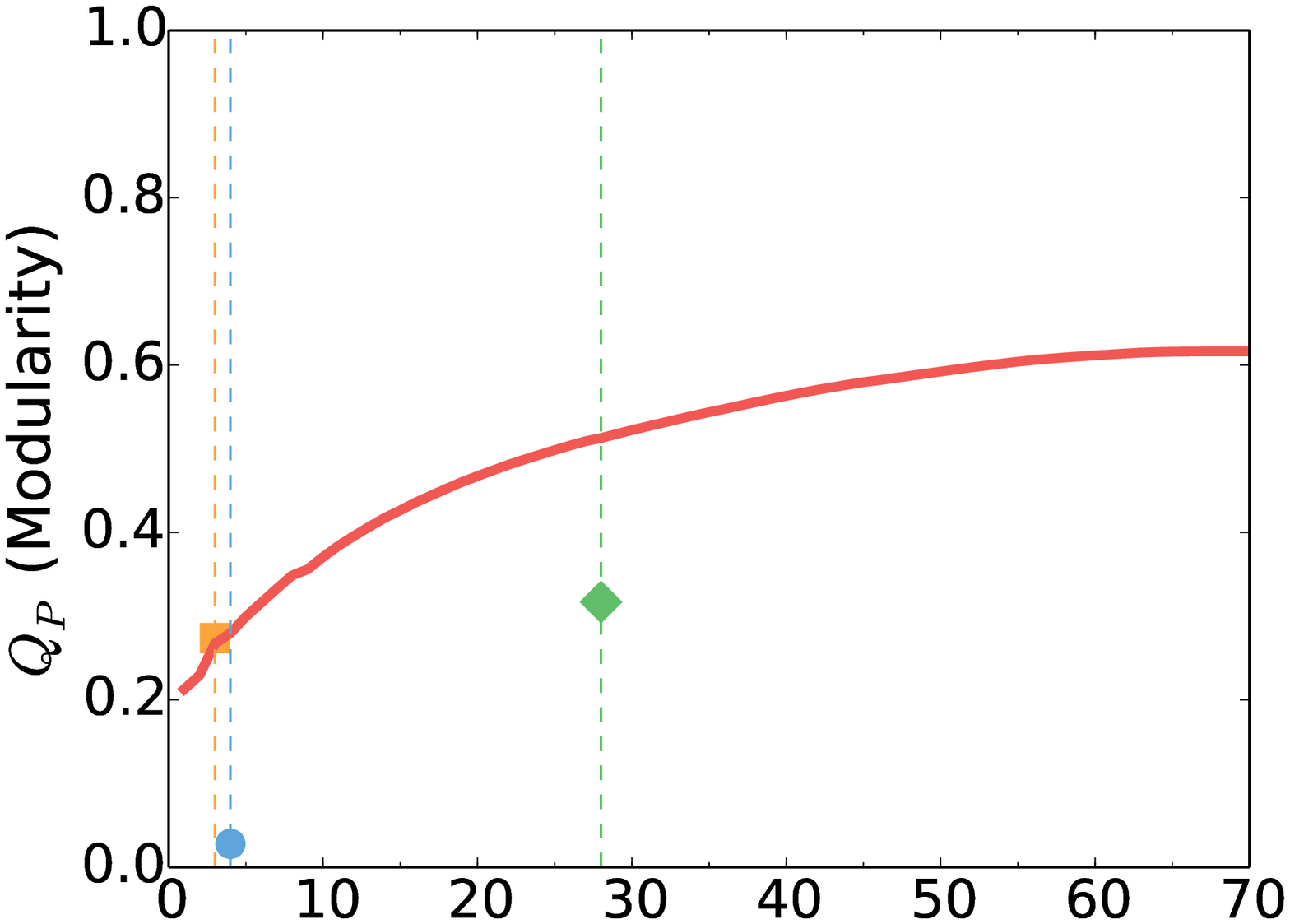}}
\hspace{0.2cm}
\subfloat[\emph{AMD Hope}]{\label{fig:supplement:real_world_amd_hope_avg_sn__mod_cscd}\includegraphics[width=0.42\linewidth]{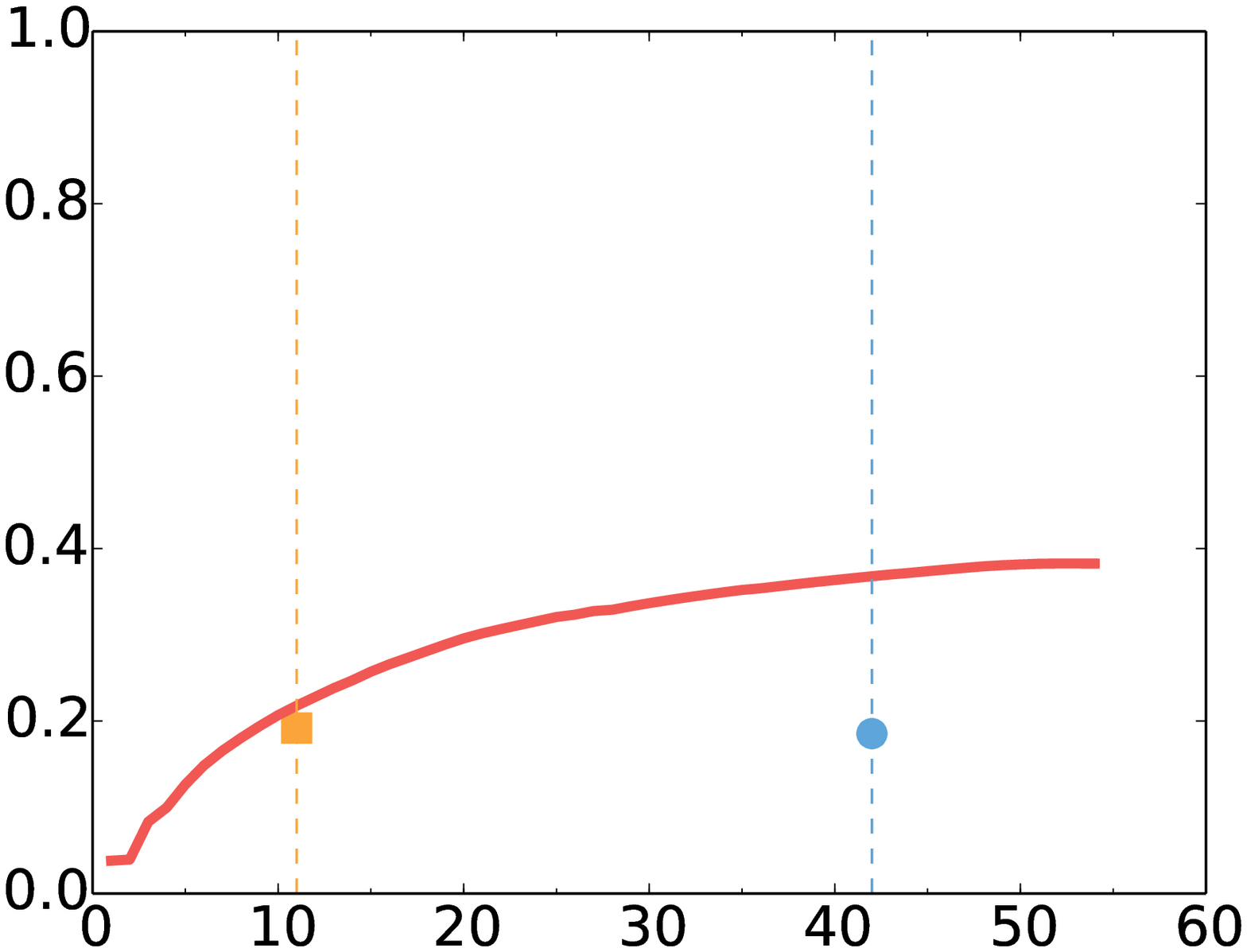}}

\subfloat[\emph{Reality mining}]{\label{fig:supplement:real_world_reality_mining_avg_sn__mod_cscd}\includegraphics[width=0.447\linewidth]{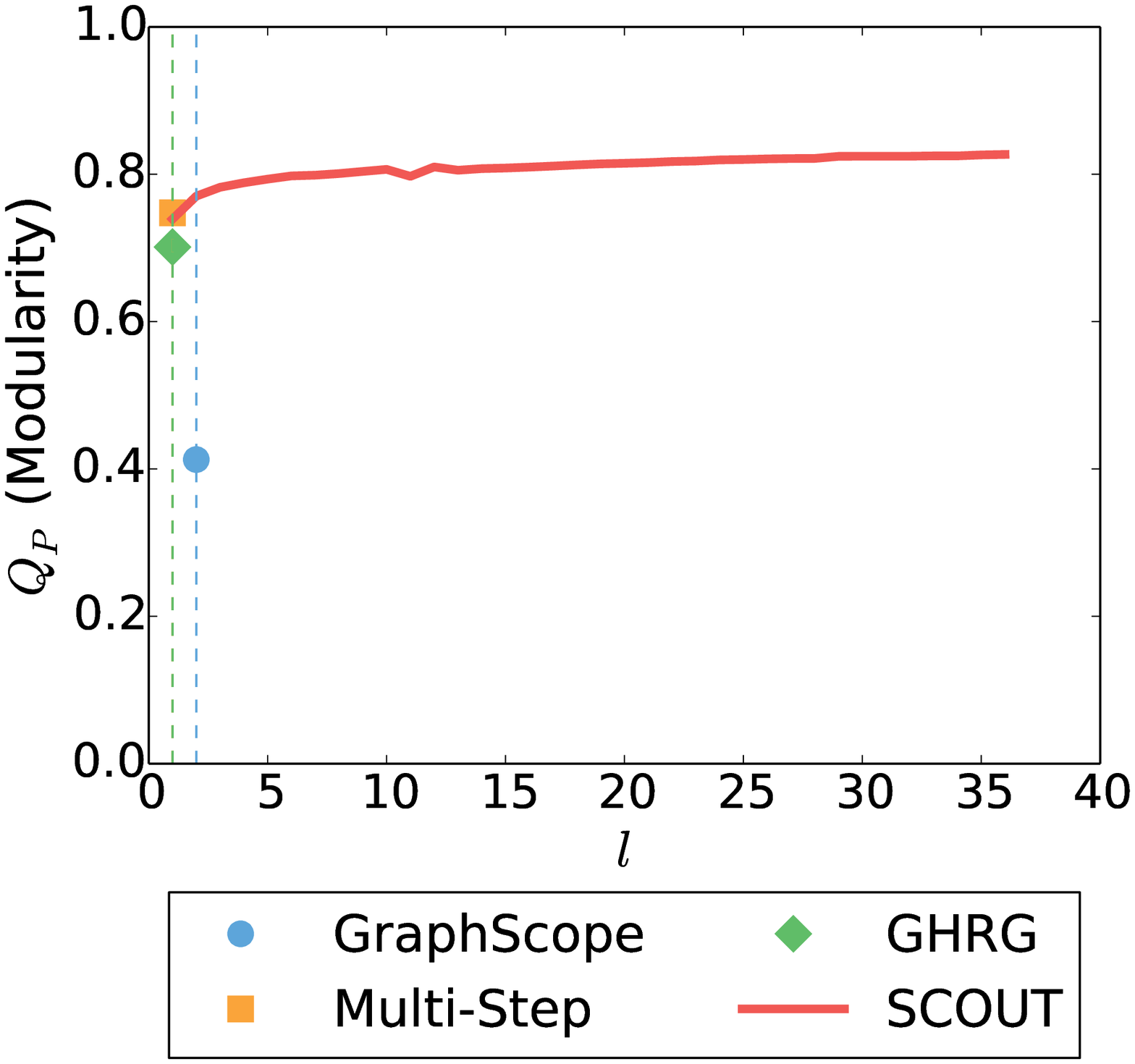}}
\hspace{0.2cm}
\subfloat[\emph{Enron}]{\label{fig:supplement:real_world_enron_avg_sn__mod_cscd}\includegraphics[width=0.42\linewidth]{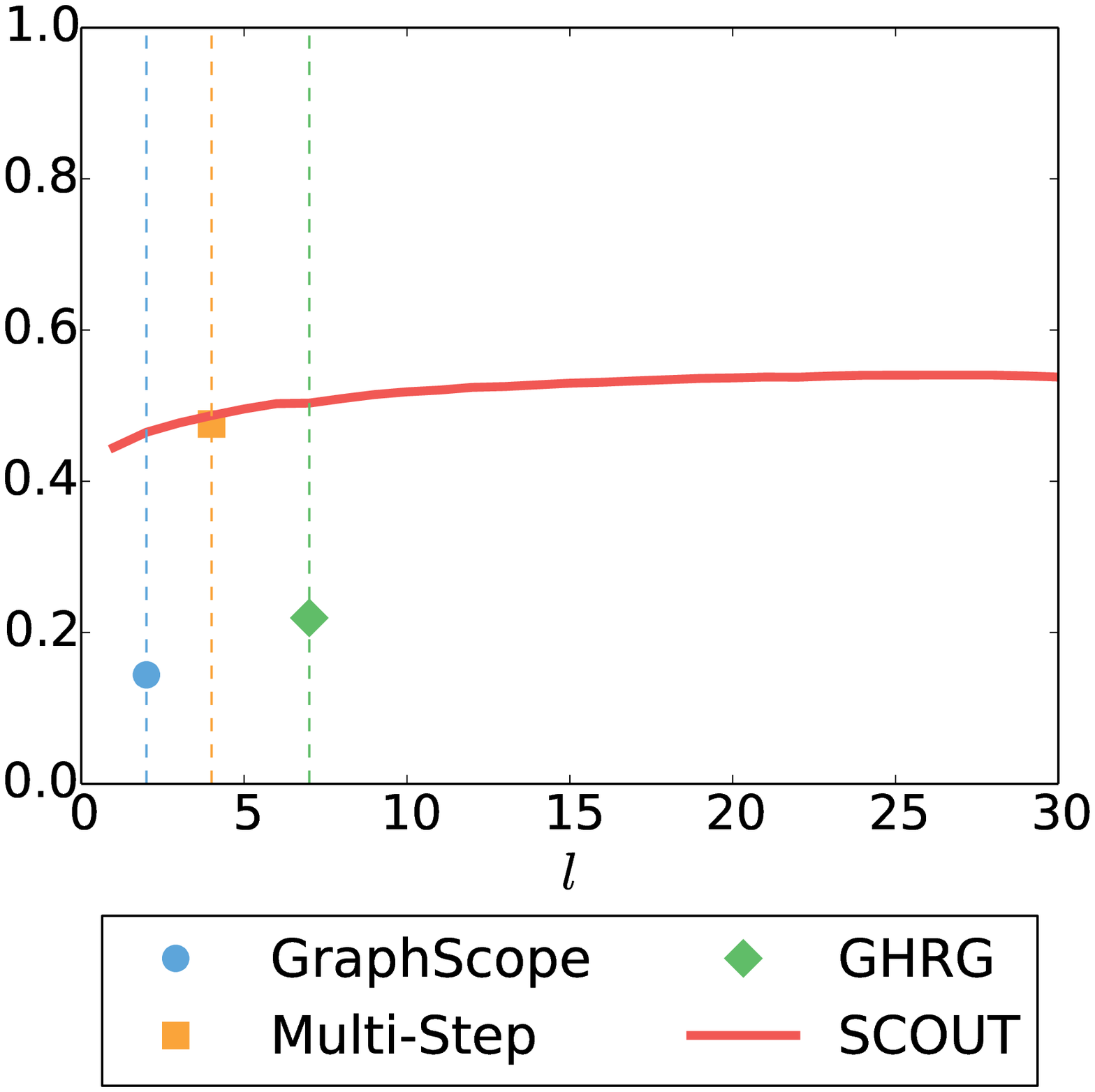}}

\caption[$Q_P$ scores of SCOUT's solutions for different numbers of segments $l$ and the solutions of the existing methods]{$Q_P$ scores of  1) SCOUT's solutions for different numbers of segments $l$ and 2) the solutions of the existing methods for
\textbf{\protect\subref{fig:supplement:real_world_hypertext_avg_sn__mod_cscd}} \emph{Hypertext} network,
\textbf{\protect\subref{fig:supplement:real_world_amd_hope_avg_sn__mod_cscd}} \emph{AMD Hope} network,
\textbf{\protect\subref{fig:supplement:real_world_reality_mining_avg_sn__mod_cscd}} \emph{Reality Mining} network, and
\textbf{\protect\subref{fig:supplement:real_world_enron_avg_sn__mod_cscd}} \emph{Enron} network.
For SCOUT, the line shows its $Q_P$ score when solving the CSCD problem while varying the number of segments. 
For each of the existing methods, the mark shows $Q_P$ score of its solution, with the position of the mark along the $x$-axis corresponding to the number of segments $l$ in the solution.
Equivalent results for the remaining real-world networks are shown in Figure~\ref{fig:real_world_avg_sn__mod_cscd} in the main paper.
}
\label{fig:supplement:real_world_avg_sn__mod_cscd}
\end{figure*}

\clearpage
\section*{Supplementary Tables}
\addcontentsline{toc}{section}{Supplementary Tables} 

\begin{table}[h]
\centering
{\renewcommand{\arraystretch}{1.25}
\caption[Real-world networks that we use in our study]{Real-world networks that we use in our study.
``Network'' columns show general information about a given network. 
``Snapshots'' columns show properties of the network snapshots averaged over all snapshots.
}
\resizebox{\linewidth}{!}{
\begin{tabular}{@{\extracolsep{4pt}}@{\hskip5pt}lccccccc} 
\toprule
\multicolumn{5}{c}{Network} & \multicolumn{3}{c}{Snapshots} \\
\cline{1-5} \cline{6-8}
\multicolumn{1}{c}{Name} & \multicolumn{1}{c}{\# of nodes} & \multicolumn{1}{c}{\# of edges} & \multicolumn{1}{c}{Time span} & \multicolumn{1}{c}{Edge type} & \multicolumn{1}{c}{\# of nodes} & \multicolumn{1}{c}{\# of edges} & \multicolumn{1}{c}{Duration} \\
 \midrule
\emph{Hypertext} & 113 & 21K & 3 days & Proximity & $47 \pm 20$ & $73 \pm 53$ & 30 min \\
\emph{AMD Hope} & 409 & 1.26M & 3 days & Co-location & $161 \pm 101$ & $4.5$K $\pm$ $3.9$K & 1 hour \\
\emph{High School} & 327 & 189K & 5 days & Proximity & $227 \pm 39$ & $499 \pm 237$ & 1 hour \\
\emph{Reality Mining} & 78 & 5K & 10 months & Phone call & $28 \pm 11$ & $25 \pm 12$ & 1 week \\
\emph{Enron} & 184 & 121K & 2.5 years & Email & $99 \pm 40$ & $267 \pm 154$ & 1 month \\
\emph{Senate} & 51 & 28K & 227 years & Voting similarity & $35 \pm 15$ & $248 \pm 268$ & 2 years \\
\bottomrule
\end{tabular}
\label{tab:supplement:real_world_data}
}
}
\end{table}

\bookmarksetup{startatroot}

\bibliographystyle{IEEEtran}
\bibliography{bibliography}

\begin{thebibliography}{10}
\providecommand{\url}[1]{#1}
\csname url@samestyle\endcsname
\providecommand{\newblock}{\relax}
\providecommand{\bibinfo}[2]{#2}
\providecommand{\BIBentrySTDinterwordspacing}{\spaceskip=0pt\relax}
\providecommand{\BIBentryALTinterwordstretchfactor}{4}
\providecommand{\BIBentryALTinterwordspacing}{\spaceskip=\fontdimen2\font plus
\BIBentryALTinterwordstretchfactor\fontdimen3\font minus
  \fontdimen4\font\relax}
\providecommand{\BIBforeignlanguage}[2]{{%
\expandafter\ifx\csname l@#1\endcsname\relax
\typeout{** WARNING: IEEEtran.bst: No hyphenation pattern has been}%
\typeout{** loaded for the language `#1'. Using the pattern for}%
\typeout{** the default language instead.}%
\else
\language=\csname l@#1\endcsname
\fi
#2}}
\providecommand{\BIBdecl}{\relax}
\BIBdecl

\bibitem{newman2010networks}
M.~Newman, \emph{{Networks: An Introduction}}.\hskip 1em plus 0.5em minus
  0.4em\relax Oxford University Press, 2010.

\bibitem{berger2013computational}
B.~Berger, J.~Peng, and M.~Singh, ``Computational solutions for omics data,''
  \emph{{Nature Reviews Genetics}}, vol.~14, no.~5, pp. 333--346, 2013.

\bibitem{mitra2013integrative}
K.~Mitra, A.-R. Carvunis, S.~K. Ramesh, and T.~Ideker, ``Integrative approaches
  for finding modular structure in biological networks,'' \emph{{Nature Reviews
  Genetics}}, vol.~14, no.~10, pp. 719--732, 2013.

\bibitem{holme2015modern}
P.~Holme, ``Modern temporal network theory: a colloquium,'' \emph{The European
  Physical Journal B}, vol.~88, no.~9, pp. 1--30, 2015.

\bibitem{faisal2014dynamic}
F.~E. Faisal and T.~Milenkovi{\'c}, ``Dynamic networks reveal key players in
  aging,'' \emph{Bioinformatics}, vol.~30, no.~12, pp. 1721--1729, 2014.

\bibitem{romero2011differences}
D.~M. Romero, B.~Meeder, and J.~Kleinberg, ``Differences in the mechanics of
  information diffusion across topics: idioms, political hashtags, and complex
  contagion on twitter,'' in \emph{Proceedings of the 20th international
  conference on World wide web}.\hskip 1em plus 0.5em minus 0.4em\relax ACM,
  2011, pp. 695--704.

\bibitem{fortunato2010community}
S.~Fortunato, ``Community detection in graphs,'' \emph{Physics Reports}, vol.
  486, no.~3, pp. 75--174, 2010.

\bibitem{newman2004finding}
M.~E. Newman and M.~Girvan, ``Finding and evaluating community structure in
  networks,'' \emph{{Physical Review E}}, vol.~69, no.~2, p. 026113, 2004.

\bibitem{yang2015defining}
J.~Yang and J.~Leskovec, ``Defining and evaluating network communities based on
  ground-truth,'' \emph{Knowledge and Information Systems}, vol.~42, no.~1, pp.
  181--213, 2015.

\bibitem{milenkovic2008uncovering}
T.~Milenkovi{\'c} and N.~Pr{\v{z}}ulj, ``Uncovering biological network function
  via graphlet degree signatures,'' \emph{{Cancer Informatics}}, no.~6, pp.
  257--273, 2008.

\bibitem{solava2012graphlet}
R.~W. Solava, R.~P. Michaels, and T.~Milenkovi{\'c}, ``Graphlet-based edge
  clustering reveals pathogen-interacting proteins,'' \emph{{Bioinformatics}},
  vol.~28, no.~18, pp. 480--486, 2012.

\bibitem{hulovatyy2015exploring}
Y.~Hulovatyy, H.~Chen, and T.~Milenkovi{\'c}, ``Exploring the structure and
  function of temporal networks with dynamic graphlets,''
  \emph{Bioinformatics}, vol.~31, no.~12, pp. i171--i180, 2015.

\bibitem{wiwie2015comparing}
C.~Wiwie, J.~Baumbach, and R.~R{\"o}ttger, ``Comparing the performance of
  biomedical clustering methods,'' \emph{{Nature Methods}}, vol.~12, no.~11,
  pp. 1033--1038, 2015.

\bibitem{hric2014community}
D.~Hric, R.~K. Darst, and S.~Fortunato, ``Community detection in networks:
  Structural communities versus ground truth,'' \emph{{Physical Review E}},
  vol.~90, no.~6, p. 062805, 2014.

\bibitem{hartmann2014clustering}
T.~Hartmann, A.~Kappes, and D.~Wagner, ``Clustering evolving networks,''
  \emph{arXiv preprint arXiv:1401.3516}, 2014.

\bibitem{meng2016multiplex}
L.~Meng, Y.~Hulovatyy, A.~Striegel, and T.~Milenkovi{\'c}, ``On the interplay
  between individuals' evolving interaction patterns and traits in dynamic
  multiplex social networks,'' \emph{Network Science and Engineering, IEEE
  Transactions on}, vol.~PP, no.~99, pp. 1--1, 2016.

\bibitem{chi2007evolutionary}
Y.~Chi, X.~Song, D.~Zhou, K.~Hino, and B.~L. Tseng, ``Evolutionary spectral
  clustering by incorporating temporal smoothness,'' in \emph{{Proceedings of
  the 13th ACM SIGKDD International Conference on Knowledge Discovery and Data
  Mining}}.\hskip 1em plus 0.5em minus 0.4em\relax ACM, 2007, pp. 153--162.

\bibitem{fu2009dynamic}
W.~Fu, L.~Song, and E.~P. Xing, ``Dynamic mixed membership blockmodel for
  evolving networks,'' in \emph{Proceedings of the 26th annual international
  conference on machine learning}.\hskip 1em plus 0.5em minus 0.4em\relax ACM,
  2009, pp. 329--336.

\bibitem{mucha2010community}
P.~J. Mucha, T.~Richardson, K.~Macon, M.~A. Porter, and J.-P. Onnela,
  ``Community structure in time-dependent, multiscale, and multiplex
  networks,'' \emph{{Science}}, vol. 328, no. 5980, pp. 876--878, 2010.

\bibitem{kawadia2012sequential}
V.~Kawadia and S.~Sreenivasan, ``Sequential detection of temporal communities
  by estrangement confinement,'' \emph{{Scientific Reports}}, vol.~2, 2012.

\bibitem{berger2006framework}
T.~Y. Berger-Wolf and J.~Saia, ``A framework for analysis of dynamic social
  networks,'' in \emph{Proceedings of the 12th ACM SIGKDD international
  conference on Knowledge discovery and data mining}.\hskip 1em plus 0.5em
  minus 0.4em\relax ACM, 2006, pp. 523--528.

\bibitem{palla2007quantifying}
G.~Palla, A.-L. Barab{\'a}si, and T.~Vicsek, ``Quantifying social group
  evolution,'' \emph{Nature}, vol. 446, no. 7136, pp. 664--667, 2007.

\bibitem{rosvall2010mapping}
M.~Rosvall and C.~T. Bergstrom, ``Mapping change in large networks,''
  \emph{{PLOS ONE}}, vol.~5, no.~1, p. e8694, 2010.

\bibitem{greene2010tracking}
D.~Greene, D.~Doyle, and P.~Cunningham, ``Tracking the evolution of communities
  in dynamic social networks,'' in \emph{Advances in social networks analysis
  and mining (ASONAM), 2010 international conference on}.\hskip 1em plus 0.5em
  minus 0.4em\relax IEEE, 2010, pp. 176--183.

\bibitem{pandit2011detecting}
S.~Pandit, Y.~Yang, V.~Kawadia, S.~Sreenivasan, and N.~V. Chawla, ``Detecting
  communities in time-evolving proximity networks,'' in \emph{Network Science
  Workshop (NSW), 2011 IEEE}.\hskip 1em plus 0.5em minus 0.4em\relax IEEE,
  2011, pp. 173--179.

\bibitem{aynaud2011multi}
T.~Aynaud and J.-L. Guillaume, ``Multi-step community detection and
  hierarchical time segmentation in evolving networks,'' in \emph{Proceedings
  of the 5th SNA-KDD workshop}, 2011.

\bibitem{lancichinetti2012consensus}
A.~Lancichinetti and S.~Fortunato, ``Consensus clustering in complex
  networks,'' \emph{{Scientific Reports}}, vol.~2, 2012.

\bibitem{zhang2012common}
S.~Zhang, J.~Zhao, and X.-S. Zhang, ``Common community structure in
  time-varying networks,'' \emph{{Physical Review E}}, vol.~85, no.~5, p.
  056110, 2012.

\bibitem{ranshous2015anomaly}
S.~Ranshous, S.~Shen, D.~Koutra, S.~Harenberg, C.~Faloutsos, and N.~F.
  Samatova, ``Anomaly detection in dynamic networks: a survey,'' \emph{Wiley
  Interdisciplinary Reviews: Computational Statistics}, vol.~7, no.~3, pp.
  223--247, 2015.

\bibitem{barnett2016change}
I.~Barnett and J.-P. Onnela, ``Change point detection in correlation
  networks,'' \emph{{Scientific Reports}}, vol.~6, 2016.

\bibitem{mintz1993life}
S.~Mintz, ``Life stages,'' \emph{{Encyclopedia of American Social History}},
  vol.~3, pp. 7--33, 1993.

\bibitem{sun2007graphscope}
J.~Sun, C.~Faloutsos, S.~Papadimitriou, and P.~S. Yu, ``{GraphScope:
  parameter-free mining of large time-evolving graphs},'' in \emph{{Proceedings
  of the 13th ACM SIGKDD International Conference on Knowledge Discovery and
  Data Mining}}.\hskip 1em plus 0.5em minus 0.4em\relax ACM, 2007, pp.
  687--696.

\bibitem{peel2015detecting}
L.~Peel and A.~Clauset, ``Detecting change points in the large-scale structure
  of evolving networks,'' in \emph{Twenty-Ninth AAAI Conference on Artificial
  Intelligence}, 2015.

\bibitem{yoo2015improving}
B.~Yoo, F.~Faisal, H.~Chen, and T.~Milenkovi{\'c}, ``Improving identification
  of key players in aging via network de-noising and core inference.''
  \emph{{IEEE/ACM Transactions on Computational Biology and Bioinformatics}},
  2015.

\bibitem{newman2015structure}
M.~Newman and A.~Clauset, ``Structure and inference in annotated networks,''
  \emph{arXiv preprint arXiv:1507.04001}, 2015.

\bibitem{isella2011s}
L.~Isella, J.~Stehl{\'e}, A.~Barrat, C.~Cattuto, J.-F. Pinton, and W.~Van~den
  Broeck, ``What's in a crowd? analysis of face-to-face behavioral networks,''
  \emph{{Journal of Theoretical Biology}}, vol. 271, no.~1, pp. 166--180, 2011.

\bibitem{amdhope}
``{AMD Hope RFID Data},''
  \url{http://networkdata.ics.uci.edu/data.php?d=amdhope}, 2008.

\bibitem{mastrandrea2015contact}
R.~Mastrandrea, J.~Fournet, and A.~Barrat, ``Contact patterns in a high school:
  a comparison between data collected using wearable sensors, contact diaries
  and friendship surveys,'' \emph{{PLOS ONE}}, vol.~10, no.~9, p. e0136497,
  2015.

\bibitem{eagle2006reality}
N.~Eagle and A.~Pentland, ``{Reality Mining: Sensing Complex Social Systems},''
  \emph{{Personal and Ubiquitous Computing}}, vol.~10, no.~4, pp. 255--268,
  2006.

\bibitem{priebe2005scan}
C.~E. Priebe, J.~M. Conroy, D.~J. Marchette, and Y.~Park, ``{Scan statistics on
  Enron graphs},'' \emph{{Computational \& Mathematical Organization Theory}},
  vol.~11, no.~3, pp. 229--247, 2005.

\bibitem{vinh2010information}
N.~X. Vinh, J.~Epps, and J.~Bailey, ``Information theoretic measures for
  clusterings comparison: Variants, properties, normalization and correction
  for chance,'' \emph{The Journal of Machine Learning Research}, vol.~11, pp.
  2837--2854, 2010.

\bibitem{rosenberg2007v}
A.~Rosenberg and J.~Hirschberg, ``V-measure: A conditional entropy-based
  external cluster evaluation measure.'' in \emph{EMNLP-CoNLL}, vol.~7, 2007,
  pp. 410--420.

\bibitem{ferlez2008monitoring}
J.~Ferlez, C.~Faloutsos, J.~Leskovec, D.~Mladenic, and M.~Grobelnik,
  ``{Monitoring network evolution using MDL},'' in \emph{Data Engineering,
  2008. ICDE 2008. IEEE 24th International Conference on}.\hskip 1em plus 0.5em
  minus 0.4em\relax IEEE, 2008, pp. 1328--1330.

\bibitem{duan2009community}
D.~Duan, Y.~Li, Y.~Jin, and Z.~Lu, ``Community mining on dynamic weighted
  directed graphs,'' in \emph{{Proceedings of the 1st ACM International
  Workshop on Complex Networks Meet Information \& Knowledge
  Management}}.\hskip 1em plus 0.5em minus 0.4em\relax ACM, 2009, pp. 11--18.

\bibitem{cheng2010uncovering}
X.-Q. Cheng and H.-W. Shen, ``Uncovering the community structure associated
  with the diffusion dynamics on networks,'' \emph{Journal of Statistical
  Mechanics: Theory and Experiment}, vol. 2010, no.~04, p. P04024, 2010.

\bibitem{shi2000normalized}
J.~Shi and J.~Malik, ``Normalized cuts and image segmentation,'' \emph{Pattern
  Analysis and Machine Intelligence, IEEE Transactions on}, vol.~22, no.~8, pp.
  888--905, 2000.

\bibitem{flake2000efficient}
G.~W. Flake, S.~Lawrence, and C.~L. Giles, ``Efficient identification of web
  communities,'' in \emph{Proceedings of the sixth ACM SIGKDD international
  conference on Knowledge discovery and data mining}.\hskip 1em plus 0.5em
  minus 0.4em\relax ACM, 2000, pp. 150--160.

\bibitem{kadane2004methods}
J.~B. Kadane and N.~A. Lazar, ``Methods and criteria for model selection,''
  \emph{{Journal of the American Statistical Association}}, vol.~99, no. 465,
  pp. 279--290, 2004.

\bibitem{grunwald2000model}
P.~Gr{\"u}nwald, ``Model selection based on minimum description length,''
  \emph{Journal of Mathematical Psychology}, vol.~44, no.~1, pp. 133--152,
  2000.

\bibitem{akaike1998information}
H.~Akaike, ``Information theory and an extension of the maximum likelihood
  principle,'' in \emph{Selected Papers of Hirotugu Akaike}.\hskip 1em plus
  0.5em minus 0.4em\relax Springer, 1998, pp. 199--213.

\bibitem{schwarz1978estimating}
G.~Schwarz \emph{et~al.}, ``Estimating the dimension of a model,'' \emph{{The
  Annals of Statistics}}, vol.~6, no.~2, pp. 461--464, 1978.

\bibitem{karrer2011stochastic}
B.~Karrer and M.~E. Newman, ``Stochastic blockmodels and community structure in
  networks,'' \emph{{Physical Review E}}, vol.~83, no.~1, p. 016107, 2011.

\bibitem{clauset2004finding}
A.~Clauset, M.~E. Newman, and C.~Moore, ``Finding community structure in very
  large networks,'' \emph{{Physical Review E}}, vol.~70, no.~6, p. 066111,
  2004.

\bibitem{raghavan2007near}
U.~N. Raghavan, R.~Albert, and S.~Kumara, ``Near linear time algorithm to
  detect community structures in large-scale networks,'' \emph{{Physical Review
  E}}, vol.~76, no.~3, p. 036106, 2007.

\bibitem{newman2006finding}
M.~E. Newman, ``Finding community structure in networks using the eigenvectors
  of matrices,'' \emph{{Physical Review E}}, vol.~74, no.~3, p. 036104, 2006.

\bibitem{rosvall2010map}
M.~Rosvall, D.~Axelsson, and C.~T. Bergstrom, ``The map equation,'' \emph{The
  European Physical Journal Special Topics}, vol. 178, no.~1, pp. 13--23, 2010.

\bibitem{pons2005computing}
P.~Pons and M.~Latapy, ``Computing communities in large networks using random
  walks,'' in \emph{Computer and Information Sciences-ISCIS 2005}.\hskip 1em
  plus 0.5em minus 0.4em\relax Springer, 2005, pp. 284--293.

\bibitem{blondel2008fast}
V.~D. Blondel, J.-L. Guillaume, R.~Lambiotte, and E.~Lefebvre, ``Fast unfolding
  of communities in large networks,'' \emph{Journal of Statistical Mechanics:
  Theory and Experiment}, vol. 2008, no.~10, p. P10008, 2008.

\bibitem{aynaud2010static}
T.~Aynaud and J.-L. Guillaume, ``Static community detection algorithms for
  evolving networks,'' in \emph{Modeling and Optimization in Mobile, Ad Hoc and
  Wireless Networks (WiOpt), 2010 Proceedings of the 8th International
  Symposium on}.\hskip 1em plus 0.5em minus 0.4em\relax IEEE, 2010, pp.
  513--519.

\bibitem{sarovios2014graphscope}
S.~Beis, ``{Implementation of GraphScope algorithm for clustering time evolving
  graphs},'' \url{https://github.com/sarovios/social-graph-cluster}, 2014.

\end{thebibliography}
\addcontentsline{toc}{section}{\refname}

\end{document}